\documentclass[english]{article}
\usepackage[utf8]{inputenc}
\usepackage{amsmath}
\usepackage{amssymb}
\usepackage{tensor}
\usepackage{mathrsfs}
\usepackage{cite}
\usepackage{multirow}
\usepackage{framed}
\usepackage{graphicx}
\usepackage{shuffle}
\usepackage{breqn}
\usepackage{float}
\usepackage[colorlinks=true,urlcolor=blue,anchorcolor=blue,citecolor=blue,filecolor=blue,linkcolor=blue,menucolor=blue,linktocpage=true,pdfproducer=medialab,pdfa=true]{hyperref}
\usepackage{calc}
\usepackage{bm}
\usepackage{color}

\allowdisplaybreaks

\newcommand{\nocontentsline}[3]{}
\let\origcontentsline\addcontentsline
\newcommand\stoptoc{\let\addcontentsline\nocontentsline}
\newcommand\resumetoc{\let\addcontentsline\origcontentsline}

\renewcommand{\theequation}{\thesection.\arabic{equation}}

\makeatletter
\def\theequation{\thesection.\arabic{equation}}
\@addtoreset{equation}{section}

\newcommand\eqnss[2]   {eqs.\,(\ref{#1})--(\ref{#2})}

\newcommand\nn         {\nonumber}

\def\beq{\begin{equation}}
\def\eeq{\end{equation}}
\def\bsp#1\esp{\begin{split}#1\end{split}}
\def\bal#1\eal{\begin{align}#1\end{align}}

\def\bleq#1{\begin{equation}
\label{eq:#1}
\end{equation}
\vspace{-40pt}
\begin{quote}\raggedright}
\def\eleq{\end{quote}}

\def\beeq{\begin{eqnarray}}
\def\eeeq{\end{eqnarray}}
\newcommand\bom[1]     {{\mbox{\boldmath $#1$}}}

\newcommand{\CF}       {C_{\mathrm{F}}}
\newcommand{\CA}       {C_{\mathrm{A}}}
\newcommand{\TR}       {T_{\mathrm{R}}}
\newcommand{\bT}       {\bom{T}}
\newcommand{\Nc}       {N_{\mathrm{c}}}

\newcommand\qb         {{\bar q}}

\newcommand{\ep}      {\varepsilon}
\newcommand{\eps}      {\varepsilon}

\newcommand\Li         {\mathop{\mathrm{Li}}\nolimits}

\newcommand{\rd}       {{\mathrm{d}}}
\newcommand{\PS}[1]    {\rd\phi_{#1}}

\newcommand\dsig[1]    {\rd\sigma^{{\rm #1}}}
\newcommand\dsiga[2]   {\rd\sigma^{{\rm #1,A}_{\scriptscriptstyle #2}}}

\newcommand\la         {\langle}
\newcommand\ra         {\rangle}

\newcommand{\cA}       {{\cal A}}

\newcommand\SME[3]     {|{\cal M}_{#1}^{#2}{#3}|^2}

\newcommand{\mom}[1]   {\{p\}^{#1}}

\newcommand{\momt}[1]   {\{\ti{p}\}^{#1}}
\newcommand{\momh}[1]   {\{\ha{p}\}^{#1}}
\newcommand{\momhh}[1]   {\{\ha{\ha{p}}\}^{#1}}
\newcommand{\momht}[1]   {\{\ha{\ti{p}}\}^{#1}}
\newcommand{\momth}[1]   {\{\ti{\ha{p}}\}^{#1}}
\newcommand{\momtt}[1]   {\{\ti{\ti{p}}\}^{#1}}

\newcommand{\cmap}[2]   {\xrightarrow{{C}_{#1}^{#2}}}
\newcommand{\smap}[1]   {\xrightarrow{{S}_{#1}}}

\def\hP{\hat{P}}
\newcommand{\calS}     {{\cal S}}
\newcommand{\cC}[2]    {{\cal C}_{#1}^{#2}}
\newcommand{\cS}[2]    {{\cal S}_{#1}^{#2}}

\newcommand{\La}    {\bom{L_{a}}}
\newcommand{\Lb}    {\bom{L_{b}}}
\newcommand{\Lab}    {\bom{L_{ab}}}
\newcommand{\Leta}  {\bom{L_{\eta}}}
\newcommand{\LbTOa} {\bom{L_{b\to a}}}

\newcommand{\LaTOb}  {\bom{ L_{a\to b}}}

\newcommand{\IcC}[2]   {{\mathrm C}_{#1}^{#2}}
\newcommand{\IcS}[2]   {{\mathrm S}_{#1}^{#2}}
\newcommand{\bI}       {\bom{I}}

\newcommand{\ba}[1]    {\bar{#1}}
\newcommand{\ti}[1]    {\tilde{#1}}

\newcommand{\tha}[1]    {\tilde{\hat{#1}}}

\newcommand{\ha}[1]    {\hat{#1}}
\newcommand{\wha}[1]   {\widehat{#1}}

\newcommand\kTm[1]     {k_{\perp}^{#1}}
\newcommand\kT[1]      {k_{\perp #1}}
\newcommand\kTt[1]     {\tilde{k}_{\perp #1}}
\newcommand\kTh[1]     {\hat{k}_{\perp #1}}

\def\s12{s_{ab}}

\newcommand\al	{\alpha}

\newcommand\lam	{\lambda}

\newcommand\CT {\text{CT}}
\newcommand\bCT {\textbf{CT}}

\newcommand{\colorful}{CoLoRFulNNLO }

\definecolor{burntorange}{rgb}{0.8, 0.33, 0.0}

\allowdisplaybreaks

\begin{document}

\numberwithin{equation}{section}

\begin{titlepage}
\noindent
\hfill December 2025\\
\vspace{0.6cm}
\begin{center}
{\LARGE \bf 
    \colorful for hadron collisions: integrating the \\[0.5em] 
    iterated single-unresolved subtraction terms
}\\ 
\vspace{1.4cm}

\large
L.~Fek\'esh\'azy$^{\, a,b}$, G.~Somogyi$^{\, c}$ and S.~Van Thurenhout$^{\, c}$\\
\vspace{1.4cm}
\normalsize
{\it $^{\, a}$II. Institut für Theoretische Physik, Universität Hamburg, Luruper Chaussee
149, 22761, Hamburg, Germany}\\
{\it $^{\, b}$Institute for Theoretical Physics, ELTE Eötvös Loránd University, Pázmány Péter sétány
1/A, 1117, Budapest, Hungary}\\
{\it $^{\, c}$HUN-REN Wigner Research Centre for Physics, 1121 Budapest, Konkoly-Thege Mikl\'os \'ut 29-33, Hungary}\\
\vspace{1.4cm}

{\large \bf Abstract}
\vspace{-0.2cm}
\end{center}
We present the analytic integration of the iterated single-unresolved subtraction terms in the extension of the \colorful subtraction scheme to color-singlet production in hadron collisions. We exploit the fact that, in this scheme, subtraction terms are defined through momentum mappings which lead to exact phase space convolutions for real emissions. This allows us to write the integrated subtraction terms as parametric integrals, which can be evaluated using standard tools. Finally, we show that the integrated iterated single-unresolved approximate cross section can be written as a convolution of the Born cross section with an appropriately defined insertion operator.
\vspace*{0.3cm}
\end{titlepage}
\clearpage

\tableofcontents

\section{Introduction}
\label{sec:Introduction}
The Standard Model of particle physics (SM) provides an excellent description of particle interactions over a wide range of energies. Nevertheless, despite its tremendous successes, it is more-or-less accepted that extensions of the SM will become important beyond a certain energy scale. While experimentalists occasionally find anomalies that are in tension with the predictions of the SM, such tensions have a tendency to disappear over time\footnote{A recent example of this is the evolution of the anomalous magnetic moment of the muon from anomaly to yet another confirmation of the Standard Model \cite{Aoyama:2020ynm,Muong-2:2006rrc,Muong-2:2021ojo,Muong-2:2023cdq,Muong-2:2025xyk,Muong-2:2024hpx,Muong-2:2021vma,Muong-2:2021ovs,Muong-2:2021xzz,Aliberti:2025beg}.}, and no convincing evidence of physics beyond the Standard Model has been found so far. This forces us to, bluntly put, look harder. For our experimental friends, this means building ever-more-efficient machines and devising, or improving, tools for particle identification. To keep up with their progress, us theorists need to provide high-precision predictions. An important part of this is the computation of perturbative cross sections to sufficient orders in the relevant coupling. This is straightforward in principle. However, in practice the calculational workflow is complicated due various divergences, leading to infinities in intermediate steps. To get sensible (i.e., finite) physical predictions, such infinities should be dealt with.\newline

In general, two types of divergences can appear during the evaluation of Feynman diagrams. First, there are the ultraviolet (UV) divergences, originating from momenta running through loops becoming large. The resulting infinities are treated once and for all by renormalization. Second, there are infrared (IR) divergences, coming from particles becoming soft and/or collinear to one another. This is relevant for both virtual and real particles and, unfortunately, there is no clear-cut method to treat the corresponding infinities. The two main approaches followed in the literature are phase space slicing \cite{Catani:2007vq,Gaunt:2015pea} and the construction of a (local) subtraction scheme \cite{Gehrmann-DeRidder:2005btv,Czakon:2014oma,Magnea:2018hab,Caola:2017dug,Cacciari:2015jma,Herzog:2018ily,NNLOJET:2025rno}. In this work, we focus our attention on the completely local subtraction scheme \colorful \cite{Somogyi:2006da}. The latter was initially developed for jet production from a colorless initial state \cite{Somogyi:2005xz,Somogyi:2006db,Somogyi:2006da,DelDuca:2015zqa,DelDuca:2016csb,DelDuca:2016ily,Somogyi:2020mmk}. Recently, the scheme has been extended to color-singlet production in hadron collisions to next-to-next-to-leading order (NNLO) accuracy \cite{DelDuca:2024ovc,DelDuca:2025yph}. In particular, the latter paper rigorously defined all subtraction terms necessary to regularize singularities coming from double-real initial-state emissions. In this paper, we start setting up the analytic integration of these counterterms. This, however, turns out to be a daunting task and, as such, the complete integration will be presented over separate publications. One of the main reasons for this division is that, depending on the exact definitions of the subtraction terms, different integration methods are used. Specifically, we either employ reverse unitarity \cite{Anastasiou:2002yz} in combination with integration-by-parts reduction \cite{Chetyrkin:1981qh} and the method of differential equations \cite{Henn:2013pwa}, or we perform the integration directly by setting up a parametric representation for the integrated counterterm.\newline

This paper presents the analytic integration of the subtraction terms regularizing iterated single-unresolved emission. Regularizing loop- and phase space integrals using $d=4-2\eps$ dimensional regularization, these integrals need to be computed to $\mathcal{O}(\eps^0)$. The integration is performed directly by setting up a parametric representation of the phase space integrals. This is based on the important fact that the choice of momentum mappings in the \colorful scheme leads to an exact convolution structure for the real-emission phase space in terms of the reduced phase space of mapped momenta and an integration measure for the unresolved emissions. The resulting integrands involve multivariate rational functions raised to $\eps$-dependent powers whose $\eps$-expansions, as will be shown explicitly, can be integrated in terms of generalized polylogarithms (GPLs). The result of the integration can be written as a convolution of an insertion operator with the Born-level cross section. Contrary to jet production from color-singlet initial states, in which case the insertion operator was simply a \textit{function} of the convolution variables, the insertion operator now needs to be interpreted as a \textit{distribution} acting on the parton density functions (PDFs). As such, one needs to carefully set up the distributional expansion. In practice we avoid writing any explicit distributions by defining an appropriate subtraction to regularize any singularities, which is more convenient for turning the results into a numeric code.\newline

The purposes of the present paper are then two-fold. The first is to present the explicit results for the integration of the subtraction terms that regularize the singularities coming from iterated single-unresolved emissions. The second is to lay out the direct integration method, which is also used for the integration of some of the other subtraction terms. In light of this, we provide an in-depth discussion of the used methods and tools in Appendix \ref{appx:ICasrCas}, which we hope to be useful for, say, the adventurous graduate student getting into this type of calculation.\newline

The remainder of the paper is organized as follows. In Section \ref{sec:Scheme} we set up some notation and briefly review how double-real singularities are regularized in the \colorful subtraction scheme. Next, in Section \ref{sec:intA12generic}, we provide a generic overview of our integration method, which is then applied to the subtraction terms of interest in Section \ref{sec:ICTs}. Section \ref{sec:A12_Iop} then sets up the final result in the form of the insertion operator. Finally, in Section \ref{sec:conclusion}, we provide a brief summary and an outlook on future steps and applications.

\section{Review of the \colorful subtraction scheme}
\label{sec:Scheme}
Consider a collision of hadrons $h_A(p_A)$ and $h_B(p_B)$ leading to the production of a color-singlet final state $X$ in association with $m$ jets,
\begin{equation}
    h_A(p_A)+h_B(p_B) \rightarrow X + m \text{ jets}.
\end{equation}
Because of QCD factorization, the cross section of such a process can be written as
\begin{equation}
\label{eq:fact}
    \sigma_{AB}(p_A,p_B) =\,\sum_{a,b}
	\int_0^1 \rd x_a \, f_{a/A}(x_a,\mu_F^2) \int_0^1 \rd x_b \, f_{b/B}(x_b,\mu_F^2)\, 
    \sigma_{ab}(p_a,p_b;\mu_F^2)\,.
\end{equation}
Here, $f_{a/A}$ and $f_{b/B}$ are the PDFs, which depend on the hadronic momentum fractions $x_a$ and $x_b$ and on the factorization scale $\mu_F^2$. $\sigma_{ab}$ is the standard partonic cross section and depends on the partonic momenta $p_{a} =\,x_{a}p_{A}$ and $p_{b} =\,x_{b}p_{B}$ as well as on the factorization scale. It can be computed in perturbation theory, i.e.,
\begin{equation}
    \sigma_{ab}(p_a,p_b;\mu_F^2) =\,
    \sum_{k=0}^{\infty}\sigma^{\text{N$^k$LO}}_{ab}(p_a,p_b;\mu_R^2,\mu_F^2)\,.
\end{equation}
Note that the fixed-order cross sections also depend on the renormalization scale $\mu_R^2$. In the following, we suppress the dependence on $\mu_R^2$ and $\mu_F^2$. The NNLO correction to the partonic cross section receives five distinct contributions,
\begin{align}
\label{eq:NNLOxs}
    \sigma^{\text{NNLO}}_{ab}(p_a,p_b) &=\,\int_{X+m+2}\rd \sigma_{ab}^{\text{RR}}(p_a,p_b)J_{X+m+2}+\int_{X+m+1}\rd \sigma_{ab}^{\text{RV}}(p_a,p_b)J_{X+m+1}\nonumber\\&+\int_{X+m}\rd \sigma_{ab}^{\text{VV}}(p_a,p_b)J_{X+m}  + \int_{X+m+1}\rd \sigma_{ab}^{C_1}(p_a,p_b)J_{X+m+1}+\int_{X+m}\rd \sigma_{ab}^{C_2}(p_a,p_b)J_{X+m}\,.
\end{align}
The first three terms correspond to the double-real (RR, two real emissions), real-virtual (RV, one real emission and one loop) and double-virtual (VV, two loops) contributions. The final two terms, $C_1$ and $C_2$, originate from the UV renormalization of the PDFs. $J_{X+m}$ is the measurement or jet function, i.e., some physical jet observable in terms of the momenta of the $m$ final-state partons. For an IR safe observable, the sum in eq.~(\ref{eq:NNLOxs}) is finite. However, each separate term is IR divergent and hence requires regularization. The latter is implemented by way of subtraction. In particular, we construct subtraction terms that match the point-wise singularity structure of the partonic cross sections. These subtraction terms are then organized into approximate cross sections according to the nature of the IR divergence they regularize. For example, the double-real contribution is written as
\begin{align}
\label{eq:NNLORR}
    \sigma_{ab, \mathrm{reg.}}^{\text{RR}}(p_a,p_b) =\,\int_{X+m+2}\Bigg\{&\rd\sigma_{ab}^{\text{RR}}(p_a,p_b)J_{X+m+2}-\rd\sigma_{ab}^{\text{RR},A_1}(p_a,p_b)J_{X+m+1}-\rd\sigma_{ab}^{\text{RR},A_2}(p_a,p_b)J_{X+m}\nonumber\\&+\rd\sigma_{ab}^{\text{RR},A_{12}}(p_a,p_b)J_{X+m}\Bigg\}
\end{align}
which is, by construction, IR finite in four dimensions. Each term on the right-hand side of eq.~(\ref{eq:NNLORR}) has a specific purpose,
\begin{itemize}
    \item $\rd\sigma_{ab}^{\text{RR},A_1}(p_a,p_b)$ cancels the singularities coming from a single unresolved emission,
    \item $\rd\sigma_{ab}^{\text{RR},A_2}(p_a,p_b)$ cancels the singularities coming from a double unresolved emission and
    \item $\rd\sigma_{ab}^{\text{RR},A_{12}}(p_a,p_b)$ cancels the singularities coming from iterated single-unresolved emissions.
\end{itemize}
Their explicit construction was discussed in \cite{DelDuca:2025yph} and will not be repeated here. Of course, the subtracted approximate cross sections need to be added back, integrated over the phase space of the unresolved emissions. In the present work we focus on the integration of the iterated single-unresolved subtraction terms that appear in $\dsiga{RR}{12}_{ab}$ in the color-singlet case, $m=0$. For ease of reference, we will simply call these terms the $A_{12}$ subtraction terms . The approximate cross section, $\dsiga{RR}{12}_{ab}$, can be written as \cite{DelDuca:2024ovc}
\beq
\dsiga{RR}{12}_{ab} = 
	\PS{X+2}(\mom{}_{X+2}) \cA_{12}^{(0)}\,,
\label{eq:dsigRRA12}
\eeq
with
\beq
\cA_{12}^{(0)} =
	\sum_{s \in F}\bigg[
	\cA_{2}^{(0)}\, \cS{s}{} 
	+ \sum_{\substack{r \in F \\ r \ne s}} \bigg(\frac12 \cA_{2}^{(0)}\, \cC{rs}{FF}
	- \cA_{2}^{(0)}\, \cC{rs}{FF}\cS{s}{}\bigg)
	+ \sum_{c \in I} \bigg( \cA_{2}^{(0)}\, \cC{cs}{IF}
	- \cA_{2}^{(0)}\, \cC{cs}{IF}\cS{s}{} \bigg)\bigg]
\label{eq:A12}
\eeq
and
\bal
\cA_{2}^{(0)}\, \cS{s}{} &=
	\sum_{\substack{r \in F \\ r \ne s}}\bigg[
	\cS{rs}{(0,0)}\, \cS{s}{}
	+ \sum_{c \in I}\bigg(	
	\cC{crs}{IFF (0,0)}\, \cS{s}{}	
	- \cC{crs}{IFF}\cS{rs}{(0,0)}\, \cS{s}{}
	\bigg)
	\bigg]	
	\,, \label{eq:SsA2new}
\\
\cA_{2}^{(0)}\, \cC{rs}{FF} &=
	\cS{rs}{(0,0)}\, \cC{rs}{FF}
	+ \sum_{c \in I}\bigg(
	\cC{crs}{IFF (0,0)}\, \cC{rs}{FF}
	-\cC{crs}{IFF}\cS{rs}{(0,0)}\, \cC{rs}{FF}
	\bigg)
	\,, \label{eq:CrsA2}
\\
\cA_{2}^{(0)}\, \cC{rs}{FF}\cS{s}{} &=
    \sum_{c \in I}
	\cC{crs}{IFF (0,0)}\, \cC{rs}{FF}\cS{s}{}
	\,, \label{eq:CrsFFSsA2new}  
\\
\cA_{2}^{(0)}\, \cC{cs}{IF}  &=
	\sum_{\substack{r \in F \\ r \ne s}} \bigg( \cC{csr}{IFF (0,0)}\, \cC{cs}{IF}
	+ \sum_{\substack{d \in I \\ d \ne c}}
	\cC{cs,dr}{IF,IF (0,0)}\, \cC{cs}{IF}
	 \bigg)
	\,, \label{eq:CasA2new}
\\
\cA_{2}^{(0)}\, \cC{cs}{IF}\cS{s}{} &=
	\sum_{\substack{r \in F \\ r \ne s}}\bigg(
	\cS{rs}{(0,0)}\, \cC{cs}{IF}\cS{s}{}
	+ \cC{csr}{IFF (0,0)}\,  \cC{cs}{IF}\cS{s}{}
	- \cC{csr}{IFF}\cS{rs}{(0,0)}\, \cC{cs}{IF}\cS{s}{}
	\bigg)
	\,.
\label{eq:CasIFSsA2new}
\eal
Here, $I\,(F)$ represents the set of initial-(final-)state partons. As was explained in detail in a previous paper \cite{DelDuca:2025yph}, these expressions come about by taking the single-unresolved limits of the counterterms that regularize the double-unresolved emissions. Furthermore, we have exploited several cancellations in writing eqs.~(\ref{eq:SsA2new})-(\ref{eq:CasIFSsA2new}). Some of these cancellations already occur at the level of the iterated IR factorization formul\ae \cite{Somogyi:2005xz}, while others are specific to color-singlet production. In particular, our framework is such that, in the color-singlet case, all double-unresolved subtraction terms involving a soft-collinear limit cancel among themselves. This then implies that eqs.~(\ref{eq:SsA2new})-(\ref{eq:CasIFSsA2new}) do not involve any terms that would correspond to a single-unresolved limit of a soft-collinear type double subtraction term.

As it turns out, this set of subtraction terms leads to a total of 104 basic integrals to be computed. In section \ref{sec:intA12generic} we provide a generic overview of the main technical steps of these calculations. Section \ref{sec:ICTs} then presents the integration of each counterterm listed above.

\subsection{Momentum fractions}
The subtraction terms are written in terms of specific \textit{momentum fractions}, which, for $i\in I$ and $f,g\in F$, are defined as follows (see \cite{DelDuca:2025yph} for more details)
\begin{align}
    \label{eq:defX2}
    &x_{f,i} =\, \frac{s_{f Q}}{s_{i Q}}\,,\qquad x_{i,f} =\, 1-x_{f,i}\,, \qquad x_{i,fg} =\, 1-x_{f,i}-x_{g,i}\,,\\
    \label{eq:defZ}
    &z_{f,g} =\, \frac{s_{f Q}}{s_{(f g)Q}}\,,\qquad z_{g,f} =\, 1-z_{f,g}\,.
\end{align}
Here $Q$ is the total incoming partonic momentum, $Q = p_a+p_b$, while $s_{m n}=2p_m\cdot p_n$ and $s_{(l m)n} = 2(p_l+p_m)\cdot p_n$. We note that the above definitions are generic. In particular, they are also valid if any of the momenta have undergone some mapping. For example, from eq.~(\ref{eq:defX2}) it follows that
\begin{equation}
    x_{\wha{rs},\ti{\ha{a}}} =\, \frac{s_{\wha{rs}Q}}{s_{\ti{\ha{a}}Q}}\,.
\end{equation}

\subsection{Partonic cross section}
In this work we will only need the Born-level partonic cross section, which we define as
\beq
\bsp
\label{eq:fullXSB}
    \rd\sigma_{ab}^{B}(p_a,p_b) =\,&\, \sum_{\{0\}}\frac{1}{S_{\{0\}}}\rd\sigma_{ab,X}^{B}(p_a,p_b)\\=\,&\, \frac{{\cal N}}{\omega(a)\omega(b)\Phi(p_a\cdot p_b)}\sum_{\{0\}}\frac{1}{S_{\{0\}}} 
\PS{X}(\mom{}_{X};Q) 
\SME{ab,X}{(0)}{(p_a,p_b;\mom{}_{X})}
\esp
\eeq
with $S_{\{n\}}$ the Bose symmetry factor for $n$ identical partons in the final state. As we only consider color-singlet production, we simply have $n=0$ for the Born process. Nevertheless, it is worth emphasizing that in general, $\rd\sigma_{ab,X}^{B}(p_a,p_b)$ denotes the cross section for a specific partonic subprocess and the full cross section corresponds to a sum over all subprocesses $a+b \to X+ n\mbox{ partons}$. This summation over subprocesses is generically denoted by $\displaystyle \sum_{\{n\}}$\,. For the Born cross section in color-singlet production ($n=0$), this sum of course has only a single term. However, this will no longer be the case when we consider the double-real emission cross section in Section \ref{sec:A12_Iop} below. Hence, we have chosen to set up our notation here, even though, it is somewhat redundant for the Born cross section. Finally, $\mathcal{N}$ collects all non-QCD factors while 
\begin{equation}
    \Phi(p_a\cdot p_b)=\,2p_a\cdot p_b=\,s_{ab}    
\end{equation}
is the partonic flux factor. The $\omega$-function takes into account the averaging over initial-state colors and spins. In particular we have
    \begin{equation}
        \omega(q) =\, \omega(\qb) =\, 4T_R C_A = 2\Nc
    \end{equation}
    for quarks or antiquarks and 
    \begin{equation}
        \omega(g) =\, 4C_A C_F (1-\eps) =\, 2(1-\eps)(\Nc^2-1) 
    \end{equation}
for gluons. $C_A$ ($C_F$) is the usual quadratic Casimir of the SU($\Nc$) color group in the adjoint (fundamental) representation and we use the standard convention $\TR=1/2$. In our calculations, the Born cross section $\rd\sigma_{ab,X}^{B}(p_a,p_b)$ will always be accompanied by a factor of
\begin{equation}
    \left[\frac{\al_s}{2\pi}S_{\eps}\left(\frac{\mu^2}{s_{ab}}\right)^{\eps}\right]^2
\end{equation}
with $\mu$ the arbitrary energy scale introduced by $d=4-2\eps$ dimensional regularization and
\begin{equation}
    S_{\eps} =\, \frac{(4\pi)^{\eps}}{\Gamma(1-\eps)}\,.
\end{equation}
As such, we find it convenient to define
\beq
\bsp
\label{eq:XSB}
    \rd\sigma_{ab}(p_a,p_b) &=\, \left[\frac{\al_s}{2\pi}S_{\eps}\left(\frac{\mu^2}{s_{ab}}\right)^{\eps}\right]^2\rd\sigma_{ab,X}^{B}(p_a,p_b)\,.
\esp
\eeq
We emphasize once more that this quantity characterizes a single partonic subprocess. Below, we will explicitly assume that the prefactor in eq.~(\ref{eq:XSB}) is \textit{unaffected} by rescaling of the initial-state momenta. So, for some generic momentum mapping $(p_a,p_b)\to(\Bar{p}_a,\Bar{p}_b)$ we define the mapped cross section as
\begin{equation}
\label{eq:red_XS}
    \rd\sigma_{\Bar{a}\Bar{b}}(\Bar{p}_a,\Bar{p}_b) =\, \left[\frac{\al_s}{2\pi}S_{\eps}\left(\frac{\mu^2}{s_{ab}}\right)^{\eps}\right]^2\rd\sigma^B_{\Bar{a}\Bar{b},X}(\Bar{p}_a,\Bar{p}_b)\,.
\end{equation}

\section{Integration procedure}
\label{sec:intA12generic}
In this section, we provide a general description of our integration procedure. Although our main interest in this paper is the integration of the approximate cross section $\rd\sigma^{\mathrm{RR,A_{12}}}$ in the case of color-singlet production, cf. eqs.~(\ref{eq:dsigRRA12})-(\ref{eq:CasIFSsA2new}), our workflow is in fact \textit{completely generic}. To emphasize this, we keep the number of jets $m$ arbitrary throughout this section.

\subsection{Hadronic cross sections and dealing with endpoint singularities}
Suppose we have some $A_{12}$ counterterm that we symbolically denote by $CT$. Following the notation introduced in \cite{DelDuca:2024ovc}, the latter will generically be of the form
\begin{equation}
\label{eq:toyCT}
    CT =\, (8\pi\al_s\mu^{2\eps})^2\,\mathrm{Sing}^{(0)}_2\SME{\Bar{a}\Bar{b},X+m}{}{(\Bar{p}_a,\Bar{p}_b;\{\Bar{p}\}_{X+m})}\,.
\end{equation}
Here $\mathrm{Sing}^{(0)}_2$ describes the universal singular structure of the IR limit under consideration (involving a splitting function in a collinear limit and an eikonal factor in a soft one). The momenta $\Bar{p}$ that enter the factorized matrix element are obtained from the original partonic momenta by a successive application of two single-unresolved momentum mappings,
\begin{equation}
\label{eq:mommapToy}
    (p_a,p_b;\{p\}_{X+m+2})\xrightarrow{\mathrm{Map}_1}(\hat{p}_a,\ha{p}_b;\{\hat{p}\}_{X+m+1})\xrightarrow{\mathrm{Map}_2}(\Bar{p}_a,\Bar{p}_b;\{\Bar{p}\}_{X+m})\,.
\end{equation}
Specifically, $\mathrm{Map}_1$ and $\mathrm{Map}_2$ can correspond to a soft, initial-state collinear or final-state collinear mapping, see Appendix \ref{appx:mommaps} for more details. Since all these mappings involve a rescaling of the initial-state momenta, we generically write
\beq
\bsp
\label{eq:momsmapToy}
    &\Bar{p}_a =\, \Bar{x}_1 \Bar{y}_1 p_a\quad\text{\:and\:}\quad
    \Bar{p}_b =\, \Bar{x}_2 \Bar{y}_2 p_b\,.
\esp
\eeq
For the purpose of integration, the transformation of the final-state momenta is irrelevant, and hence we leave this unspecified. Note that, in general, the mapping can also affect the parton flavors. The counterterm in eq.~(\ref{eq:toyCT}) is then subtracted off the partonic cross section to regularize some specific IR limit, after which it needs to be integrated over the final-state phase space $\rd\phi_{X+m+2}(\{p\}_{X+m+2})$ and added back. This integration can generically be written as
\beq
\bsp
    \mathcal{N}\int_2\rd\phi_{X+m+2}(\{p\}_{X+m+2};Q)\,\frac{1}{\omega(a)\omega(b)\Phi(p_a\cdot p_b)}CT\,.
\esp
\eeq
The momentum mapping defined by eq.~(\ref{eq:mommapToy}) now leads to a factorization of the $\rd\phi_{X+m+2}(\{p\}_{X+m+2};Q)$ phase space which we write as
\begin{equation}
    \rd\phi_{X+m+2}(\{p\}_{X+m+2};Q) =\, \rd\phi_2\,\rd\phi_{X+m}(\{\Bar{p}\}_{X+m};Q)\,.
\end{equation}
So, we get a product (or, more generally, a \textit{convolution}) of the two-particle unresolved phase space $\rd\phi_2$ and the combined phase space of the ($X + m$ jets) final state, $\rd\phi_{X+m}(\{\Bar{p}\}_{X+m};Q)$. Following the notation introduced in \cite{DelDuca:2025yph}, the integrated subtraction term then takes on the form
\beq
\bsp
\label{eq:ICT-partonic}
     \left[ \CT\right]\otimes\rd\sigma_{\Bar{a}\Bar{b},m} =\, \int_{0}^{1}\rd x_1\,\rd x_2\,\int_{0}^{1}\rd y_1\,\rd y_2\, [\CT(x_1,x_2,y_1,y_2;\eps)]\rd\sigma_{\Bar{a}\Bar{b},m}(x_1 y_1 p_a,x_2 y_2 p_b)
\esp
\eeq
where, in analogy to eq.~(\ref{eq:XSB}) above, we defined
\beq
\bsp
    \rd\sigma_{ab,m}(p_a,p_b) &=\, \left[\frac{\al_s}{2\pi}S_{\eps}\left(\frac{\mu^2}{s_{ab}}\right)^{\eps}\right]^2\rd\sigma_{ab,X+m}^{B}(p_a,p_b)\,.
\esp
\eeq
In order to write eq.~(\ref{eq:ICT-partonic}) in terms of the \textit{mapped} cross section, we divided our expression by $\omega(\Bar{a})\omega(\Bar{b})$. Of course, we then need to multiply this quantity back, and we assume that the resulting factor $\frac{\omega(\Bar{a})}{\omega(a)}\frac{\omega(\Bar{b})}{\omega(b)}$ is included in the integrand $[\CT(x_1,x_2,y_1,y_2;\eps)]$. The replacement of the integration over the unresolved emissions to an integration over $\{x_1,x_2,y_1,y_2\}$ follows from setting up a \textit{parametric representation} of the phase space measure. In practice, this comes about by first choosing some convenient reference frame, like the rest frame of the incoming partons. Because of spherical symmetry, it is often useful to employ $d$-dimensional polar coordinates. Then, the corresponding angular integral factorizes and can be performed once and for all \cite{Somogyi:2011ir}, while the radial part is set by the various Dirac-delta distributions of the phase space measure. The latter are typically of the form $\delta(\Bar{x}_1-x_1)$, with $x_1$ an integration variable and $\Bar{x}_1$ a parameter of the momentum mapping, cf. eq.~(\ref{eq:momsmapToy}). Hence, with a slight abuse of notation, we will simply \textbf{write the momentum mapping in terms of the \textit{unbarred} variables $x_1,\dots$ when integrating}. This reasoning will be followed for the integration of all $A_{12}$ subtraction terms throughout this text. For example, we will write the flux factor in terms of mapped momenta using
\begin{equation}
    \Phi(\Bar{p}_a\cdot\Bar{p}_b) =\, x_1 y_1 x_2 y_2 \Phi(p_a\cdot p_b)\,.
\end{equation}
We assume the additional factor of $x_1 y_1 x_2 y_2$ to already be absorbed into the definition of the integrand $[\CT(x_1,x_2,y_1,y_2;\eps)]$. The square brackets around the latter denote the fact that a parametric representation of the phase space measure, as discussed above, has already been set up.\newline 

The expression in eq.~(\ref{eq:ICT-partonic}) corresponds to the integrated counterterm at the \textit{partonic} level. To obtain the corresponding expression at the \textit{hadronic} level, we take the convolution with the PDFs and sum over parton flavors, cf.~eq.~(\ref{eq:fact}),
\beq
\bsp
     \left[ \CT\right]\otimes\rd\sigma_{AB} =\, \sum_{a,b}\int_{0}^{1}\rd x_a\,\rd x_b\,f_{a/A}(x_a)f_{b/B}(x_b)&\int_{0}^{1}\rd x_1\,\rd x_2\,\int_{0}^{1}\rd y_1\,\rd y_2\, [\CT(x_1,x_2,y_1,y_2;\eps)]\\&\times\rd\sigma_{\Bar{a}\Bar{b},m}(x_1 y_1 x_a p_A,x_2 y_2 x_b p_B)\,.
\esp
\eeq
Note that the cross section, which is now written in terms of the hadronic momenta using $p_{a/b} =\, x_{a/b}p_{A/B}$, depends on all six integration variables. To simplify the structure of the convolution, we perform a change of variables
\begin{equation}
    y_1\to\frac{z_1}{x_1}\,,\qquad y_2\to\frac{z_2}{x_2}
\end{equation}
leading to
\beq
\bsp
     \left[ \CT\right]\otimes\rd\sigma_{AB}=\,\sum_{a,b}\int_{0}^{1}\rd x_a\,\rd x_b\,f_{a/A}(x_a)f_{b/B}(x_b)&\int_{0}^{1}\rd x_1\,\rd x_2\,\int_{0}^{x_1}\frac{\rd z_1}{x_1}\,\int_{0}^{x_2}\frac{\rd z_2}{x_2}\,\rd\sigma_{\Bar{a}\Bar{b},m}(z_1 x_a p_A,z_2 x_b p_B)\\&\times[\CT(x_1,x_2,z_1/x_1,z_2/x_2;\eps)]\,.
\esp
\eeq
In its current form, we would be required to evaluate the reduced cross section in the integrated subtraction term in several \textit{different} phase space points. However, in practice it is more convenient to only evaluate it in a \textit{single} point. For this reason, we perform an additional change of variables
\begin{equation}
\label{eq:xabTrans}
    x_a\to\frac{x_a}{z_1}\,,\qquad x_b\to\frac{x_b}{z_2}
\end{equation}
such that now we have
\beq
\bsp
     \left[ \CT\right]\otimes\rd\sigma_{AB}=\,\sum_{a,b}&\int_{0}^{1}\frac{\rd x_1}{x_1}\,\frac{\rd x_2}{x_2}\,\int_{0}^{x_1}\frac{\rd z_1}{z_1}\,\int_{0}^{x_2}\frac{\rd z_2}{z_2}\,\int_{0}^{z_1}\rd x_a\,\int_{0}^{z_2}\rd x_b\,\rd\sigma_{\Bar{a}\Bar{b},m}(x_a p_A,x_b p_B)\\&\times[\CT(x_1,x_2,z_1/x_1,z_2/x_2;\eps)] (z_1 z_2)^{2\eps}\,f_{a/A}(x_a/z_1)f_{b/B}(x_b/z_2)\,.
\esp
\eeq
Note the appearance of the factor $(z_1 z_2)^{2\eps}$, which originates from the transformation in eq.~(\ref{eq:xabTrans}) hitting the partonic energy $s_{ab}=\,x_a x_b s_{AB}$ in the reduced cross section, cf.~eq.~(\ref{eq:red_XS}). Next, we use that the argument of the PDFs should be between zero and one to write
\begin{equation}
    \int_{0}^{z_1}\rd x_a\,f_{a/A}(x_a/z_1) =\, \int_{0}^{1}\rd x_a\,f_{a/A}(x_a/z_1) \quad\text{and}\quad\int_{0}^{z_2}\rd x_b\,f_{b/B}(x_b/z_2) =\, \int_{0}^{1}\rd x_b\,f_{b/B}(x_b/z_2)\,.
\end{equation}
Then, exchanging the order of the $(x_1,x_2)$ and $(z_1,z_2)$ integrations, we find
\beq
\bsp
     \left[ \CT\right]\otimes\rd\sigma_{AB}=\,\sum_{a,b}&\int_{0}^{1}\rd x_a\,\rd x_b\,\rd\sigma_{\Bar{a}\Bar{b},m}(x_a p_A,x_b p_B)\int_{0}^{1}\frac{\rd z_1}{z_1}\,\frac{\rd z_2}{z_2}\,\int_{z_1}^{1}\frac{\rd x_1}{x_1}\,\int_{z_2}^{1}\frac{\rd x_2}{x_2}\,\\&\times[\CT(x_1,x_2,z_1/x_1,z_2/x_2;\eps)] (z_1 z_2)^{2\eps}\,f_{a/A}(x_a/z_1)f_{b/B}(x_b/z_2)\,.
\esp
\eeq
As the PDFs do not depend on $x_1$ and $x_2$, the integration over the latter can be done once and for all. Denoting the result by
\begin{equation}
    \int_{z_1}^{1}\frac{\rd x_1}{x_1}\,\int_{z_2}^{1}\frac{\rd x_2}{x_2}\,[\CT(x_1,x_2,z_1/x_1,z_2/x_2;\eps)] \equiv\, {\bom{[\bCT(z_1,z_2;\eps)]}}
\end{equation}
we thus have
\beq
\bsp
     \left[ \CT\right]\otimes\rd\sigma_{AB}=\,\sum_{a,b}\int_{0}^{1}\rd x_a\,\rd x_b\,\rd\sigma_{\Bar{a}\Bar{b},m}(x_a p_A,x_b p_B)\,&\int_{0}^{1}\rd z_1\,\rd z_2\,(z_1 z_2)^{2\eps}{\bom{[\bCT(z_1,z_2;\eps)]}} \\&\times\frac{f_{a/A}(x_a/z_1)}{z_1}\frac{f_{b/B}(x_b/z_2)}{z_2}\,.
\esp
\eeq
The computation of ${\bom{[\bCT(z_1,z_2;\eps)]}}$ up to finite ($\mathcal{O}(\eps^0)$) terms is performed \textit{analytically} and constitutes the main subject of this work. For convenience, we will often call this object the integrated subtraction term. Having an analytic expression instead of just a numerical evaluation has several advantages. First, it allows us to verify the validity of our subtraction scheme explicitly by checking analytic pole cancellation between the partonic cross sections and the approximate ones. Second, it allows for better control over the final convolution integrals involving the PDFs, which are of course evaluated numerically. However, care needs to be taken with the integration over $z_1$ and $z_2$, as ${\bom{[\bCT(z_1,z_2;\eps)]}}$ generically develops \textit{endpoint singularities}. Usually this entails $z_1$ and/or $z_2$ approaching one, though more complicated cases can also occur. As such, ${\bom{[\bCT(z_1,z_2;\eps)]}}$ needs to be interpreted as a \textit{distribution} acting on the PDFs, which we emphasize with the bold notation. In particular, ${\bom{[\bCT(z_1,z_2;\eps)]}}$ should really be thought of as a combination of (double) plus-distributions, Dirac-delta distributions and regular terms. While it is of course perfectly possible mathematically to write out the different combinations, in practice it is not very useful, especially with the construction of a numerical code in mind. Instead, we choose to regularize the integration over $z_1$ and $z_2$ by setting up an appropriate \textit{subtraction}, i.e., we subtract all offending limits and add back the integrated versions. For our symbolic example we have
\beq
\bsp
\label{eq:distexpTOY}
     \left[ \CT\right]\otimes\rd\sigma_{AB}&=\,\sum_{a,b}\int_{0}^{1}\rd x_a\,\rd x_b\,\rd\sigma_{\Bar{a}\Bar{b},m}(x_a p_A,x_b p_B)\,\int_{0}^{1}\rd z_1\,\rd z_2\,\\&\times\Bigg\{(z_1 z_2)^{2\eps}[\CT(z_1,z_2;\eps)]\frac{f_{a/A}(x_a/z_1)}{z_1}\frac{f_{b/B}(x_b/z_2)}{z_2}\\&+(z_2)^{2\eps}(-\bom{L}_1+[\bom{L}_1])[\CT(z_1,z_2;\eps)]f_{a/A}(x_a)\frac{f_{b/B}(x_b/z_2)}{z_2}\\&+(z_1)^{2\eps}(-\bom{L}_2+[\bom{L}_2])[\CT(z_1,z_2;\eps)]\frac{f_{a/A}(x_a/z_1)}{z_1}f_{b/B}(x_b)\\&+(-\bom{L}_{12}+\bom{L}_1\bom{L}_{12}+\bom{L}_2\bom{L}_{12}+[\bom{L}_{12}]-[\bom{L}_1\bom{L}_{12}]-[\bom{L}_2\bom{L}_{12}])[\CT(z_1,z_2;\eps)]f_{a/A}(x_a)f_{b/B}(x_b)\Bigg\}\,.
\esp
\eeq
Here $[\CT(z_1,z_2;\eps)]$ represents the integrated counterterm, now interpreted as a \textit{function} (as opposed to a \textit{distribution}). Furthermore, $\bom{L}$ denotes a formal limit operator. For example, $\bom{L}_2[\CT(z_1,z_2;\eps)]$ selects the leading singularity in $d$ dimensions as $z_2\to 1$, dropping both subleading and non-singular terms,
\begin{equation}
\label{eq:L2}
    \bom{L}_2[\CT(z_1,z_2;\eps)] =\, \sum_i (1-z_2)^{c_i+d_i\eps}g_i(z_1;\eps)\,.
\end{equation}
Note that the functions $g_i(z_1;\eps)$ are independent of $z_2$, and that all $c_i$ should be negative. Likewise, $\bom{L}_{12}[\CT(z_1,z_2;\eps)]$ takes care of the singular behavior as both $z_1\to 1$ \textit{and} $z_2\to 1$,
\begin{equation}
\label{eq:L12}
    \bom{L}_{12}[\CT(z_1,z_2;\eps)] =\, \sum_i (1-z_1)^{a_i+b_i\eps}(1-z_2)^{c_i+d_i\eps}g_i(z_1/z_2;\eps)\,.
\end{equation}
In this case the functions $g_i$ still depend on both $z_1$ and $z_2$, but only through their ratio. Furthermore, all $a_i$ and $c_i$ should be negative. Finally, the iterated limits $\bom{L}_1\bom{L}_{12}[\CT(z_1,z_2;\eps)]$ and $\bom{L}_2\bom{L}_{12}[\CT(z_1,z_2;\eps)]$ take care of the singularities of $\bom{L}_{12}[\CT(z_1,z_2;\eps)]$ as $z_1$ or $z_2$ approaches one, e.g.,
\begin{equation}
\label{eq:L2L12}
    \bom{L}_2\bom{L}_{12}[\CT(z_1,z_2;\eps)] =\, \sum_i (1-z_1)^{a_i+b_i\eps}(1-z_2)^{c_i+d_i\eps}g_i(\eps)\,.
\end{equation}
The functions $g_i$ are now completely independent of $z_1$ and $z_2$ while, as before, $a_i$ and $c_i$ should be negative. Once all the limits are computed, the resulting expressions need to be integrated over the appropriate variable(s) and added back to complete the subtraction, which is denoted by $[\bom{L}]$ in eq.~(\ref{eq:distexpTOY}). We define
\begin{align}
    &[\bom{L}_1][\text{CT}(z_1,z_2;\eps)] =\, \int_0^1\rd z_1\, \bom{L}_1[\text{CT}(z_1,z_2;\eps)]\,,\\
    &[\bom{L}_2][\text{CT}(z_1,z_2;\eps)] =\, \int_0^1\rd z_2\, \bom{L}_2[\text{CT}(z_1,z_2;\eps)]\,,\\
    &[\bom{L}_{12}][\text{CT}(z_1,z_2;\eps)] =\, \int_0^1\rd z_1\,\int_0^1\rd z_2\, \bom{L}_{12}[\text{CT}(z_1,z_2;\eps)]\,,\\
    &[\bom{L}_1\bom{L}_{12}][\text{CT}(z_1,z_2;\eps)] =\, \int_0^1\rd z_1\, \bom{L}_1\bom{L}_{12}[\text{CT}(z_1,z_2;\eps)]\,,\\
    &[\bom{L}_2\bom{L}_{12}][\text{CT}(z_1,z_2;\eps)] =\, \int_0^1\rd z_2\, \bom{L}_2\bom{L}_{12}[\text{CT}(z_1,z_2;\eps)]\,.
\end{align}
From the discussion above, it is clear that the integration of the single limits is trivial. For example, from eqs.~(\ref{eq:L2}) and (\ref{eq:L2L12}) we see that
\beq
\bsp
    &[\bom{L}_2][\text{CT}(z_1,z_2;\eps)] =\, \sum_i\frac{g_{i}(z_1;\eps)}{1+c_i+d_i\eps}\,,\\
    &[\bom{L}_2\bom{L}_{12}][\text{CT}(z_1,z_2;\eps)] =\,\sum_i\frac{g_i(\eps)}{1+c_i+d_i\eps}(1-z_1)^{a_i+b_i\eps}\,.
\esp
\eeq
The integration of the double limit, eq.~(\ref{eq:L12}), is non-trivial and in practice requires an actual computation. We can compactify the expression for the integrated subtraction term in eq.~(\ref{eq:distexpTOY}) by introducing its \textit{coefficient functions}, which we define as
\begin{align}
\label{eq:toyCFin}
    &\CT(z_1,z_2;\eps\,|\,z_1,z_2) =\, (z_1 z_2)^{2\eps}[\CT(z_1,z_2;\eps)]\,,\\
    &\CT(z_1,z_2;\eps\,|\,1,z_2) =\, (z_2)^{2\eps}(-\bom{L}_1+[\bom{L}_1])[\CT(z_1,z_2;\eps)]\,,\\
    &\CT(z_1,z_2;\eps\,|\,z_1,1) =\, (z_1)^{2\eps}(-\bom{L}_2+[\bom{L}_2])[\CT(z_1,z_2;\eps)]\,,\\
    &\CT(z_1,z_2;\eps\,|\,1,1) =\, (-\bom{L}_{12}+\bom{L}_1\bom{L}_{12}+\bom{L}_2\bom{L}_{12}+[\bom{L}_{12}]-[\bom{L}_1\bom{L}_{12}]-[\bom{L}_2\bom{L}_{12}])[\CT(z_1,z_2;\eps)]\,.
\label{eq:toyCFout}
\end{align} 
Hence
\beq
\bsp
\label{eq:distexpCFsTOY}
     \left[ \CT\right]\otimes\rd\sigma_{AB}&=\,\sum_{a,b}\int_{0}^{1}\rd x_a\,\rd x_b\,\rd\sigma_{\Bar{a}\Bar{b},m}(x_a p_A,x_b p_B)\,\int_{0}^{1}\rd z_1\,\rd z_2\,\\&\times\Bigg\{\CT(z_1,z_2;\eps\,|\,z_1,z_2)\frac{f_{a/A}(x_a/z_1)}{z_1}\frac{f_{b/B}(x_b/z_2)}{z_2}+\CT(z_1,z_2;\eps\,|\,1,z_2)f_{a/A}(x_a)\frac{f_{b/B}(x_b/z_2)}{z_2}\\&+\CT(z_1,z_2;\eps\,|\,z_1,1)\frac{f_{a/A}(x_a/z_1)}{z_1}f_{b/B}(x_b)+\CT(z_1,z_2;\eps\,|\,1,1)f_{a/A}(x_a)f_{b/B}(x_b)\Bigg\}\,.
\esp
\eeq
As such, the computation of the coefficient functions requires the determination of (a) the integrated counterterm and (b) the asymptotic behavior of the integrated counterterm in all relevant limits. Furthermore, if we want the subtraction to be fully consistent at NNLO to finite terms in $\eps$, the integrated subtraction term itself should be computed to $\mathcal{O}(\eps^0)$, while the single and double limits should be computed to $\mathcal{O}(\eps)$ and $\mathcal{O}(\eps^2)$ respectively, since integrating the limit formul\ae\, introduces additional poles. All of this of course needs to be considered for each $A_{12}$ subtraction term separately. Luckily, it turns out that the computations for different counterterms are not completely different from one another and share some generic features. In fact, we can write down a recipe to compute the coefficient functions which can be followed for \textit{all} $A_{12}$ counterterms.\footnote{As will be discussed in a future publication, this recipe will be useful \textit{beyond} $A_{12}$ as well. For example, the integration of the subtraction to the integrated $A_1$ counterterms follows the same steps.}

\subsection{The integration recipe}
\label{sec:A12ints}
The computation of the integrated subtraction terms and the regularization of their endpoint singularities can be systematized through the following steps. 
\begin{enumerate}
    \item Write the integrated counterterm in the form of a \textbf{parametric integral}. This is achieved by choosing some explicit parametrization of the phase space measures for unresolved emissions and expressing the singular structure $\mathrm{Sing}^{(0)}_{2}$ with the chosen variables. The resulting parametric integral representations can involve non-trivial $d$-dimensional angular integrals, the treatment of which is however well-understood, see e.g.~\cite{Somogyi:2011ir}. 
    \item The resulting integrand is typically a complicated function of integration variables involving products of rational functions raised to $\eps$-dependent powers. In particular, the denominators can have non-trivial zeroes, leading to overlapping singularities. The latter can be treated by setting up a \textbf{sector decomposition} \cite{Heinrich:2008si}, which in our computations we do using an in-house routine. Once no overlapping singularities remain, the $\eps$-poles can be extracted in a straightforward manner. In this way, we obtain a representation where each coefficient in the $\eps$-expansion is given by a finite parametric integral.
    \item The integrands representing the expansion coefficients are given by products of rational functions and logarithms of rational functions. Thus, we can attempt to evaluate the integrals in terms of generalized polylogarithms (GPLs) \cite{Goncharov:1998kja}, performing the integrations over all variables one after the other.\footnote{A brief summary of useful properties of GPLs can be found in Appendix \ref{appx:GPLs}. These (and many more) are nicely implemented in the package {\tt PolyLogTools} \cite{Duhr:2019tlz}, which we use extensively.} Since the integration kernels of GPLs are linear, higher-order polynomials of the current integration variable that appear in denominators should be fully  \textbf{factorized}. This generically leads to expressions with an algebraic dependence on the rest of the integration variables. To stay within the realm of GPLs, one should then set up transformations of variables to \textbf{rationalize} such expressions. This is automated, e.g., in the {\tt Mathematica} package {\tt RationalizeRoots} \cite{Besier:2019kco}.
        \item Perform the \textbf{integration} in the chosen variable in terms of GPLs. For this, the weight vectors of the GPLs should be independent of the integration variable. As such, one needs to ensure that the integrand is written in a \textbf{fibration basis} with respect to the integration variable.\footnote{This is a slight abuse of notation introduced for notational simplicity. Of course a fibration basis is determined by choosing a specific ordering of \textit{all} variables that appear in the GPL. However, for the purpose of integration it is most important that the integration variable is the first variable of the fibration, while the ordering of the other variables typically has little to no influence. For example, if the integration variable is $x$, choosing a fibration basis with respect to $x$ really means with respect to $\{x,\dots\}$ with the ordering of the other variables left implicit.}
    \item Repeat steps 3-4 for all integration variables.
    \item The result obtained in step 5 still needs to be integrated over the variables that explicitly appear in the PDFs. However, these integrals typically suffer from \textit{endpoint singularities}, which are treated by setting up an appropriate \textbf{subtraction}. For this, one needs to determine the asymptotic behavior of the integrand in all relevant limits, which can be done using the method of \textbf{expansion by regions} \cite{Beneke:1997zp}. The determination of all relevant regions is non-trivial but automated in, e.g., the {\tt Mathematica} package {\tt asy2.m} \cite{Pak:2010pt,Jantzen:2012mw}. The resulting limit formul\ae\, are then subtracted and, to complete the subtraction, integrated over the appropriate variable(s) and added back.
\end{enumerate}
A somewhat surprising complication arises in step 3 of our integration recipe, in which we need to perform univariate partial fraction decompositions of intermediate results. As the latter are typically large multivariate expressions, the computation of such decompositions can become a significant bottleneck. For example, a typical function that appears is
\beq
\bsp
r(x_a,x_b;y) &=
\Big[(-4 + y) (1 - y + x_b y) (2 - y + x_b y) (4 - y + x_b y) (1 - x_a - y + x_b y)^3 
\\&\times
(-1 + x_a - y + x_b y) (-4 - 4 x_b - y + x_b y) (-4 x_b - y + x_b y) \\&\times
(-4 x_a - 4 x_b - y + x_b y) (4 x_a - 4 x_b - y + x_b y) (2 + 2 x_b - y + x_b y)^3 
\\&\times
(6 + 2 x_b - y + x_b y) (2 - 4 x_a + 2 x_b - y + x_b y) (2 + 4 x_a + 2 x_b - y + x_b y) 
\\&\times
(-1 + x_a - x_a y + x_a x_b y) (1 + x_a - x_a y + x_a x_b y) (-2 + 2 x_a - x_a y + x_a x_b y) 
\\&\times
(2 + 2 x_a - x_a y + x_a x_b y) (-x_b + x_a x_b - x_a y + x_a x_b y)^3 
\\&\times
(-4 + 2 x_a + 2 x_a x_b - x_a y + x_a x_b y) (4 + 2 x_a + 2 x_a x_b - x_a y + x_a x_b y) 
\\&\times
(1 - 2 x_a + x_a^2 - y - x_a y + x_b y + x_a x_b y) 
\\&\times
\left(2 x_b - 2 x_a x_b + x_a y - x_b y - x_a x_b y + x_b^2 y\right)^3\Big]^{-1}\,,
\esp
\label{eq:realex}
\eeq
which must be integrated symbolically over $y$. Unfortunately, setting up the partial fraction decomposition with respect to $y$ using publicly available tools, such as {\tt Apart} in {\tt Mathematica}, is intractable, as simple estimates put the required time at $\sim 10^9\,$s. We circumvented this issue with the development of a new univariate partial fractioning routine called {\tt LinApart} \cite{Chargeishvili:2024nut}. The latter is based on a closed formula for the decomposition which is rooted in the residue theorem. The main advantage of our routine is a major improvement in both time and memory consumption in the computation of partial fraction decompositions of complicated rational functions. For example, it only takes {\tt LinApart} $\sim 10^{-2}\,$s to compute the decomposition of the function in eq.~(\ref{eq:realex}).\newline

We finish this section by stressing that none of the steps presented above depend on the presence of additional jets in the process under consideration. As such, the whole procedure is also applicable \textit{beyond} the color-singlet case.

\section{
\texorpdfstring{Overview of the integrated $A_{12}$ subtraction terms}{Overview of the integrated A12 subtraction terms}}
\label{sec:ICTs}
In this section we provide an overview of the computation of the integrated $A_{12}$ subtraction terms listed in eqs.~(\ref{eq:SsA2new})-(\ref{eq:CasIFSsA2new}). The explicit integrations broadly follow the procedure outlined in sec.~\ref{sec:intA12generic} above. We will not discuss the explicit construction of the counterterms, as this was done in detail in \cite{DelDuca:2025yph}. We emphasize however that for $A_{12}$, all subtraction terms involve mapped momenta that arise by iterating the initial-final collinear (IF), final-final collinear (FF) and soft (S) mappings of Appendix~\ref{appx:mommaps}. Consequently, when integrating, it will be convenient to collect the counterterms by the particular sequence of iterated momentum mappings they correspond to. We distinguish the following five distinct cases:
\begin{itemize}
    \item IF--IF iteration,
    \item S--S iteration,
    \item IF--S iteration,
    \item S--FF iteration and
    \item IF--FF iteration.
\end{itemize}
Note that the ordering of the mappings in this notation is understood to correspond to convolutions of maps such that the rightmost mapping is performed first. E.g, the IF--S case denotes a soft mapping followed by an initial-final collinear mapping.\newline

This section exclusively treats the integrations in the case of color-singlet production. However, most of the integrals to be discussed are in fact \textit{completely generic}, in the sense that they require no modification in the case of jet production. The few exceptions come from counterterms whose definitions contain eikonal factors that depend on hard momenta, which can only happen if the soft mapping is involved. For color-singlet production, there are only two hard partons, namely the incoming ones, whose momenta are moreover back-to-back. Beyond color-singlet production, these counterterms may involve up to four hard partons with general kinematics.

\subsection{Integration of IF--IF iterated subtraction terms}
We start with the integration of the IF--IF iterated subtraction terms. The factorized matrix element is written in terms of the momenta $( \hat{\hat{p}}_a, \hat{\hat{p}}_b;\momhh{}_{X})$, obtained by iterating the initial-final collinear mapping in eq.~(\ref{eq:IFmap}),
\begin{equation}
\label{eq:IFIF-map}
    ( p_a,p_b;\mom{}_{X+2})\xrightarrow{C_{ab,s}^{II,F}}( \hat{p}_a, \hat{p}_b;\momh{}_{X+1})\xrightarrow{C_{\hat{a}\hat{b},\hat{r}}^{II,F}}( \hat{\hat{p}}_a, \hat{\hat{p}}_b;\momhh{}_{X})\,.
\end{equation}
Specifically we have\footnote{The transformation of the final-state momenta is not needed and hence omitted here and in the following sections.}
\begin{align}
    &\hat{\hat{p}}_{a}^{\mu} =\,\xi_{\hat{a},\hat{r}}\xi_{a,s}p_{a}^{\mu}\,,\qquad\hat{\hat{p}}_{b}^{\mu} =\,\xi_{\hat{b},\hat{r}}\xi_{b,s}p_{b}^{\mu}\,.
\end{align}
The $(X+2)$-particle phase space is composed of an iteration of initial-final collinear convolutions, 
\beq
\bsp
\label{eq:PS-IFIF}
    \rd\phi_{X+2}(\{p\}_{X+2};Q) =\,&\,\int_{0}^{1}\rd\xi_a\,\rd\xi_b\,\int_{0}^{1}\rd\ha{\xi}_a\,\rd\ha{\xi}_b\,\rd\phi_{X}(\{\ha{\ha{p}}\}_{X};\ha{\ha{Q}})\rd\phi_{II,F}(\ha{p}_r,\ha{\xi}_a,\ha{\xi}_b)\, \rd\phi_{II,F}(p_s,\xi_a,\xi_b)
\esp
\eeq
with\footnote{The explicit derivation of this representation of the phase space measures can be found in Appendix \ref{sec:defIC}.}
\beq
\bsp
\label{eq:PSfact}
    \rd\phi_{II,F}(\ha{p}_r,\ha{\xi}_{a},\ha{\xi}_{b}) =\,&\, \frac{S_{\epsilon}}{8\pi^2}\frac{\rd\Omega_{d-2}}{\Omega_{d-2}}\rd\xi_{\ha{a},\ha{r}}\,\rd\xi_{\ha{b},\ha{r}}\,s_{\ha{a}\ha{b}}^{1-\eps}\frac{\xi_{\ha{a},\ha{r}}\xi_{\ha{b},\ha{r}}(1+\xi_{\ha{a},\ha{r}}\xi_{\ha{b},\ha{r}})}{(\xi_{\ha{a},\ha{r}}+\xi_{\ha{b},\ha{r}})^2}\left[\frac{\xi_{\ha{a},\ha{r}}\xi_{\ha{b},\ha{r}}(1-\xi_{\ha{a},\ha{r}}^2)(1-\xi_{\ha{b},\ha{r}}^2)}{(\xi_{\ha{a},\ha{r}}+\xi_{\ha{b},\ha{r}})^2}\right]^{-\eps}\\&\times\theta(1-\xi_{\ha{a},\ha{r}}\xi_{\ha{b},\ha{r}})\delta(\xi_{\ha{a},\ha{r}}-\ha{\xi}_a)\delta(\xi_{\ha{b},\ha{r}}-\ha{\xi}_b)\,, 
    \\ \rd\phi_{II,F}(p_s,\xi_{a},\xi_{b}) =\,&\, \frac{S_{\epsilon}}{8\pi^2}\frac{\rd\Omega_{d-2}}{\Omega_{d-2}}\rd\xi_{a,s}\,\rd\xi_{b,s}\,s_{ab}^{1-\eps}\frac{\xi_{a,s}\xi_{b,s}(1+\xi_{a,s}\xi_{b,s})}{(\xi_{a,s}+\xi_{b,s})^2}\left[\frac{\xi_{a,s}\xi_{b,s}(1-\xi_{a,s}^2)(1-\xi_{b,s}^2)}{(\xi_{a,s}+\xi_{b,s})^2}\right]^{-\eps}\\&\times\theta(1-\xi_{a,s}\xi_{b,s})\delta(\xi_{a,s}-\xi_a)\delta(\xi_{b,s}-\xi_b)\,.
\esp
\eeq
There are two distinct counterterms in this class, namely $\cC{asr}{IFF(0,0)}\cC{as}{IF}$ and $\cC{as,br}{IF,IF(0,0)}\cC{as}{IF}$. Because of the shared momentum mapping and unresolved phase space, their integrated versions can be written uniformly as
\beq
\bsp
\label{eq:IFIF-distexp}
    \left[ \CT\right]\otimes\rd\sigma_{AB} =\, \sum_{a,b}&\int_{0}^{1}\rd x_a\,\rd x_b\,\rd\sigma_{\ha{\ha{a}}\ha{\ha{b}}}(x_a p_A,x_b p_B)\,\int_{0}^{1}\rd \eta_a\,\rd \eta_b\,\\&\times\Bigg\{\CT(\eta_a,\eta_b;\eps\,|\,\eta_a,\eta_b)\frac{f_{a/A}(x_a/\eta_a)}{\eta_a}\frac{f_{b/B}(x_b/\eta_b)}{\eta_b}\\&+\CT(\eta_a,\eta_b;\eps\,|\,1,\eta_b)f_{a/A}(x_a)\frac{f_{b/B}(x_b/\eta_b)}{\eta_b}\\&+\CT(\eta_a,\eta_b;\eps\,|\,\eta_a,1)\frac{f_{a/A}(x_a/\eta_a)}{\eta_a}f_{b/B}(x_b)\\&+\CT(\eta_a,\eta_b;\eps\,|\,1,1)f_{a/A}(x_a)f_{b/B}(x_b)\Bigg\}\,.
\esp
\eeq
Here the coefficient functions are defined as in eqs.~(\ref{eq:toyCFin})-(\ref{eq:toyCFout}) with $z_1$ and $z_2$ replaced by $\eta_a$ and $\eta_b$,
\begin{align}
    &\CT(\eta_a,\eta_b;\eps\,|\,\eta_a,\eta_b) =\, (\eta_a \eta_b)^{2\eps}[\CT(\eta_a,\eta_b;\eps)]\,,\\
    &\CT(\eta_a,\eta_b;\eps\,|\,1,\eta_b) =\, (\eta_b)^{2\eps}(-\bom{L}_a+[\bom{L}_a])[\CT(\eta_a,\eta_b;\eps)]\,,\\
    &\CT(\eta_a,\eta_b;\eps\,|\,\eta_a,1) =\, (\eta_a)^{2\eps}(-\bom{L}_b+[\bom{L}_b])[\CT(\eta_a,\eta_b;\eps)]\,,\\
    &\CT(\eta_a,\eta_b;\eps\,|\,1,1) =\, (-\bom{L}_{ab}+\bom{L}_a\bom{L}_{ab}+\bom{L}_b\bom{L}_{ab}+[\bom{L}_{ab}]-[\bom{L}_a\bom{L}_{ab}]-[\bom{L}_b\bom{L}_{ab}])[\CT(\eta_a,\eta_b;\eps)]\,.
\end{align} 
Since the subtraction terms of this type do not involve any eikonal factors, their integrated versions are generic and can also be used \textit{beyond} the color-singlet case.

\subsubsection{
\texorpdfstring{$\cC{asr}{IFF (0,0)}\cC{as}{IF}$}{CarsIFF00CasIF}}
The integration of the $\cC{asr}{IFF(0,0)}\cC{as}{IF}$ subtraction term gives a good idea of all techniques and complications that generically arise in the $A_{12}$ integration procedure. As such, we use it as a template example, which can be found in Appendix \ref{appx:ICasrCas}. There we present an in-depth discussion of the computational workflow, including an overview of the tools that were used. Given the amount of detail, we hope that this can serve as a pedagogical example on how this type of calculation is performed in practice.\newline

Here we will simply summarize our findings. The integrated subtraction term, or equivalently the coefficient function ${\rm{C}}_{asr}^{IFF(0,0)}{\rm{C}}_{as}^{IF}(\eta_a,\eta_b;\eps\,|\,\eta_a,\eta_b)$, was computed to $\mathcal{O}(\eps^0)$ using steps 1-5 of our recipe in sec.~\ref{sec:A12ints}. While its pole parts are relatively simple, the finite part is a complicated function of $\eta_a$ and $\eta_b$. In particular, the result contains more than 500 weight-two GPLs with a quite elaborate alphabet. For example, some of the letters are solutions of quartic equations, such that they contain nested square roots. However, it can be shown that, once the full $\bI^{(0)}_{12}$ insertion operator is constructed, all such contributions cancel, leading to significant simplifications in the analytic structure. Next, we use step 6 in sec.~\ref{sec:A12ints} to compute the asymptotic behavior of the integrated counterterm. It turns out that the latter is actually regular as $\eta_a\to 1$ and we have ${\rm{C}}_{asr}^{IFF(0,0)}{\rm{C}}_{as}^{IF}(\eta_a,\eta_b;\eps\,|\,1,\eta_b)=0$. The remaining coefficient functions in eq.~(\ref{eq:IFIF-distexp}) are non-zero. Their analytic structure is however quite simple. For example, the most complicated structures in the alphabet of the GPLs are rational functions of the convolution variables $\eta_a$ and $\eta_b$. Furthermore, all GPLs can be mapped to logarithms and classical polylogarithms up to weight four.

\subsubsection{
\texorpdfstring{$\cC{as,br}{IF,IF(0,0)}\cC{as}{IF}$}{CasbrIFIF00CasIF}}
The integrated counterterm was set up in \cite{DelDuca:2025yph},
\beq
\bsp
    \left[ \IcC{as,br}{IF,IF(0,0)} \IcC{as}{IF}\right]\otimes\rd\sigma_{\ha{\ha{a}}\ha{\ha{b}}} =\,&\, \int_{0}^{1} \rd\xi_a \rd\xi_b\, \int_{0}^{1} \rd\hat{\xi}_a\, \rd\hat{\xi}_b\,\rd\sigma_{\ha{\ha{a}}\ha{\ha{b}}}(\hat{\hat{p}}_a,\hat{\hat{p}}_b)\,
	\left[ \IcC{as,br}{IF,IF(0,0)} \IcC{as}{IF}(\xi_a,\xi_b,\ha{\xi}_a,\ha{\xi}_b;\eps)\right]
\esp
\eeq
with
\beq
\bsp
\label{eq:CasbrCas-br}
    &\left[ \IcC{as,br}{IF,IF(0,0)} \IcC{as}{IF}(\xi_a,\xi_b,\ha{\xi}_a,\ha{\xi}_b;\eps)\right] =\, 4 \frac{\omega(as)}{\omega(a)}\, \frac{\omega(br)}{\omega(b)}\, \s12^2\,
	\left[ \frac{\xi_a^2 \xi_b^2 (1-\xi_a^2) (1-\xi_b^2)}{(\xi_a + \xi_b)^2} \right]^{-\eps}\,
	\\&\times\frac{\xi_a^3 \xi_b^3 (1+\xi_a \xi_b)}{(\xi_a + \xi_b)^2} 
	\left[ \frac{\hat{\xi}_a \hat{\xi}_b (1-\hat{\xi}_a^2) (1-\hat{\xi}_b^2)}{(\hat{\xi}_a + \hat{\xi}_b)^2} \right]^{-\eps} 
	\frac{\hat{\xi}_a^2 \hat{\xi}_b^2 (1+\hat{\xi}_a \hat{\xi}_b)}{(\hat{\xi}_a + \hat{\xi}_b)^2}
	\frac{{\cal F}(x_{\ha{b},\ha{r}},\ha{\xi}_a \ha{\xi}_b)}{x_{a,s} s_{as} x_{\ha{b},\ha{r}} s_{\ha{b}\ha{r}}}\,
	\\&\times P_{(as)s}^{(0)}(x_{a,s};\ep)\, P_{(br)r}^{(0)}(x_{\ha{b},\ha{r}};\ep)\,.
\esp
\eeq
Here the function ${\cal F}(x_{\ha{b},\ha{r}},\ha{\xi}_a \ha{\xi}_b)$ is defined as
\beq
\label{eq:Fifif}
{\cal F}(x_{\ha{b},\ha{r}},\ha{\xi}_a \ha{\xi}_b) =\, \left(\frac{x_{\ha{b},\ha{r}}}{\ha{\xi}_a\ha{\xi}_b}\right)^2
\eeq
while the momentum fractions are as in eq.~(\ref{eq:defX2}). The argument of $\omega$ represents the parton flavor with $(as) = a+s$, computed using the standard rules, e.g. $\omega(gg) = \omega(g)$, $\omega(qg) = \omega(q)$ etc. Because of the structure of the splitting functions in eq.~(\ref{eq:CasbrCas-br}), see also \eqnss{eq:Pqg-ave-IF}{eq:Pgg-ave-IF}, there are 25 distinct basic integrals to compute. The integrands are very similar to those for the $\cC{asr}{IFF(0,0)}\cC{as}{IF}$ subtraction term, and hence the computation very closely follows the discussion in secs.~\ref{sec:prepint}-\ref{sec:distexp}. For this reason, we omit the explicit details here. One noteworthy difference, however, is that in this case all four coefficient functions appearing in eq.~(\ref{eq:IFIF-distexp}) are non-zero. Put another way, the integrated counterterm develops singularities as any of its variables approaches its endpoint, and hence all steps of regularization are required.

\subsection{Integration of S--S iterated subtraction terms}
The factorized matrix element for the S--S iterated counterterms is written in terms of the momenta $(\ti{\ti{p}}_a, \ti{\ti{p}}_b;\momtt{}_{X})$, which are defined by successive application of the single soft mapping in eq.~(\ref{eq:softmap}),
\beq
(p_a, p_b;\mom{}_{X+2}) \xrightarrow{S_s} (\ti{p}_a, \ti{p}_b;\momt{}_{X+1}) \xrightarrow{S_{\ti{r}}} (\ti{\ti{p}}_a, \ti{\ti{p}}_b;\momtt{}_{X})
\label{eq:itersoftsoftmap}
\eeq
with
\beq
\label{eq:SSmommap}
\ti{\ti{p}}_a^\mu =\, \lam_s\lam_{\ti{r}} p_a^\mu\,,
\qquad
\ti{\ti{p}}_b^\mu =\, \lam_s\lam_{\ti{r}} p_b^\mu\,.
\eeq
The soft momentum mapping is defined in such a way that the $(X+2)$-particle phase space convolution is
\beq
 \PS{X+2}(\mom{}_{X+2};Q) =\, \int_{0}^{1} \rd\lam\,\frac{\lam s_{ab}}{\pi}\, \PS{2}(p_s,K;Q) \int_{0}^{1} \rd\ti{\lam}\, \frac{\ti{\lam} s_{\ti{a}\ti{b}}}{\pi}\, \PS{2}(\ti{p}_r,\ti{K};\ti{Q})\,
	\PS{X}(\momtt{}_{X};\ti{\ti{Q}} )\,.
\label{eq:PS-SS}
\eeq
In total we have four counterterms in this class, labelled by $\cS{rs}{(0,0)}\cS{s}{ }$, $\cC{ars}{IFF}\cS{rs}{(0,0)}\cS{s}{}$, $\cS{rs}{(0,0)}\cC{as}{IF}\cS{s}{}$ and $\cC{ars}{IFF}\cS{rs}{(0,0)}\cC{as}{IF}\cS{s}{}$. Because of the common momentum mapping and phase space factorization, their integrated versions inherit a shared structure. Denoting a generic S--S iterated subtraction term by $CT$ we have
\beq
\bsp
\label{eq:SS-distexp}
     \left[ \CT\right]\otimes\rd\sigma_{AB} =\,\sum_{a,b}\int_{0}^{1}\rd x_a\,\rd x_b\,\rd\sigma_{\ti{\ti{a}}\ti{\ti{b}}}(x_a\, p_A,x_b\, p_B)\int_{0}^{1}&\rd\eta\,\Bigg\{\CT(\eta;\eps\,|\,\eta)\frac{f_{a/A}(x_a/\eta)}{\eta}\frac{f_{b/B}(x_b/\eta)}{\eta}\\&+\CT(\eta;\eps\,|\,1)f_{a/A}(x_a)f_{b/B}(x_b)\Bigg\}
\esp
\eeq
with
\begin{align}
    &\CT(\eta;\eps\,|\,\eta) =\, \eta^{4\eps}[\CT(\eta;\eps)]\quad\text{and}\quad
    \CT(\eta;\eps\,|\,1) =\, (-\Leta+[\Leta])[\CT(\eta;\eps)]\,.
\end{align}
The precise definitions and some details about the explicit computations for each of the four counterterms will be presented below.

\subsubsection{
\texorpdfstring{$\cS{rs}{(0,0)}\cS{s}{ }$}{Srs00Ss}}
\label{sec:defISrsSs0}
The integrated subtraction term is \cite{DelDuca:2025yph}
\beq
\bsp
\label{eq:Int_SSS}
& \left[ \IcS{rs}{(0,0)} \IcS{s}{}\right]\otimes\rd\sigma_{\ti{\ti{a}}\ti{\ti{b}}}=\,\int_{0}^{1} \rd\lam\, \int_{0}^{1} \rd\ti{\lam}\,
	\rd\sigma_{\ti{\ti{a}}\ti{\ti{b}}}(\ti{\ti{p}}_a,\ti{\ti{p}}_b)\,
	\Bigg[\frac18 
	\sum_{\substack{i,k,j,\ell \in I}} 
	\left[ \IcS{rs}{(0,0)} \IcS{s}{}(\lam,\ti{\lam};\eps)\right]^{(ik,j\ell)}  \,
	\{\bT_{i} \bT_{k},\bT_{j} \bT_{\ell}\}
	\\&\qquad\qquad\qquad\qquad\qquad-\frac14 \CA
	\sum_{\substack{i,k \in I}} 
	\left[ \IcS{rs}{(0,0)} \IcS{s}{}(\lam,\ti{\lam};\eps)\right]^{(i,k)}
	\bT_{i} \bT_{k}  \Bigg]
\esp
\eeq
with
\beq
\bsp
[\IcS{rs}{(0,0)} \IcS{s}{}(\lam,\ti{\lam};\eps)]^{(ik,j\ell)} =\,&
	\left( \frac{\s12}{\pi} \right)^2
	\left( \frac{(4\pi)^2 }{S_\ep} \s12^\ep \right)^2 
	\int_1 \PS{2}(p_s,K;Q)\, \int_1 \PS{2}(\ti{p}_r,\ti{K};\ti{Q})\,
	   \lam^5 \ti{\lam}^3 \, \calS_{\ti{\ti{i}}\ti{\ti{k}}}(\ti{r}) \calS_{\ti{j}\ti{\ell}}(s) \,, \\
[\IcS{rs}{(0,0)} \IcS{s}{}(\lam,\ti{\lam};\eps)]^{(i,k)} =\,&
	\left( \frac{\s12}{\pi} \right)^2
	\left( \frac{(4\pi)^2 }{S_\ep} \s12^\ep \right)^2   \int_1 \PS{2}(p_s,K;Q)\, \\&\times\int_1 \PS{2}(\ti{p}_r,\ti{K};\ti{Q})\,
	   \lam^5 \ti{\lam}^3 \, \calS_{\ti{\ti{i}}\ti{\ti{k}}}(\ti{r})
	\left(
		\calS_{\ti{\ti{i}}\ti{r}}(s)
		+ \calS_{\ti{\ti{k}}\ti{r}}(s)
		- \calS_{\ti{\ti{i}}\ti{\ti{k}}}(s)
	\right) \,.
\esp
\label{eq:ISSS}
\eeq
Here $\calS_{ij}(k)$ denotes the standard soft eikonal factor,
\begin{equation}
\label{eq:eikonal}
    \calS_{ij}(k) =\, \frac{2s_{ij}}{s_{ik}s_{jk}}\,.
\end{equation}
Note that the definition in eq.~(\ref{eq:Int_SSS}) is specific to color-singlet production. In particular, if colored particles also appear in the final state, the sums in eq.~(\ref{eq:Int_SSS}) must also run over these, and the discussion below would have to be updated accordingly.

Let us first consider the integration over the $\PS{2}(p_s,K;Q)$ phase space, which involves four basic integrals
\beq
\bsp
    \frac{\s12}{\pi} 
	\left( \frac{(4\pi)^2 }{S_\ep} \s12^\ep \right)   \int_1 \PS{2}(p_s,K;Q)\left\{\calS_{\ti{j}\ti{l}}(s),\calS_{\ti{\ti{i}}\ti{r}}(s),\calS_{\ti{\ti{k}}\ti{r}}(s),\calS_{\ti{\ti{i}}\ti{\ti{k}}}(s)\right\}\,.
\esp
\eeq
However, since the eikonal is homogeneous we have\footnote{Note that this trick only works in the color-singlet case, as here $i,k,j,\ell \in I$ and initial-state momenta are simply rescaled, cf.~eq.~(\ref{eq:SSmommap}). Beyond color-singlet, any of the $i,k,j,\ell$ could be in the final state, and their corresponding mapped momenta would not simply be rescaled but Lorentz-boosted.}
\beq
    \calS_{\ti{\ti{i}}\ti{r}}(s)=\calS_{\ti{i}\ti{r}}(s),\qquad \calS_{\ti{\ti{k}}\ti{r}}(s)=\calS_{\ti{k}\ti{r}}(s),\qquad \calS_{\ti{\ti{i}}\ti{\ti{k}}}(s)=\calS_{\ti{i}\ti{k}}(s)\,
\eeq
and hence the integrals reduce to
\beq
\bsp
    \frac{\s12}{\pi} 
	\left( \frac{(4\pi)^2 }{S_\ep} \s12^\ep \right)   \int_1 \PS{2}(p_s,K;Q)\left\{\calS_{\ti{j}\ti{\ell}}(s),\calS_{\ti{i}\ti{r}}(s),\calS_{\ti{k}\ti{r}}(s),\calS_{\ti{i}\ti{k}}(s)\right\}\,
\esp
\eeq
which are now all of the same type. For the sake of explicitness, let us pick out
\beq
\bsp
\label{eq:SSint}
    \frac{\s12}{\pi}
	\left( \frac{(4\pi)^2 }{S_\ep} \s12^\ep \right)   \int_1 \PS{2}(p_s,K;Q)\calS_{\ti{j}\ti{\ell}}(s)\equiv\,\mathcal{I}_{\ti{j}\ti{\ell}}\,.
\esp
\eeq
To compute this integral, we start by choosing a reference frame. We find it convenient to work in the rest frame of $Q^\mu$ and we orient the frame such that $\ti{p}_j^\mu$ is pointing in the $z$-direction while $\ti{p}_\ell^\mu$ lies in the $z$--$y$ plane,
\beq
\bsp
&Q^{\mu} =\,\sqrt{s_{ab}}(1,\mathbf{0}_{d-1}),\\
&\ti{p}_{j}^{\mu} =\,\ti{E}_{j}(1,\mathbf{0}_{d-2},1),\\
&\ti{p}_{\ell}^{\mu} =\,\ti{E}_{\ell}(1,\mathbf{0}_{d-3},\sin\chi_{j\ell},\cos\chi_{j\ell})\\
&p_s^{\mu} =\,E_s(1,\ldots,\sin\vartheta\sin\phi,\sin\vartheta\cos\phi,\cos\vartheta)\,.
\esp
\eeq
The $\ldots$ denote angular variables of which the integrand is independent, such that their integration is trivial. The two-particle phase space $\PS{2}(p_s,K;Q)$ then reads
\beq
\bsp
\PS{2}(p_s,K;Q) =\,\frac{2\pi}{s_{ab}}\frac{S_{\eps}}{(4\pi)^2}s_{ab}^{1-\eps}\frac{\Gamma^2(1-\eps)}{2\pi\Gamma(1-2\eps)}\epsilon_s^{1-2\eps} \rd\epsilon_s\,\rd\Omega_2\,\delta(1-\epsilon_s-\lam^2)
\esp
\eeq
with
\begin{equation}
    \rd\Omega_2 =\,\rd(\cos\vartheta)(\sin\vartheta)^{-2\eps}\rd(\cos\phi)(\sin\phi)^{-1-2\eps}
\end{equation}
and
\begin{equation}
    \epsilon_s =\,\frac{2E_s}{\sqrt{s_{ab}}}\,.
\end{equation}
Substituting into eq.~(\ref{eq:SSint}) we find, after some algebra,
\beq
\bsp
\mathcal{I}_{\ti{j}\ti{\ell}} =\,\frac{2\Gamma^2(1-2\eps)}{\pi\Gamma(1-2\eps)}(1-\cos\chi_{j\ell})(1-\lam^2)^{-1-2\eps}\int\rd\Omega_2\,(1-\cos\vartheta)^{-1}(1-\sin\chi_{j\ell}\sin\vartheta\cos\phi-\cos\chi_{j\ell}\cos\vartheta)^{-1}\,.
\esp
\eeq
The angular integral is well-known and reads \cite{Somogyi:2011ir}
\begin{equation}
    \int\rd\Omega_2\,(1-\cos\vartheta)^{-1}(1-\sin\chi_{j\ell}\sin\vartheta\cos\phi-\cos\chi_{j\ell}\cos\vartheta)^{-1} =\,\frac{\pi}{\eps}\,{}_{2}F_{1}(1,1,1-\eps,1-Y_{\ti{j}\ti{\ell}})
\end{equation}
with
\begin{equation}
    Y_{\ti{j}\ti{\ell}} =\,\frac{1-\cos\chi_{j\ell}}{2}\,.
\end{equation}
Hence we finally find
\beq
\bsp
 \frac{\s12}{\pi}
	\left( \frac{(4\pi)^2 }{S_\ep} \s12^\ep \right)   \int_1 \PS{2}(p_s,K;Q)\calS_{\ti{j}\ti{\ell}}(s) =&\, -4Y_{\ti{j}\ti{\ell}}\frac{1}{\eps}\frac{\Gamma^2(1-2\eps)}{\Gamma(1-2\eps)}(1-\lam^2)^{-1-2\eps} {}_{2}F_{1}(1,1,1-\eps,1-Y_{\ti{j}\ti{\ell}})\,.
\esp
\eeq
Next we need to perform the integration over the $\PS{2}(\ti{p}_r,\ti{K};\ti{Q})$ phase space in eq.~(\ref{eq:ISSS}). This is similar to the computation presented above and, setting the frame to
\beq
\bsp
&\ti{Q}^{\mu} =\,\lam\sqrt{s_{ab}}(1,\mathbf{0}_{d-1})\,,\\
&\ti{\ti{p}}_{i}^{\mu} =\,\ti{\ti{E}}_i(1,\mathbf{0}_{d-1},1)\,,\\
&\ti{\ti{p}}_{k}^{\mu} =\,\ti{\ti{E}}_k(1,\mathbf{0}_{d-1},-1)\,,\\
&\ti{p}_r^{\mu} =\,\ti{E}_r(1,\ldots,\sin\vartheta,\cos\vartheta)\,,
\esp
\eeq
we find that the integrated counterterm becomes
\beq
\bsp
\label{eq:ISrsSs0}
\left[ \IcS{rs}{(0,0)} \IcS{s}{}\right]\otimes\rd\sigma_{\ti{\ti{a}}\ti{\ti{b}}}=\,&\,\int_{0}^{1} \rd\lam\, \int_{0}^{1} \rd\ti{\lam}\,
	\rd\sigma_{\ti{\ti{a}}\ti{\ti{b}}}(\ti{\ti{p}}_a,\ti{\ti{p}}_b)\,
	\Bigg\{[S_{rs}^{(0,0)}S_s(\lam,\ti{\lam};\eps)]^{(ab,ab)}(\bT_a^2)^2\\&+\frac{C_A}{2}\left(2[S_{rs}^{(0,0)}S_s(\lam,\ti{\lam};\eps)]^{(ab,ar)}-[S_{rs}^{(0,0)}S_s(\lam,\ti{\lam};\eps)]^{(ab,ab)}\right)\bT_a^2\Bigg\}
\esp
\eeq
with
\beq
\bsp
\label{eq:SrsSs-abab}
[S_{rs}^{(0,0)}S_s(\lam,\ti{\lam};\eps)]^{(ab,ab)} =\,-\frac{2^{4+2\eps}}{\eps}\frac{\Gamma^2(1-\eps)}{\Gamma(1-2\eps)}&\int_{-1}^{1}\rd(\cos\vartheta)\,\frac{(\sin\vartheta)^{-2\eps}}{1-\cos^2\vartheta}(1-\lam^2)^{-1-2\eps}\lam^{3-2\eps}(1-\ti{\lam}^2)^{-1-2\eps}\ti{\lam}^{3}
\esp
\eeq
and
\beq
\bsp
\label{eq:SrsSs-abar}
[S_{rs}^{(0,0)}S_s(\lam,\ti{\lam};\eps)]^{(ab,ar)} =\,-\frac{2^{3+2\eps}}{\eps}\frac{\Gamma^2(1-\eps)}{\Gamma(1-2\eps)}\int_{-1}^{1}&\rd(\cos\vartheta)\,\,{}_{2}F_{1}\left(1,1,1-\eps,\frac{1+\cos\vartheta}{2}\right)\frac{(\sin\vartheta)^{-2\eps}}{1+\cos\vartheta}\\&\times(1-\lam^2)^{-1-2\eps}\lam^{3-2\eps}(1-\ti{\lam}^2)^{-1-2\eps}\ti{\lam}^{3}\,.
\esp
\eeq
Note that the former involves the product $\calS_{ab}(\ti{r})\calS_{ab}(s)$ and the latter $\calS_{ab}(\ti{r})\calS_{ar}(s)$. To simplify the discussion to follow we set
\beq
\bsp
\label{eq:SrsSsints}
    [S_{rs}^{(0,0)}S_s(\lam,\ti{\lam};\eps)] \equiv\,&\, [S_{rs}^{(0,0)}S_s(\lam,\ti{\lam};\eps)]^{(ab,ab)}(\bT^2_{\mathrm{ini}})^2+\frac{C_A}{2}\Big(2[S_{rs}^{(0,0)}S_s(\lam,\ti{\lam};\eps)]^{(ab,ar)}\\&-[S_{rs}^{(0,0)}S_s(\lam,\ti{\lam};\eps)]^{(ab,ab)}\Big)\bT^2_{\mathrm{ini}}\,.
\esp
\eeq
In writing eq.~(\ref{eq:SrsSsints}), we have used that for color-singlet production the color-charge algebra is trivial, since color conservation ($\bT_a+\bT_b=0$) implies 
\beq
\bT_a^2 = \bT_b^2 \equiv \bT^2_{\mathrm{ini}}\,,
\qquad
\bT_a\bT_b = -\bT^2_{\mathrm{ini}}
\label{eq:TaTb}
\eeq
and
\beq
\{\bT_a\bT_b,\bT_a\bT_b\} = 2\left(\bT^2_{\mathrm{ini}}\right)^2\,.
\label{eq:TaTbTaTb}
\eeq
Finally, replacing $\lam$ by $\eta/\ti{\lam}$ removes the $\lam$-dependence in the reduced cross section, such that we finally find
\beq
\bsp
\label{eq:ISrsSs2}
&\left[ \IcS{rs}{(0,0)} \IcS{s}{}\right]\otimes\rd\sigma_{\ti{\ti{a}}\ti{\ti{b}}}=\,\int_{0}^{1} \rd\eta\, 
	\rd\sigma_{\ti{\ti{a}}\ti{\ti{b}}}(\eta\, p_a,\eta\, p_b)\,
	[S_{rs}^{(0,0)}S_s(\eta;\eps)]
\esp
\eeq
with
\beq
\bsp
\left[S_{rs}^{(0,0)}S_s(\eta;\eps)\right] =\, \int_{\eta}^{1}\frac{\rd\ti{\lam}}{\ti{\lam}}\,[S_{rs}^{(0,0)}S_s(\eta/\ti{\lam},\ti{\lam};\eps)]\,.
\esp
\eeq
It is now straightforward to compute both the $(ab,ab)$ integral, cf.~eq.~(\ref{eq:SrsSs-abab}), and its asymptotic behavior as $\eta\to 1$, using the steps outlined in sec.~\ref{sec:A12ints}. The result is a relatively simple function of $\eta$, involving logarithms and classical polylogarithms up to weight three with the following arguments
\beq
\bsp
\label{eq:SrsSseps0}
&\Bigg\{\pm\eta,1\pm\eta,2\eta,\eta^2,\frac{2\eta}{\eta-1}\Bigg\}\,.
\esp
\eeq 
For the $(ab,ar)$ integral, cf.~eq.~(\ref{eq:SrsSs-abar}), we need to deal with the hypergeometric function. We find it convenient to go to an integral representation using
\begin{equation}
    {}_2F_{1}(a,b,c;z) =\,\frac{ \Gamma (c)  }{\Gamma (b) \Gamma (c-b)}\int_{0}^{1}\rd t\,t^{b-1}(1-t z)^{-a}(1-t)^{-b+c-1}
\end{equation}
and explicitly integrate over the additional variable $t$. 
This again makes the integration and the limit computation as $\eta\to 1$ a matter of following the recipe of sec.~\ref{sec:A12ints}. We find that the functional form is very similar to that of the $(ab,ab)$ integral. In fact, the result contains logarithms and classical polylogarithms evaluated in the same arguments as in eq.~(\ref{eq:SrsSseps0}).

\subsubsection{
\texorpdfstring{$\cC{ars}{IFF}\cS{rs}{(0,0)}\cS{s}{}$}{CarsIFFSrs00Ss}}
The integrated subtraction term takes on the form~\cite{DelDuca:2025yph}
\beq
\bsp
\label{eq:Int_SsCarsIFFSrs0sing}
&\left[ \IcC{ars}{IFF} \IcS{rs}{(0,0)} \IcS{s}{}\right]\otimes\rd\sigma_{\ti{\ti{a}}\ti{\ti{b}}}=\,\int_{0}^{1} \rd\lam\, \int_{0}^{1} \rd\ti{\lam}\, \rd\sigma_{\ti{\ti{a}}\ti{\ti{b}}}(\ti{\ti{p}}_a,\ti{\ti{p}}_b)
	 \left[ \IcC{ars}{} \IcS{rs}{(0,0)} \IcS{s}{}(\lam,\ti{\lam};\eps)\right]
\esp
\eeq
with
\beq
\bsp
\label{eq:ISsCarsIFFSrs0}
\left[ \IcC{ars}{IFF} \IcS{rs}{(0,0)} \IcS{s}{}(\lam,\ti{\lam};\eps)\right] =\,&\,\left( \frac{\s12}{\pi} \right)^2
	\left( \frac{(4\pi)^2 }{S_\ep} \s12^\ep \right)^2 
	\int_1 \PS{2}(p_s,K;Q)\, \int_1 \PS{2}(\ti{p}_r,\ti{K};\ti{Q})\,
	  \frac{2\lam^5 \ti{\lam}^3}{ x_{\ti{r},\ti{\ti{a}}} s_{\ti{\ti{a}}\ti{r}}}\,\bT_a^2 \\&\qquad\qquad\qquad\qquad\qquad\times
	 P^{(\mathrm S), (0)}_{(ar)rs}(x_{\ti{a},\ti{r}s},x_{\ti{r},\ti{a}},x_{s,\ti{a}},s_{\ti{a}\ti{r}}, s_{\ti{a}s}, s_{\ti{r}s};\ep) \,.
\esp
\eeq
Here $P^{(\mathrm S), (0)}_{(ar)rs}$ represents the soft limit of the collinear splitting function, cf.~eqs.~(\ref{eq:SsP1})-(\ref{eq:SsP5}). Since parton $s$ should be soft ($\cS{s}{ }$), it should be a gluon\footnote{Of course a quark can also become soft, but this does not lead to any divergences.}, which in turn implies that parton $r$ should also be a gluon ($\cS{rs}{(0,0)}$). Parton $a$ however can either be a quark or a gluon. From eqs.~(\ref{eq:SsP2}) and (\ref{eq:SsP5}) it then follows that
\begin{equation}
   P^{(\mathrm S), (0)}_{(ar)rs}\to P^{(\mathrm S), (0)}_{a g_r g_s} =\, \frac{2}{x_{s,\ti{a}} s_{\ti{a}s}}\,\bT_a^2+C_A\left(\frac{s_{\ti{a}\ti{r}}}{s_{\ti{a}s}s_{\ti{r}s}}-\frac{1}{x_{s,\ti{a}} s_{\ti{a}s}}+\frac{x_{\ti{r},\ti{a}}}{x_{s,\ti{a}} s_{\ti{r}s}}\right)\,.
\end{equation}
The computation of $\left[ \IcC{ars}{IFF} \IcS{rs}{(0,0)} \IcS{s}{}(\lam,\ti{\lam};\eps)\right]$ in eq.~(\ref{eq:ISsCarsIFFSrs0}) now follows the same steps as the computation of $\left[ \IcS{rs}{(0,0)} \IcS{s}{}(\lam,\ti{\lam};\eps)\right]$ above and, as such, we omit the details of the derivation and just present the results. The integrated counterterm can be written as
\beq
\bsp
    \left[ \IcC{ars}{IFF} \IcS{rs}{(0,0)} \IcS{s}{}\right]\otimes\rd\sigma_{\ti{\ti{a}}\ti{\ti{b}}}=\,&\,\int_{0}^{1} \rd\lam\, \int_{0}^{1} \rd\ti{\lam}\,
	\rd\sigma_{\ti{\ti{a}}\ti{\ti{b}}}(\ti{\ti{p}}_a,\ti{\ti{p}}_b)\Bigg\{2[\IcC{ars}{IFF} \IcS{rs}{(0,0)} \IcS{s}{}(\lam,\ti{\lam};\eps)]^{(ab,ab)}(\bT_a^2)^2\\&+2[\IcC{ars}{IFF} \IcS{rs}{(0,0)} \IcS{s}{}(\lam,\ti{\lam};\eps)]^{(ab,ar)}C_A\,\bT_a^2\Bigg\}
\esp
\eeq
where, in analogy to the case above, we defined\footnote{The $(ab,ab)$ and $(ab,ar)$ labels are simply introduced to emphasize the similarity to the analysis of $\left[\cS{rs}{(0,0)}\cS{s}{ }(\lam,\ti{\lam};\eps)\right]$.}
\beq
\bsp
\label{eq:CarsSrsSs-abab}
    [\IcC{ars}{IFF} \IcS{rs}{(0,0)} \IcS{s}{}(\lam,\ti{\lam};\eps)]^{(ab,ab)} =\,-\frac{2^{3+2\eps}}{\eps}\frac{\Gamma^2(1-\eps)}{\Gamma(1-2\eps)}&\int_{-1}^{1}\rd(\cos\vartheta)\,\frac{(\sin\vartheta)^{-2\eps}}{1-\cos\vartheta}\lam^{3-2\eps}(1-\lam^2)^{-1-2\eps}\\&\times(1-\ti{\lam}^2)^{-1-2\eps}\ti{\lam}^{3}
\esp
\eeq
and
\beq
\bsp
\label{eq:CarsSrsSs-abar}
    [\IcC{ars}{IFF} \IcS{rs}{(0,0)} \IcS{s}{}(\lam,\ti{\lam};\eps)]^{(ab,ar)} =\, -\frac{2^{2+2\eps}}{\eps}\frac{\Gamma^2(1-\eps)}{\Gamma(1-2\eps)}&\int_{-1}^{1}\rd(\cos\vartheta)\,\,{}_{2}F_{1}\left(1,1,1-\eps,\frac{1+\cos\vartheta}{2}\right)\\&\times(\sin\vartheta)^{-2\eps}(1-\lam^2)^{-1-2\eps}\lam^{3-2\eps}(1-\ti{\lam}^2)^{-1-2\eps}\ti{\lam}^{3}\,.
\esp
\eeq
As before, we collect the integrals as
\beq
\bsp
    [\IcC{ars}{IFF} \IcS{rs}{(0,0)} \IcS{s}{}(\lam,\ti{\lam};\eps)] =\,&\, 2[\IcC{ars}{IFF} \IcS{rs}{(0,0)} \IcS{s}{}(\lam,\ti{\lam};\eps)]^{(ab,ab)}(\bT_a^2)^2+2[\IcC{ars}{IFF} \IcS{rs}{(0,0)} \IcS{s}{}(\lam,\ti{\lam};\eps)]^{(ab,ar)}C_A\,\bT_a^2
\esp
\eeq
and introduce $\eta=\lambda\ti{\lambda}$,
\beq
\bsp
\left[\IcC{ars}{IFF} \IcS{rs}{(0,0)} \IcS{s}{}(\eta;\eps)\right] =\, \int_{\eta}^{1}\frac{\rd\ti{\lam}}{\ti{\lam}}\,[\IcC{ars}{IFF} \IcS{rs}{(0,0)} \IcS{s}{}(\eta/\ti{\lam},\ti{\lam};\eps)]\,.
\esp
\eeq
The explicit computation of the integrals is now completely analogous to the evaluation of those for $\cS{rs}{(0,0)}\cS{s}{ }$ above. In fact, the final results contain logarithms and classical polylogarithms with the same arguments as in eq.~(\ref{eq:SrsSseps0}).

Although the computation of the integrated subtraction term closely follows the discussion of $\cS{rs}{(0,0)}\cS{s}{ }$, we do not have any eikonal factors in the integrand. As such, the result we find is \textit{generic} and remains valid once final-state jets are included.

\subsubsection{
\texorpdfstring{$\cS{rs}{(0,0)}\cC{as}{IF}\cS{s}{}$}{Srs00CasSs}}
The integrated subtraction term takes on the form~\cite{DelDuca:2025yph}
\beq
\bsp
\label{eq:ISrsCasSs}
& \left[ \IcS{rs}{(0,0)} \IcC{as}{IF} \IcS{s}{}\right]\otimes\rd\sigma_{\ti{\ti{a}}\ti{\ti{b}}}=\, 
	  -\bT_a^2\int_{0}^{1} \rd\lam\, \int_{0}^{1} \rd\ti{\lam}\,\rd\sigma_{\ti{\ti{a}}\ti{\ti{b}}}(\ti{\ti{p}}_a, \ti{\ti{p}}_b)\,
	\left[ \IcS{rs}{(0,0)} \IcC{as}{IF} \IcS{s}{}(\lam,\ti{\lam};\eps)\right]^{(a,b)} \,
\esp
\eeq
with
\beq
\bsp
    \left[ \IcS{rs}{(0,0)} \IcC{as}{IF} \IcS{s}{}(\lam,\ti{\lam};\eps)\right]^{(a,b)}
	=\,&\, - \left( \frac{\s12}{\pi} \right)^2
	\left( \frac{(4\pi)^2 }{S_\ep} \s12^\ep \right)^2
	\int_1 \PS{2}(p_s,K;Q)\, \int_1 \PS{2}(\ti{p}_r,\ti{K};\ti{Q})
	 \, \frac{2\lam^5 \ti{\lam}^3}{s_{\ti{a}s} x_{s,\ti{a}} }\, 
     \bT_a^2
	\calS_{\ti{\ti{a}}\ti{\ti{b}}}(\ti{r}) \,.
\esp
\label{eq:ISrs00CasIFSs}
\eeq
We immediately wrote the eikonal above in terms of the initial-state partons $a$ and $b$, as appropriate for color-singlet production.\footnote{So, beyond the color-singlet case, our analysis would need to be adapted.} As mentioned before, the eikonal is also homogeneous in this case, $\calS_{\ti{\ti{a}}\ti{\ti{b}}}(\ti{r})=\calS_{\ti{a}\ti{b}}(\ti{r})=\calS_{ab}(\ti{r})$. It turns out that this integral is exactly equivalent to the $(ab,ab)$-type integral of Section~\ref{sec:defISrsSs0} above (up to the extra factor of $\bT_a^2$ in eq.~(\ref{eq:ISrs00CasIFSs}) above). To see this, note that the integral of $\frac{2}{s_{\ti{a}s} x_{s,\ti{a}}}=\frac{2}{s_{as} x_{s,a}}$ over the $\PS{2}(p_s,K;Q)$ phase space is independent of $p_a^\mu$. As such we can write
\begin{equation}
    \int_1\PS{2}(p_s,K;Q)\, \frac{2}{s_{as} x_{s,a}} =\, \int_1\PS{2}(p_s,K;Q)\, \frac{2}{s_{bs} x_{s,b}}\,.    
\end{equation}
Hence
\beq
\bsp
\int_1\PS{2}(p_s,K;Q)\, \frac{2}{s_{as} x_{s,a}} =\,&\, 
\int_1\PS{2}(p_s,K;Q)\, \left(\frac{s_{aQ}}{s_{as} s_{sQ}} + \frac{s_{bQ}}{s_{bs} s_{sQ}}\right)\,.
\esp
\eeq
Since $a,b\in I$ we obviously have $s_{aQ}=\,s_{bQ}=\,s_{ab}$, so simply performing the sum in the parenthesis gives
\beq
\bsp
   \int_1\PS{2}(p_s,K;Q)\, \frac{2}{s_{as} x_{s,a}} =\, 
\int_1\PS{2}(p_s,K;Q)\, \frac{s_{ab}}{s_{as}s_{bs}}\,.
\esp
\eeq
So for the integration we can make the replacement
\beq
\frac{2}{s_{as} x_{s,a}} \calS_{ab}(\ti{r}) \rightarrow
\calS_{ab}(s)\calS_{ab}(\ti{r})
\eeq
in which the right-hand side indeed corresponds to the kernel of $[\IcS{rs}{(0,0)} \IcS{s}{}(\lam,\ti{\lam};\eps)]^{(ab,ab)}$, cf.~eq.~(\ref{eq:SrsSs-abab}). Hence the integrated counterterm becomes of the form eq.~(\ref{eq:SS-distexp})
with
\begin{equation}
    [\IcS{rs}{(0,0)} \IcC{as}{IF} \IcS{s}{}(\eta;\eps)]^{(a,b)} =\, 2\,\bT_a^2 [\IcS{rs}{(0,0)} \IcS{s}{}(\eta;\eps)]^{(ab,ab)}\,
\end{equation}
and similarly for the (integrated) limit formula.

\subsubsection{
\texorpdfstring{$\cC{ars}{IFF}\cS{rs}{(0,0)}\cC{as}{IF}\cS{s}{}$}{CarsIFFSrs00CasSs}}
The integrated subtraction term takes on the form~\cite{DelDuca:2025yph}
\beq
\bsp
\label{eq:ICarsSrsCasSs}
&\left[ \IcC{ars}{IFF} \IcS{rs}{(0,0)} \IcC{as}{IF} \IcS{s}{}\right]\otimes\rd\sigma_{\ti{\ti{a}}\ti{\ti{b}}}=\,\int_{0}^{1} \rd\lam\, \int_{0}^{1} \rd\ti{\lam}\,\rd\sigma_{\ti{\ti{a}}\ti{\ti{b}}}(\ti{\ti{p}}_a,\ti{\ti{p}}_b) 
	\,
	\left[ \IcC{ars}{IFF} \IcS{rs}{(0,0)} \IcC{as}{IF} \IcS{s}{}(\lam,\ti{\lam};\eps)\right] 
\esp
\eeq
with
\beq
\bsp
    \left[ \IcC{ars}{IFF} \IcS{rs}{(0,0)} \IcC{as}{IF} \IcS{s}{}(\lam,\ti{\lam};\eps)\right]
	=\,&\, \left( \frac{\s12}{\pi} \right)^2
	\left( \frac{(4\pi)^2 }{S_\ep} \s12^\ep \right)^2
	\int_1 \PS{2}(p_s,K;Q)\, \int_1 \PS{2}(\ti{p}_r,\ti{K};\ti{Q})
	 \\ \,&\,\times \frac{2\lam^5 \ti{\lam}^3}{ x_{s,\ti{a}} s_{\ti{a}s}}\, \frac2{ x_{\ti{r},\ti{\ti{a}}} s_{\ti{\ti{a}}\ti{r}}}(\bT_a^2)^2\,.
\esp
\eeq
The integral of interest was already computed above. In particular, up to a factor of two and $(\bT_a^2)^2$, it corresponds to $[\IcC{ars}{IFF} \IcS{rs}{(0,0)} \IcS{s}{}(\lam,\ti{\lam};\eps)]^{(ab,ab)}$, cf.~eq.~(\ref{eq:CarsSrsSs-abab}). As such the integrated subtraction term becomes of the type eq.~(\ref{eq:SS-distexp})
with
\begin{equation}
    [\IcC{ars}{IFF} \IcS{rs}{(0,0)} \IcC{as}{IF} \IcS{s}{}(\eta;\eps)] =\, 2\,(\bT_a^2)^2 [\IcC{ars}{IFF} \IcS{rs}{(0,0)} \IcS{s}{}(\eta;\eps)]^{(ab,ab)}\,.
\end{equation}
This result is generic and does not require any modification once final-state jets are included.

\subsection{Integration of IF--S iterated subtraction terms}
In this case, the factorized matrix element is written in terms of the momenta $(\ha{\ti{p}}_a, \ha{\ti{p}}_b;\momht{}_{X})$, which come about by first applying the single soft mapping in eq.~(\ref{eq:softmap}) and then the initial-state collinear mapping of eq.~(\ref{eq:IFmap}),
\beq
(p_a, p_b;\mom{}_{X+2}) \smap{s} (\ti{p}_a, \ti{p}_b;\momt{}_{X+1}) 
	\cmap{\ti{a}\ti{b},\ti{r}}{II,F} (\ha{\ti{p}}_a, \ha{\ti{p}}_b;\momht{}_{X})\,.
	\label{eq:SIF-map}
\eeq
In particular we have
\beq
\bsp
\ha{\ti{p}}_a^\mu &=\,\lam_s\xi_{\ti{a},\ti{r}} p_a^\mu\,,
\qquad
\ha{\ti{p}}_b^\mu =\,\lam_s\xi_{\ti{b},\ti{r}} p_b^\mu\,.
\esp
\label{eq:itersoftIFmap2}
\eeq
The $(X+2)$-particle phase space is composed from a soft convolution followed by an initial-final collinear convolution,
\beq
\bsp
    &\PS{X+2}(\mom{}_{X+2};Q)  =\, 
	\frac{s_{ab}}{\pi}\int_{0}^{1} \rd \lam\,\lam
	 \int_{0}^{1} \rd{\xi}_a\, \rd{\xi}_b\,
	\PS{X}(\momht{}_{X}; \ha{\ti{Q}})
	\PS{2}(p_s,K;Q)\PS{II,F}(\ti{p}_{r},{\xi}_a,{\xi}_b)
	\,.
\esp
\eeq
We now have a total of three subtraction terms, namely $\cC{ars}{IFF(0,0)}\cS{s}{}$, $\cC{ars}{IFF(0,0)}\cC{rs}{FF}\cS{s}{}$ and $\cC{asr}{IFF(0,0)}\cC{as}{IF}\cS{s}{}$. As above, their integrated versions take on a similar form, which can be written as
\beq
\bsp
\label{eq:SIF-distexp}
    &\left[\CT\right]\otimes\rd\sigma_{AB}=\,\sum_{a,b}\int_0^1 \rd x_a\, \rd x_b\, \int_0^1 \rd \eta_a\, \int_0^{1} \rd \eta_b\,
\rd\sigma_{\ha{\ti{a}}\ha{\ti{b}}}(x_a p_A,x_b p_B) \\&\times\Bigg\{
 \bigg(\CT(\eta_a,\eta_b;\eps\,|\,\eta_a,\eta_b)_{\eta_b<\eta_a}+\CT(\eta_a,\eta_b;\eps\,|\,\eta_a,\eta_b)_{\eta_a<\eta_b}\bigg)
\frac{f_{a/A}(x_a/\eta_a)}{\eta_a}\frac{f_{b/B}(x_b/\eta_b)}{\eta_b}
\\&+\CT(\eta_a,\eta_b;\eps\,|\,1,\eta_b)_{\eta_b<\eta_a} f_{a/A}(x_a) \frac{f_{b/B}(x_b/\eta_b)}{\eta_b}+\CT(\eta_a,\eta_b;\eps\,|\,\eta_a,1)_{\eta_a<\eta_b} \frac{f_{a/A}(x_a/\eta_a)}{\eta_a} f_{b/B}(x_b)
\\&
+\CT(\eta_a,\eta_b;\eps\,|\,\eta_a,\eta_a)_{\eta_b<\eta_a} \frac{f_{a/A}(x_a/\eta_a)}{\eta_a} \frac{f_{b/B}(x_b/\eta_a)}{\eta_a} +\CT(\eta_a,\eta_b;\eps\,|\,\eta_b,\eta_b)_{\eta_a<\eta_b} \frac{f_{a/A}(x_a/\eta_b)}{\eta_b} \frac{f_{b/B}(x_b/\eta_b)}{\eta_b} 
\\&+\bigg(\CT(\eta_a,\eta_b;\eps\,|\,1,1)_{\eta_b<\eta_a}+\CT(\eta_a,\eta_b;\eps\,|\,1,1)_{\eta_a<\eta_b}\bigg) f_{a/A}(x_a) f_{b/B}(x_b)
 \Bigg\}\,.
\esp
\eeq
The coefficient functions are defined as follows
\begin{align}
    &\CT(\eta_a,\eta_b;\eps\,|\,\eta_a,\eta_b)_{\eta_b<\eta_a} =\, (\eta_a \eta_b)^{2\ep}[\CT(\eta_a,\eta_b;\eps)]_{\eta_b<\eta_a}\,,\\
    &\CT(\eta_a,\eta_b;\eps\,|\,\eta_a,\eta_b)_{\eta_a<\eta_b} =\, (\eta_a \eta_b)^{2\ep}[\CT(\eta_a,\eta_b;\eps)]_{\eta_a<\eta_b}\,,\\
     &\CT(\eta_a,\eta_b;\eps\,|\,1,\eta_b)_{\eta_b<\eta_a} =\, \eta_b^{2\ep}\Big([\La] - \La\Big)[\CT(\eta_a,\eta_b;\eps)]_{\eta_b<\eta_a}\,,\\
    &\CT(\eta_a,\eta_b;\eps\,|\,\eta_a,1)_{\eta_a<\eta_b} =\, \eta_a^{2\ep}\Big([\Lb] - \Lb\Big)[\CT(\eta_a,\eta_b;\eps)]_{\eta_a<\eta_b}\,,\\
    &\CT(\eta_a,\eta_b;\eps\,|\,\eta_a,\eta_a)_{\eta_b<\eta_a} =\, \eta_a^{4\ep}\Big([\LbTOa] - \LbTOa\Big)[\CT(\eta_a,\eta_b;\eps)]_{\eta_b<\eta_a}\,,\\
     &\CT(\eta_a,\eta_b;\eps\,|\,\eta_b,\eta_b)_{\eta_a<\eta_b} =\, \eta_b^{4\ep}\Big([\LaTOb] - \LaTOb\Big)[\CT(\eta_a,\eta_b;\eps)]_{\eta_a<\eta_b}\,,\\
    &\CT(\eta_a,\eta_b;\eps\,|\,1,1)_{\eta_b<\eta_a} =\, \Big([\Lab]-[\La\Lab]
-[\LbTOa\Lab]\nn\\&\qquad\qquad\qquad\qquad\qquad-\Lab+\La\Lab+\LbTOa\Lab\Big)[\CT(\eta_a,\eta_b;\eps)]_{\eta_b<\eta_a}\,,\\
    &\CT(\eta_a,\eta_b;\eps\,|\,1,1)_{\eta_a<\eta_b} =\, \Big([\Lab]-[\Lb\Lab]
-[\LaTOb\Lab]\nn\\&\qquad\qquad\qquad\qquad\qquad-\Lab+\Lb\Lab+\LaTOb\Lab\Big)[\CT(\eta_a,\eta_b;\eps)]_{\eta_a<\eta_b}\,.
\end{align}
Note the appearance of two new asymptotic limits, namely $\eta_b\to \eta_a$ (denoted by $\LbTOa$) and $\eta_a\to \eta_b$ (denoted by $\LaTOb$). Furthermore, the integrated counterterm now needs to be split into two regions, ${\eta_b<\eta_a}$ and ${\eta_a<\eta_b}$.\footnote{It is assumed that each coefficient function only has support in its corresponding region. For example, we define $\CT(\eta_a,\eta_b;\eps\,|\,\eta_a,\eta_b)_{\eta_b<\eta_a}=\,0$ if $\eta_b>\eta_a$.} We will discuss these points further below. As none of the IF--S iterated subtraction terms involve eikonal factors, the results in this section carry over to processes with jet production without any modification.

\subsubsection{
\texorpdfstring{$\cC{ars}{IFF(0,0)}\cS{s}{}$}{CarsIFFSs}}

The parton-level integrated counterterm can be written as~\cite{DelDuca:2025yph}
\beq
\bsp
\label{eq:ICarsSs0}
   \left[ \IcC{ars}{IFF(0,0)} \IcS{s}{}\right]\otimes\rd\sigma_{\ha{\ti{a}}\ha{\ti{b}}}=\,\int_{0}^{1}\rd\lam\,\int_{0}^{1}\rd\xi_a\,\rd\xi_b\,\rd\sigma_{\ha{\ti{a}}\ha{\ti{b}}}(\ha{\ti{p}}_a,\ha{\ti{p}}_b)\,\left[ \IcC{ars}{IFF(0,0)} \IcS{s}{}(\lambda,\xi_a,\xi_b;\eps)\right] 
\esp
\eeq
with
\beq
\bsp
    \left[ \IcC{ars}{IFF(0,0)} \IcS{s}{}(\lambda,\xi_a,\xi_b;\eps)\right] =\,&\, 2\frac{(4\pi)^2}{S_{\eps}}s_{ab}^{1+\eps}\int_1\PS{2}(p_s,K;Q)\,\frac{s_{ab}}{\pi}\lam^{5-2\eps}\left[\frac{\xi_a\xi_b(1-\xi_a^2)(1-\xi_b^2)}{(\xi_a+\xi_b)^2}\right]^{-\eps}\frac{\xi_a^2\xi_b^2(1+\xi_a\xi_b)}{(\xi_a+\xi_b)^2}\\&\times\frac{1}{x_{\ti{a},\ti{r}}s_{\ti{a}\ti{r}}}\frac{\omega(ar)}{\omega(a)}P^{(\mathrm S), (0)}_{(ar) r s}
	(x_{\ti{a},\ti{r}s}, x_{\ti{r},\ti{a}}, x_{s,\ti{a}},
	s_{\ti{a}\ti{r}}, s_{\ti{a}s}, s_{\ti{r}s};\ep)
	 P_{(ar) r}^{(0)}(x_{\ti{a},\ti{r}};\ep)\,.
\esp
\eeq
The momentum fractions are defined as in eq.~(\ref{eq:defX2}). Note that parton $s$ is necessarily a gluon. The soft functions $P^{(\mathrm S), (0)}_{(ar) r g_s}$ for $I\to IF(F)$ splitting can then be obtained from the corresponding soft functions for $F\to FF(F)$ splitting given in \eqnss{eq:SsP1}{eq:SsP5} by the crossing relation~\cite{DelDuca:2025yph}
\beq
P^{(\mathrm S), (0)}_{(ar) r s}
	(x_a,x_r,x_s,s_{ar}, s_{as}, s_{rs};\eps) =
    -(-1)^{F(a)+F(ar)}P^{(\mathrm S),(0)}_{\ba{a} r s}
	\left(\frac{1}{x_a}, 
	-\frac{x_r}{x_a}, 
	-\frac{x_s}{x_a},
	-s_{ar}, -s_{as}, s_{rs};\eps\right)\,,
	\label{eq:sapspliIFF}
\eeq
with
\beq
F(q) =\,F(\qb) =\,-1\,,\qquad F(g) =\,0\,.
\eeq
The integration over the $\PS{2}(p_s,K;Q)$ phase space is similar to that for the $\cS{rs}{(0,0)}\cS{s}{}$ counterterm. In particular, we again find two types of basic integrals,
\begin{itemize}
    \item $(ab,ar)$-type coming from the $\mathcal{S}_{\ti{a}\ti{r}}(s)$ eikonal factor in the soft splitting function,
    \item $(ab,ab)$-type coming from the other structures in the soft splitting functions.
\end{itemize}
As before, the $(ab,ar)$-type integral comes with a ${}_2F_1$ hypergeometric function. With all these considerations in mind we write 
\beq
\bsp
\label{eq:CarsSsints}
    \left[ \IcC{ars}{IFF(0,0)} \IcS{s}{}(\lambda,\xi_a,\xi_b;\eps)\right] \equiv\,&\, 2\,\frac{\omega(ar)}{\omega(a)}\bT_{ar}^2\left[ \IcC{ars}{IFF(0,0)} \IcS{s}{}(\lambda,\xi_a,\xi_b;\eps)\right]^{(ab,ab)}\\&+\frac{\omega(ar)}{\omega(a)}\left(\bT_{a}^2+\bT_{r}^2-\bT_{ar}^2\right)\left[ \IcC{ars}{IFF(0,0)} \IcS{s}{}(\lambda,\xi_a,\xi_b;\eps)\right]^{(ab,ar)}
\esp
\eeq
with
\beq
\bsp
\label{eq:CarsSs-abab}
    \left[ \IcC{ars}{IFF(0,0)} \IcS{s}{}(\lambda,\xi_a,\xi_b;\eps)\right]^{(ab,ab)} =\,&\, -\frac{4\,\Gamma^2(1-\eps)}{\eps\,\Gamma(1-2 \eps)}\left(\frac{\xi_a \left(1-\xi_a^2\right) \xi_b \left(1-\xi_b^2\right)}{(\xi_a+\xi_b)^2}\right)^{-\eps}\\&\times\frac{ (1-\lambda^2)^{-1-2 \eps} \lambda^{3-2\eps}  (1+\xi_a \xi_b)}{\xi_a+\xi_b}  \frac{\xi_b}{1-\xi_b^2} P^{(0)}_{(ar)r}(\xi_a \xi_b;\eps)
\esp
\eeq
and
\beq
\bsp
\label{eq:CarsSs-abar}
    \left[ \IcC{ars}{IFF(0,0)} \IcS{s}{}(\lambda,\xi_a,\xi_b;\eps)\right]^{(ab,ar)} =\,&\, -\frac{8\,\Gamma^2(1-\eps)}{\eps\,\Gamma(1-2 \eps)} \left(\frac{\xi_a \left(1-\xi_a^2\right) \xi_b \left(1-\xi_b^2\right)}{(\xi_a+\xi_b)^3 (1-\xi_a \xi_b)}\right)^{-\eps}\\&\times\frac{\left(1-\lambda^2\right)^{-1-2 \eps}\lambda^{3-2\eps}(1+\xi_a \xi_b)}{\xi_a+\xi_b}
\frac{\xi_b\left(\xi_a(1-\xi_b^2)\right)^{-\eps}}{1-\xi_b^2}\,
\\&\times
{}_2F_1\left(-\eps,-\eps;1-\eps;1-\frac{\xi_a \left(1-\xi_b^2\right)}{(\xi_a+\xi_b) (1-\xi_a \xi_b)}\right) P^{(0)}_{(ar)r}(\xi_a \xi_b;\eps)\,.
\esp
\eeq
Note that the reduced cross section in eq.~(\ref{eq:ICarsSs0}) depends on the three convolution variables $\lambda$, $\xi_a$ and $\xi_b$. This structure can be simplified by setting
\begin{equation}
    \xi_a =\, \frac{\eta_a}{\lambda}\,,\qquad \xi_b =\, \frac{\eta_b}{\lambda}\,.
\end{equation}
We then find the following form for the integrated subtraction term
\beq
\bsp
\label{eq:ICarsSs1}
   &\left[ \IcC{ars}{IFF(0,0)} \IcS{s}{}\right]\otimes\rd\sigma_{\ha{\ti{a}}\ha{\ti{b}}}=\,\int_{0}^{1}\frac{\rd\lambda}{\lambda^2}\,\int_{0}^{\lambda}\rd\eta_a\,\int_{0}^{\lambda}\rd\eta_b\,\rd\sigma_{\ha{\ti{a}}\ha{\ti{b}}}(\eta_a p_a,\eta_b p_b)\left[ \IcC{ars}{IFF(0,0)} \IcS{s}{}(\lambda,\eta_a/\lambda,\eta_b/\lambda;\eps)\right]\,.
\esp
\eeq
For practical purposes, it would be more convenient to have the integration over the convolution variables $\eta_a$ and $\eta_b$, which appear in the reduced differential cross section, to be unconstrained. Note however that this is complicated due to the non-trivial geometry of the integration region, originating from the condition that both $\eta_a<\lambda$ \textit{and} $\eta_b<\lambda$. We solve this issue by splitting the integration range in two regions, one with $\eta_a<\eta_b$ and one with $\eta_b<\eta_a$. This can be achieved by inserting the identity into eq.~(\ref{eq:ICarsSs1}) in the form
\begin{equation}
    \int_{0}^{1}\rd\eta_a\,\int_{0}^{1}\rd\eta_b\,\left[\theta(\eta_a-\eta_b)+\theta(\eta_b-\eta_a)\right] =\, 1\,.
\end{equation}
As the reduced cross section is now independent of $\lambda$, the integration over the latter can be done analytically. Denoting
\begin{equation}
\label{eq:CarsSs-bLTa}
    \left[ \IcC{ars}{IFF(0,0)} \IcS{s}{}(\eta_a,\eta_b;\eps)\right]_{\eta_b<\eta_a} =\, \theta(\eta_a-\eta_b)\int_{\eta_a}^{1}\frac{\rd\lambda}{\lambda^2}\,\left[ \IcC{ars}{IFF(0,0)} \IcS{s}{}(\lambda,\eta_a/\lambda,\eta_b/\lambda;\eps)\right]
\end{equation}
and
\begin{equation}
\label{eq:CarsSs-aLTb}
    \left[ \IcC{ars}{IFF(0,0)} \IcS{s}{}(\eta_a,\eta_b;\eps)\right]_{\eta_a<\eta_b} =\, \theta(\eta_b-\eta_a)\int_{\eta_b}^{1}\frac{\rd\lambda}{\lambda^2}\,\left[ \IcC{ars}{IFF(0,0)} \IcS{s}{}(\lambda,\eta_a/\lambda,\eta_b/\lambda;\eps)\right]
\end{equation}
we find
\beq
\bsp
\label{eq:ICarsSs4}
   \left[ \IcC{ars}{IFF(0,0)} \IcS{s}{}\right]\otimes\rd\sigma_{\ha{\ti{a}}\ha{\ti{b}}}=\,&\,\int_{0}^{1}\rd\eta_a\,\rd\eta_b\,\rd\sigma_{\ha{\ti{a}}\ha{\ti{b}}}(\eta_a p_a,\eta_b p_b)\Bigg(\left[ \IcC{ars}{IFF(0,0)} \IcS{s}{}(\eta_a,\eta_b;\eps)\right]_{\eta_b<\eta_a}\\&\quad+\left[ \IcC{ars}{IFF(0,0)} \IcS{s}{}(\eta_a,\eta_b;\eps)\right]_{\eta_a<\eta_b}\Bigg)\,.
\esp
\eeq
Finally, we regularize the endpoint singularities by setting up a subtraction. This is more complicated than in previous cases, as now there will also be $\eta_a\to\eta_b$ and $\eta_b\to\eta_a$ singularities. Hence, we need to introduce the formal limit operators $\LbTOa$ and $\LaTOb$, together with their integrated versions. With all these considerations in mind, we find that the final form of the integrated subtraction term indeed becomes of the form presented in eq.~(\ref{eq:SIF-distexp}).\newline

Note that there are 20 basic integrals to compute as we have two types, $(ab,ab)$ and $(ab,ar)$, each with five different structures coming from the splitting function, cf.~eqs.~(\ref{eq:CarsSs-abab})-(\ref{eq:CarsSs-abar}). Furthermore, each needs to be computed in the two regions $\eta_a<\eta_b$ and $\eta_b<\eta_a$, cf.~eqs.~(\ref{eq:CarsSs-bLTa})-(\ref{eq:CarsSs-aLTb}). Finally, we also require the limit formul\ae\, appearing in the endpoint regularization of eq.~(\ref{eq:SIF-distexp}) and their integrals. However, the computation of each of these ingredients nicely follows the procedure outlined in sec.~\ref{sec:A12ints}. We find that the analytic structure of the coefficient function $\IcC{ars}{IFF(0,0)} \IcS{s}{}(\eta_a,\eta_b;\eps\,|\,\eta_a,\eta_b)$ (in both regions) is relatively simple. In particular, the weight-two GPLs that depend on the convolution variables $\eta_a$ and $\eta_b$ have arguments in $\Bigg\{\pm \eta_a,\pm\eta_b, 1/\eta_a, 1/\eta_b, \pm\sqrt{\eta_a\eta_b}\Bigg\}$. Also, the integrated counterterm is actually finite as $\eta_a\to 1$, such that $\IcC{ars}{IFF(0,0)} \IcS{s}{}(\eta_a,\eta_b;\eps\,|\,1,\eta_b)_{\eta_b<\eta_a}=0$.

\subsubsection{
\texorpdfstring{$\cC{ars}{IFF(0,0)}\cC{rs}{FF}\cS{s}{}$}{CarsIFF00CrsSs}}
The integrated counterterm is~\cite{DelDuca:2025yph}
\beq
\bsp
&\left[ \IcC{ars}{IFF(0,0)} \IcC{rs}{FF} \IcS{s}{}\right]\otimes\rd\sigma_{\ha{\ti{a}}\ha{\ti{b}}}=\,\int_{0}^{1} \rd\lam\, \int_{0}^{1} \rd\xi_a\, \rd\xi_b\,\rd\sigma_{\ha{\ti{a}}\ha{\ti{b}}}(\ha{\ti{p}}_a, \ha{\ti{p}}_b)\,
	\left[ \IcC{ars}{IFF(0,0)} \IcC{rs}{FF} \IcS{s}{}(\lambda,\xi_a,\xi_b;\eps)\right]\,
\esp
\eeq
with
\beq
\bsp
\left[ \IcC{ars}{IFF(0,0)} \IcC{rs}{FF} \IcS{s}{}(\lambda,\xi_a,\xi_b;\eps)\right] =\,&\,
	2 \frac{(4\pi)^2 }{S_\ep} s_{ab}^{1+\ep} 
	\int_1 \PS{2}(p_s,K;Q)\,\frac{\s12}{\pi} \lam^{5-2\eps}
	\left[ \frac{\xi_a \xi_b (1-\xi_a^2) (1-\xi_b^2)}{(\xi_a + \xi_b)^2} \right]^{-\eps} 
	\\
	&\times\frac{\xi_a^2 \xi_b^2 (1+\xi_a \xi_b)}{(\xi_a + \xi_b)^2}  
	\frac{1}{x_{\ti{a},\ti{r}} s_{\ti{a}\ti{r}}}\,
	\frac{2z_{\ti{r},s}}{s_{\ti{r}s} z_{s,\ti{r}}} \bT_r^2
	\frac{\omega(ar)}{\omega(a)}\, P_{(ar)r}^{(0)}(x_{\ti{a},\ti{r}};\ep) \,.
\esp
\eeq
It turns out that the required integrals were already computed above. In particular, using
\begin{equation}
    \frac{2z_{\ti{r},s}}{s_{\ti{r}s} z_{s,\ti{r}}} =\, \frac{2}{s_{\ti{r}s}} \frac{x_{\ti{r},\ti{a}}}{x_{s,\ti{a}}}\,,
\end{equation}
we see that they correspond to $\left[ \IcC{ars}{IFF(0,0)} \IcS{s}{}(\lambda,\xi_a,\xi_b;\eps)\right]^{(ab,ab)}$, cf.~eq.~(\ref{eq:CarsSs-abab}), up to a factor of $2\,\bT_r^2$. As such, we find that the final result reduces to eq.~(\ref{eq:SIF-distexp})
with
\begin{equation}
    [\IcC{ars}{IFF(0,0)}\IcC{rs}{FF}\IcS{s}{}(\eta_a,\eta_b;\eps)] =\, 2\,\bT_r^2[\IcC{ars}{IFF(0,0)}\IcS{s}{}(\eta_a,\eta_b;\eps)]^{(ab,ab)}\,.
\end{equation}

\subsubsection{
\texorpdfstring{$\cC{asr}{IFF(0,0)}\cC{as}{IF}\cS{s}{}$}{CarsIFF00CasSs}}
The integrated counterterm is~\cite{DelDuca:2025yph}
\beq
\bsp
&\left[ \IcC{asr}{IFF(0,0)} \IcC{as}{IF} \IcS{s}{}\right]\otimes\rd\sigma_{\ha{\ti{a}}\ha{\ti{b}}}=\,\int_{0}^{1} \rd\lam\, \int_{0}^{1} \rd\xi_a\, \rd\xi_b\,\rd\sigma_{\ha{\ti{a}}\ha{\ti{b}}}(\ha{\ti{p}}_a, \ha{\ti{p}}_b)\,
	\left[ \IcC{asr}{IFF(0,0)} \IcC{as}{IF} \IcS{s}{}(\lambda,\xi_a,\xi_b;\eps)\right]\,
\esp
\eeq
with
\beq
\bsp
 \left[ \IcC{asr}{IFF(0,0)} \IcC{as}{IF} \IcS{s}{}(\lambda,\xi_a,\xi_b;\eps)\right] =\,&\,
	2 \frac{(4\pi)^2 }{S_\ep} \s12^{1+\ep} 
	\int_1 \PS{2}(p_s,K;Q)\,\frac{\s12}{\pi} \lam^{5-2\eps} 
	\left[\frac{\xi_a \xi_b (1-\xi_a^2) (1-\xi_b^2)}{(\xi_a + \xi_b)^2} \right]^{-\eps} \\
	&\times
	\frac{\xi_a^2 \xi_b^2 (1+\xi_a \xi_b)}{(\xi_a + \xi_b)^2}  
	\frac{1}{x_{\ti{a},\ti{r}}s_{\ti{a}\ti{r}}}\,
	\frac{2}{s_{\ti{a}s}x_{s,\ti{a}}} \bT_a^2
	 \frac{\omega(ar)}{\omega(a)}\, P_{(ar) r}^{(0)}(x_{\ti{a},\ti{r}};\ep)\,.
\esp
\eeq
The required integrals were already computed above. In particular, we see that they correspond to $\left[ \IcC{ars}{IFF(0,0)} \IcS{s}{}(\lambda,\xi_a,\xi_b;\eps)\right]^{(ab,ab)}$, cf.~eq.~(\ref{eq:CarsSs-abab}), up to a factor of $2\,\bT_a^2$. As such we find that the integrated subtraction term becomes as in eq.~(\ref{eq:SIF-distexp}) with
\begin{equation}
    [\IcC{asr}{IFF(0,0)}\IcC{as}{IF}\IcS{s}{}(\eta_a,\eta_b;\eps)] =\, 2\,\bT_a^2[\IcC{ars}{IFF(0,0)}\IcS{s}{}(\eta_a,\eta_b;\eps)]^{(ab,ab)}\,.
\end{equation}

\subsection{Integration of S--FF iterated subtraction terms}
\label{sec:IFFS}
The S--FF iterated factorized matrix element is written in terms of the momenta $(\ti{\ha{p}}_a, \ti{\ha{p}}_b;\momth{}_{X})$, which come about by first applying the final-state collinear mapping of eq.~(\ref{eq:FFmap}) and then the single soft mapping in eq.~(\ref{eq:softmap}),
\beq
(p_a, p_b;\mom{}_{X+2}) \cmap{rs}{FF} (\ha{p}_a, \ha{p}_b;\momh{}_{X+1}) \smap{\widehat{rs}} (\ti{\ha{p}}_a, \ti{\ha{p}}_b;\momth{}_{X}) \,.
	\label{eq:iterFFsoftmap}
\eeq
In particular we have
\beq
\bsp
\ti{\ha{p}}_a^\mu &=\,\lambda_{\wha{rs}}(1-\al_{rs}) p_a^\mu\,,
\qquad
\ti{\ha{p}}_b^\mu =\,\lambda_{\wha{rs}} (1-\al_{rs})p_b^\mu\,.
\esp
\label{eq:iterFFsoftmap2}
\eeq
The $(X+2)$-particle phase space is composed as a final-final collinear convolution followed by a single soft convolution,
\beq
\bsp
    \PS{X+2}(\mom{}_{X+2};Q)  =\,&\,
	\int_{0}^{1} \rd\al\, \frac{s_{\wha{rs} Q }}{2\pi}\,
	\frac{(1-\alpha)^2\s12}{\pi} \int_{0}^{1} \rd\lam\lam
	\PS{X}(\momth{}_{X};\ti{\ha{Q}})
	\PS{2}(\ha{p}_{rs},\ha{K};\ha{Q})
	\PS{2}(p_r,p_s; \hat{p}_{rs}+\al Q)\,.
\label{eq:PS-FFS}
\esp
\eeq
We now have two different subtraction terms, namely $\cS{rs}{(0,0)}\cC{rs}{FF}$ and $\cC{ars}{IFF}\cS{rs}{(0,0)}\cC{rs}{FF}$. Their integrated versions have the same structure as the integrated S--S iterated counterterms, cf.~eq.~(\ref{eq:SS-distexp}).

\subsubsection{
\texorpdfstring{$\cS{rs}{(0,0)}\cC{rs}{FF}$}{Srs00Crs}}
The integrated counterterm can be written as~\cite{DelDuca:2025yph}
\beq
\bsp
\label{eq:SrsCrs-def}
    &\left[ \IcS{rs}{(0,0)} \IcC{rs}{FF}\right]\otimes\rd\sigma_{\ti{\ha{a}}\ti{\ha{b}}}=\,\bT^2_{\mathrm{ini}}\int_{0}^{1} \rd\al\, \int_{0}^{1} \rd\lam\,\rd\sigma_{\ti{\ha{a}}\ti{\ha{b}}}(\ti{\ha{p}}_a,\ti{\ha{p}}_b)\,
	\Big\{ \left[ \IcS{rs}{(0,0)} \IcC{rs}{FF}(\al,\lam;\eps)\right]^{(a,a)} + \left[ \IcS{rs}{(0,0)} \IcC{rs}{FF}(\al,\lam;\eps)\right]^{(b,b)}
	\\&\qquad\qquad\qquad\qquad\qquad\qquad\qquad-2 \left[ \IcS{rs}{(0,0)} \IcC{rs}{FF}(\al,\lam;\eps)\right]^{(a,b)} \Big\}
\esp
\eeq
with
\beq
\bsp
\label{eq:SrsCrs-br}
     \left[ \IcS{rs}{(0,0)} \IcC{rs}{FF}(\al,\lam;\eps)\right]^{(i,k)} =\,&\,  \frac{(4\pi)^2 }{S_\ep} \s12^\ep
	\int_1 \, \PS{2}(p_r,p_s; \hat{p}_{rs}+\al Q)\,\frac{s_{\ha{a}\ha{b}}}{\pi}\frac{\lam^3(1-\alpha)^2}{s_{rs}} 
\\&\times\frac{(4\pi)^2 }{S_\ep} \s12^\ep
	\int_1 \,\PS{2}(\ha{p}_{rs},\ha{K};\ha{Q})
	\frac{s_{\wha{rs}Q}}{2\pi}\, \calS_{\ti{\ha{i}}\ti{\ha{k}}}^{\mu \nu}(\wha{rs}) \,
	 \la \mu |\hP^{(0)}_{rs}(z_{r,s},\kT{r,s};\ep)|\nu \ra
	\,
\esp
\eeq
for $i,k\in I$. Here $\calS_{\ti{\ha{i}}\ti{\ha{k}}}^{\mu \nu}(\wha{rs})$ is the uncontracted eikonal function,
\beq
\calS_{\ti{\ha{i}}\ti{\ha{k}}}^{\mu \nu}(\wha{rs}) =\,
	\frac{4\ti{\ha{p}}_i^\mu \ti{\ha{p}}_k^\nu}{s_{\ti{\ha{i}}\wha{rs}} s_{\ti{\ha{k}}\wha{rs}}}\,.
	\label{eq:uneik}
\eeq
The definition in eq.~(\ref{eq:SrsCrs-def}) is specific for the case of color-singlet production. As such, the analysis below will need to be adapted when considering jets. The transverse vector $\kT{r,s}^\mu$ in the spin-dependent splitting function, cf.~eqs.~(\ref{eq:Pqq0FF})-(\ref{eq:Pgg0FF}), is orthogonal to $\ha{p}_{rs}^\mu$ and $Q^\mu$ and can be written as
\beq
\bsp
    \kT{r,s}^{\mu} =\,&\, \left(\frac{s_{rQ}}{s_{(rs)Q}} - \frac{s_{rs}}{\al s_{(rs)Q}}\right) p_s^\mu - \left(\frac{s_{sQ}}{s_{(rs)Q}} - \frac{s_{rs}}{\al s_{(rs)Q}}\right) p_r^\mu + \left(\frac{s_{rs}}{\al s_{\wha{rs}Q}} \left(\frac{s_{sQ}}{s_{(rs)Q}} - \frac{s_{rQ}}{s_{(rs)Q}}\right)\right) \ha{p}_{rs}^\mu\,.
\esp
\eeq
To proceed, we decompose the splitting function into its different Lorentz structures,
\beq
\bsp
   \la\mu|\ha{P}^{(0)}_{rs}(z_{r,s},\kT{r,s};\ep)|\nu\ra\ =
-g^{\mu\nu} A_{rs}(z_{r,s};\eps)
+B_{rs}(z_{r,s};\eps) \frac{\kT{r,s}^\mu \kT{r,s}^\nu}{\kT{r,s}^2}
\esp
\eeq
with
\bal
\label{eq:Afrfs}
A_{gg}(z;\eps) &=\,2C_A \left(\frac{z}{1-z} + \frac{1-z}{z}\right)\,, \quad
A_{q\qb}(z;\eps) =\,\TR
\eal
and
\bal
\label{eq:Bfrfs}
B_{gg}(z;\eps) &=\,-4C_A(1-\eps)z(1-z)\,, \quad
B_{q\qb}(z;\eps) =\,4z(1-z)\TR\,.
\eal
Next we consider the tensor integral involving the transverse momentum,
\beq
\bsp
   &I^{\mu\nu}=\,\int_1 \PS{2}(p_r,p_s; \hat{p}_{rs}+\al Q) \frac{\kT{r,s}^\mu \kT{r,s}^\nu}{\kT{r,s}^2} \,.
\esp
\eeq
For this, we first set up an ansatz in terms of all possible tensor structures built up out of the metric, $\ha{p}_{rs}^\mu$ and $Q^\mu$,
\beq
\bsp
\label{eq:tensAns}
    I^{\mu\nu} =\,\int_1 \PS{2}(p_r,p_s; \hat{p}_{rs}+\al Q)( a_1 g^{\mu\nu}+a_2 \ha{p}^{\mu}_{rs}\ha{p}^{\nu}_{rs}+a_3 Q^{\mu}Q^{\nu}+a_4 \ha{p}^{\mu}_{rs}Q^{\nu}+a_5 Q^{\mu}\ha{p}^{\nu}_{rs})\,.
\esp
\eeq
The unknown $a_i$ can now be fixed by contracting $I^{\mu\nu}$ with each tensor on the right-hand side of eq.~(\ref{eq:tensAns}). Using that $\kT{r,s}^\mu$ was chosen to be orthogonal to both $Q^\mu$ and $\ha{p}_{rs}^\mu$, we find
\beq
\bsp
\label{eq:kTrep}
    I^{\mu\nu}=\, \int_1 \PS{2}(p_r,p_s; \hat{p}_{rs}+\al Q) \frac{1}{2(1-\eps)}\left[g^{\mu\nu}+\frac{Q^2}{(\ha{p}_{rs}\cdot Q)^2}\,\ha{p}_{rs}^\mu \ha{p}_{rs}^\nu-\frac{\ha{p}_{rs}^\mu Q^\nu+\ha{p}_{rs}^\nu Q^\mu}{\ha{p}_{rs}\cdot Q}\right].
\esp
\eeq
Hence the integrand in eq.~(\ref{eq:SrsCrs-br}) becomes
\beq
\bsp
    &\frac{s_{\wha{rs}Q}}{s_{rs}}\, \calS_{\ti{\ha{i}}\ti{\ha{k}}}^{\mu \nu}(\wha{rs}) \,
	 \la \mu |\hP^{(0)}_{rs}(z_{r,s},\kT{r,s};\ep)|\nu \ra \\&=\, -\frac{s_{\wha{rs}Q}}{s_{rs}}\,\calS_{\ti{\ha{i}}\ti{\ha{k}}}^{\mu \nu}(\wha{rs}) g^{\mu\nu}\left(A_{rs}(z_{r,s};\eps)-\frac{1}{2(1-\eps)}B_{rs}(z_{r,s};\eps)\right) \\&=\, -\frac{s_{\wha{rs}Q}}{s_{rs}}\,\calS_{\ti{\ha{i}}\ti{\ha{k}}}(\wha{rs}) \,P^{(0)}_{rs}(z_{r,s};\ep)\,.
\esp
\eeq
Note the appearance of the azimuthally averaged splitting function, which was obtained using
\beq
\label{eq:ABtoP}
    A_{rs}(z_{r,s};\eps)-\frac{1}{2(1-\eps)}B_{rs}(z_{r,s};\eps) =\, P^{(0)}_{rs}(z_{r,s};\ep)\,,
\eeq
cf.~eqs.~(\ref{eq:Pqq-ave})-(\ref{eq:Pgg-ave}). Collecting everything we then find
\beq
\bsp
    \left[ \IcS{rs}{(0,0)} \IcC{rs}{FF}(\al,\lam;\eps)\right]^{(i,k)} =\,&\, -\frac{(4\pi)^2}{S_{\eps}}s_{ab}^{\eps}\int_1 \PS{2}(\ha{p}_{rs},\ha{K};\ha{Q})\,\frac{s_{\wha{rs} Q }}{2\pi}\frac{\lam^3(1-\al)^2}{s_{rs}}\calS_{\ti{\ha{i}}\ti{\ha{k}}}(\wha{rs})\\&\times\frac{(4\pi)^2}{S_{\eps}}s_{ab}^{\eps}\int_1 \PS{2}(p_r,p_s; \hat{p}_{rs}+\al Q)\,\frac{s_{\ha{a}\ha{b}}}{\pi}P^{(0)}_{rs}(z_{r,s};\ep)\,.
\esp
\eeq
Now, the integral over the $\PS{2}(p_r,p_s; \hat{p}_{rs}+\al Q)$ phase space actually exactly matches the definition of $\left[\IcC{rs}{FF}(\al;\eps)\right]$ in the integrated single-unresolved approximate cross section $\int_1\dsiga{RR}{1}_{ab}$, whose evaluation will be discussed in detail elsewhere. Here, we simply quote the result, which reads
\beq
\bsp
\label{eq:dphi2Crs}
    &\left[\frac{(4\pi)^2}{S_{\eps}}s_{ab}^{\eps}\right]\int_1 \PS{2}(p_r,p_s; \hat{p}_{rs}+\al Q)\,\frac{s_{\wha{rs} Q }}{2\pi}\frac{1}{s_{rs}}P^{(0)}_{rs}(z_{r,s};\ep) \\&=\, x_{\wha{rs},a}\int_{0}^{1}\rd v\,\al^{-1-\eps}(\al+x_{\wha{rs},a})^{-1-\eps}v^{-\eps}(1-v)^{-\eps}P^{(0)}_{rs}\left(\frac{\al+x_{\wha{rs},a}v}{2\al+x_{\wha{rs},a}};\ep\right)\,.
\esp
\eeq
Furthermore, the analysis of the $\PS{2}(\ha{p}_{rs},\ha{K};\ha{Q})$ phase space is similar to the computation for the $\cS{rs}{(0,0)}\cS{s}{ }$ subtraction term in sec.~\ref{sec:defISrsSs0} and hence will not be presented here. Instead we just give the final result, which is
\beq
\bsp
    &\left[ \IcS{rs}{(0,0)} \IcC{rs}{FF}(\al,\lam;\eps)\right]^{(i,k)} =\, \frac{4}{\eps}(1-\cos\chi_{ik})\,{}_{2}F_{1}\left(1,1,1-\eps,\frac{1+\cos\chi_{ik}}{2}\right)\frac{\Gamma^2(1-\eps)}{\Gamma(1-2\eps)}\lam^3(1-\lam^2)^{-2\eps}\\&\times\al^{-1-\eps}(1-\al)^{3-2\eps}(\al+(1-\lam^2)(1-\al))^{-1-\eps}\int_{0}^{1}\rd v\,v^{-\eps}(1-v)^{-\eps}P^{(0)}_{rs}\left(\frac{\al+(1-\lam^2)(1-\al)v}{2\al+(1-\lam^2)(1-\al)};\ep\right)\,.
\esp
\eeq
As $(i,k)\in I$, we immediately see that $\left[ \IcS{rs}{(0,0)} \IcC{rs}{FF}(\al,\lam;\eps)\right]^{(a,a)}=\,\left[ \IcS{rs}{(0,0)} \IcC{rs}{FF}(\al,\lam;\eps)\right]^{(b,b)}=\,0$ such that only the mixed contributions remain. For these however we have $\cos\chi_{ik}=\pi$ such that, using ${}_{2}F_{1}(a,b,c,0) =\,1$, we simply get
\beq
\bsp
    \left[ \IcS{rs}{(0,0)} \IcC{rs}{FF}(\al,\lam;\eps)\right] =\,&\, \frac{8}{\eps}\,\frac{\Gamma^2(1-\eps)}{\Gamma(1-2\eps)}\lam^3(1-\lam^2)^{-2\eps}\al^{-1-\eps}(1-\al)^{3-2\eps}(\al+(1-\lam^2)(1-\al))^{-1-\eps}\\&\times\int_{0}^{1}\rd v\,v^{-\eps}(1-v)^{-\eps}P^{(0)}_{rs}\left(\frac{\al+(1-\lam^2)(1-\al)v}{2\al+(1-\lam^2)(1-\al)};\ep\right)\,.
\esp
\eeq
For ease of notation we set $\left[ \IcS{rs}{(0,0)} \IcC{rs}{FF}(\al,\lam;\eps)\right]^{(a,b)} =\, \left[ \IcS{rs}{(0,0)} \IcC{rs}{FF}(\al,\lam;\eps)\right]$.
The parton-level integrated subtraction term then becomes
\beq
\bsp
    &\left[ \IcS{rs}{(0,0)} \IcC{rs}{FF}\right]\otimes\rd\sigma_{\ti{\ha{a}}\ti{\ha{b}}}=\, \int_{0}^{1} \rd\al\, \int_{0}^{1} \rd\lam\,\rd\sigma_{\ti{\ha{a}}\ti{\ha{b}}}(\lam(1-\al)p_a,\lam(1-\al)p_b)\left[ \IcS{rs}{(0,0)} \IcC{rs}{FF}(\al,\lam;\eps)\right]\,.
\esp
\eeq
Next we introduce a change of variables $\eta=\lam(1-\al)$ to reduce the number of convolution variables in the reduced differential cross section. This leads to
\beq
\bsp
    \left[ \IcS{rs}{(0,0)} \IcC{rs}{FF}\right]\otimes\rd\sigma_{\ti{\ha{a}}\ti{\ha{b}}}=\,\int_{0}^{1} \rd\eta\,\rd\sigma_{\ti{\ha{a}}\ti{\ha{b}}}(\eta\, p_a,\eta\, p_b)\,\left[ \IcS{rs}{(0,0)} \IcC{rs}{FF}(\eta;\eps)\right]
\esp
\eeq
with
\begin{equation}
\label{eq:SrsCrs-br-FIN}
    \left[ \IcS{rs}{(0,0)} \IcC{rs}{FF}(\eta;\eps)\right] =\, \int_{\eta}^{1}\frac{\rd\lam}{\lam}\,\left[ \IcS{rs}{(0,0)} \IcC{rs}{FF}(1-\eta/\lam,\lam;\eps)\right]\,.
\end{equation}
The final expression for the integrated counterterm at the hadronic level then indeed becomes as eq.~(\ref{eq:SS-distexp}).\newline

A priori we have five distinct basic integrals to compute, corresponding to the different $z$-structures in the splitting function,
\beq
\left\{\frac{1}{1-z}, \frac{1}{z}, z^0, z, z^2\right\}\,.
\eeq
However, there is a $z_{r,s}\leftrightarrow 1-z_{r,s}$ symmetry, which in practice means that the $\frac{1}{1-z}$-term in the splitting function can be replaced by $\frac{1}{z}$ without changing the value of the integral. As such, only four unique integrals remain\footnote{In fact, by the same symmetry, one can also relate the integrals with $z^0$ and $z^1$. However, in practice, we compute both and use their relation as a check of the computation.}, corresponding to the structures $\left\{\frac{1}{z},z^0,z^1,z^2\right\}$. The computation of $\left[ \IcS{rs}{(0,0)} \IcC{rs}{FF}(\eta;\eps)\right]$ in eq.~(\ref{eq:SrsCrs-br-FIN}) is now relatively straightforward. Note however that this is actually a double integral, as there is still an implicit $v$-integration of the form
\begin{equation}
    \mathcal{I}_v =\, \int_{0}^{1}\rd v\,v^{-\eps}(1-v)^{-\eps}P^{(0)}_{rs}\left(\frac{-2 \eta^2 t v+(\eta-1) \eta t^2 v+(\eta+1) \eta v+t}{\eta^2 (t-1)^2-\eta t^2+\eta+2 t};\ep\right)\,.
\end{equation}
The $\lam$-integration was replaced by an integration over $t$, which now runs between zero and one, using
\begin{equation}
    t =\, \frac{\eta-\lam}{\eta-1}\,.
\end{equation}
The evaluation of these integrals is straightforward and generically leads to some hypergeometric solutions. For example, for the most complicated case, corresponding to the $\frac{1}{z}$-term in the splitting function, we find
\beq
\bsp
\label{eq:Iv1}
    \mathcal{I}_v =\,&\, \frac{8\Gamma^4 (1-\eps)}{\eps\, \Gamma (1-2 \eps) \Gamma (2-2 \eps)}\, \eta^3 (1-\eta)^{-1-4 \eps} t^{-2-\eps}  \left(\eta^2 (t-1)^2-\eta t^2+\eta+2 t\right) (\eta-\eta t)^{-2 \eps} (-\eta t+\eta+1)^{-1-\eps}\\&\times (-\eta t+\eta+t)^{3 \eps} (-\eta t+\eta+t+1)^{-2 \eps}{}_{2}F_{1}\left(1,1-\eps,2-2 \eps,-\frac{\eta (\eta (t-1)-t-1) (t-1)}{t}\right)\,.
\esp
\eeq
Note that the last argument of the hypergeometric function in eq.~(\ref{eq:Iv1}) is proportional to $1/t$, which will cause problems when integrating over $t$, since the singularity at $t=0$ in this form is not factorized. We can circumvent this issue by using the hypergeometric identity \cite{gradshteyn2007}
\beq
\bsp
    {}_2F_1(a,b,c,z) =\,&\,\frac{(-z)^{-a}\, \Gamma (c) \Gamma (b-a) }{\Gamma (b) \Gamma (c-a)}\, {}_2F_1\left(a,a-c+1,a-b+1,\frac{1}{z}\right)\\&+\frac{(-z)^{-b}\, \Gamma (c) \Gamma (a-b) }{\Gamma (a) \Gamma (c-b)}\, {}_2F_1\left(b,b-c+1,-a+b+1,\frac{1}{z}\right)\,.
\esp
\eeq
Hence we get two different hypergeometric functions, whose last argument is $\frac{t}{\eta (t-1) (\eta (t-1)-t-1)}$.
This however is still problematic, now because of the overall $\frac{1}{t-1}$ in the last argument (this interferes with the straightforward extraction of the singularity at $t=1$.) Again a hypergeometric identity comes to the rescue, this time
\begin{equation}
    \, {}_2F_1(a,b,c,t)=\, (1-t)^{-b} \, {}_2F_1\left(c-a,b,c,\frac{t}{t-1}\right)\,,
\end{equation}
and the remaining ${}_2F_1$ is now regular in the integration limits.\footnote{Of course, now the offending factors involving $t^{-1}$ and $(t-1)^{-1}$ are moved outside of the hypergeometric function. These however can then be treated by subtraction.} Next we integrate over $t$, which closely follows the steps outlined in sec.~\ref{sec:A12ints}, and compute the asymptotic behavior. We omit the explicit results here but note that they have a relatively simple structure with classical polylogarithms in $\eta$ up to weight three.

\subsubsection{
\texorpdfstring{$\cC{ars}{IFF}\cS{rs}{(0,0)}\cC{rs}{FF}$}{CarsIFF00SrsCrs}}
\label{sec:CarsIFF00SrsCrs}
The parton-level integrated counterterm is~\cite{DelDuca:2025yph}
\beq
\bsp
   &\left[ \IcC{ars}{IFF} \IcS{rs}{(0,0)} \IcC{rs}{FF}\right]\otimes\rd\sigma_{\ti{\ha{a}}\ti{\ha{b}}}=\,\bT_a^2\int_{0}^{1} \rd\al\, \int_{0}^{1} \rd\lam\, \rd\sigma_{\ti{\ha{a}}\ti{\ha{b}}}(\ti{\ha{p}}_a,\ti{\ha{p}}_b)\,
	\left[ \IcC{ars}{IFF} \IcS{rs}{(0,0)} \IcC{rs}{FF}(\al,\lam;\eps) \right]
\esp
\eeq
with
\beq
\bsp
    \label{eq:CarsSrsCrs-brA}
    &\left[ \IcC{ars}{IFF} \IcS{rs}{(0,0)} \IcC{rs}{FF}(\al,\lam;\eps) \right] =\,\frac{(4\pi)^2 }{S_\ep} \s12^\ep
	\int_1 \,\PS{2}(\ha{p}_{rs},\ha{K};\ha{Q})
	\frac{s_{\wha{rs}Q}}{2\pi} \frac{\lam^3(1-\alpha)^2}{s_{rs}}
\\&\times\frac{(4\pi)^2 }{S_\ep} \s12^\ep
	\int_1 \,\PS{2}(p_r,p_s; \hat{p}_{rs}+\al Q)\,\frac{s_{\ha{a}\ha{b}}}{\pi}\,\frac{2}{s_{\ti{\ha{a}}\wha{rs}}}\left(\frac{1}{x_{\wha{rs},\ti{\ha{a}}}}A_{rs}(z_{r,s};\eps)+\frac{s_{\ti{\ha{a}}\kT{r,s}}^2}{4\kT{r,s}^2s_{\ti{\ha{a}}\wha{rs}}}B_{rs}(z_{r,s};\eps)\right)\,.
\esp
\eeq
The functions $A_{rs}(z_{r,s};\eps)$ and $B_{rs}(z_{r,s};\eps)$ were defined in eqs.~(\ref{eq:Afrfs})-(\ref{eq:Bfrfs}) above. To proceed, we must first handle the structure $s_{\ti{\ha{a}}\kT{r,s}}$. The analysis closely follows the discussion for $\left[\IcS{rs}{(0,0)} \IcC{rs}{FF}(\al,\lam;\eps) \right]$. In particular, using eq.~(\ref{eq:kTrep}), we can write
\beq
\bsp
    &\int_1 \PS{2}(p_r,p_s; \hat{p}_{rs}+\al Q)\, \frac{s_{\ti{\ha{a}}\kT{r,s}}^2}{\kT{r,s}^2s_{\ti{\ha{a}}\wha{rs}}} =\, \int_1 \PS{2}(p_r,p_s; \hat{p}_{rs}+\al Q) \,\frac{4\,\ti{\ha{p}}_{a\mu}\ti{\ha{p}}_{a\nu}}{s_{\ti{\ha{a}}\wha{rs}}}\frac{\kT{r,s}^{\mu}\kT{r,s}^{\nu}}{\kT{r,s}^2}\\&\to\,\int_1 \PS{2}(p_r,p_s; \hat{p}_{rs}+\al Q) \,\frac{4\,\ti{\ha{p}}_{a\mu}\ti{\ha{p}}_{a\nu}}{s_{\ti{\ha{a}}\wha{rs}}}\frac{1}{2(1-\eps)}\left[g^{\mu\nu}+\frac{Q^2}{(\ha{p}_{rs}\cdot Q)^2}\,\ha{p}_{rs}^\mu \ha{p}_{rs}^\nu-\frac{\ha{p}_{rs}^\mu Q^\nu+\ha{p}_{rs}^\nu Q^\mu}{\ha{p}_{rs}\cdot Q}\right]\,.
\esp
\eeq
After some algebra this leads to
\beq
\bsp
    \int_1 \PS{2}(p_r,p_s; \hat{p}_{rs}+\al Q)\, \frac{s_{\ti{\ha{a}}\kT{r,s}}^2}{\kT{r,s}^2s_{\ti{\ha{a}}\wha{rs}}}\to\,\int_1 \PS{2}(p_r,p_s; \hat{p}_{rs}+\al Q)\,\frac{2}{1-\eps}\left(\frac{s_{\ti{\ha{a}}\wha{rs}}s_{ab}}{s_{\wha{rs}Q}^2}-\frac{s_{\ti{\ha{a}}Q}}{s_{\wha{rs}Q}}\right)\,.
\esp
\eeq
Substituting into eq.~(\ref{eq:CarsSrsCrs-brA}) we find that the integrand reduces to
\beq
\bsp
\label{eq:CarsSrsCrs-brB}
    &\left[ \IcC{ars}{IFF} \IcS{rs}{(0,0)} \IcC{rs}{FF}(\al,\lam;\eps) \right] =\, 2\left[\frac{(4\pi)^2}{S_{\eps}}s_{ab}^{\eps}\right]^2\,\,\int_1 \PS{2}(\ha{p}_{rs},\ha{K};\ha{Q})\frac{s_{\wha{rs} Q }}{2\pi}\frac{\lam^3(1-\al)^2}{s_{rs}}\\&\times\int_1 \PS{2}(p_r,p_s;\hat{p}_{rs}+\al Q)\frac{s_{\ha{a}\ha{b}}}{\pi}\,\Bigg\{\frac{2}{s_{\ti{\ha{a}}\wha{rs}}x_{\wha{rs},\ti{\ha{a}}}}\,P^{(0)}_{rs}(z_{r,s};\ep)+\frac{s_{ab}}{s_{\wha{rs} Q }^2}\,\frac{B_{rs}(z_{r,s};\eps)}{2(1-\eps)}\Bigg\}
\esp
\eeq
in which the azimuthally averaged splitting function was introduced using eq.~(\ref{eq:ABtoP}). Now, after setting $\eta=\lam(1-\al)$, the full integrated subtraction term at the hadronic level again becomes as in eq.~(\ref{eq:SS-distexp}). Furthermore, the required integrals were actually already computed above. In particular, the computation can be reduced to evaluating integrals involving
\beq
\label{eq:ints}
\frac{1}{s_{\ti{\ha{a}}\wha{rs}}x_{\wha{rs},\ti{\ha{a}}}} \frac{1}{s_{rs}}z_{r,s}^k
\qquad\mbox{and}\qquad
\frac{s_{ab}}{s_{\wha{rs} Q }^2} \frac{1}{s_{rs}}z_{r,s}^k
\eeq
with $k\in\{-1,0,1,2\}$. Now, for the integration of the $\cS{rs}{(0,0)}\cC{rs}{FF}$ subtraction term above, we already evaluated the integral of 
\beq
\frac{-2s_{\tha{a}\tha{b}}}{s_{\tha{a}\wha{rs}}s_{\tha{b}\wha{rs}}} \frac{1}{s_{rs}} z_{r,s}^k =\, \frac{-2s_{ab}}{s_{a\wha{rs}}s_{b\wha{rs}}} \frac{1}{s_{rs}} z_{r,s}^k\,.
\eeq
But, using the same type of argument as used in the computation of $\left[ \IcS{rs}{(0,0)} \IcC{as}{IF} \IcS{s}{}(\lam,\ti{\lam};\eps)\right]$, during integration we can replace
\beq
\frac{1}{s_{\ti{\ha{a}}\wha{rs}} x_{\wha{rs},\ti{\ha{a}}}} \frac{1}{s_{rs}}z_{r,s}^k \to\,-\frac{1}{4} \left(\frac{-2s_{ab}}{s_{a\wha{rs}}s_{b\wha{rs}}}\right) \frac{1}{s_{rs}} z_{r,s}^k
\eeq
for the first type of integral in eq.~(\ref{eq:ints}). Similarly, for the second type in eq.~(\ref{eq:ints}) we have\footnote{The additional factor of $\eps/(1-2\eps)$ appears due to differences in the angular integration.}
\beq
\frac{s_{ab}}{s_{\wha{rs} Q }^2} \frac{1}{s_{rs}}z_{r,s}^k \to\,\frac{\ep}{4(1-2\ep)}\left(\frac{-2s_{ab}}{s_{a\wha{rs}}s_{b\wha{rs}}}\right) \frac{1}{s_{rs}} z_{r,s}^k\,.
\eeq
Hence the necessary integrals exactly correspond to those for $\cS{rs}{(0,0)}\cC{rs}{FF}$ above. In particular we have
\beq
\bsp
    \left[ \IcC{ars}{IFF} \IcS{rs}{(0,0)} \IcC{rs}{FF}(\eta;\eps) \right] =\,&\, -\frac{1}{2}\Bigg\{\left[ \IcS{rs}{(0,0)} \IcC{rs}{FF}(\eta;\eps)\right]+B_{rs}(z_{r,s};\eps)\vert_{z^1}\left[ \IcS{rs}{(0,0)} \IcC{rs}{FF}(\eta;\eps)\right]_{z^1}\\&+B_{rs}(z_{r,s};\eps)\vert_{z^2}\left[ \IcS{rs}{(0,0)} \IcC{rs}{FF}(\eta;\eps)\right]_{z^2}\Bigg\}
\esp
\eeq
with $B_{rs}(z_{r,s};\eps)\vert_{z^1/z^2}$ the coefficient of $z_{r,s}/z_{r,s}^2$ in $B_{rs}(z_{r,s};\eps)$. A similar notation was introduced for terms in $\left[ \IcS{rs}{(0,0)} \IcC{rs}{FF}(\eta;\eps)\right]$ originating from specific structures in the splitting function. Furthermore, none of the steps outlined above were specific to color-singlet production. Hence the result can be used as is when considering an extension to colored final states.

\subsection{Integration of IF--FF iterated subtraction terms}
Finally, we discuss the IF--FF iterated subtraction terms, for which the factorized matrix element is written in terms of the momenta $( \hat{\hat{p}}_a, \hat{\hat{p}}_b;\momhh{}_{X} )$. The latter are obtained by first applying the final-final collinear mapping in eq.~(\ref{eq:FFmap}) and then the initial-final collinear mapping in eq.~(\ref{eq:IFmap}),
\begin{equation}
\label{IFFF-map}
    ( p_a,p_b;\mom{}_{X+2} ) \cmap{rs}{FF} ( \hat{p}_a, \hat{p}_b;\momh{}_{X+1} ) \cmap{\ha{a}\ha{b},\wha{rs}}{II,F}( \hat{\hat{p}}_a, \hat{\hat{p}}_b;\momhh{}_{X} )\,.
\end{equation}
Specifically we have
\beq
\bsp
\label{eq:carIFmaphh}
\hat{\hat{p}}_a^\mu =\, \xi_{\ha{a},\wha{rs}}(1-\al_{rs})p_a^\mu\,,
\qquad
\hat{\hat{p}}_b^\mu =\, \xi_{\ha{b},\wha{rs}}(1-\al_{rs})p_b^\mu\,.
\esp
\eeq
The subtraction term now needs to be integrated over the $(X+2)$-particle phase space, which is composed from the final-final collinear convolution followed by an initial-final collinear convolution,
\beq
\bsp
\PS{X+2}(\mom{}_{X+2};Q) =\,&\, \int_{0}^{1} \rd\al\, \frac{s_{\wha{rs} Q }}{2\pi}\,
	\int_{0}^{1} \rd\xi_a\, \rd\xi_b\,
	\PS{X}(\momhh{}_{X};\ha{\ha{Q}}) \,
	\PS{II,F}(\ha{p}_{rs},\xi_a,\xi_b)\,
	\PS{2}(p_r,p_s; \hat{p}_{rs}+\al Q)\,.
\label{eq:PS-IFFF}
\esp
\eeq
This class only contains one counterterm, namely $\cC{ars}{IFF(0,0)}\cC{rs}{FF}$, whose integrated version takes on the same form as the integrated IF--S iterated subtraction terms, cf.~eq.~(\ref{eq:SIF-distexp}). Moreover, the result is generic and valid for general processes \textit{beyond} color-singlet production as well.

\subsubsection{
\texorpdfstring{$\cC{ars}{IFF(0,0)}\cC{rs}{FF}$}{CarsIFF00Crs}}
The integrated counterterm is~\cite{DelDuca:2025yph}
\beq
\bsp
    \left[ \IcC{ars}{IFF(0,0)} \IcC{rs}{FF}\right]\otimes\rd\sigma_{\ha{\ha{a}}\ha{\ha{b}}}=\,\int_{0}^{1} \rd\al\, 
\int_{0}^{1} \rd\xi_a\, \rd\xi_b\,\rd\sigma_{\ha{\ha{a}}\ha{\ha{b}}}(\hat{\hat{p}}_a,\hat{\hat{p}}_b)\, \left[ \IcC{ars}{IFF(0,0)} \IcC{rs}{FF}(\al,\xi_a,\xi_b;\eps)\right] 
\esp
\eeq
with
\beq
\bsp
\label{eq:CarsCrs-br1}
    &\left[ \IcC{ars}{IFF(0,0)} \IcC{rs}{FF}(\al,\xi_a,\xi_b;\eps)\right] =\, \frac{(4\pi)^2 }{S_\ep} \s12^{1+\ep} 
	\int_1 \PS{2}(p_r,p_s; \hat{p}_{rs}+\al Q)\, \frac{s_{\wha{rs} Q }}{\pi}\, (1-\alpha)^{4-2\eps}
\\&\times\left[ \frac{\xi_a \xi_b (1-\xi_a^2) (1-\xi_b^2)}{(\xi_a + \xi_b)^2} \right]^{-\eps} \frac{\xi_a^2 \xi_b^2 (1+\xi_a \xi_b)}{(\xi_a + \xi_b)^2}\,	\frac{1}{s_{rs}\, x_{\ha{a},\wha{rs}}\, s_{\ha{a}\wha{rs}}}\, \frac{\omega(ars)}{\omega(a)}
	 P^{\mathrm{(C)}, (0)}_{(ars) r s}(z_{r,s},x_{\ha{a},\wha{rs}}, s_{\ha{a}\kT{r,s}} ;\ep)  \,.
\esp
\eeq
The treatment of $\PS{2}(p_r,p_s; \hat{p}_{rs}+\al Q)$ is similar to the setup of $\left[\IcC{rs}{FF}(\al;\eps)\right]$ in $A_1$, see also eq.~(\ref{eq:dphi2Crs}). As such we skip the explicit details and just quote the final result, which reads
\beq
\bsp
\label{eq:CarsCrs-br}
     &\left[ \IcC{ars}{IFF(0,0)} \IcC{rs}{FF}(\al,\xi_a,\xi_b;\eps)\right] =\, 2 \frac{\omega(ars)}{\omega(a)}
	(1-\alpha)^3\left[ (1-\alpha)^2\, \frac{\xi_a \xi_b (1-\xi_a^2) (1-\xi_b^2)}{(\xi_a + \xi_b)^2} \right]^{-\eps}
\frac{(1+\xi_a \xi_b)(1-\xi_a\xi_b)}{\xi_a + \xi_b}\\&\qquad\qquad\qquad\qquad\qquad\qquad\quad\times 
\frac{\xi_b}{1-\xi_b^2}
	\int_{0}^{1}\rd v\,\al^{-1-\eps}(\al+(1-\al)(1-\xi_a\xi_b))v^{-\eps}(1-v)^{-\eps}\\&\qquad\qquad\qquad\qquad\qquad\qquad\quad\qquad\qquad\qquad\times P^{\mathrm{(C)}, (0)}_{(ars) r s}(z_{r,s},x_{\ha{a},\wha{rs}} ;\ep)  \,.
\esp
\eeq
Here, with a slight abuse of notation, we introduced
\beq
\bsp
\label{eq:Ph0}
    P^{\mathrm{(C)}, (0)}_{(ars) r s}(z_{r,s},x_{\ha{a},\wha{rs}} ;\ep) =\,&\, P^{(0)}_{rs}\left(z_{r,s};\ep\right) P^{(0)}_{(ars)(rs)}\left(x_{\ha{a},\wha{rs}};\ep\right)-\delta_{rs,g}\,\bT_{ars}^2\, z_{r,s}z_{s,r} b_{rs}^{(0)}b_{(ars)(rs)}^{(0)} R
\esp
\eeq
with
\begin{equation}
    R =\, x_{\ha{a},\wha{rs}}\frac{s_{\ha{a}\wha{rs}}s_{ab}}{s_{\wha{rs} Q }^2}
\end{equation}
and
\beq
b^{(0)}_{qg} =\, b^{(0)}_{gq} =\, b^{(0)}_{gg} =\, 2\,C_A\,,\qquad
b^{(0)}_{q\qb} =\, -\frac{2}{1-\ep}\TR\,.
\eeq
In writing eq.~(\ref{eq:CarsCrs-br}), the terms involving $s_{\ha{a}\kT{r,s}}$ in eq.~(\ref{eq:CarsCrs-br1}) were treated in the same way as explained in Section~\ref{sec:CarsIFF00SrsCrs}, hence eq.~(\ref{eq:Ph0}) gives a form of the strongly-ordered splitting function whose integral is equal to the integral of the original one. Note that, in terms of the integration variables, the argument of the azimuthally averaged final-final splitting function, cf.~eqs.~(\ref{eq:Pqg-ave})-(\ref{eq:Pgg-ave}), is
\begin{equation}
    z_{r,s} =\, \frac{\al+v (1-\al)(1-\xi_a\xi_b)}{2 \al+(1-\al)(1-\xi_a\xi_b)}
\end{equation}
while the one in the azimuthally averaged initial-final splitting function, cf.~eqs.~(\ref{eq:Pqg-ave-IF})-(\ref{eq:Pgg-ave-IF}), is
\begin{equation}
    x_{\ha{a},\wha{rs}} =\, \xi_a\xi_b\,.
\end{equation}
Similarly,
\begin{equation}
    R =\, \frac{\xi_a^2\xi_b(1-\xi_b^2)}{(1-\xi_a\xi_b)^2(\xi_a+\xi_b)}\,.
\end{equation}
Next we simplify the convolutional nature of the integrated subtraction term by transforming the integrations over $\xi_a$ and $\xi_b$ to integrals over $\eta_a =\, (1-\al) \xi_a$ and $\eta_b =\, (1-\al) \xi_b$. This leads to
\beq
\bsp
   \left[ \IcC{ars}{IFF(0,0)} \IcC{rs}{FF}\right]\otimes\rd\sigma_{\ha{\ha{a}}\ha{\ha{b}}}=\,&\,\int_{0}^{1}\frac{\rd \al}{(1-\al)^2}\, \int_0^{1-\al} \rd \eta_a\, \int_0^{1-\al} \rd \eta_b\, \rd\sigma_{\ha{\ha{a}}\ha{\ha{b}}}(\eta_a p_a,\eta_b p_b)\\&\times[\IcC{ars}{IFF(0,0)}\IcC{rs}{FF}(\al,\eta_a/(1-\al),\eta_b/(1-\al))]\,.
\esp
\eeq
The integrated counterterm is now of a similar form as eq.~(\ref{eq:ICarsSs1}) for $\cC{ars}{IFF(0,0)}\cS{s}{}$. Following the discussion presented there we then find, at the level of hadronic variables, that the result is of the form eq.~(\ref{eq:SIF-distexp})
with
\begin{equation}
    [\IcC{ars}{IFF(0,0)} \IcC{rs}{FF}(\eta_a,\eta_b;\eps)]_{\eta_b<\eta_a}  =\, \int_{0}^{1-\eta_a}\frac{\rd\al}{(1-\al)^2}\, [\IcC{ars}{IFF(0,0)}\IcC{rs}{FF}(\al,\eta_a/(1-\al),\eta_b/(1-\al))]
\end{equation}
and
\begin{equation}
    [\IcC{ars}{IFF(0,0)} \IcC{rs}{FF}(\eta_a,\eta_b;\eps)]_{\eta_a<\eta_b}  =\, \int_{0}^{1-\eta_b}\frac{\rd\al}{(1-\al)^2}\, [\IcC{ars}{IFF(0,0)}\IcC{rs}{FF}(\al,\eta_a/(1-\al),\eta_b/(1-\al))]\,.
\end{equation}
Because of the structure of the splitting function in eq.~(\ref{eq:Ph0}) and the need to introduce the regions $\eta_a<\eta_b$ and $\eta_b<\eta_a$, we now have $2(5\times 4+1) =\, 42$ basic integrals to evaluate. However, the explicit integration closely follows the discussion presented for the $\cS{rs}{(0,0)}\cC{rs}{FF}$ counterterm above, see sec.~\ref{sec:IFFS}. As such, we omit the explicit details here. One noteworthy difference is that in this case the $\eta_{a/b}\to\eta_{b/a}$ limits turn out to be regular, such that no subtractions are actually necessary here.

\section{
\texorpdfstring{The $\bI_{12}^{(0)}$ insertion operator}{The I12 insertion operator}}
\label{sec:A12_Iop}

\subsection{Constructing the insertion operator}
Consider some arbitrary $A_{12}$ subtraction term $CT$. As discussed, the result of integrating $CT$ over the phase space of unresolved emissions can symbolically be written as
\begin{equation}
    \int_2 CT =\, \left[\CT\right]\otimes\rd\sigma_{ab}\,,
\end{equation}
which, using eq.~(\ref{eq:XSB}), can be written as a convolution with the Born cross section,
\begin{equation}
    \int_2 CT =\, \left[\frac{\al_s}{2\pi}S_{\eps}\left(\frac{\mu^2}{s_{ab}}\right)^{\eps}\right]^2 \left[\CT\right]\otimes\rd\sigma_{ab,X}^B\,.
\end{equation}
Next, we combine all subtraction terms in the appropriate way as to construct the approximate cross section $\rd\sigma_{ab}^{RR,A_{12}}$, cf.~eq.~(\ref{eq:dsigRRA12}), and replace each subtraction term $CT$ by its integrated version,
\beq
    CT \to\,\left[\frac{\al_s}{2\pi}S_{\eps}\left(\frac{\mu^2}{s_{ab}}\right)^{\eps}\right]^2 \left[\CT\right]\otimes\rd\sigma_{ab,X}^B\,.
\eeq
The resulting expression is only appropriate for one specific partonic subprocess. The full integrated approximate cross section is then obtained by summing over all possible such subprocesses. Hence, denoting the integrated counterterms with square brackets for simplicity, we symbolically write
\beq
\bsp
    \int_2 \dsiga{RR}{12}_{ab}=\,&\,\left[\frac{\al_s}{2\pi}S_{\eps}\left(\frac{\mu^2}{s_{ab}}\right)^{\eps}\right]^2\sum_{\{2\}}\frac{1}{S_{\{2\}}} 
\sum_{s \in F}\Bigg\{\sum_{\substack{r \in F \\ r \ne s}}\Bigg[
	\left[\cS{rs}{(0,0)}\, \cS{s}{}\right]
	\\&+ \sum_{c \in I}\bigg(	
	\left[\cC{crs}{IFF (0,0)}\, \cS{s}{}\right]
	- \left[\cC{crs}{IFF}\cS{rs}{(0,0)}\, \cS{s}{}\right]
	\bigg)
	\Bigg]+\frac{1}{2}\sum_{\substack{r \in F \\ r \ne s}}\Bigg[\left[\cS{rs}{(0,0)}\, \cC{rs}{FF}\right]
	\\&+ \sum_{c \in I}\bigg(
	\left[\cC{crs}{IFF (0,0)}\, \cC{rs}{FF}\right]
	-\left[\cC{crs}{IFF}\cS{rs}{(0,0)}\, \cC{rs}{FF}\right]
	\bigg)\Bigg]-\sum_{\substack{r \in F \\ r \ne s}}\sum_{c \in I}
	\left[\cC{crs}{IFF (0,0)}\, \cC{rs}{FF}\cS{s}{}\right]\\&+\sum_{c \in I}\sum_{\substack{r \in F \\ r \ne s}} \bigg( \left[\cC{csr}{IFF (0,0)}\, \cC{cs}{IF}\right]
	+ \sum_{\substack{d \in I \\ d \ne c}}
	\left[\cC{cs,dr}{IF,IF (0,0)}\, \cC{cs}{IF}\right]
	 \bigg)\\&-\sum_{c \in I}\sum_{\substack{r \in F \\ r \ne s}}\bigg(
	\left[\cS{rs}{(0,0)}\, \cC{cs}{IF}\cS{s}{}\right]
	+ \left[\cC{csr}{IFF (0,0)}\,  \cC{cs}{IF}\cS{s}{}\right]
	- \left[\cC{csr}{IFF}\cS{rs}{(0,0)}\, \cC{cs}{IF}\cS{s}{}\right]
	\bigg)\Bigg\}\otimes\rd\sigma_{ab,X}^B\,.
\esp
\eeq
Finally, we need to sum over the flavors of the unresolved partons. While tedious, the procedure is straightforward and directly generalizes the discussion presented in appendix A of \cite{Bolzoni:2010bt}. After this summation is performed, we can write the integrated approximate cross section as a convolution of a single object, the $\bI_{12}^{(0)}$ \textit{insertion operator}, with the Born cross section of eq.~(\ref{eq:fullXSB}),
\beq
\bsp
    \int_2 \dsiga{RR}{12}_{ab}& =\, \left(\bI_{12}^{(0)}(\ep) \otimes \dsig{B}\right)_{ab}\\&=\, \int_0^1 \rd x_a\, \rd x_b\, f_{a/A}(x_a) f_{b/B}(x_b) \left[\int_0^1 \rd \eta_a\, \rd \eta_b\, \bI_{12;ac,bd}^{(0)}(\eta_a,\eta_b;\eps)\, \dsig{B}_{cd}(\eta_a x_a p_A,\eta_b x_b p_B)\right]\,.
\esp
\eeq
At this point, the operator is still to be understood as a distribution acting on the Born cross section. However, as explained in the previous section at the level of the separate integrated subtraction terms, for coding purposes it is more convenient to use an expansion in terms of coefficient functions. A slight complication here is that we actually have three distinct types of expansion, which we can denote by IFIF, SS and IFS; cf. eqs.~(\ref{eq:IFIF-distexp}), (\ref{eq:SS-distexp}) and (\ref{eq:SIF-distexp}) respectively. If we collect all counterterms of the same type into their appropriate $\bI$ operator we can symbolically write
\beq
\bsp
\label{eq:I12}
    \bI_{12}^{(0)}(\ep) =\, \frac{1}{S_{\{2\}}}\left(\bI^{(0),\text{IFIF}}_{12}(\ep)+\bI^{(0),\text{SS}}_{12}(\ep)+\bI^{(0),\text{IFS}}_{12}(\ep)\right)\,.
\esp
\eeq
Based on the results of the previous section, we can write the action of each of these on the Born cross section in the following way
\beq
\bsp
    \left(\bI_{12}^{(0),\text{IFIF}}(\ep) \otimes \dsig{B}\right)_{ab} =\,\int_{0}^{1}&\rd x_a\,\rd x_b\,\dsig{B}_{cd}(x_a p_A,x_b p_B)\,\int_{0}^{1}\rd \eta_a\,\rd \eta_b\,\\&\times\Bigg\{I_{12;ac,bd}^{(0),\text{IFIF}}(\eta_a,\eta_b;\eps\,|\,\eta_a,\eta_b)\frac{f_{a/A}(x_a/\eta_a)}{\eta_a}\frac{f_{b/B}(x_b/\eta_b)}{\eta_b}\\&+I_{12;ac,bd}^{(0),\text{IFIF}}(\eta_a,\eta_b;\eps\,|\,1,\eta_b)f_{a/A}(x_a)\frac{f_{b/B}(x_b/\eta_b)}{\eta_b}\\&+I_{12;ac,bd}^{(0),\text{IFIF}}(\eta_a,\eta_b;\eps\,|\,\eta_a,1)\frac{f_{a/A}(x_a/\eta_a)}{\eta_a}f_{b/B}(x_b)\\&+I_{12;ac,bd}^{(0),\text{IFIF}}(\eta_a,\eta_b;\eps\,|\,1,1)f_{a/A}(x_a)f_{b/B}(x_b)\Bigg\}\,,
\esp
\eeq

\beq
\bsp
    \left(\bI_{12}^{(0),\text{SS}}(\ep) \otimes \dsig{B}\right)_{ab} =\,\int_{0}^{1}\rd x_a\,\rd x_b\,\dsig{B}_{cd}(x_a\, p_A,x_b\, p_B)\int_{0}^{1}&\rd\eta\,\Bigg\{I_{12;ac,bd}^{(0),\text{SS}}(\eta;\eps\,|\,\eta)\frac{f_{a/A}(x_a/\eta)}{\eta}\frac{f_{b/B}(x_b/\eta)}{\eta}\\&+I_{12;ac,bd}^{(0),\text{SS}}(\eta;\eps\,|\,1)f_{a/A}(x_a)f_{b/B}(x_b)\Bigg\}
\esp
\eeq
and
\beq
\bsp
    &\left(\bI_{12}^{(0),\text{IFS}}(\ep) \otimes \dsig{B}\right)_{ab} =\,\int_0^1 \rd x_a\, \rd x_b\, \int_0^1 \rd \eta_a\, \int_0^{\eta_a} \rd \eta_b\,
\dsig{B}_{cd}(x_a p_A,x_b p_B) \\&\times\Bigg\{
 \bigg(I_{12;ac,bd}^{(0),\text{IFS}}(\eta_a,\eta_b;\eps\,|\,\eta_a,\eta_b)_{\eta_b<\eta_a}+I_{12;ac,bd}^{(0),\text{IFS}}(\eta_a,\eta_b;\eps\,|\,\eta_a,\eta_b)_{\eta_a<\eta_b}\bigg)
\frac{f_{a/A}(x_a/\eta_a)}{\eta_a}\frac{f_{b/B}(x_b/\eta_b)}{\eta_b}
\\&+I_{12;ac,bd}^{(0),\text{IFS}}(\eta_a,\eta_b;\eps\,|\,1,\eta_b)_{\eta_b<\eta_a} f_{a/A}(x_a) \frac{f_{b/B}(x_b/\eta_b)}{\eta_b}+I_{12;ac,bd}^{(0),\text{IFS}}(\eta_a,\eta_b;\eps\,|\,\eta_a,1)_{\eta_a<\eta_b} \frac{f_{a/A}(x_a/\eta_a)}{\eta_a} f_{b/B}(x_b)
\\&
+I_{12;ac,bd}^{(0),\text{IFS}}(\eta_a,\eta_b;\eps\,|\,\eta_a,\eta_a)_{\eta_b<\eta_a} \frac{f_{a/A}(x_a/\eta_a)}{\eta_a} \frac{f_{b/B}(x_b/\eta_a)}{\eta_a} \\&+I_{12;ac,bd}^{(0),\text{IFS}}(\eta_a,\eta_b;\eps\,|\,\eta_b,\eta_b)_{\eta_a<\eta_b} \frac{f_{a/A}(x_a/\eta_b)}{\eta_b} \frac{f_{b/B}(x_b/\eta_b)}{\eta_b} 
\\&+\bigg(I_{12;ac,bd}^{(0),\text{IFS}}(\eta_a,\eta_b;\eps\,|\,1,1)_{\eta_b<\eta_a}+I_{12;ac,bd}^{(0),\text{IFS}}(\eta_a,\eta_b;\eps\,|\,1,1)_{\eta_a<\eta_b}\bigg) f_{a/A}(x_a) f_{b/B}(x_b)
 \Bigg\}\,.
\esp
\eeq
Now, to write the action of the full $\bI_{12}^{(0)}$ insertion operator on the Born cross section in a uniform way, we rewrite the SS-type expansion as
\beq
\bsp
    \left(\bI_{12}^{(0),\text{SS}}(\ep) \otimes \dsig{B}\right)_{ab} =\,\frac{1}{2}&\int_0^1 \rd x_a\, \rd x_b\, \int_0^1 \rd \eta_a\, \int_0^{\eta_a} \rd \eta_b\,\dsig{B}_{cd}(x_a p_A,x_b p_B)\\&\times\Bigg\{I_{12;ac,bd}^{(0),\text{SS}}(\eta_a;\eps\,|\,\eta_a)\frac{f_{a/A}(x_a/\eta_a)}{\eta_a}\frac{f_{b/B}(x_b/\eta_a)}{\eta_a}\\&\quad+I_{12;ac,bd}^{(0),\text{SS}}(\eta_b;\eps\,|\,\eta_b)\frac{f_{a/A}(x_a/\eta_b)}{\eta_b}\frac{f_{b/B}(x_b/\eta_b)}{\eta_b}\\&\quad+\left[I_{12;ac,bd}^{(0),\text{SS}}(\eta_a;\eps\,|\,1)+I_{12;ac,bd}^{(0),\text{SS}}(\eta_b;\eps\,|\,1)\right] f_{a/A}(x_a) f_{b/B}(x_b)\Bigg\}\,.
\esp
\eeq
Collecting we then find
\beq
\bsp
    \int_2 \dsiga{RR}{12}_{ab} =\, \frac{1}{S_{\{2\}}}\int_0^1 \rd x_a\, \rd x_b\, &\int_0^1 \rd \eta_a\, \int_0^{\eta_a} \rd \eta_b\,\dsig{B}_{cd}(x_a p_A,x_b p_B)\\&\times\Bigg\{I_{12;ac,bd}^{(0)}(\eta_a,\eta_b;\eps\,|\,\eta_a,\eta_b)\frac{f_{a/A}(x_a/\eta_a)}{\eta_a}\frac{f_{b/B}(x_b/\eta_b)}{\eta_b}\\&\quad+I_{12;ac,bd}^{(0)}(\eta_a,\eta_b;\eps\,|\,1,\eta_b)f_{a/A}(x_a)\frac{f_{b/B}(x_b/\eta_b)}{\eta_b}\\&\quad+I_{12;ac,bd}^{(0)}(\eta_a,\eta_b;\eps\,|\,\eta_a,1)\frac{f_{a/A}(x_a/\eta_a)}{\eta_a}f_{b/B}(x_b)\\&\quad+I_{12;ac,bd}^{(0)}(\eta_a,\eta_b;\eps\,|\,\eta_a,\eta_a)\frac{f_{a/A}(x_a/\eta_a)}{\eta_a}\frac{f_{b/B}(x_b/\eta_a)}{\eta_a}\\&\quad+I_{12;ac,bd}^{(0)}(\eta_a,\eta_b;\eps\,|\,\eta_b,\eta_b)\frac{f_{a/A}(x_a/\eta_b)}{\eta_b}\frac{f_{b/B}(x_b/\eta_b)}{\eta_b}\\&\quad+I_{12;ac,bd}^{(0)}(\eta_a,\eta_b;\eps\,|\,1,1)f_{a/A}(x_a)f_{b/B}(x_b)\Bigg\}
\esp
\eeq
where
\begin{align}
\label{eq:I12-in}
    &I_{12;ac,bd}^{(0)}(\eta_a,\eta_b;\eps\,|\,\eta_a,\eta_b)=\,I^{\text{IFIF}}_{12;ac,bd}(\eta_a,\eta_b;\eps\,|\,\eta_a,\eta_b)+I^{\text{SIF}}_{12;ac,bd}(\eta_a,\eta_b;\eps\,|\,\eta_a,\eta_b)_{\eta_a<\eta_b}\nn\\&\qquad\qquad\qquad\qquad\qquad\quad+I^{\text{SIF}}_{12;ac,bd}(\eta_a,\eta_b;\eps\,|\,\eta_a,\eta_b)_{\eta_b<\eta_a}\,,\\
    &I_{12;ac,bd}^{(0)}(\eta_a,\eta_b;\eps\,|\,1,\eta_b)=\,I^{\text{IFIF}}_{12;ac,bd}(\eta_a,\eta_b;\eps\,|\,1,\eta_b)+I^{\text{SIF}}_{12;ac,bd}(\eta_a,\eta_b;\eps\,|\,1,\eta_b)_{\eta_b<\eta_a}\,,\\
    &I_{12;ac,bd}^{(0)}(\eta_a,\eta_b;\eps\,|\,\eta_a,1)=\,I^{\text{IFIF}}_{12;ac,bd}(\eta_a,\eta_b;\eps\,|\,\eta_a,1)+I^{\text{SIF}}_{12;ac,bd}(\eta_a,\eta_b;\eps\,|\,\eta_a,1)_{\eta_a<\eta_b}\,,\\
    &I_{12;ac,bd}^{(0)}(\eta_a,\eta_b;\eps\,|\,\eta_a,\eta_a)=\,\frac{1}{2}I^{\text{SS}}_{12;ac,bd}(\eta_a;\eps\,|\,\eta_a)+I^{\text{SIF}}_{12;ac,bd}(\eta_a,\eta_b;\eps\,|\,\eta_a,\eta_a)_{\eta_b<\eta_a}\,,\\
    &I_{12;ac,bd}^{(0)}(\eta_a,\eta_b;\eps\,|\,\eta_b,\eta_b)=\,\frac{1}{2}I^{\text{SS}}_{12;ac,bd}(\eta_b;\eps\,|\,\eta_b)+I^{\text{SIF}}_{12;ac,bd}(\eta_a,\eta_b;\eps\,|\,\eta_b,\eta_b)_{\eta_a<\eta_b}\,,\\
    &I_{12;ac,bd}^{(0)}(\eta_a,\eta_b;\eps\,|\,1,1)=\,I^{\text{IFIF}}_{12;ac,bd}(\eta_a,\eta_b;\eps\,|\,1,1)+\frac{1}{2}I^{\text{SS}}_{12;ac,bd}(\eta_a;\eps\,|\,1)+\frac{1}{2}I^{\text{SS}}_{12;ac,bd}(\eta_b;\eps\,|\,1)\nn\\&\qquad\qquad\qquad\qquad\qquad\quad+I^{\text{SIF}}_{12;ac,bd}(\eta_a,\eta_b;\eps\,|\,1,1)_{\eta_a<\eta_b}+I^{\text{SIF}}_{12;ac,bd}(\eta_a,\eta_b;\eps\,|\,1,1)_{\eta_b<\eta_a}\,.
\label{eq:I12-out}
\end{align}

\subsection{Application to Higgs boson production in HEFT}
\begin{figure}[b]
    \centering
    \includegraphics{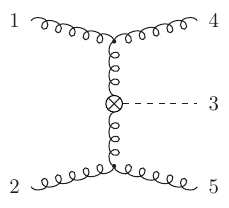}
    \caption{Sample diagram for the purely gluonic double-real emission subprocess for Higgs production in the HEFT approximation.}
    \label{fig:Hgg}
\end{figure}

To illustrate the construction and functional form of the coefficient functions of the $\bI_{12}^{(0)}$ insertion operator, we consider the production of a Higgs boson in the HEFT approximation. We focus on the purely gluonic subprocess and discard light-quark contributions, cf. fig.~\ref{fig:Hgg}. Note that this subprocess leads to the most complicated IR structure for color-singlet production.
While it is possible to write down the full expression of the insertion operator in a completely generic way (following a generalization of the procedure outlined in Appendix A of \cite{Bolzoni:2010bt}), in practice this is a cumbersome and, in fact, unnecessary, job. A much more convenient way to go about this is to simply list out all subtractions required for the regularization of the IR behavior of the cross section. For each subtraction, we add back the integrated version, and collecting the latter reconstructs the insertion operator. Consider, for example, the IF--IF iterated subtraction terms, $\cC{asr}{IFF(0,0)}\cC{as}{IF}$ and $\cC{as,br}{IF,IF(0,0)}\cC{as}{IF}$. Using the labelling of fig.~\ref{fig:Hgg}, it is clear that we need to account for the following contributions:
\beq
\bsp
    &\Bigg\{\cC{145}{IFF(0,0)}\cC{14}{IF}, \cC{145}{IFF(0,0)}\cC{15}{IF},\cC{245}{IFF(0,0)}\cC{24}{IF}, \cC{245}{IFF(0,0)}\cC{25}{IF},\cC{14,25}{IF,IF(0,0)}\cC{14}{IF},\cC{14,25}{IF,IF(0,0)}\cC{25}{IF},\\&\quad\cC{15,24}{IF,IF(0,0)}\cC{15}{IF},\cC{15,24}{IF,IF(0,0)}\cC{24}{IF}\Bigg\}\,.
\esp
\eeq
So, denoting their integrated version with square brackets, the $\bI_{12}^{(0),\text{IFIF}}(\eps)$ operator becomes
\beq
\bsp
   \bI_{12}^{(0),\text{IFIF}}(\eps) =\,&\, [\cC{145}{IFF(0,0)}\cC{14}{IF}]+[\cC{145}{IFF(0,0)}\cC{15}{IF}]+[\cC{245}{IFF(0,0)}\cC{24}{IF}]+[\cC{245}{IFF(0,0)}\cC{25}{IF}]\\&+[\cC{14,25}{IF,IF(0,0)}\cC{14}{IF}]+[\cC{14,25}{IF,IF(0,0)}\cC{25}{IF}]+[\cC{15,24}{IF,IF(0,0)}\cC{15}{IF}]+[\cC{15,24}{IF,IF(0,0)}\cC{24}{IF}]\,.
\esp
\eeq
This exercise can easily be repeated for the remaining subtraction terms. We find
\beq
\bsp
    \bI_{12}^{(0),\text{SS}}(\eps) &=\, [\cS{45}{(0,0)}\cS{4}{ }]+[\cS{45}{(0,0)}\cS{5}{ }]-[\cC{145}{IFF}\cS{45}{(0,0)}\cS{4}{}]-[\cC{145}{IFF}\cS{45}{(0,0)}\cS{5}{}]-[\cC{245}{IFF}\cS{45}{(0,0)}\cS{4}{}]-[\cC{245}{IFF}\cS{45}{(0,0)}\cS{5}{}]\\&-[\cS{45}{(0,0)}\cC{14}{IF}\cS{4}{}]-[\cS{45}{(0,0)}\cC{15}{IF}\cS{5}{}]-[\cS{45}{(0,0)}\cC{24}{IF}\cS{4}{}]-[\cS{45}{(0,0)}\cC{25}{IF}\cS{5}{}]+[\cC{145}{IFF}\cS{45}{(0,0)}\cC{14}{IF}\cS{4}{}]\\&+[\cC{145}{IFF}\cS{45}{(0,0)}\cC{15}{IF}\cS{5}{}]+[\cC{245}{IFF}\cS{45}{(0,0)}\cC{24}{IF}\cS{4}{}]+[\cC{245}{IFF}\cS{45}{(0,0)}\cC{25}{IF}\cS{5}{}]+[\cS{45}{(0,0)}\cC{45}{FF}]-[\cC{145}{IFF}\cS{45}{(0,0)}\cC{45}{FF}]\\&-[\cC{245}{IFF}\cS{45}{(0,0)}\cC{45}{FF}]
\esp
\eeq
and
\beq
\bsp
    \bI_{12}^{(0),\text{IFS}}(\eps) &=\, [\cC{145}{IFF(0,0)}\cS{4}{}]+[\cC{145}{IFF(0,0)}\cS{5}{}]+[\cC{245}{IFF(0,0)}\cS{4}{}]+[\cC{245}{IFF(0,0)}\cS{5}{}]-[\cC{145}{IFF(0,0)}\cC{45}{FF}\cS{4}{}]\\&-[\cC{145}{IFF(0,0)}\cC{45}{FF}\cS{5}{}]-[\cC{245}{IFF(0,0)}\cC{45}{FF}\cS{4}{}]-[\cC{245}{IFF(0,0)}\cC{45}{FF}\cS{5}{}]-[\cC{145}{IFF(0,0)}\cC{14}{IF}\cS{4}{}]\\&-[\cC{145}{IFF(0,0)}\cC{15}{IF}\cS{5}{}]-[\cC{245}{IFF(0,0)}\cC{24}{IF}\cS{4}{}]-[\cC{245}{IFF(0,0)}\cC{25}{IF}\cS{5}{}]+[\cC{145}{IFF(0,0)}\cC{45}{FF}]+[\cC{245}{IFF(0,0)}\cC{45}{FF}]\,.
\esp
\eeq
Hence, the full $\bI_{12}^{(0)}(\eps)$ operator, as defined in eq.~(\ref{eq:I12}) with $S_{\{2\}}=2$ for the two gluons in the final state, receives 39 distinct contributions. For concreteness, we present the pole parts of the corresponding coefficient functions in Appendix \ref{appx:Ipoles}. We note that these are much more elaborate than the poles of the full $\bI$ operator, which were presented in \cite{DelDuca:2024ovc}. One of the main reasons for this difference in complexity is that, at the level of the insertion operators, there are large cancellations between $A_{12}$ and the subtractions to the integrated $A_1$ counterterms. The latter are a necessary ingredient in the regularization of IR singularities in the real-virtual cross section, which will be discussed in a future publication.

\section{Conclusions}
\label{sec:conclusion}
In this paper, we addressed the integration of the iterated single-unresolved subtraction terms of the \colorful scheme for hadron collisions. In particular, we concentrated on the case of color-singlet production, which involves a self-contained subset of all subtraction terms necessary to deal with generic hadron-initiated processes. We demonstrated that the integrated subtraction terms must be understood as distributions. We explained in detail how to rewrite the hadronic integrated approximate cross section in such a way that these distributions act via convolutions on the parton density functions. We then showed that the corresponding distributions can be represented by their coefficient functions. In fact, this way of presenting the integrated counterterms is the most convenient from the point of view of a numerical implementation.

Then we addressed the concrete computation of the various coefficient functions. For the $A_{12}$ subtractions, we found it most convenient to consider an explicit parametric representation of the integrated counterterms, which we were then able to evaluate in terms of generalized polylogarithms. We presented our integration recipe in detail and included an elaborate appendix where all necessary steps are explicitly discussed for one of the most cumbersome cases. We hope that this level of detail will be useful for the adventurous graduate student or postdoc taking on similar calculations.

Finally, we presented the organization of the integrated counterterms into the $\bI^{(0)}_{12}$ insertion operator, whose structure we showcased for the fully gluonic subprocess of Higgs boson production in the HEFT approximation. We note that in terms of its infrared structure, this is the most complicated subprocess which can arise in color-singlet production. In order to give a flavor of our final expressions, the pole parts of the insertion operator were also presented in an appendix.

Although our discussion has largely focused on color-singlet processes, we stress that most of our results are in fact fully generic. First, our integration recipe does not depend in any way on the absence of colored particles in the final state. As such, the same method can be applied when extending our formalism to account for jet production. Second, the majority of the results we presented are fully generic and require no modification once processes beyond color-singlet production are considered. Indeed, as highlighted in the main text, only a few of our integrated subtraction terms are truly specific to the color-singlet case. These terms involve a soft mapping and contain eikonal factors with hard momenta. The kinematics of the hard momenta are considerably simpler for color-singlet production than for generic processes and, as such, the corresponding integrals for the latter will have to be adapted accordingly. Furthermore, the set of (integrated) subtraction terms will have to be extended with terms that do not contribute to color-singlet production. A detailed investigation of these extensions is left for future studies.

\subsection*{Acknowledgments}
\vspace*{-1mm}
We are grateful to V.~Del Duca, C.~Duhr, F.~Guadagni, P.~Mukherjee and F.~Tramontano for many useful discussions. This work has been supported by grant K143451 of the National Research, Development and Innovation Fund in Hungary and the Bolyai Fellowship program of the Hungarian Academy of Sciences. The work of L.F.\ was supported by the German Academic Exchange Service (DAAD) through its Bi-Nationally Supervised Scholarship program.

\appendix

\renewcommand{\theequation}{\ref{appx:dglap}.\arabic{equation}}
\setcounter{equation}{0}
\renewcommand{\thefigure}{\ref{appx:dglap}.\arabic{figure}}
\setcounter{figure}{0}
\renewcommand{\thetable}{\ref{appx:dglap}.\arabic{table}}
\setcounter{table}{0}

\section{Collinear splitting functions}
\label{appx:dglap}
The tree-level Altarelli-Parisi splitting functions for the final-final splitting are~\cite{Altarelli:1977zs}
\bal
\la s|\hP^{(0)}_{q g}(z,\kTm{};\eps)|s'\ra &=\,
	\CF \left[\frac{1 + z^2}{1 - z} - \eps (1 - z)\right] \delta_{ss'}\,,
\label{eq:Pqg0FF}
\\
\la s|\hP^{(0)}_{g q}(z,\kTm{};\eps)|s'\ra &=\,
	\CF \left[\frac{1 + (1 - z)^2}{z} - \eps z\right] \delta_{ss'}\,,	
\label{eq:Pgq0FF}
\\
\la \mu|\hP^{(0)}_{q \qb}(z,\kTm{};\eps)|\nu\ra &=\,
	\TR \left[-g^{\mu\nu} + 4 z(1 - z) 
	\frac{\kTm{\mu} \kTm{\nu}}{\kTm{2}}\right]\,,	
\label{eq:Pqq0FF}
\\
\la \mu|\hP^{(0)}_{g g}(z,\kTm{};\eps)|\nu\ra &=\,
	2\CA \left[
	-g^{\mu\nu}\left(\frac{z}{1 - z} + \frac{1 - z} {z}\right) 
	-2(1 - \eps)z(1 - z) 
	\frac{\kTm{\mu} \kTm{\nu}}{\kTm{2}}\right]\,,
\label{eq:Pgg0FF}
\eal
from which we may derive the azimuthally averaged splitting functions,
\bal
P_{qg}^{(0)}(z;\eps) &=\,
	\CF\left[\frac{1+z^2}{1-z} - \eps(1-z)\right]\,,
\label{eq:Pqg-ave}
\\
P_{gq}^{(0)}(z;\eps) &=\,
	\CF\left[\frac{1+(1-z)^2}{z} - \eps z\right]\,,
\label{eq:Pgq-ave}
\\
P_{q\qb}^{(0)}(z;\eps) &=\,
	\TR\left[1 - \frac{2z(1-z)}{1-\eps}\right]\,,
\label{eq:Pqq-ave}
\\
P_{gg}^{(0)}(z;\eps) &=\,
	2\CA\left[\frac{z}{1-z} + \frac{1-z}{z} + z(1-z)\right]\,.
\label{eq:Pgg-ave}
\eal
The Altarelli-Parisi kernels for initial-final splitting are obtained from \eqnss{eq:Pqg0FF}{eq:Pgg0FF} through the crossing relation
\beq
\hP_{(ar) r}^{(0)}(x,k_\perp;\ep) = 
	-(-1)^{F(a)+F(ar)} x \hP_{\ba{a}r}(1/x,k_\perp/x;\ep)\,,
	\label{eq:ifcross}
\eeq
and read 
\bal
\la s|\hP^{(0)}_{q_a g_r}(x,\kTm{};\eps)|s'\ra &=\,
	\CF \left[\frac{1 + x^2}{1 - x} - \eps (1 - x)\right] \delta_{ss'}\,,
\label{eq:Pqg0IF}
\\
\la \mu|\hP^{(0)}_{g_a q_r}(x,\kTm{};\eps)|\nu\ra &=\,
	\TR \left[-g^{\mu\nu} x - 4\frac{1-x}{x} 
	\frac{\kTm{\mu} \kTm{\nu}}{\kTm{2}}\right] \,,	
\\
\la s|\hP^{(0)}_{q_a \qb_r}(x,\kTm{};\eps)|s'\ra &=\,
	\CF \left[1 - \eps - 2 x(1  -x)\right] \delta_{ss'}\,,	
\\
\la \mu|\hP^{(0)}_{g_a g_r}(x,\kTm{};\eps)|\nu\ra &=\,
	2\CA \left[
	-g^{\mu\nu}\left(x(1 - x) + \frac{x} {1-x}\right) 
	-2(1 - \eps)\frac{1 - x}{x} 
	\frac{\kTm{\mu} \kTm{\nu}}{\kTm{2}}\right]\,.
\label{eq:Pgg0IF}
\eal
The azimuthally averaged initial-final splitting functions are
\bal
P_{qg}^{(0)}(x;\eps) &=\,
	\CF\left[\frac{1+x^2}{1-x} - \eps(1-x)\right]\,,
\label{eq:Pqg-ave-IF}
\\
P_{gq}^{(0)}(x;\eps) &=\,
	\TR\left[x + 2\frac{1-x}{(1-\eps)x}\right]\,,
\label{eq:Pgq-ave-IF}
\\
P_{q\qb}^{(0)}(x;\eps) &=\,
	\CF\left[1 - \eps - 2x(1-x)\right]\,,
\label{eq:Pqq-ave-IF}
\\
P_{gg}^{(0)}(x;\eps) &=\,
	2\CA\left[x(1-x) + \frac{x}{1-x} + \frac{1-x}{x}\right]\,.
\label{eq:Pgg-ave-IF}
\eal

\renewcommand{\theequation}{\ref{appx:SO-AP-functions}.\arabic{equation}}
\setcounter{equation}{0}
\renewcommand{\thefigure}{\ref{appx:SO-AP-functions}.\arabic{figure}}
\setcounter{figure}{0}
\renewcommand{\thetable}{\ref{appx:SO-AP-functions}.\arabic{table}}
\setcounter{table}{0}
\section{Strongly-ordered collinear splitting functions}
\label{appx:SO-AP-functions}

In this appendix, we recall the tree-level strongly-ordered splitting functions for final-state splitting~\cite{Somogyi:2005xz}.
For quark splitting we have
\beeq
\lefteqn{ \la s|\hP^{\mathrm{(C)}, (0)}_{q_i (\qb_r' q')}
	(z_{r,s}, \kT{r,s}, \kTt{rs}, z_{\ha{i},\wha{rs}},  
	\kT{\ha{i}\wha{rs}}, \ha{p}_i; \eps)|s'\ra } \nn\\
	&& =\,\TR \Bigg[
	P^{(0)}_{q_{\ha{i}} g_{\wha{rs}}}(z_{\ha{i},\wha{rs}};\eps)
	- 2\CF z_{r,s} (1-z_{r,s})\left(1 - z_{\ha{i},\wha{rs}} - 
	\frac{s_{\ha{i}\kT{r,s}}^2}{\kT{r,s}^2 s_{\ha{i}\wha{rs}}}\right)\Bigg] 
	\delta_{ss'}\,,
\label{eq:CFFPqqbq0FFF}
\eeeq
\beq
\la s|\hP^{\mathrm{(C)}, (0)}_{g_i (q_r g_s)}
	(z_{r,s}, \kT{r,s}, \kTt{rs}, z_{\ha{i},\wha{rs}},  
	\kT{\ha{i}\wha{rs}}, \ha{p}_i; \eps)|s'\ra =\,
	P^{(0)}_{q_{\wha{rs}} g_{\ha{i}}}(z_{\wha{rs},\ha{i}};\eps)
	P^{(0)}_{q_r g_s}(z_{r,s};\eps)
	\delta_{ss'}\,,
\label{eq:CFFPgqg0FFF}
\eeq
\beeq
\lefteqn{ \la s|\hP^{\mathrm{(C)}, (0)}_{q_i (g_r g_s)}
	(z_{r,s}, \kT{r,s}, \kTt{rs}, z_{\ha{i},\wha{rs}},  
	\kT{\ha{i}\wha{rs}}, \ha{p}_i; \eps)|s'\ra } \label{eq:CFFPqgg0FFF}\\
	&& \!\!\!\!\!\!\!\!\!\!\!\! =\,2\CA \Bigg[
	P^{(0)}_{q_{\ha{i}} g_{\wha{rs}}}(z_{\ha{i},\wha{rs}};\eps)
	\left(\frac{z_{r,s}}{1-z_{r,s}} + \frac{1-z_{r,s}}{z_{r,s}}\right)
	+\CF (1-\eps) z_{r,s} (1-z_{r,s})\left(1 - z_{\ha{i},\wha{rs}} - 
	\frac{s_{\ha{i}\kT{r,s}}^2}{\kT{r,s}^2 s_{\ha{i}\wha{rs}}}\right)\Bigg] 
	\delta_{ss'}\,.  \nn
\eeeq
For gluon splitting we have
\beq
\la \mu|\hP^{\mathrm{(C)}, (0)}_{\qb_i (q_r g_s)}
	(z_{r,s}, \kT{r,s}, \kTt{rs}, z_{\ha{i},\wha{rs}},  
	\kT{\ha{i}\wha{rs}}, \ha{p}_i; \eps)|\nu\ra =\,
	\la \mu |\hP^{(0)}_{\qb_{\ha{i}} q_{\wha{rs}}}
	(z_{\wha{rs},\ha{i}}, \kT{\wha{rs}\ha{i}}; \eps)| \nu \ra 
	P^{(0)}_{q_r g_s}(z_{r,s}; \eps)\,,
	\label{eq:CFFPqbqg0FFF}
\eeq
\beeq
\lefteqn{ \la \mu|\hP^{\mathrm{(C)}, (0)}_{g_i (q_r \qb)}
	(z_{r,s}, \kT{r,s}, \kTt{rs}, z_{\ha{i},\wha{rs}},  
	\kT{\ha{i}\wha{rs}}, \ha{p}_i; \eps)| \nu \ra } \nn\\
	&=&
	2\CA \TR \Bigg[-g^{\mu\nu}\left(\frac{z_{\ha{i},\wha{rs}}}{1-z_{\ha{i},\wha{rs}}}
	+ \frac{1-z_{\ha{i},\wha{rs}}}{z_{\ha{i},\wha{rs}}}
	+ z_{r,s}(1-z_{r,s})\frac{s_{\ha{i}\kT{r,s}}^2}{\kT{r,s}^2 s_{\ha{i}\wha{rs}}}\right)
	+ 4 z_{r,s}(1-z_{r,s}) \frac{1-z_{\ha{i},\wha{rs}}}{z_{\ha{i},\wha{rs}}}
	\frac{\kTt{rs}^\mu \kTt{rs}^\nu}{\kTt{rs}^2}\Bigg] \nn\\
	&-& 4 \CA (1-\eps) z_{\ha{i},\wha{rs}}(1-z_{\ha{i},\wha{rs}}) P^{(0)}_{q_r \qb}(z_{r,s};\eps)
	\frac{\kT{\ha{i}\wha{rs}}^\mu \kT{\ha{i}\wha{rs}}^\nu}{\kT{\ha{i}\wha{rs}}^2}\,,
\label{eq:CFFPgqqb0FFF}
\eeeq
\beeq
\lefteqn{ \la \mu|\hP^{\mathrm{(C)}, (0)}_{g_i (g_r g_s)}
	(z_{r,s}, \kT{r,s}, \kTt{rs}, z_{\ha{i},\wha{rs}},  
	\kT{\ha{i}\wha{rs}}, \ha{p}_i; \eps)| \nu \ra } \nn\\
	&=\,&
	4\CA^2 \Bigg[-g^{\mu\nu}\left(\frac{z_{\ha{i},\wha{rs}}}{1-z_{\ha{i},\wha{rs}}}
	+ \frac{1-z_{\ha{i},\wha{rs}}}{z_{\ha{i},\wha{rs}}}\right)
	\left(\frac{z_{r,s}}{1-z_{r,s}} + \frac{1-z_{r,s}}{z_{r,s}}\right) \nn\\
	&& \qquad + g^{\mu\nu} z_{r,s}(1-z_{r,s}) \frac{1-\eps}{2}
	\frac{s_{\ha{i}\kT{r,s}}^2}{\kT{r,s}^2 s_{\ha{i}\wha{rs}}}
	-2 (1-\eps) z_{r,s}(1-z_{r,s})
	\frac{1-z_{\ha{i},\wha{rs}}}{z_{\ha{i},\wha{rs}}}
	\frac{\kTt{rs}^\mu \kTt{rs}^\nu}{\kTt{rs}^2}\Bigg] \nn\\
	&-& 4 \CA (1-\eps) z_{\ha{i},\wha{rs}}(1-z_{\ha{i},\wha{rs}}) P^{(0)}_{g_r g_s}(z_{r,s};\eps)
	\frac{\kT{\ha{i}\wha{rs}}^\mu \kT{\ha{i}\wha{rs}}^\nu}{\kT{\ha{i}\wha{rs}}^2}\,,
\label{eq:CFFPggg0FFF}
\eeeq
where the azimuthally averaged tree-level splitting functions $P^{(0)}$ are given in \eqnss{eq:Pqg-ave}{eq:Pgg-ave}.
The strongly-ordered splitting functions for initial-state splitting can be obtained from those for final-state splitting by crossing relations. These were derived in~\cite{DelDuca:2025yph} and here we limit ourselves to recalling the final results. For $a\to a+(rs)\to a+r+s$ splitting the strongly-ordered splitting function is obtained by the 
crossing relation, 
\beq
\bsp
&
\hP^{\mathrm{(C)}, (0)}_{(ars) r s}
(z_r, \kT{r}, \kTh{rs}, x_{\ha{a}}, \kT{\wha{rs}}, \ha{p}_a;\ep)
\\ &\qquad= 
	-(-1)^{F(a)+F(ars)} x_{\ha{a}}
	\hP^{\mathrm{(C)}, (0)}_{\ba{a} rs}
	(z_r, \kT{r}, \kTh{rs}, 1/x_{\ha{a}}, 
    \kT{\wha{rs}}/x_{\ha{a}}, -\ha{p}_a;\ep)\,,
\label{eq:PCars-crossing}
\esp
\eeq
while the $a\to (as)+r\to a+r+s$ strongly-ordered splitting function is given by the following crossing formula, 
\beq
\bsp&
\hP^{\mathrm{(C)}, (0)}_{(ars) (as) s}
(x_a, \kT{s}, \kTh{s}, x_{\ha{a}}, \kT{\ha{r}}, \ha{p}_r;\ep) 
\\&\qquad= 
	(-1)^{F(as)+F(ars)} x_a x_{\ha{a}}
	\hP^{\mathrm{(C)}, (0)}_{r \ba{a} s}(1/x_a, \kT{s}/x_a, 
    \kTh{s}/x_a, 1/x_{\ha{a}}, \kT{\ha{r}}/x_{\ha{a}}, 
    \ha{p}_r;\ep)\,.
\esp	
\label{eq:PCars-crossing-IF}
\eeq
Note that the flavors of the partons involved in the most collinear splitting always appear as the second and third index 
in the subscript.

\renewcommand{\theequation}{\ref{appx:Soft-AP-functions}.\arabic{equation}}
\setcounter{equation}{0}
\renewcommand{\thefigure}{\ref{appx:Soft-AP-functions}.\arabic{figure}}
\setcounter{figure}{0}
\renewcommand{\thetable}{\ref{appx:Soft-AP-functions}.\arabic{table}}
\setcounter{table}{0}
\section{The soft limit of collinear splitting functions}
\label{appx:Soft-AP-functions}
Here we present the soft functions $P^{(\mathrm S),(0)}_{irs}$ that appear in the single soft limit of the tree-level splitting functions for final-state splitting. They represent the single soft limit of a set of $n$ collinear partons~\cite{DelDuca:2019ggv}, for $n=3$ and read~\cite{Somogyi:2005xz},
\bal
P^{(\mathrm S) (0)}_{f_i f_r (q_s)}(z_i,z_r,z_s,s_{ir},s_{is},s_{rs}) &=\,0\,,
\label{eq:SsP1}
\\
P^{(\mathrm S) (0)}_{q_i g_r (g_s)}(z_i,z_r,z_s,s_{ir},s_{is},s_{rs}) &=\,
	\CF \frac{2}{s_{is}}\frac{z_i}{z_s}
	+ \CA\left(
		\frac{s_{ir}}{s_{is} s_{rs}} 
		- \frac{1}{s_{is}} \frac{z_i}{z_s}
		+ \frac{1}{s_{rs}} \frac{z_r}{z_s}
		\right)\,,
\label{eq:SsP2}
\\
P^{(\mathrm S) (0)}_{g_i q_r (g_s)}(z_i,z_r,z_s,s_{ir},s_{is},s_{rs}) &=\,
	\CF \frac{2}{s_{rs}}\frac{z_r}{z_s}
	+ \CA\left(
		\frac{s_{ir}}{s_{is} s_{rs}} 
		+ \frac{1}{s_{is}} \frac{z_i}{z_s}
		- \frac{1}{s_{rs}} \frac{z_r}{z_s}
		\right)\,,
\label{eq:SsP3}
\\
P^{(\mathrm S) (0)}_{q_i \qb_r (g_s)}(z_i,z_r,z_s,s_{ir},s_{is},s_{rs}) &=\,
	\CF \frac{2 s_{ir}}{s_{is} s_{rs}}
	+ \CA\left(
		-\frac{s_{ir}}{s_{is} s_{rs}} 
		+ \frac{1}{s_{is}} \frac{z_i}{z_s}
		+ \frac{1}{s_{rs}} \frac{z_r}{z_s}
		\right)\,,
\label{eq:SsP4}
\\
P^{(\mathrm S) (0)}_{g_i g_r (g_s)}(z_i,z_r,z_s,s_{ir},s_{is},s_{rs}) &=\,
	\CA\left(
		\frac{s_{ir}}{s_{is} s_{rs}} 
		+ \frac{1}{s_{is}} \frac{z_i}{z_s}
		+ \frac{1}{s_{rs}} \frac{z_r}{z_s}
		\right)\,.
\label{eq:SsP5}
\eal
The index of the soft parton $s$ appears last in the subscript. Finally, notice that \eqnss{eq:SsP1}{eq:SsP5} can be written in a uniform way~\cite{Bolzoni:2010bt},
\beq
\bsp
&
P^{(\mathrm S), (0)}_{f_i f_r g_s}
(z_i,z_r,z_s,s_{ir},s_{is},s_{rs}) 
\\&\qquad=
(\bT_{i}^2+\bT_r^2-\bT_{ir}^2) \frac{s_{ir}}{s_{is} s_{rs}} 
+(\bT_{ir}^2+\bT_i^2-\bT_r^2) \frac{1}{s_{is}} \frac{z_i}{z_s}
+(\bT_{ir}^2-\bT_i^2+\bT_r^2) \frac{1}{s_{rs}} \frac{z_r}{z_s}\,.
\esp
\eeq
The corresponding soft functions for initial-state splitting can be obtained by the crossing relation given in eq.~(\ref{eq:sapspliIFF}).

\renewcommand{\theequation}{\ref{appx:ICasrCas}.\arabic{equation}}
\setcounter{equation}{0}
\renewcommand{\thefigure}{\ref{appx:ICasrCas}.\arabic{figure}}
\setcounter{figure}{0}
\renewcommand{\thetable}{\ref{appx:ICasrCas}.\arabic{table}}
\setcounter{table}{0}

\section{
\texorpdfstring{Integrating the $\cC{asr}{IFF(0,0)}\cC{as}{IF}$ subtraction term}{Integrating the CasrIFF00CasIF subtraction term}}
\label{appx:ICasrCas}
As mentioned in the main text, the integration of all $A_{12}$ subtraction terms follows more or less the same methodology. In this appendix, we provide an in-depth presentation of all computational steps applied to the $\cC{asr}{IFF(0,0)}\cC{as}{IF}$ subtraction term. For completeness, we repeat the explicit definition of both the counterterm in sec.~\ref{sec:CarsCas-ct} and its integral over the unresolved phase space in sec.~\ref{sec:defIC}. In the latter section, we also present in detail the determination of the parametric representation of the phase space measure. The actual computation of the integrated counterterm is then presented in secs.~\ref{sec:prepint}--\ref{sec:hadronint}.

\stoptoc
\subsection{The $\cC{asr}{IFF(0,0)}\cC{as}{IF}$ subtraction term}
\label{sec:CarsCas-ct}
The subtraction term is defined to be~\cite{DelDuca:2025yph}
\beq
\bsp
\cC{asr}{IFF (0,0)}\cC{as}{IF}(p_a,p_b;\mom{}_{X+m+2}) &= 
	(8\pi\alpha_s\mu^{2\ep})^2
	\frac{1}{x_{a,s} s_{as} x_{\ha{a},\ha{r}} s_{\ha{a}\ha{r}}}
	\hP^{\mathrm{(C)},(0)}_{(ars) (as)s}
	(x_{a,s}, \kT{s,a}, \kTh{s,a}, x_{\ha{a},\ha{r}}, 
    \kT{\ha{r},\ha{a}}, \ha{p}_r; \ep)
\\&\times
    \SME{(ars)b,X+m}{(0)}
    {(\ha{\ha{p}}_a,\ha{\ha{p}}_b;\momhh{}_{X+m})} {\cal F}
    (x_{\ha{a},\ha{r}},\xi_{\ha{a},\ha{r}}\xi_{\ha{b},\ha{r}}) 
	\,,
\esp
\label{eq:carscas}
\eeq
with $\mu$ the arbitrary scale introduced by dimensional regularization and\footnote{The function ${\cal F}$ originates from the $\cC{asr}{IFF(0,0)}$ counterterm in $A_2$, where it is introduced to cancel spurious poles introduced by the definitions of the momentum fractions that appear in the subtraction. The fact that it also needs to be included here is tied to the delicate structure of cancellations between the subtraction terms which needs to be preserved.}
\begin{align}
    \label{eq:Fiff}
    {\cal F}(x_{\ha{a},\ha{r}},\xi_{\ha{a},\ha{r}}\xi_{\ha{b},\ha{r}})  &=\,	
	\left[1-\frac{s_{\ha{r} Q }}{s_{\ha{a} Q }}\right]^2	
	\left[1-\frac{s_{\ha{r}\ha{Q}}}{s_{\ha{a}\ha{Q}}}\right]^{-2}\,,\\
    x_{a,s} &=\,1-\frac{s_{s Q }}{s_{ab}}\,,\\
    \label{eq:xahrh}
    x_{\hat{a},\hat{r}} &=\,1-\frac{s_{\hat{r} Q }}{s_{\hat{a} Q }} \,,\\
    s_{k(ij)} &=\,s_{ik}+s_{jk}\,,\\
    s_{ij} &=\,2p_i\cdot p_j\,,\\
    Q^{\mu} &=\, p_a^{\mu}+p_b^{\mu}\,,\\
    \kT{s,a}^\mu &=\, p_s^\mu -  \frac{s_{bs}}{s_{ab}} p_a^\mu 
	- \frac{s_{as}}{s_{ab}} p_b^\mu\,.
\end{align}
The subscripts of the splitting function and reduced matrix element represent the flavors of partons, with $(as)\equiv a+s$ and $(ars)\equiv a+r+s$. Flavor combinations are computed in the standard way,
\begin{align}
    &q + \overline{q} =\,g,\nn\qquad 
    \overline{q} + q =\,g,\nn\qquad
q + g =\,q,\nn\\
& g + q =\,q,\qquad
\overline{q} + g =\,\overline{q},\qquad
g + \overline{q} =\,\overline{q},\\
&g + g =\,g,\nn\qquad
q + q =\,g,\nn\qquad
\overline{q} + \overline{q} =\,g\,.\nn
\end{align}
The momentum mapping, which takes $(p_a,p_b;\{p\}_{X+2})$ to $(\ha{\ha{p}}_a,\ha{\ha{p}}_b;\{\hat{\hat{p}}\}_{X})$, is defined as follows. First, there is a collinear mapping $(p_a,p_b;\{p\}_{X+2})\xrightarrow{C_{ab,s}^{I,F}}(\ha{p}_a,\ha{p}_b;\{\hat{p}\}_{X+1})$,
\begin{align}
\label{eq:map1IN}
    &\hat{p}_{a}^{\mu} =\,\xi_{a,s}p_{a}^{\mu}\,,\\
    &\hat{p}_{b}^{\mu} =\,\xi_{b,s}p_{b}^{\mu}\,,\\
    &\hat{p}_{n}^{\mu} =\,\Lambda(P,\hat{P})^\mu_\nu\,p_{n}^{\nu} \hspace{50pt} (n\in F \text{ and } \,n\neq s)\,,\\
    &\hat{p}_{X}^{\mu} =\,\Lambda(P,\hat{P})^\mu_\nu\,p_{X}^{\nu}\,.
\label{eq:map1FIN}
\end{align}
Here $F$ denotes the set of final-state partons while $\Lambda(P,\hat{P})^\mu_\nu$ represents a proper Lorentz transformation that takes the massive momentum $P$ into another momentum $\hat{P}$ of the same mass with
\begin{align}
    &P^{\mu} =\,Q^{\mu}-p_s^{\mu}\,,\qquad
    \hat{P}^{\mu} =\,\ha{Q}^{\mu} =\,\xi_{a,s}p_a^{\mu}+\xi_{b,s}p_b^{\mu}\,.
\end{align}
Requiring $P^2=\hat{P}^2$ fixes
\begin{equation}
\label{eq:conXis}
    \xi_{a,s}\xi_{b,s} =\,1-\frac{s_{s Q }}{s_{ab}}\,.
\end{equation}
A particular choice that satisfies eq.~(\ref{eq:conXis}) is
\begin{align}
\label{eq:xiar}
    &\xi_{a,s} =\,\sqrt{\frac{s_{ab}-s_{bs}}{s_{ab}-s_{as}}\frac{s_{ab}-s_{s Q }}{s_{ab}}}\,,\quad \xi_{b,s} =\,\sqrt{\frac{s_{ab}-s_{as}}{s_{ab}-s_{bs}}\frac{s_{ab}-s_{s Q }}{s_{ab}}}\,.
\end{align}
Then, the set of double-hatted momenta appearing in eq.~(\ref{eq:carscas}) is obtained by re-applying the mapping defined in eqs.~(\ref{eq:map1IN})-(\ref{eq:map1FIN}), i.e. $(\ha{p}_a,\ha{p}_b;\{\hat{p}\}_{X+1})\xrightarrow{C_{\hat{a}\hat{b},\hat{r}}^{I,F}}(\ha{\ha{p}}_a,\ha{\ha{p}}_b;\{\hat{\hat{p}}\}_{X})$ with
\begin{align}
\label{eq:map2IN}
    &\hat{\hat{p}}_{a}^{\mu} =\,\xi_{\ha{a},\ha{r}}\hat{p}_{a}^{\mu}\,,\\
    &\hat{\hat{p}}_{b}^{\mu} =\,\hat{\xi}_{b,r}\hat{p}_{b}^{\mu}\,,\\
    &\hat{\hat{p}}_{n}^{\mu} =\,\Lambda(\hat{P},\hat{\hat{P}})^\mu_\nu\,\hat{p}_{n}^{\nu} \hspace{50pt}(n\in F \text{ and } \,n\neq r)\,,\\
    &\hat{\hat{p}}_{X}^{\mu} =\,\Lambda(\hat{P},\hat{\hat{P}})^\mu_\nu\,\hat{p}_{X}^{\nu}\,.
\label{eq:map2FIN}
\end{align}
As above, $\Lambda(\hat{P},\hat{\hat{P}})^\mu_\nu$ is a proper Lorentz transformation which takes $\hat{P}$ to $\hat{\hat{P}}$ with $\hat{P}^2=\hat{\hat{P}}^2$. The latter furthermore fixes
\begin{equation}
    \xi_{\ha{a},\ha{r}}\hat{\xi}_{b,r} =\,1-\frac{s_{\hat{r}\hat{Q}}}{s_{\hat{a}\hat{b}}}\,.
\end{equation}
Since for color-singlet production $\hat{\hat{P}}^2=M_X^2$, we also have
\begin{equation}
    \xi_{\ha{a},\ha{r}}\hat{\xi}_{b,r} =\,\frac{M_X^2}{\xi_{a,s}\xi_{b,s}s_{ab}}\,.
\end{equation}
One particular choice is
\begin{align}
\label{eq:xiHar}
    &\xi_{\ha{a},\ha{r}} =\,\sqrt{\frac{s_{\hat{a}\hat{b}}-s_{\hat{b}\hat{r}}}{s_{\hat{a}\hat{b}}-s_{\hat{a}\hat{r}}}\frac{s_{\hat{a}\hat{b}}-s_{\hat{r}\hat{Q}}}{s_{\hat{a}\hat{b}}}}\,,\qquad\hat{\xi}_{b,r} =\,\sqrt{\frac{s_{\hat{a}\hat{b}}-s_{\hat{a}\hat{r}}}{s_{\hat{a}\hat{b}}-s_{\hat{b}\hat{r}}}\frac{s_{\hat{a}\hat{b}}-s_{\hat{r}\hat{Q}}}{s_{\hat{a}\hat{b}}}}\,.
\end{align}
Finally, $\hP^{\mathrm{(C)},(0)}_{(ars) (as)s}(x_{a,s}, \kT{s,a}, \kTh{s,a}, x_{\ha{a},\ha{r}}, \kT{\ha{r},\ha{a}}, \ha{p}_r; \ep)$ in eq.~(\ref{eq:carscas}) is the strongly-ordered splitting function for $I\rightarrow (IF)F$ splitting. It can be related to the one for $F\rightarrow F(FF)$ splitting, cf.~eqs.~(\ref{eq:CFFPqqbq0FFF})-(\ref{eq:CFFPggg0FFF}), by the crossing relation in eq.~(\ref{eq:PCars-crossing-IF}). Note that we choose $\kT{s,a}$ to be orthogonal to both $p_a$ and $p_b$.

\subsection{Definition of the integrated counterterm}
\label{sec:defIC}
The counterterm defined by eq.~(\ref{eq:carscas}) needs to be integrated over the $(X+2)$-particle phase space,
\beq
\bsp
\label{eq:IC0}
    \int\cC{asr}{IFF(0,0)}&\cC{as}{IF}=\,\mathcal{N}\int_2\rd\phi_{X+2}(\{p\}_{X+2};Q)\,\frac{1}{\omega(a)\omega(b)\Phi(p_a\cdot p_b)}\cC{asr}{IFF(0,0)}\cC{as}{IF}(p_a,p_b;\{p\}_{X+2})\,.
\esp
\eeq
Here $\mathcal{N}$ collects all non-QCD factors while the partonic flux factor is
\begin{equation}
    \Phi(p_a\cdot p_b) =\, 2p_a\cdot p_b =\, s_{ab}\,.
\end{equation}
The $\omega$ function accounts for the numbers of colors and spins of the incoming partons in $d$ dimensions,
\beq
\omega(q) =\,2 \Nc\,, \qquad\qquad \omega(g) =\,2 (\Nc^2-1)(1-\eps)\,.
\label{eq:omegafac}
\eeq
Because of the momentum mapping, the $(X+2)$-particle phase space can be written as
\beq
\bsp
\label{eq:PS-IFIFapp}
    \rd\phi_{X+2}(\{p\}_{X+2};Q) =\,&\,\int_{0}^{1}\rd\xi_a\,\rd\xi_b\,\int_{0}^{1}\rd\ha{\xi}_a\,\rd\ha{\xi}_b\,\rd\phi_{X}(\{\ha{\ha{p}}\}_{X};\ha{\ha{Q}})\rd\phi_{II,F}(\ha{p}_r,\ha{\xi}_a,\ha{\xi}_b)\, \rd\phi_{II,F}(p_s,\xi_a,\xi_b)
\esp
\eeq
with
\beq
\bsp
\label{eq:PSfactapp}
    &\rd\phi_{II,F}(\ha{p}_r,\ha{\xi}_a,\ha{\xi}_b) =\, \frac{\rd^d \ha{p}_r}{(2\pi)^{d-1}}\delta_{+}(\ha{p}_r^2)\delta(\xi_{\hat{a},\hat{r}}-\ha{\xi}_a)\delta(\xi_{\hat{b},\hat{r}}-\ha{\xi}_b)\,, \\&\rd\phi_{II,F}(p_s,\xi_a,\xi_b)  =\,\frac{\rd^d p_s}{(2\pi)^{d-1}}\delta_{+}(p_s^2)\delta(\xi_{a,s}-\xi_a)\delta(\xi_{b,s}-\xi_b)\,.
\esp
\eeq
Hence we have
\beq
\bsp
\label{eq:IC1}
    \int\cC{asr}{IFF(0,0)}\cC{as}{IF}=\,&\,\mathcal{N}\int_{0}^{1}\rd\xi_a\,\rd\xi_b\,\int_{0}^{1}\rd\ha{\xi}_a\,\rd\ha{\xi}_b\,\rd\phi_{X}(\{\ha{\ha{p}}\}_{X};\ha{\ha{Q}})\,\rd\phi_{II,F}(\ha{p}_r,\ha{\xi}_{a},\ha{\xi}_{b})\, \rd\phi_{II,F}(p_s,\xi_{a},\xi_{b})\\&\times\frac{(8\pi\alpha_s\mu^{2\eps})^2}{\omega(a)\omega(b)\Phi(p_a\cdot p_b)}\frac{1}{x_{a,s}s_{as}x_{\hat{a},\hat{r}}s_{\hat{a}\hat{r}}}\hP^{\mathrm{(C)},(0)}_{(ars) (as)s}(x_{a,s}, \kT{s,a}, \kTh{s,a}, x_{\ha{a},\ha{r}}, \kT{\ha{r},\ha{a}}, \ha{p}_r; \ep)\\&\times\SME{(ars)b,X+m}{(0)}{(\ha{\ha{p}}_a,\ha{\ha{p}}_b;\momhh{}_{X+m})} {\cal F} (x_{\ha{a},\ha{r}},\xi_{\ha{a},\ha{r}}\xi_{\ha{b},\ha{r}})\,.
\esp
\eeq
To proceed, we now need to answer two questions:
\begin{itemize}
    \item[(a)] How do we treat the phase space integration involving the transverse momentum $\kT{s,a}^\mu$?
    \item[(b)] How do we write down a parametric representation of the phase space measures $\rd\phi_{II,F}(p_s,\xi_{a},\xi_{b})$ and $\rd\phi_{II,F}(\ha{p}_r,\ha{\xi}_{a},\ha{\xi}_{b})$?
\end{itemize}
Let us provide some details on how to answer these questions. First consider the phase space integration of the transverse momentum. To be specific, we are interested in the following integral
\begin{equation}
    I_T =\, \int_1\rd\phi_{II,F}(p_s,\xi_{a},\xi_{b}) \frac{s_{\ha{r}\kT{s,a}}^2}{\kT{s,a}^2 s_{\ha{r}\wha{as}}}\,.
\end{equation}
Using the definition of the invariant $s_{\ha{r}\kT{s,a}}$, we can write $I_T$ as
\begin{equation}
\label{eq:IT}
    I_T =\, 4\,\frac{\ha{p}_{r,\mu}\ha{p}_{r,\nu}}{s_{\ha{r}\wha{as}}} I^{\mu\nu}
\end{equation}
with
\begin{equation}
    I^{\mu\nu} =\, \int_1\rd\phi_{II,F}(p_s,\xi_{a},\xi_{b})\frac{\kT{s,a}^{\mu}\kT{s,a}^{\nu}}{\kT{s,a}^2}\,.
\end{equation}
Now, to work out the tensor integral $I^{\mu\nu}$ we proceed as follows. First, we set up an ansatz for a rank-two tensor built out of the metric, $\ha{p}$ and $Q$ to write
\begin{equation}
\label{eq:ansTens}
    I^{\mu\nu} =\,\int_1 \rd\phi_{II,F}(p_s,\xi_{a},\xi_{b})( a_1 g^{\mu\nu}+a_2 \ha{p}^{\mu}_{a}\ha{p}^{\nu}_{a}+a_3 Q^{\mu}Q^{\nu}+a_4 \ha{p}^{\mu}_{a}Q^{\nu}+a_5 Q^{\mu}\ha{p}^{\nu}_{a})\,.
\end{equation}
Next, we fix all unknowns of the ansatz by contracting $I^{\mu\nu}$ with all tensor structures that appear on the right-hand side of eq.~(\ref{eq:ansTens}), keeping in mind that $\kT{s,a}$ was defined to be orthogonal to both $\ha{p}_a$ and $Q$. We find
\begin{equation}
    I^{\mu\nu} =\,\int_1 \rd\phi_{II,F}(p_s,\xi_{a},\xi_{b}) \frac{1}{2(1-\eps)}\left(g^{\mu\nu}+\frac{Q^2}{(\ha{p}_a\cdot Q)^2}\ha{p}_a^{\mu}\ha{p}_a^{\nu}-\frac{\ha{p}_a^{\mu}Q^{\nu}+\ha{p}_a^{\nu}Q^{\mu}}{\ha{p}_a\cdot Q}\right)\,.
\end{equation}
Substituting into eq.~(\ref{eq:IT}) and performing the contractions, our integral of interest reduces to
\begin{equation}
    \int_1\rd\phi_{II,F}(p_s,\xi_{a},\xi_{b}) \frac{s_{\ha{r}\kT{s,a}}^2}{\kT{s,a}^2 s_{\ha{r}\wha{as}}} =\, \int_1\rd\phi_{II,F}(p_s,\xi_{a},\xi_{b})\left(\frac{s_{ab}s_{\ha{r}\wha{as}}}{s_{\wha{as}Q}^2}-x_{\ha{r},\wha{as}}\right)\,.
\end{equation}
Next, we discuss how to turn the phase space integrations over $\rd\phi_{II,F}(p_s,\xi_{a},\xi_{b})$ and $\rd\phi_{II,F}(\ha{p}_r,\ha{\xi}_{a},\ha{\xi}_{b})$ into (multi-dimensional) parametric integrals. We focus our attention on $\rd\phi_{II,F}(p_s,\xi_{a},\xi_{b})$, but of course the treatment of $\rd\phi_{II,F}(\ha{p}_r,\ha{\xi}_{a},\ha{\xi}_{b})$ is completely analogous. The phase space measure is defined as
\begin{equation}
\label{eq:ps}
    \rd\phi_{II,F}(p_s,\xi_{a},\xi_{b}) =\, \frac{\rd^d p_s}{(2\pi)^{d-1}}\delta_+(p_s^2)\delta(\xi_a-\xi_{a,s})\delta(\xi_b-\xi_{b,s})
\end{equation}
with $\xi_{a,s}$ and $\xi_{b,s}$ as in eq.~(\ref{eq:xiar}). It will be convenient in what follows to work in the rest frame of $Q$, oriented such that $p_a$ points in the $z$-direction. So, in $d$-dimensional polar coordinates, we write, assuming all partons to be massless,
\begin{align}
    &p_{a}^{\mu} =\, \frac{\sqrt{s_{ab}}}{2}(1,\mathbf{0}_{d-2},1)\,,\\
    &p_{b}^{\mu} =\, \frac{\sqrt{s_{ab}}}{2}(1,\mathbf{0}_{d-2},-1)\,,\\
    &p_{s}^{\mu} =\, E_s(1,\_\_\_\,,\cos\vartheta)\,.
\end{align}
The dashes denote angular variables of which the integrand is independent, such that their integration is trivial. In this frame, the corresponding invariants are
\begin{align}
    &s_{as} =\, \sqrt{s_{ab}}E_s(1-\cos\vartheta)\,,\\
    &s_{bs} =\, \sqrt{s_{ab}}E_s(1+\cos\vartheta)\,,\\
    &s_{Qs} =\, 2\sqrt{s_{ab}}E_s\,.
\end{align}
Hence the parameters of the momentum mapping can be written as
\begin{equation}
\label{eq:xias}
    \xi_{a,s} =\, \sqrt{\frac{s_{ab}-\sqrt{s_{ab}}E_s(1+\cos\vartheta)}{s_{ab}-\sqrt{s_{ab}}E_s(1-\cos\vartheta)}\frac{s_{ab}-2\sqrt{s_{ab}}E_s}{s_{ab}}}
\end{equation}
and
\begin{equation}
\label{eq:xibs}
    \xi_{b,s} =\, \sqrt{\frac{s_{ab}-\sqrt{s_{ab}}E_s(1-\cos\vartheta)}{s_{ab}-\sqrt{s_{ab}}E_s(1+\cos\vartheta)}\frac{s_{ab}-2\sqrt{s_{ab}}E_s}{s_{ab}}}\,.
\end{equation}
Let us now work out the invariant measure in this frame. At the risk of being overly-explicit,\footnote{But, as the saying goes: Nothing risked, nothing gained...} we remind the reader that generically, in $d$ dimensions, we have
\begin{equation}
    \rd^d p_s =\, \frac{\rd^{d-1}\mathbf{p}_s}{2E_s(2\pi)^{d-1}}\,,
\end{equation}
or, using polar coordinates,
\begin{equation}
\label{eq:psMeas}
    \rd^d p_s =\, \frac{1}{2E_s(2\pi)^{d-1}}|\mathbf{p}_s|^{d-2}\rd|\mathbf{p}_s|\rd\Omega_{d-1}\,.
\end{equation}
The $(d-1)$-dimensional angular measure can be compactly written as
\begin{equation}
\label{eq:angMeas}
    \rd\Omega_{d-1} =\, \prod_{k=1}^{d-2}\rd\vartheta_{k}\sin^{d-k-2}\vartheta_k\,.
\end{equation}
Next we turn the momentum integration into an integration over the energy using $E_s^2=\,|\mathbf{p}_s|^2$,
\begin{equation}
    |\mathbf{p}_s|\rd|\mathbf{p}_s| =\, E_s\rd E_s\,.
\end{equation}
Substituting into eq.~(\ref{eq:psMeas}) we find, after some trivial algebra,
\begin{equation}
    \rd^d p_s =\, \frac{E_s^{d-3}}{2(2\pi)^{d-1}}\rd E_s\rd\Omega_{d-1}\,.
\end{equation}
Because of the particular frame we chose, the integration involving $\vartheta$ will be non-trivial. As such, we pinch it off from the angular measure, writing
\begin{equation}
    \rd\Omega_{d-1}\to\,\rd\vartheta\sin^{d-3}\vartheta\,\rd\Omega_{d-2}\,,
\end{equation}
cf. also eq.~(\ref{eq:angMeas}). Finally, we rewrite the $\vartheta$-integration as an integration over $\cos\vartheta$,
\begin{equation}
    \rd\Omega_{d-1}\to\,\rd(\cos\vartheta)\sin^{d-4}\vartheta\,\rd\Omega_{d-2}\,.
\end{equation}
With all these considerations in mind, our phase space measure in eq.~(\ref{eq:ps}) becomes
\beq
\bsp
\label{eq:ps2}
    \rd\phi_{II,F}(p_s,\xi_{a},\xi_{b}) =\,&\,\frac{\rd E_s\,E_s^{1-2\eps}}{2(2\pi)^{3-2\eps}}\rd\Omega_{d-2}\,\rd(\cos\vartheta)\sin^{-2\eps}\vartheta\,\theta(E_s)\,\delta\left(\xi_a-\sqrt{\frac{\sqrt{s_{ab}}-E_s(1+\cos\vartheta)}{\sqrt{s_{ab}}-E_s(1-\cos\vartheta)}\frac{\sqrt{s_{ab}}-2E_s}{\sqrt{s_{ab}}}}\right)\\&\times\delta\left(\xi_b-\sqrt{\frac{\sqrt{s_{ab}}-E_s(1-\cos\vartheta)}{\sqrt{s_{ab}}-E_s(1+\cos\vartheta)}\frac{\sqrt{s_{ab}}-2E_s}{\sqrt{s_{ab}}}}\right)\,.
\esp
\eeq
Next, we want to turn the integration over $(E_s,\cos\vartheta)$ into an integration over $(\xi_{a,s},\xi_{b,s})$. For this, we use eqs.~(\ref{eq:xias}) and (\ref{eq:xibs}) to write
\begin{equation}
    E_s =\, \frac{\sqrt{s_{ab}}}{2}(1-\xi_{a,s}\xi_{b,s})
\end{equation}
and
\begin{equation}
\label{eq:cost}
    \cos\vartheta =\, \frac{(\xi_{a,s}-\xi_{b,s})(1+\xi_{a,s}\xi_{b,s})}{(\xi_{a,s}+\xi_{b,s})(1-\xi_{a,s}\xi_{b,s})}\,.
\end{equation}
Note that the latter equation implies that
\begin{equation}
    \sin^2\vartheta =\, \frac{4\xi_{a,s}\xi_{b,s}(1-\xi_{a,s}^2)(1-\xi_{b,s}^2)}{(\xi_{a,s}+\xi_{b,s})^2(1-\xi_{a,s}\xi_{b,s})^2}\,.
\end{equation}
Furthermore, the transformation introduces the Jacobian
\begin{equation}
    J =\, -2\sqrt{s_{ab}}\frac{\xi_{a,s}\xi_{b,s}(1+\xi_{a,s}\xi_{b,s})}{(\xi_{a,s}+\xi_{b,s})^2(1-\xi_{a,s}\xi_{b,s})}\,.
\end{equation}
Putting all of this together, we find that the phase space in eq.~(\ref{eq:ps2}) can be written as
\beq
\bsp
    \rd\phi_{II,F}(p_s,\xi_{a},\xi_{b}) =\,&\, \frac{\rd\Omega_{d-2}}{(2\pi)^{3-2\eps}}\rd\xi_{a,s}\,\rd\xi_{b,s}\,\sqrt{s_{ab}}\frac{\xi_{a,s}\xi_{b,s}(1+\xi_{a,s}\xi_{b,s})}{(\xi_{a,s}+\xi_{b,s})^2(1-\xi_{a,s}\xi_{b,s})}\left[\frac{\sqrt{s_{ab}}}{2}(1-\xi_{a,s}\xi_{b,s})\right]^{1-2\eps}\\&\times\left[\frac{4\xi_{a,s}\xi_{b,s}(1-\xi_{a,s}^2)(1-\xi_{b,s}^2)}{(\xi_{a,s}+\xi_{b,s})^2(1-\xi_{a,s}\xi_{b,s})^2}\right]^{-\eps}\theta(1-\xi_{a,s}\xi_{b,s})\delta(\xi_{a,s}-\xi_a)\delta(\xi_{b,s}-\xi_b)\,.
\esp
\eeq
Finally, using that
\begin{equation}
    \Omega_{d-2}=\,\int\rd\Omega_{d-2}=\,(2\pi)^{1-\eps}S_{\eps}
\end{equation}
with
\begin{equation}
    S_{\eps} =\,\frac{(4\pi)^{\eps}}{\Gamma(1-\eps)}\,,
\end{equation}
and collecting factors of the same form, the phase space measure reduces to
\beq
\bsp
\label{eq:psFIN}
    \rd\phi_{II,F}(p_s,\xi_{a},\xi_{b}) =\,&\, \frac{S_{\epsilon}}{8\pi^2}\frac{\rd\Omega_{d-2}}{\Omega_{d-2}}\rd\xi_{a,s}\,\rd\xi_{b,s}\,s_{ab}^{1-\eps}\frac{\xi_{a,s}\xi_{b,s}(1+\xi_{a,s}\xi_{b,s})}{(\xi_{a,s}+\xi_{b,s})^2}\left[\frac{\xi_{a,s}\xi_{b,s}(1-\xi_{a,s}^2)(1-\xi_{b,s}^2)}{(\xi_{a,s}+\xi_{b,s})^2}\right]^{-\eps}\\&\times\theta(1-\xi_{a,s}\xi_{b,s})\delta(\xi_{a,s}-\xi_a)\delta(\xi_{b,s}-\xi_b)\,.
\esp
\eeq
This is the parametric representation we were after. Naturally, the analysis of $\rd\phi_{II,F}(\ha{p}_r,\ha{\xi}_{a},\ha{\xi}_{b})$ is completely analogous, and we have
\beq
\bsp
    \rd\phi_{II,F}(\ha{p}_r,\ha{\xi}_{a},\ha{\xi}_{b}) =\,&\, \frac{S_{\epsilon}}{8\pi^2}\frac{\rd\Omega_{d-2}}{\Omega_{d-2}}\rd\xi_{\ha{a},\ha{r}}\,\rd\xi_{\ha{b},\ha{r}}\,s_{\ha{a}\ha{b}}^{1-\eps}\frac{\xi_{\ha{a},\ha{r}}\xi_{\ha{b},\ha{r}}(1+\xi_{\ha{a},\ha{r}}\xi_{\ha{b},\ha{r}})}{(\xi_{\ha{a},\ha{r}}+\xi_{\ha{b},\ha{r}})^2}\left[\frac{\xi_{\ha{a},\ha{r}}\xi_{\ha{b},\ha{r}}(1-\xi_{\ha{a},\ha{r}}^2)(1-\xi_{\ha{b},\ha{r}}^2)}{(\xi_{\ha{a},\ha{r}}+\xi_{\ha{b},\ha{r}})^2}\right]^{-\eps}\\&\times\theta(1-\xi_{\ha{a},\ha{r}}\xi_{\ha{b},\ha{r}})\delta(\xi_{\ha{a},\ha{r}}-\ha{\xi}_a)\delta(\xi_{\ha{b},\ha{r}}-\ha{\xi}_b)\,.
\esp
\eeq
Of course, whenever the integrand is independent of the angular variables appearing in $\rd\Omega_{d-2}$, we can set $\rd\Omega_{d-2}/\Omega_{d-2}\to\,1$. This will in fact be the case for any $A_{12}$ counterterm whose integration involves $\rd\phi_{II,F}$. Furthermore, the integrals over $(\xi_{a,s},\xi_{b,s})$ and $(\xi_{\ha{a},\ha{r}},\xi_{\ha{b},\ha{r}})$ are completely trivial due to the delta-distributions, and the result is to simply replace $(\xi_{a,s},\xi_{b,s},\xi_{\ha{a},\ha{r}},\xi_{\ha{b},\ha{r}})$ by $(\xi_a,\xi_b,\ha{\xi}_a,\ha{\xi}_b)$. Taking all these considerations into account, we are finally ready to present the fully parametric representation of the parton-level integrated counterterm. We write eq.~(\ref{eq:IC1}) as\footnote{Since the integrations over $\xi_a, \xi_b, \ha{\xi}_a$ and $\ha{\xi}_b$ run between zero and one, the conditions $\xi_a\xi_b\leq 1$ and $\ha{\xi}_a\ha{\xi}_b\leq 1$ are automatically satisfied, and hence we can safely omit the corresponding $\theta$-functions.}
\beq
\bsp
\label{eq:IC2}
    &\left[{\rm{C}}_{asr}^{IFF(0,0)}{\rm{C}}_{as}^{IF}\right]\otimes\rd\sigma_{\ha{\ha{a}}\ha{\ha{b}}}=\,\int_{0}^{1}\rd\xi_a\,\rd\xi_b\,\int_{0}^{1}\rd\ha{\xi}_a\,\rd\ha{\xi}_b\,\rd\sigma_{\ha{\ha{a}}\ha{\ha{b}}}(\hat{\hat{p}}_a,\hat{\hat{p}}_b)\,\left[{\rm{C}}_{asr}^{IFF(0,0)}{\rm{C}}_{as}^{IF}(\xi_a,\xi_b,\ha{\xi}_a,\ha{\xi}_b;\eps)\right]
\esp
\eeq
with\footnote{Note that, to write the integrated counterterm in terms of the reduced cross section with \textit{mapped} variables, we need to divide by $\omega(asr)$, cf. eq.~(\ref{eq:red_XS}). Of course, this needs to be multiplied back and, together with the original $\omega(a)$ coming from the cross section in terms of the original variables, this is absorbed into the integrand, cf. eq.~(\ref{eq:CasrCasBR}). Finally, we also used that the flavor of parton $b$ is unaffected by the mapping.}
\begin{equation}
\label{eq:redXSapp}
    \rd\sigma_{\ha{\ha{a}}\ha{\ha{b}}}(\ha{\ha{p}}_a,\ha{\ha{p}}_b) =\, \left[\frac{\al_s}{2\pi}S_{\eps}\left(\frac{\mu^2}{s_{ab}}\right)^{\eps}\right]^2\frac{\mathcal{N}}{\omega(asr)\omega(b)\Phi(\ha{\ha{p}}_a\cdot\ha{\ha{p}}_b)}\,\rd\phi_{X}(\{\ha{\ha{p}}\}_X;\ha{\ha{Q}})\SME{(ars)b,X+m}{(0)}{(\ha{\ha{p}}_a,\ha{\ha{p}}_b;\momhh{}_{X+m})}\,.
\end{equation}
Note that, because of the action of the Dirac-delta distributions in the phase space measures, the mapped momenta are now
\begin{equation}
\label{eq:dhat}
    \ha{\ha{p}}_{a} =\, \xi_a\ha{\xi}_a p_a\,,\qquad \ha{\ha{p}}_{b} =\, \xi_b\ha{\xi}_b p_b
\end{equation}
and hence
\begin{equation}
    \Phi(\ha{\ha{p}}_a\cdot\ha{\ha{p}}_b) =\, \xi_a\ha{\xi}_a\xi_b\ha{\xi}_b\,\Phi(p_a\cdot p_b)\,.
\end{equation}
The factor of $\xi_a\ha{\xi}_a\xi_b\ha{\xi}_b$ was moved into the integrand $\left[{\rm{C}}_{asr}^{IFF(0,0)}{\rm{C}}_{as}^{IF}(\xi_a,\xi_b,\ha{\xi}_a,\ha{\xi}_b;\eps)\right]$, which we define as
\beq
\bsp
\label{eq:CasrCasBR}
&\left[{\rm{C}}_{asr}^{IFF(0,0)}{\rm{C}}_{as}^{IF}(\xi_a,\xi_b,\ha{\xi}_a,\ha{\xi}_b;\eps)\right] =\,
	4 \frac{\omega(asr)}{\omega(a)} 
	s_{ab}^2
	\left[
	\frac{\xi_{a}^2 \xi_{b}^2 (1-\xi_{a}^2)(1-\xi_{b}^2)}
	{(\xi_{a} + \xi_{b})^2}\right]^{-\eps}
	\frac{\xi_{a}^3 \xi_{b}^3 (1+\xi_{a} \xi_{b})}
	{(\xi_{a} + \xi_{b})^2}
	\\ &\quad\times\left[
	\frac{\hat{\xi}_{a} \hat{\xi}_{b} (1-\hat{\xi}_{a}^2)(1-\hat{\xi}_{b}^2)}
	{(\hat{\xi}_{a} + \hat{\xi}_{b})^2}\right]^{-\eps}
	\frac{\hat{\xi}_{a}^2 \hat{\xi}_{b}^2(1+\hat{\xi}_{a} \hat{\xi}_{b})}
	{(\hat{\xi}_{a} + \hat{\xi}_{b})^2}
	\frac{{\cal F}}{s_{as} x_{a,s} s_{\ha{a}\ha{r}} x_{\ha{a},\ha{r}}} 
	 P^{\mathrm{(C)} (0)}_{(asr) (as) s}(x_{a,s} 
	x_{\ha{a},\ha{r}};\eps) \,.
\esp
\eeq
Following eq.~(\ref{eq:dhat}), we also set $s_{\ha{a}\ha{b}}=\xi_a\xi_b\,s_{ab}$. Finally we have performed the azimuthal integration, which amounts to passing 
from the full splitting function $\hP^{\mathrm{(C)} (0)}_{(asr) (as) s}$ to 
its azimuthally averaged form, $P^{\mathrm{(C)} (0)}_{(asr) (as) s}$. For 
the latter, we find
\beq
\label{eq:PCave}
P^{\mathrm{(C)} (0)}_{(asr) (as) s}(x_{a,s} 
	x_{\ha{a},\ha{r}};\eps) =\,
	P_{(as) s}^{(0)}(x_{a,s};\eps) P_{(asr) r}^{(0)}(x_{\ha{a},\ha{r}};\eps)
	- F(asr)\delta_{(as),g} \bT^2_{asr} \frac{1-x_{a,s}}{x_{a,s}}
	b^{(0)}_{a s} b^{(0)}_{(asr) (as)} R\,
\eeq
with
\beq
\label{eq:bfacs}
b^{(0)}_{qg} =\, b^{(0)}_{gq} =\, b^{(0)}_{gg} =\, 2\CA\,,\qquad
b^{(0)}_{q\qb} =\, -\frac{2}{1-\ep}\TR
\eeq
and
\beq
\label{eq:defR}
R =\,x_{\ha{a},\ha{r}} \frac{s_{\ha{a}\ha{r}} s_{ab}}{s_{\ha{a} Q }^2}\,.
\eeq
From \eqnss{eq:Pqg-ave-IF}{eq:Pgg-ave-IF} it is clear that 
$P_{(as) s}^{(0)}(x_{a,s};\ep) P_{f_{crs} r}^{(0)}(x_{\ha{a},\ha{r}};\ep)$
is simply a linear combination of terms of the form
\beq
\bigg\{
\frac{1}{x_{a,s}}\,, \frac{1}{1-x_{a,s}}\,, 1\,, x_{a,s}\,, x_{a,s}^2 
\bigg\}
\times
\bigg\{
\frac{1}{x_{\ha{a},\ha{r}}}\,, \frac{1}{1-x_{\ha{a},\ha{r}}}\,, 1\,, 
x_{\ha{a},\ha{r}}\,, x_{\ha{a},\ha{r}}^2 
\bigg\}\,.
\eeq
Hence we must evaluate integrals whose integrands contain
\beq
\label{eq:intStr}
\bigg\{
\frac{1}{1-x_{\ha{a},\ha{r}}}
\frac{1}{1-x_{a,s}}\,,\frac{x_{a,s}^k}{1-x_{\ha{a},\ha{r}}}\,,
\frac{x_{\ha{a},\ha{r}}^l}{1-x_{a,s}}\,, 
x_{a,s}^k x_{\ha{a},\ha{r}}^l\,, R\bigg\}\,,\qquad k,l =\,-1,0,1,2\,.
\eeq
In order to write the integrals explicitly in terms of the integration variables, note that we have 
\bal
s_{as} &=\,s_{ab} \frac{\xi_a(1-\xi_b^2)}{\xi_a+\xi_b}\,, &
s_{s Q } &=\,s_{ab}(1-\xi_a \xi_b)\,,
\eal
which implies 
\beq
x_{a,s} =\,\xi_a \xi_b
\eeq
and
\bal
s_{\ha{a}\ha{b}} &=\,\xi_a \xi_b s_{ab}\,, &
s_{\ha{a}\ha{r}} &=\,s_{\ha{a}\ha{b}} 
	\frac{\hat{\xi}_{a}(1-\hat{\xi}_{b}^2)}{\hat{\xi}_{a}+\hat{\xi}_{b}}\,, &
s_{\ha{b}\ha{r}} &=\,s_{\ha{a}\ha{b}} 
	\frac{\hat{\xi}_{b}(1-\hat{\xi}_{a}^2)}{\hat{\xi}_{a}+\hat{\xi}_{b}}\,.
\eal
Moreover, using $\ha{p}_a =\,\xi_a p_a$ and $\ha{p}_b =\,\xi_a p_b$, we find
\beq
s_{\ha{a} Q } =\,\frac{1}{\xi_b}s_{\ha{a}\ha{b}}\,,\quad
s_{\ha{b} Q } =\,\frac{1}{\xi_a}s_{\ha{a}\ha{b}}\,
\eeq
and
\beq
\bsp
s_{\ha{r} Q } =\,
    \frac{\hat{\xi}_{a}\xi_b(1-\hat{\xi}_{b}^2) 
    + \xi_{a}\hat{\xi}_{b}(1-\hat{\xi}_{a}^2)}
    {\xi_a \xi_b (\hat{\xi}_{a} + \hat{\xi}_{b})} s_{\ha{a}\ha{b}}\,.
\esp
\eeq
From eq.~(\ref{eq:Fiff}) it then follows that
\beq
{\cal F} =\,\frac{1}{\hat{\xi}_{a}^2 \hat{\xi}_{b}^2}
    \frac{[\xi_a (\hat{\xi}_{a} + \hat{\xi}_{b}) - \hat{\xi}_{a}\xi_b(1-\hat{\xi}_{b}^2) 
    - \xi_{a}\hat{\xi}_{b}(1-\hat{\xi}_{a}^2)]^2}
    {\xi_a^2 (\hat{\xi}_{a} + \hat{\xi}_{b})^2}\,
\eeq
while eq.~(\ref{eq:xahrh}) gives
\beq
x_{\ha{a},\ha{r}} =\,
    \frac{\xi_a (\hat{\xi}_{a} + \hat{\xi}_{b}) - \hat{\xi}_{a}\xi_b(1-\hat{\xi}_{b}^2) 
    - \xi_{a}\hat{\xi}_{b}(1-\hat{\xi}_{a}^2)}
    {\xi_a (\hat{\xi}_{a} + \hat{\xi}_{b})}\,.
\eeq
Finally, from eq.~(\ref{eq:defR}),
\begin{equation}
   R =\, \frac{\hat{\xi}_{a}^2 \hat{\xi}_{b} (1-\hat{\xi}_{b}^2)}
	{(1 - \hat{\xi}_{a} \hat{\xi}_{b})^2(\hat{\xi}_{a} + \hat{\xi}_{b})}\,.
\end{equation}
Note now that the integration in eq.~(\ref{eq:IC2}) is over the four convolution variables that appear in the factorized matrix element. To simplify the convolution structure, we replace the $\ha{\xi}_a$ and $\ha{\xi}_b$ variables by
\begin{equation}
    \ha{\xi}_a \rightarrow \frac{\eta_a}{\xi_a}, \qquad \ha{\xi}_b \rightarrow \frac{\eta_b}{\xi_b}\
\end{equation}
such that
\beq
\bsp
  &\left[{\rm{C}}_{asr}^{IFF(0,0)}{\rm{C}}_{as}^{IF}\right]\otimes\rd\sigma_{\ha{\ha{a}}\ha{\ha{b}}}\\&\qquad=\,\int_{0}^{1} \rd \eta_a\, \int_0^{1} \rd \eta_b\,\rd\sigma_{\ha{\ha{a}}\ha{\ha{b}}}(\eta_a p_a,\eta_b p_b)\, \int_{\eta_a}^{1}\frac{\rd \xi_a}{\xi_a} \int_{\eta_b}^{1}\frac{\rd \xi_b}{\xi_b}
  [\IcC{asr}{IFF(0,0)}\IcC{as}{IF}(\xi_a,\xi_b,\eta_a/\xi_a,\eta_b/\xi_b;\eps)]\,.
\esp
\eeq
The utility of this transformation is that the reduced cross section now only depends on two convolution variables, namely $\eta_a$ and $\eta_b$. In particular, as the factorized matrix element no longer depends on $\xi_a$ and $\xi_b$, the integration over the latter can be done analytically once and for all. Denoting the result of this integral as
\begin{equation}
    [\IcC{asr}{IFF(0,0)}\IcC{as}{IF}(\eta_a,\eta_b;\eps)] =\,\int_{\eta_a}^{1}\frac{\rd \xi_a}{\xi_a}\, \int_{\eta_b}^{1}\frac{\rd \xi_b}{\xi_b}\,[\IcC{asr}{IFF(0,0)}\IcC{as}{IF}(\xi_a,\xi_b,\eta_a/\xi_a,\eta_b/\xi_b;\eps)]
\end{equation}
we have
\beq
\bsp
\label{eq:IC3}
    &\left[{\rm{C}}_{asr}^{IFF(0,0)}{\rm{C}}_{as}^{IF}\right]\otimes\rd\sigma_{\ha{\ha{a}}\ha{\ha{b}}}=\,\int_{0}^{1} \rd \eta_a\, \int_0^{1} \rd \eta_b\,\rd\sigma_{\ha{\ha{a}}\ha{\ha{b}}}(\eta_a p_a,\eta_b p_b)\,[\IcC{asr}{IFF(0,0)}\IcC{as}{IF}(\eta_a,\eta_b;\eps)]\,.
\esp
\eeq
However, care has to be taken with the interpretation of this integral. In particular, there will be singularities as any of the variables of the integrand approaches one.\footnote{A priori, any of the integration variables approaching zero could also be problematic. However, once the parton-level integrated counterterm is convoluted with the PDFs, this option is eliminated.} Hence we need an additional step of regularization, which is achieved by setting up a subtraction. For this we compute the asymptotic behavior of $[\IcC{asr}{IFF(0,0)}\IcC{as}{IF}(\eta_a,\eta_b;\eps)]$ in all relevant limits using the method of \textit{expansion by regions} \cite{Beneke:1997zp} as will be discussed in detail in sec.~\ref{sec:distexp}. In particular, we need to determine the asymptotic behavior of $[\IcC{asr}{IFF(0,0)}\IcC{as}{IF}(\eta_a,\eta_b;\eps)]$ in three distinct limits, which we symbolically denote as $\La$, $\Lb$ and $\Lab$.
For example, the operator $\La$ selects the leading singularity in $d$ dimensions as $\eta_a\to 1$, dropping both subleading and non-singular terms. Also, while the double limit $\Lab$ takes into account the divergences as both $\eta_a$ \textit{and} $\eta_b$ approach one at the same time, the resulting expression can still diverge as either $\eta_a$ \textit{or} $\eta_b$ approaches one, meaning we also need to compute the overlaps $\La\Lab$ and $\Lb\Lab$.
The regularized integrated subtraction term is then obtained by subtracting off each limit and adding back the integrated versions, leading to
\beq
\bsp
\label{eq:fulldistexp}
\left[{\rm{C}}_{asr}^{IFF(0,0)}{\rm{C}}_{as}^{IF}\right]\otimes\rd\sigma_{\ha{\ha{a}}\ha{\ha{b}}}=\,&\,\int_{0}^{1} \rd \eta_a\, \int_0^{1} \rd \eta_b\,\bigg\{
[\IcC{asr}{IFF(0,0)}\IcC{as}{IF}(\eta_a,\eta_b;\eps)]\rd\sigma_{\ha{\ha{a}}\ha{\ha{b}}}(\eta_a p_a,\eta_b p_b)
\\&-
\La[ \IcC{asr}{IFF(0,0)}\IcC{as}{IF}(\eta_a,\eta_b;\eps)]\rd\sigma_{\ha{\ha{a}}\ha{\ha{b}}}( p_a,\eta_b p_b)
\\&-
\Lb[ \IcC{asr}{IFF(0,0)}\IcC{as}{IF}(\eta_a,\eta_b;\eps)]\rd\sigma_{\ha{\ha{a}}\ha{\ha{b}}}(\eta_a p_a, p_b)
\\&-
\Big(\Lab[ \IcC{asr}{IFF(0,0)}\IcC{as}{IF}(\eta_a,\eta_b;\eps)]-\La\Lab[ \IcC{asr}{IFF(0,0)}\IcC{as}{IF}(\eta_a,\eta_b;\eps)]\\&-\Lb\Lab[ \IcC{asr}{IFF(0,0)}\IcC{as}{IF}(\eta_a,\eta_b;\eps)]\Big)\rd\sigma_{\ha{\ha{a}}\ha{\ha{b}}}(p_a, p_b)
\\&+
[\La][  \IcC{asr}{IFF(0,0)}\IcC{as}{IF}(\eta_a,\eta_b;\eps)]\rd\sigma_{\ha{\ha{a}}\ha{\ha{b}}}( p_a,\eta_b p_b)
\\&+
[\Lb][  \IcC{asr}{IFF(0,0)}\IcC{as}{IF}(\eta_a,\eta_b;\eps)]\rd\sigma_{\ha{\ha{a}}\ha{\ha{b}}}(\eta_a p_a, p_b)
\\&+
\Big([\Lab][ \IcC{asr}{IFF(0,0)}\IcC{as}{IF}(\eta_a,\eta_b;\eps)]-[\La \Lab][ \IcC{asr}{IFF(0,0)}\IcC{as}{IF}(\eta_a,\eta_b;\eps)]\\&-[\Lb \Lab][ \IcC{asr}{IFF(0,0)}\IcC{as}{IF}(\eta_a,\eta_b;\eps)]\Big)\rd\sigma_{\ha{\ha{a}}\ha{\ha{b}}}( p_a,p_b)
\bigg\}\,.
\esp
\eeq
Here
\begin{align}
\label{eq:intlimsin}
    &[\La][\IcC{asr}{IFF(0,0)}\IcC{as}{IF}(\eta_a,\eta_b;\eps)] =\, \int_0^1\rd \eta_a\, \La[\IcC{asr}{IFF(0,0)}\IcC{as}{IF}(\eta_a,\eta_b;\eps)]\,,\\
    &[\Lb][\IcC{asr}{IFF(0,0)}\IcC{as}{IF}(\eta_a,\eta_b;\eps)] =\, \int_0^1\rd \eta_b\, \Lb[\IcC{asr}{IFF(0,0)}\IcC{as}{IF}(\eta_a,\eta_b;\eps)]\,,\\
    &[\Lab][\IcC{asr}{IFF(0,0)}\IcC{as}{IF}(\eta_a,\eta_b;\eps)] =\, \int_0^1\rd \eta_a\,\int_0^1\rd \eta_b\, \Lab[\IcC{asr}{IFF(0,0)}\IcC{as}{IF}(\eta_a,\eta_b;\eps)]\,,\\
    &[\La\Lab][\IcC{asr}{IFF(0,0)}\IcC{as}{IF}(\eta_a,\eta_b;\eps)] =\, \int_0^1\rd \eta_a\, \La\Lab[\IcC{asr}{IFF(0,0)}\IcC{as}{IF}(\eta_a,\eta_b;\eps)]\,,\\
    &[\Lb\Lab][\IcC{asr}{IFF(0,0)}\IcC{as}{IF}(\eta_a,\eta_b;\eps)] =\, \int_0^1\rd \eta_b\, \Lb\Lab[\IcC{asr}{IFF(0,0)}\IcC{as}{IF}(\eta_a,\eta_b;\eps)]
\label{eq:intlimsfin}
\end{align}
denote the integrals of the limit formul\ae\, which need to be added back.\newline

In the following, we will illustrate the computation of $[\IcC{asr}{IFF(0,0)}\IcC{as}{IF}(\eta_a,\eta_b;\eps)]$ and its asymptotic behavior. Note however that from eq.~(\ref{eq:intStr}) it follows that the full integrated counterterm requires us to compute 26 distinct basic integrals. For the sake of explicitness, we will focus our attention on $\frac{1}{1-x_{\ha{a},\ha{r}}}\frac{1}{1-x_{a,s}}$, which leads to the most involved integral for this counterterm. So we consider
\beq
\bsp
\label{eq:omxm1omxHm1}
&\mathcal{I} =\,s_{ab}^2 \int_{0}^{1} \rd \xi_a\, \rd \xi_b\,\int_{0}^{1} \rd \hat{\xi}_{a} \rd \hat{\xi}_{b}\,\rd\sigma_{\ha{\ha{a}}\ha{\ha{b}}}(\xi_a\ha{\xi}_a p_a,\xi_b\ha{\xi}_b p_b)
	\left[
	\frac{\xi_{a}^2 \xi_{b}^2 (1-\xi_{a}^2)(1-\xi_{b}^2)}
	{(\xi_{a} + \xi_{b})^2}\right]^{-\ep}
	\frac{\xi_{a}^3 \xi_{b}^3 (1+\xi_{a} \xi_{b})}
	{(\xi_{a} + \xi_{b})^2}
 \\ &\times
	\left[
	\frac{\hat{\xi}_{a} \hat{\xi}_{b} (1-\hat{\xi}_{a}^2)(1-\hat{\xi}_{b}^2)}
	{(\hat{\xi}_{a} + \hat{\xi}_{b})^2}\right]^{-\ep}
	\frac{\hat{\xi}_{a}^2 \hat{\xi}_{b}^2(1+\hat{\xi}_{a} \hat{\xi}_{b})}
	{(\hat{\xi}_{a} + \hat{\xi}_{b})^2}
	\frac{1}{s_{as} x_{a,s} s_{\ha{a}\ha{r}} x_{\ha{a},\ha{r}}} 
 \frac{1}{1-x_{\ha{a},\ha{r}}}\frac{1}{1-x_{a,s}}
	{\cal F}\,.
\esp
\eeq
Note that we omit the overall factor $4\,\omega(asr)/\omega(as)$
as it does not influence the integration. Rewriting all factors in terms of the integration variables leads to the following charming expression
\beq
\bsp
\label{eq:intini}
&\mathcal{I} =\,\int_{0}^{1} \rd \xi_a\, \rd \xi_b\,\int_{0}^{1} \rd \hat{\xi}_{a} \rd \hat{\xi}_{b}\,\frac{\rd\sigma_{\ha{\ha{a}}\ha{\ha{b}}}(\xi_a\ha{\xi}_a p_a,\xi_b\ha{\xi}_b p_b)}{(-1+\xi_{a}\xi_{b})(-\hat{\xi}_{a}\xi_{b}-\xi_{a}\hat{\xi}_{b}+\xi_{a}\hat{\xi}_{a}^2\hat{\xi}_{b}+\hat{\xi}_{a}\xi_{b}\hat{\xi}_{b}^2)}\\&\times(1-\xi_{a})^{-\eps}\xi_{a}^{-2\eps}(1+\xi_{a})^{-\eps}(1-\hat{\xi}_{a})^{-\eps}\hat{\xi}_{a}^{-\eps}(1+\hat{\xi}_{a})^{-\eps}(1-\xi_{b})^{-1-\eps}\xi_{b}^{1-2\eps}(1+\xi_{b})^{-1-\eps}\\&\times(\xi_{a}+\xi_{b})^{-1+2\eps}(1+\xi_{a}\xi_{b})(1-\hat{\xi}_{b})^{-1-\eps}\hat{\xi}_{b}^{-\eps}(1+\hat{\xi}_{b})^{-1-\eps}(\hat{\xi}_{a}+\hat{\xi}_{b})^{-1+2\eps}(1+\hat{\xi}_{a}\hat{\xi}_{b})\\&\times(\xi_{a}+\xi_{a}\hat{\xi}_{a}\hat{\xi}_{b}+\xi_{b}(-1+\hat{\xi}_{b}^2))\,.
\esp
\eeq
Oh the fun we will have...

\subsection{Preparation of the integrand}
\label{sec:prepint}
As explained above, we reduce the number of convolution variables by two by introducing a change of variables
\begin{align}
    &\ha{\xi}_a \rightarrow \frac{\eta_a}{\xi_a}, \qquad\qquad\qquad\qquad\qquad \ha{\xi}_b \rightarrow \frac{\eta_b}{\xi_b}\,,\\
    &\xi_a \rightarrow (1 - \eta_a) (1 - \zeta_a) + \eta_a, \qquad \xi_b \rightarrow (1 - \eta_b) (1 - \zeta_b) + \eta_b\,.
\end{align}
The first set of transformations is such that only two integration variables remain in the matrix element, while the second set fixes the integration ranges to be between zero and one. The integral then takes the following form
\beq
\bsp
\label{eq:fullint}
&\mathcal{I} =\,-\int_{0}^{1} \rd \eta_a\, \rd \eta_b\,\rd\sigma_{\ha{\ha{a}}\ha{\ha{b}}}(\eta_a p_a,\eta_b p_b)\int_{0}^{1} \rd \zeta_{a} \rd \zeta_{b}\,\frac{1}{((-1 + \eta_{b}) \zeta_{b} + (-1 + \eta_{a}) \zeta_{a} (1 + (-1 + \eta_{b}) \zeta_{b}))}\\&\times\frac{1}{(\eta_{b} (-1 + \zeta_{a})^2 +\eta_{a}^2 \eta_{b} (-1 + \zeta_{a}^2) +\eta_{a} (1 - \eta_{b} (\eta_{b} + 2 (-1 + \zeta_{a}) \zeta_{a}) - 2 \zeta_{b} + 2 \eta_{b} \zeta_{b} + (-1 + \eta_{b})^2 \zeta_{b}^2))
}\\&\times(1 - \eta_{a})^{1 - 2 \eps} \eta_{a}^{-\eps} (1 - \eta_{b})^{-1 - 2 \eps}\eta_{b}^{-\eps} (1 - \zeta_{a})^{-\eps} \zeta_{a}^{-\eps} (1 - \zeta_{a} + \eta_{a} \zeta_{a})^{-\eps} (2 - \zeta_{a} +\eta_{a} \zeta_{a})^{-\eps}\\&\times (1 + \eta_{a} - \zeta_{a} + \eta_{a} \zeta_{a})^{-\eps} (1 - \zeta_{b})^{-1 - \eps}\zeta_{b}^{-1 - \eps} (1 - \zeta_{b} + \eta_{b} \zeta_{b})^{ 2 - \eps} (2 - \zeta_{b} + \eta_{b} \zeta_{b})^{-1 - \eps}\\&\times (1 + \eta_{b} - \zeta_{b} + \eta_{b} \zeta_{b})^{-1 - \eps} (2 - \zeta_{a} + \eta_{a} \zeta_{a} - \zeta_{b} + \eta_{b} \zeta_{b})^{-1 + 2 \eps}\\&\times (\eta_{a} + \eta_{b} - \eta_{b} \zeta_{a} + \eta_{a} \eta_{b} \zeta_{a} - \eta_{a} \zeta_{b} + \eta_{a} \eta_{b} \zeta_{b})^{-1 + 2 \eps} (2 - (1 - \eta_{b}) \zeta_{b} - (1 - \eta_{a}) \zeta_{a} (1 - (1 - \eta_{b}) \zeta_{b}))\\&\times ((1 - \zeta_{a}) (1 - (1 - \eta_{b}) \zeta_{b}) +\eta_{a} (\eta_{b} + \zeta_{a} - \zeta_{a} \zeta_{b} + \eta_{b} \zeta_{a} \zeta_{b}))\\&\times (-(1 - \zeta_{b}) (\zeta_{a} - \zeta_{b}) +\eta_{b} \zeta_{b} (-1 - \zeta_{a} + 2 \zeta_{b}) + \eta_{a} (\eta_{b} + \zeta_{a} - \zeta_{a} \zeta_{b} + \eta_{b} \zeta_{a} \zeta_{b}) -\eta_{b}^2 (-1 + \zeta_{b}^2))\,.
\esp
\eeq
Let us now focus on the inner $(\zeta_a,\zeta_b)$ integration, i.e., consider
\begin{equation}
    I(\eta_{a}, \eta_{b}; \ep) =\,\int_{0}^{1} \rd \zeta_{a}\, \rd \zeta_{b}\, I(\eta_{a}, \eta_{b},\zeta_a,\zeta_b; \ep)
\label{eq:intZetaaZetab}
\end{equation}
with
\beq
\bsp
\label{eq:unexpandedInt}
&I(\eta_{a}, \eta_{b},\zeta_a,\zeta_b; \ep) =\,\frac{- (\eta_{a} + \eta_{b} - \eta_{b} \zeta_{a} + \eta_{a} \eta_{b} \zeta_{a} - \eta_{a} \zeta_{b} + \eta_{a} \eta_{b} \zeta_{b})^{-1 + 2 \eps}}{((-1 + \eta_{b}) \zeta_{b} + (-1 + \eta_{a}) \zeta_{a} (1 + (-1 + \eta_{b}) \zeta_{b}))}\\&\times[(\eta_{b} (-1 + \zeta_{a})^2 +\eta_{a}^2 \eta_{b} (-1 + \zeta_{a}^2) +\eta_{a} (1 - \eta_{b} (\eta_{b} + 2 (-1 + \zeta_{a}) \zeta_{a}) - 2 \zeta_{b} + 2 \eta_{b} \zeta_{b} \\& + (-1 + \eta_{b})^2 \zeta_{b}^2))
]^{-1}(1 - \eta_{a})^{1 - 2 \eps} \eta_{a}^{-\eps} (1 - \eta_{b})^{-1 - 2 \eps}\eta_{b}^{-\eps} (1 - \zeta_{a})^{-\eps} \zeta_{a}^{-\eps} (1 - \zeta_{a} + \eta_{a} \zeta_{a})^{-\eps} \\&\times(2 - \zeta_{a} +\eta_{a} \zeta_{a})^{-\eps} (1 + \eta_{a} - \zeta_{a} + \eta_{a} \zeta_{a})^{-\eps} (1 - \zeta_{b})^{-1 - \eps}\zeta_{b}^{-1 - \eps} (1 - \zeta_{b} + \eta_{b} \zeta_{b})^{ 2 - \eps}\\&\times (2 - \zeta_{b} + \eta_{b} \zeta_{b})^{-1 - \eps} (1 + \eta_{b} - \zeta_{b} + \eta_{b} \zeta_{b})^{-1 - \eps} (2 - \zeta_{a} + \eta_{a} \zeta_{a} - \zeta_{b} + \eta_{b} \zeta_{b})^{-1 + 2 \eps}\\&\times(2 - (1 - \eta_{b}) \zeta_{b} - (1 - \eta_{a}) \zeta_{a} (1 - (1 - \eta_{b}) \zeta_{b})) ((1 - \zeta_{a}) (1 - (1 - \eta_{b}) \zeta_{b}) \\&+\eta_{a} (\eta_{b} + \zeta_{a} - \zeta_{a} \zeta_{b} + \eta_{b} \zeta_{a} \zeta_{b})) (-(1 - \zeta_{b}) (\zeta_{a} - \zeta_{b}) +\eta_{b} \zeta_{b} (-1 - \zeta_{a} + 2 \zeta_{b})\\& + \eta_{a} (\eta_{b} + \zeta_{a} - \zeta_{a} \zeta_{b} + \eta_{b} \zeta_{a} \zeta_{b}) -\eta_{b}^2 (-1 + \zeta_{b}^2))\,.
\esp
\eeq
The integrand of $I(\eta_{a}, \eta_{b}; \ep)$ is, quite obviously, a mean-looking rational function of the two integration variables $(\zeta_{a}, \zeta_{b})$, the additional parameters $(\eta_a,\eta_b)$ and the dimensional regulator $\eps$. To study its structure, we can perform a \textbf{multivariate partial fraction decomposition} of the integrand with respect to $(\zeta_{a}, \zeta_{b})$. Of course, it is not clear what exactly this means with symbolic powers of $\eps$ around. As such, we formally set $\eps=0$ and manipulate $I(\eta_{a}, \eta_{b},\zeta_a,\zeta_b; 0)$, which we are allowed to do as long as we are not actually performing the integration. The $\eps$-dependence will be reintroduced at a later stage. The partial fraction decomposition of $I(\eta_{a}, \eta_{b},\zeta_a,\zeta_b; 0)$ then reveals the presence of singularities as $\zeta_b\rightarrow 0,1$. To simplify the integration, it will be useful to disassemble $I(\eta_{a}, \eta_{b},\zeta_a,\zeta_b; 0)$ into terms with this singular behavior made explicit. As such, we write
\begin{equation}
\label{eq:I0}
    I(\eta_{a}, \eta_{b},\zeta_a,\zeta_b; 0) =\,I^{(1)}(\eta_{a}, \eta_{b},\zeta_a,\zeta_b; 0)+I^{(2)}(\eta_{a}, \eta_{b},\zeta_a,\zeta_b; 0)+I^{(3)}(\eta_{a}, \eta_{b},\zeta_a,\zeta_b; 0)\,
\end{equation}
with
\begin{align}
\label{eq:I01}
   &I^{(1)}(\eta_{a}, \eta_{b},\zeta_a,\zeta_b; 0) =\,\frac{1}{\zeta_b}\,\Bigg\{\frac{c_1^{(1)}}{(\eta_a-1) \zeta_a ((\eta_b-1) \zeta_b+1)+(\eta_b-1) \zeta_b}\nn\\&+\frac{c_2^{(1)}+\zeta_a \,c_3^{(1)}}{\eta_a^2 \eta_b \left(\zeta_a^2-1\right)+\eta_a \left(-\eta_b (\eta_b+2 (\zeta_a-1) \zeta_a)+(\eta_b-1)^2 \zeta_b^2+2 \eta_b \zeta_b-2 \zeta_b+1\right)+\eta_b (\zeta_a-1)^2}\Bigg\}\,,
\end{align}
\begin{align}
\label{eq:I02}
    &I^{(2)}(\eta_{a}, \eta_{b},\zeta_a,\zeta_b; 0) =\,\frac{1}{1-\zeta_b}\,\Bigg\{\frac{c^{(2)}_{1}}{(\eta_a-1) \zeta_a+(\eta_b-1) \zeta_b+2}\nn\\&+\frac{c^{(2)}_{2}}{(\eta_a-1) \zeta_a ((\eta_b-1) \zeta_b+1)+(\eta_b-1) \zeta_b}\nn\\&+\frac{c^{(2)}_{3}+\zeta_a \,c^{(2)}_{4} }{\eta_a^2 \eta_b \left(\zeta_a^2-1\right)+\eta_a \left(-\eta_b (\eta_b+2 (\zeta_a-1) \zeta_a)+(\eta_b-1)^2 \zeta_b^2+2 \eta_b \zeta_b-2 \zeta_b+1\right)+\eta_b (\zeta_a-1)^2}\Bigg\}
\end{align}
and
\begin{align}
\label{eq:I03}
    &I^{(3)}(\eta_{a}, \eta_{b},\zeta_a,\zeta_b; 0) =\,\frac{c^{(3)}_{1}}{(\zeta_b \eta_b+\eta_b-\zeta_b+1) (\eta_a \zeta_a-\zeta_a+\eta_b \zeta_b-\zeta_b+2)}\nn\\&+\frac{c^{(3)}_{2}}{(\eta_a \zeta_a-\zeta_a+\eta_b \zeta_b-\zeta_b+2) \left(\eta_b^2 \zeta_b^2-2 \eta_b \zeta_b^2+\zeta_b^2+2 \eta_b \zeta_b-2 \zeta_b+2\right)}\nn\\&+\frac{\zeta_b c^{(3)}_{3}}{(\eta_a \zeta_a-\zeta_a+\eta_b \zeta_b-\zeta_b+2) \left(\eta_b^2 \zeta_b^2-2 \eta_b \zeta_b^2+\zeta_b^2+2 \eta_b \zeta_b-2 \zeta_b+2\right)}\nn\\&+\frac{c^{(3)}_{4}}{(\eta_b \zeta_b-\zeta_b+2) (\eta_a \zeta_a-\eta_a \zeta_b \zeta_a+\eta_a \eta_b \zeta_b \zeta_a-\eta_b \zeta_b \zeta_a+\zeta_b \zeta_a-\zeta_a+\eta_b \zeta_b-\zeta_b)}\nn\\&+\frac{c^{(3)}_{5}}{(\zeta_b \eta_b+\eta_b-\zeta_b+1) (\eta_a \zeta_a-\eta_a \zeta_b \zeta_a+\eta_a \eta_b \zeta_b \zeta_a-\eta_b \zeta_b \zeta_a+\zeta_b \zeta_a-\zeta_a+\eta_b \zeta_b-\zeta_b)}\nn\\&+\frac{c^{(3)}_{6}}{\mathcal{D}_{1}(\eta_a,\eta_b,\zeta_a,\zeta_b)}+\frac{c^{(3)}_{7}}{\mathcal{D}_{2}(\eta_a,\eta_b,\zeta_a,\zeta_b)}+\frac{c^{(3)}_{10}+\zeta_b c^{(3)}_{11}}{\mathcal{D}_{3}(\eta_a,\eta_b,\zeta_a,\zeta_b)}\nn\\&+\frac{c^{(3)}_{8}}{2 (\eta_a+\eta_b) \zeta_b (\eta_b \zeta_a \eta_a+\eta_b \zeta_b \eta_a-\zeta_b \eta_a+\eta_a+\eta_b-\eta_b \zeta_a)}\nn\\&+\frac{c^{(3)}_{9}}{(\eta_b \zeta_b-\zeta_b+2) (\eta_b \zeta_a \eta_a+\eta_b \zeta_b \eta_a-\zeta_b \eta_a+\eta_a+\eta_b-\eta_b \zeta_a)}\nn\\&+\frac{c^{(3)}_{12}}{\mathcal{D}_{4}(\eta_a,\eta_b,\zeta_a,\zeta_b)}+\frac{c^{(3)}_{13}+\zeta_a c^{(3)}_{15}}{\mathcal{D}_{5}(\eta_a,\eta_b,\zeta_a,\zeta_b)}+\frac{c^{(3)}_{14}+\zeta_a c^{(3)}_{16}}{\mathcal{D}_{6}(\eta_a,\eta_b,\zeta_a,\zeta_b)}+\frac{c^{(3)}_{17}}{\mathcal{D}_{7}(\eta_a,\eta_b,\zeta_a,\zeta_b)}\,.
\end{align}
The denominators $\mathcal{D}_{i}(\eta_a,\eta_b,\zeta_a,\zeta_b)$ are in turn defined as
\begin{align}
    \mathcal{D}_{1}(\eta_a,\eta_b,\zeta_a,\zeta_b)=\,&\, (\eta_a \zeta_a-\eta_a \zeta_b \zeta_a+\eta_a \eta_b \zeta_b \zeta_a-\eta_b \zeta_b \zeta_a+\zeta_b \zeta_a-\zeta_a+\eta_b \zeta_b-\zeta_b)\nn\\&\times \left(\eta_b^2 \zeta_b^2-2 \eta_b \zeta_b^2+\zeta_b^2+2 \eta_b \zeta_b-2 \zeta_b+2\right)\,,\\
     \mathcal{D}_{2}(\eta_a,\eta_b,\zeta_a,\zeta_b)=\,&\, (\eta_a \zeta_a-\eta_a \zeta_b \zeta_a+\eta_a \eta_b \zeta_b \zeta_a-\eta_b \zeta_b \zeta_a+\zeta_b \zeta_a-\zeta_a+\eta_b \zeta_b-\zeta_b)\nn\\&\times \left(\eta_a \eta_b^2 \zeta_b^2+\eta_a \zeta_b^2-2 \eta_a \eta_b \zeta_b^2-2 \eta_a \zeta_b+2 \eta_a \eta_b \zeta_b+\eta_a+\eta_b\right)\,,\\
     \mathcal{D}_{3}(\eta_a,\eta_b,\zeta_a,\zeta_b)=\,&\,(\eta_b \zeta_a \eta_a+\eta_b \zeta_b \eta_a-\zeta_b \eta_a+\eta_a+\eta_b-\eta_b \zeta_a) (\eta_a \eta_b^2 \zeta_b^2+\eta_a \zeta_b^2-2 \eta_a \eta_b \zeta_b^2\nn\\&-2 \eta_a \zeta_b+2 \eta_a \eta_b \zeta_b+\eta_a+\eta_b) \,,\\
     \mathcal{D}_{4}(\eta_a,\eta_b,\zeta_a,\zeta_b)=\,&\,\eta_b \zeta_a^2 \eta_a^2-\eta_b \eta_a^2-\eta_b^2 \eta_a-2 \eta_b \zeta_a^2 \eta_a+\eta_b^2 \zeta_b^2 \eta_a-2 \eta_b \zeta_b^2 \eta_a+\zeta_b^2 \eta_a+2 \eta_b \zeta_a \eta_a\nn\\&+2 \eta_b \zeta_b \eta_a-2 \zeta_b \eta_a+\eta_a+\eta_b \zeta_a^2+\eta_b-2 \eta_b \zeta_a \,,\\
     \mathcal{D}_{5}(\eta_a,\eta_b,\zeta_a,\zeta_b)=\,&\,(\eta_b \zeta_b-\zeta_b+2) (\eta_b \zeta_a^2 \eta_a^2-\eta_b \eta_a^2-\eta_b^2 \eta_a-2 \eta_b \zeta_a^2 \eta_a+\eta_b^2 \zeta_b^2 \eta_a-2 \eta_b \zeta_b^2 \eta_a\nn\\&+\zeta_b^2 \eta_a+2 \eta_b \zeta_a \eta_a+2 \eta_b \zeta_b \eta_a-2 \zeta_b \eta_a+\eta_a+\eta_b \zeta_a^2+\eta_b-2 \eta_b \zeta_a) \,,\\
     \mathcal{D}_{6}(\eta_a,\eta_b,\zeta_a,\zeta_b)=\,&\,(\zeta_b \eta_b+\eta_b-\zeta_b+1) (\eta_b \zeta_a^2 \eta_a^2-\eta_b \eta_a^2-\eta_b^2 \eta_a-2 \eta_b \zeta_a^2 \eta_a+\eta_b^2 \zeta_b^2 \eta_a\nn\\&-2 \eta_b \zeta_b^2 \eta_a+\zeta_b^2 \eta_a+2 \eta_b \zeta_a \eta_a+2 \eta_b \zeta_b \eta_a-2 \zeta_b \eta_a+\eta_a+\eta_b \zeta_a^2+\eta_b-2 \eta_b \zeta_a) \,,\\
    \mathcal{D}_{7}(\eta_a,\eta_b,\zeta_a,\zeta_b)=\,&\, (\eta_a \zeta_a-\eta_a \zeta_b \zeta_a+\eta_a \eta_b \zeta_b \zeta_a-\eta_b \zeta_b \zeta_a+\zeta_b \zeta_a-\zeta_a+\eta_b \zeta_b-\zeta_b) (\eta_b \zeta_a^2 \eta_a^2\nn\\&-\eta_b \eta_a^2-\eta_b^2 \eta_a-2 \eta_b \zeta_a^2 \eta_a+\eta_b^2 \zeta_b^2 \eta_a-2 \eta_b \zeta_b^2 \eta_a+\zeta_b^2 \eta_a+2 \eta_b \zeta_a \eta_a\nn\\&+2 \eta_b \zeta_b \eta_a-2 \zeta_b \eta_a+\eta_a+\eta_b \zeta_a^2+\eta_b-2 \eta_b \zeta_a)\,.
\end{align}
The $c^{(i)}_j$ represent constants with respect to the $(\zeta_a,\zeta_b)$ integration.
As we are  already inside an appendix, we feel no shame in listing their explicit definitions right here, right now.\newline

\noindent\textbf{$I^{(1)}$ coefficients}
\begin{align}
    c^{(1)}_1 =\,&\, -\frac{(1-\eta_a)  (\eta_a \eta_b+1)\eta_b}{2 (1-\eta_b) (1+\eta_b) (\eta_a+\eta_b) (1-\eta_a \eta_b)} \\
    c^{(1)}_2 =\,&\, \frac{(1-\eta_a) [\eta_b (\eta_a \eta_b (\eta_a+\eta_b)+\eta_a+3 \eta_b)-2]}{2 (1-\eta_b) (1+\eta_b) (\eta_a+\eta_b) (1-\eta_a \eta_b)}\\
    c^{(1)}_3 =\,&\,-\frac{(1-\eta_a)^2 (\eta_b (\eta_a+2 \eta_b)-1)}{2 (1-\eta_b) (1+\eta_b) (\eta_a+\eta_b) (1-\eta_a \eta_b)}
\end{align}

\noindent\textbf{$I^{(2)}$ coefficients}
\begin{align}
    c^{(2)}_1 =\,&\, \frac{(1-\eta_a) \eta_b}{2 \left(\eta_b^2+1\right) (\eta_a+\eta_b)} \\
    c^{(2)}_2 =\,&\,\frac{(1-\eta_a) \eta_b^3}{\left(1-\eta_b^4\right) (1-\eta_a \eta_b)}\\
    c^{(2)}_3 =\,&\,-\frac{\left(1-\eta_a^2\right)  (\eta_a \eta_b+1)\eta_b^2}{2 (1-\eta_b) (1+\eta_b) (\eta_a+\eta_b) (1-\eta_a \eta_b)}\\
    c^{(2)}_4 =\,&\,\frac{(1-\eta_a)^2  (\eta_a \eta_b+1)\eta_b^2}{2 (1-\eta_b) (1+\eta_b) (\eta_a+\eta_b) (1-\eta_a \eta_b)}
\end{align}

\noindent\textbf{$I^{(3)}$ coefficients}
\begin{align}
    c^{(3)}_1 =\,&\, -\frac{(1-\eta_a) (1-\eta_b) \eta_b}{2 \left(\eta_b^2+1\right) (\eta_a+\eta_b)} \\
    c^{(3)}_2 =\,&\,\frac{(1-\eta_a) (1-\eta_b) (\eta_b (\eta_a+\eta_b)+2)}{\left(\eta_b^2+1\right) (\eta_a-\eta_b) (\eta_a+\eta_b)}\\
    c^{(3)}_3 =\,&\,-\frac{(1-\eta_a) (1-\eta_b)^2 (\eta_b (\eta_a+\eta_b)+2)}{\left(\eta_b^2+1\right) (\eta_a-\eta_b) (\eta_a+\eta_b)} \\
    c^{(3)}_4 =\,&\,-\frac{(1-\eta_a)  (\eta_a \eta_b+1)\eta_b}{2 (1+\eta_b) (\eta_a+\eta_b) (1-\eta_a \eta_b)} \\
    c^{(3)}_5 =\,&\,\frac{(1-\eta_a) \eta_b^3}{(1+\eta_b) \left(\eta_b^2+1\right) (1-\eta_a \eta_b)}\\
    c^{(3)}_6 =\,&\,\frac{(1-\eta_a) (1-\eta_b) (\eta_b (\eta_a+\eta_b)+2)}{\left(\eta_b^2+1\right) (\eta_a-\eta_b) (\eta_a+\eta_b)} \\
    c^{(3)}_7 =\,&\,-\frac{2 (1-\eta_a) (1-\eta_b) \eta_b}{(\eta_a-\eta_b) (\eta_a+\eta_b)} \\
    c^{(3)}_8 =\,&\,1-\eta_a \\
    c^{(3)}_9 =\,&\, -\frac{(1-\eta_a) (1-\eta_b)}{2 (\eta_a+\eta_b)}\\
    c^{(3)}_{10} =\,&\, -\frac{2 (1-\eta_a) \eta_a (1-\eta_b) \eta_b}{(\eta_a-\eta_b) (\eta_a+\eta_b)}\\
    c^{(3)}_{11} =\,&\,\frac{2 (1-\eta_a) \eta_a (1-\eta_b)^2 \eta_b}{(\eta_a-\eta_b) (\eta_a+\eta_b)}\\
    c^{(3)}_{12} =\,&\,-\frac{(1-\eta_a) (1-\eta_b)}{\eta_a+\eta_b}\\
    c^{(3)}_{13} =\,&\,-\frac{(1-\eta_a) \eta_b}{2 (1+\eta_b)} \\
    c^{(3)}_{14} =\,&\,\frac{(1-\eta_a)^2  (\eta_a \eta_b+1)\eta_b^2}{2 (1+\eta_b) (\eta_a+\eta_b) (1-\eta_a \eta_b)}\\
    c^{(3)}_{15} =\,&\,\frac{(1-\eta_a)^2 (\eta_b (\eta_a+2 \eta_b)-1)}{2 (1+\eta_b) (\eta_a+\eta_b) (1-\eta_a \eta_b)} \\
    c^{(3)}_{16} =\,&\,-\frac{(1-\eta_a)^2  (\eta_a \eta_b+1)\eta_b^2}{2 (1+\eta_b) (\eta_a+\eta_b) (1-\eta_a \eta_b)} \\
    c^{(3)}_{17} =\,&\,-\frac{2 (1-\eta_a) (1-\eta_b)}{(\eta_a+\eta_b) (1-\eta_a \eta_b)}
    \label{eq:c317}
\end{align}
It turns out that the integration involving $\mathcal{D}_{7}(\eta_a,\eta_b,\zeta_a,\zeta_b)$ is more challenging than the others. For this reason we rewrite eq.~(\ref{eq:I0}) as
\begin{equation}
\label{eq:Integral}
    I(\eta_{a}, \eta_{b}; \ep) =\,I_1(\eta_a,\eta_b;\eps)+I_2(\eta_a,\eta_b;\eps)
\end{equation}
with
\begin{align}
\label{eq:Integral1}
    I_1(\eta_{a}, \eta_{b}; \eps) =\,&\, \int_{0}^{1} \rd \zeta_{a}\, \rd \zeta_{b}\, \Bigg\{I^{(1)}(\eta_{a}, \eta_{b},\zeta_a,\zeta_b;\eps)+I^{(2)}(\eta_{a}, \eta_{b},\zeta_a,\zeta_b;\eps)+I^{(3)}(\eta_{a}, \eta_{b},\zeta_a,\zeta_b;\eps)\nn\\&-\frac{c^{(3)}_{17}}{\mathcal{D}_{7}(\eta_a,\eta_b,\zeta_a,\zeta_b)}\Bigg\}\,,\\
    I_2(\eta_{a}, \eta_{b}; \eps) =\,&\, \int_{0}^{1} \rd \zeta_{a}\, \rd \zeta_{b}\, \frac{c^{(3)}_{17}}{\mathcal{D}_{7}(\eta_a,\eta_b,\zeta_a,\zeta_b)}\,.
\label{eq:Integral2}
\end{align}
Note that we now also reinstated the full $\eps$ dependence. The computation of the integrals in eqs.~(\ref{eq:Integral1}) and (\ref{eq:Integral2}) will be explained in detail below. Before we can start the analytic integration however, we first need to deal with the \textit{overlapping singularities} that show up in the various denominators. For example, one factor that appears in the integrand is
\begin{equation}
    \frac{1}{\eta_a \eta_b \zeta_a \zeta_b-\eta_a \zeta_a \zeta_b+\eta_a \zeta_a-\eta_b \zeta_a \zeta_b+\eta_b \zeta_b+\zeta_a \zeta_b-\zeta_a-\zeta_b}\,.
\end{equation}
While this leads to regular expressions as either $\zeta_a\to 0$ or $\zeta_b\to 0$, it diverges as both $\zeta_a$ and $\zeta_b$ approach zero at the same time. To disentangle such overlaps in a consistent manner, one can apply \textbf{sector decomposition} (see e.g. \cite{Heinrich:2008si} and references therein), which factorizes all singularities of the integrand by the introduction of different integration sectors. To illustrate this procedure, consider the integral $\mathcal{I}_{\text{toy}}$\footnote{Coming soon to an Apple store near you.} defined as
\begin{equation}
    \mathcal{I}_{\text{toy}} =\,\int_{0}^{1}\rd x\int_{0}^{1}\rd y\, x^{-1-\eps}y^{-\eps}\frac{1}{x+y}\,.
\end{equation}
Clearly there is an overlapping singularity as $x,y\rightarrow 0$. To disentangle this overlap, we divide the integration region into two sectors $A$ and $B$ as follows
\begin{figure}[H]
\centering
\includegraphics{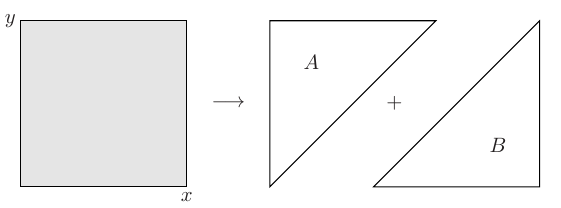}
\end{figure}
\noindent Formally this means that $\mathcal{I}_{\text{toy}}$ is rewritten as
\begin{equation}
    \mathcal{I}_{\text{toy}} =\,\int_{0}^{1}\rd x\int_{0}^{1}\rd y\, x^{-1-\eps}y^{-\eps}\frac{1}{x+y}\Big[\underbrace{\theta(x-y)}_{A}+\underbrace{\theta(y-x)}_{B}\Big]\,.
\end{equation}
Next we transform
\beq
\bsp
    y\to u x
\esp
\eeq
in region A and
\beq
\bsp
    x\to u y
\esp
\eeq
in region B leading to
\begin{equation}
    \mathcal{I}_{\text{toy}} =\,\int_{0}^{1}\rd x\, x^{-1-2\eps}\int_{0}^{1}\rd u\, \frac{u^{-\eps}}{1+u} + \int_{0}^{1}\rd y\, y^{-1-2\eps}\int_{0}^{1}\rd u\, \frac{u^{-1-\eps}}{1+u}\,.
\end{equation}
All singularities are now nicely factorized. The same reasoning is applied to our integral in eq.~(\ref{eq:Integral}) using an in-house routine. However, because of the non-trivial structure of the integrand, it does not suffice to have a single step of sector decomposition, and the process needs to be \textit{iterated}. At the end of the iterative decomposition, all singularities are explicitly factorized. Marvellous. Of course, the singularities are still $\dots$ singular. Hence, we need to regularize our integrals. This can be achieved by setting up an appropriate \textbf{subtraction}. Consider for example
\begin{equation}
    \int_{0}^{1}\rd x\,x^{-1-\eps}f(x),
\end{equation}
with $f(x)$ some function which remains finite as $x\rightarrow 0$. To regularize the $x\to 0$ singularity we write
\begin{equation}
\label{eq:subtract}
    \int_{0}^{1}\rd x\,x^{-1-\eps}f(x)\to\int_{0}^{1}\rd x\,x^{-1-\eps}[f(x)-f(0)]+\underbrace{\int_{0}^{1}\rd x\,x^{-1-\eps}f(0)}_{\frac{f(0)}{\eps}}.
\end{equation}
The first integral is regular per construction\footnote{Note that this term is often identified with a \textit{plus-distribution}, i.e., in four dimensions,
\begin{equation*}
    \int_{0}^{1}\rd x\,\left(\frac{1}{x}\right)_{+}f(x) =\, \int_{0}^{1}\rd x\,\frac{1}{x}\,[f(x)-f(0)]\,.
\end{equation*}}, while the second one produces an explicit factor of $1/\eps$. Hence, this procedure leads to a representation of eq.~(\ref{eq:Integral}) which has explicit powers of $1/\eps$ but otherwise just contains finite integrals. As such, one can now do a \textit{numeric} computation, which comes in handy for cross-checking any analytic results we obtain later on. Since the integration problem at hand is multi-dimensional, we use the {\tt CUBA} library~\cite{Hahn:2004fe}, which collects several general-purpose multi-dimensional integration algorithms. Specifically we employ {\tt Vegas}~\cite{Lepage:1977sw,Lepage:123074} for our numeric evaluations, which implements an iterative and adaptive \textbf{Monte Carlo algorithm}. For example, for $\eta_a=1/10$ and $\eta_b=2/10$ we find
\begin{equation}
\label{eq:num1}
    I(1/10,2/10;\eps) =\,\frac{0.361395}{\eps^2}+\frac{8.11945}{\eps}+45.1866\,.
\end{equation}
Note the appearance of explicit factors of $1/\eps^2$ and $1/\eps$. Also note that their coefficients are actual numbers, which is usually what you want after performing a numeric integration.

With the sector decomposed representation of our integral at hand, we can now proceed with its analytical computation. Because of a significant difference in complexity of the two integrals appearing in eq.~(\ref{eq:Integral}), we present their computations separately.

\subsection{Analytic computation of $I_1(\eta_{a}, \eta_{b}; \eps)$}
\label{sec:compI1}
Based on general grounds, we expect the result of the integration to depend on \textbf{generalized polylogarithms} (GPLs)\footnote{Also commonly referred to as \textit{multiple polylogarithms} (MPLs).}, which are recursively defined as~\cite{Goncharov:1998kja}\footnote{A brief overview of some useful properties of GPLs can be found in Appendix \ref{appx:GPLs}.}
\begin{equation}
    G(a_1,\dots,a_n;z) =\,\int_{0}^{z}\frac{\rd t}{t-a_1}\,G(a_2,\dots,a_n;t)\,, \qquad G(z) \equiv G(;z) =1\,.
\end{equation}
As such, we will make extensive use of the {\tt Mathematica} package {\tt PolyLogTools}~\cite{Duhr:2019tlz}, which allows for the integration and manipulation of such functions. An important property of GPLs is that their integration kernels are \textit{linear}. However, the denominators that appear in $I_1(\eta_{a}, \eta_{b}; \eps)$ contain higher-order polynomials in the integration variables. In particular, we have three quadratics in $\zeta_a$
\beq
\bsp
\label{eq:quadsa}
 &\Big\{\eta_b \zeta_a (\eta_a-1) \left((\eta_a-1) \zeta_a+2\right)-(\eta_a+\eta_b) (\eta_a \eta_b-1)\,,\eta_b \zeta_a(\eta_a-1)  \left((1-\eta_a) \zeta_a+2\right)\\&+\zeta_b \left((\eta_a-1)^2 \eta_b \left(\zeta_a-2\right) \zeta_a+\eta_a \eta_b (\eta_a+\eta_b-4)+\eta_a+\eta_b\right)+(\eta_a+\eta_b) (\eta_a \eta_b-1)\,,\\
    &(\eta_a-1)^2 \eta_b \zeta_a^2+2 (\eta_a-1) \eta_b \zeta_a+\eta_a (\eta_b-1) \zeta_b \left((\eta_b-1) \zeta_b+2\right)-(\eta_a+\eta_b) (\eta_a \eta_b-1)\Big\}
\esp
\eeq
and four quadratics in $\zeta_b$
\beq
\bsp
\label{eq:quadsb}
&\Big\{(\eta_b-1) \zeta_b \left((\eta_b-1) \zeta_b+2\right)+2\,,\eta_a \zeta_b(\eta_b-1)  \left((\eta_b-1) \zeta_b+2\right)+\eta_a+\eta_b\,,(\eta_a-1)^2 \eta_b \zeta_a^2\\&+2 (\eta_a-1) \eta_b \zeta_a+\eta_a\zeta_b (\eta_b-1)  \left((\eta_b-1) \zeta_b+2\right)-(\eta_a+\eta_b) (\eta_a \eta_b-1)\,,\\&\zeta_a \zeta_b\left(\eta_a (\eta_b-1)^2 \left(\zeta_b-2\right) +\eta_a \eta_b (\eta_a+\eta_b-4)+\eta_a+\eta_b\right)+\eta_a (\eta_b-1) \zeta_b \left((1-\eta_b) \zeta_b-2\right)\\&+(\eta_a+\eta_b) (\eta_a \eta_b-1)\Big\}\,.
\esp
\eeq
Hence, we start by \textbf{factorizing} these polynomials. Assuming we first integrate over $\zeta_a$, we first consider the set in eq.~(\ref{eq:quadsa}) for which we find the following six roots\footnote{As the roots obviously come in pairs, we use the notation $r_{i/j}$ to collectively denote roots $i$ and $j$. We use the convention that the top sign in $\mp$ or $\pm$ corresponds to root $i$ while the bottom one corresponds to root $j$.}
\beq
\bsp
\label{eq:roots1}
&\Bigg\{r_{1/2}=\frac{\mp\sqrt{\eta_a} \sqrt{\eta_a \eta_b+\eta_b^2-1}-\sqrt{\eta_b}}{(\eta_a-1) \sqrt{\eta_b}},\\&r_{3/4}=\frac{\eta_a \sqrt{\eta_b} \zeta_b\mp\sqrt{\eta_a} \sqrt{\eta_a \eta_b-(\eta_b-1) \left(\zeta_b-1\right) \left(\eta_b \zeta_b+\eta_b-\zeta_b+1\right)}-\sqrt{\eta_b} \zeta_b+\sqrt{\eta_b}}{(\eta_a-1) \sqrt{\eta_b} \left(\zeta_b-1\right)},\\&r_{5/6}=\frac{\mp\sqrt{\eta_a} \sqrt{\eta_a \eta_b-(\eta_b-1) \left(\zeta_b-1\right) \left(\eta_b \zeta_b+\eta_b-\zeta_b+1\right)}-\sqrt{\eta_b}}{(\eta_a-1) \sqrt{\eta_b}}\Bigg\}\,.
\esp
\eeq
Note that $\zeta_b$ explicitly appears under the square root. While this does not bode well for the $\zeta_b$-integration, we can safely ignore it for now. At this point, all denominators are at most (powers of) linear factors of $\zeta_a$ and hence we can proceed with the integration. For application of the integration routine of {\tt PolyLogTools}, called {\tt GIntegrate}, all GPLs need to be expressed in terms of a \textbf{fibration basis} with respect to the integration variable, cf.~eq.~(\ref{eq:fibre}).\footnote{This is a slight abuse of notation introduced for notational simplicity. Of course a fibration basis is determined by choosing a specific ordering of \textit{all} variables that appear in the GPL. However, for the integration it is most important that the integration variable is the first variable of the fibration, while the ordering of the other variables typically does not influence the complexity of the integration. For example, if the integration variable is $x$, choosing a fibration basis with respect to $x$ really means with respect to $\{x,\dots\}$ with the ordering of the other variables left implicit.} As currently all GPLs that appear in the integrand are logarithms with simple arguments, coming from integrations in the iterated sector decomposition, this step is trivial. Finally, the integrand should be partial fractioned with respect to $\zeta_a$. As mentioned above, we integrate using {\tt GIntegrate}. The latter does not actually compute the integral itself, but rather the \textit{primitive function} of the integrand. As such, one still needs to evaluate the result in the integration limits and take the difference. In doing so, it is possible to generate terms that formally diverge, such as $G(a;a)$. These of course should cancel in the final result. However, often this cancellation is not immediate and requires the use of GPL identities, such as the \textbf{shuffle relation} in eq.~(\ref{eq:shuffle}). Alternatively, the prefactor of the offending diverging GPL can simply be evaluated numerically, making use of the {\tt ginsh} command of {\tt PolyLogTools}, which is based on the {\tt GiNaC} framework \cite{Bauer:2000cp}. For the sake of explicitness, we now present the pole part of the solution. We find
\begin{equation}
    I_1(\eta_a,\eta_b;\eps) =\,\int_{0}^{1}\rd\zeta_b\;\Tilde{I}_1(\eta_a,\eta_b,\zeta_b;\eps)
\end{equation}
with
\bleq{res1}
    $\Tilde{I}_1(\eta_a,\eta_b,\zeta_b;\eps)=\,${\large$\frac{1}{\eps^2}\frac{\eta_a \eta_b^2+\eta_b}{2  (\eta_b-1) (1+\eta_b) (\eta_a+\eta_b) (\eta_a \eta_b-1)}+\frac{1}{\eps}\Bigg\{\frac{\eta_b (\eta_a \eta_b+1) (\eta_a-1)^2}{8 \eta_a (\eta_b-1) (1+\eta_b) (\eta_a+\eta_b) (\eta_a \eta_b-1) (\zeta_b-1) (\eta_a+\eta_b-\eta_b \zeta_b+\zeta_b-2)}+\frac{\eta_b (\eta_a \eta_b+1) G(0;\eta_b) (\eta_a-1)}{8 \eta_a (\eta_b-1) (1+\eta_b) (\eta_a+\eta_b) (\eta_a \eta_b-1)}+\frac{\eta_b (\eta_a \eta_b+1) G\left(-\frac{-\eta_a-\eta_b+2}{\eta_a-1};1\right) (\eta_a-1)}{8 \eta_a (\eta_b-1) (1+\eta_b) (\eta_a+\eta_b) (\eta_a \eta_b-1)}+\frac{\eta_b (\eta_a \eta_b+1) G(-\eta_b;\eta_a) (\eta_a-1)}{4 \eta_a (\eta_b-1) (1+\eta_b) (\eta_a+\eta_b) (\eta_a \eta_b-1)}-\frac{\eta_b (\eta_a \eta_b+1) G(1;\eta_a) (\eta_a-1)}{4 \eta_a (\eta_b-1) (1+\eta_b) (\eta_a+\eta_b) (\eta_a \eta_b-1)}-\frac{\eta_b (\eta_a \eta_b+1) G(1;\eta_b) (\eta_a-1)}{4 \eta_a (\eta_b-1) (1+\eta_b) (\eta_a+\eta_b) (\eta_a \eta_b-1)}-\frac{\eta_b (\eta_a \eta_b+1) G(-1;\eta_a) (\eta_a-1)}{8 \eta_a (\eta_b-1) (1+\eta_b) (\eta_a+\eta_b) (\eta_a \eta_b-1)}-\frac{\eta_b (\eta_a \eta_b+1) G(-1;\eta_b) (\eta_a-1)}{8 \eta_a (\eta_b-1) (1+\eta_b) (\eta_a+\eta_b) (\eta_a \eta_b-1)}-\frac{\eta_b (\eta_a \eta_b+1) G(0;\eta_a) (\eta_a-1)}{8 \eta_a (\eta_b-1) (1+\eta_b) (\eta_a+\eta_b) (\eta_a \eta_b-1)}-\frac{\eta_b (\eta_a \eta_b+1) (\eta_a-1)}{8 \eta_a (\eta_b-1) (1+\eta_b) (\eta_a+\eta_b) (\eta_a \eta_b-1) (\zeta_b-1)}+\frac{\eta_b (\eta_a \eta_b+1) (\eta_a-1)}{4 (1-\eta_b) (1+\eta_b) (\eta_a+\eta_b) (\eta_a \eta_b-1) \zeta_b (\eta_a+\eta_b \zeta_b-\zeta_b-1)}+\frac{\eta_b (\eta_a \eta_b+1) G(-1;\eta_a)}{4 (1-\eta_b) (1+\eta_b) (\eta_a+\eta_b) (\eta_a \eta_b-1)}+\frac{\eta_b (\eta_a \eta_b+1) G(-1;\eta_b)}{4 (1-\eta_b) (1+\eta_b) (\eta_a+\eta_b) (\eta_a \eta_b-1)}+\frac{\eta_b (\eta_a \eta_b+1) G(0;\eta_a)}{4 (1-\eta_b) (1+\eta_b) (\eta_a+\eta_b) (\eta_a \eta_b-1)}+\frac{(1+\eta_a) \eta_b (\eta_a \eta_b+1) G(0;\eta_b)}{8 \eta_a (\eta_b-1) (1+\eta_b) (\eta_a+\eta_b) (\eta_a \eta_b-1)}+\frac{\eta_b (\eta_a \eta_b+1) G(1;\eta_a)}{2 (1-\eta_b) (1+\eta_b) (\eta_a+\eta_b) (\eta_a \eta_b-1)}+\frac{\eta_b (\eta_a \eta_b+1) G(1;\eta_b)}{2 (1-\eta_b) (1+\eta_b) (\eta_a+\eta_b) (\eta_a \eta_b-1)}+\frac{\eta_b (\eta_a \eta_b+1) G\left(-\frac{\eta_a+1}{\eta_a-1};1\right)}{4 (\eta_b-1) (1+\eta_b) (\eta_a+\eta_b) (\eta_a \eta_b-1)}+\frac{\eta_b (\eta_a \eta_b+1) G\left(-\frac{\eta_a+1}{\eta_a-1};1\right)}{4 \eta_a (\eta_b-1) (1+\eta_b) (\eta_a+\eta_b) (\eta_a \eta_b-1)}+\frac{(1+\eta_a) \eta_b (\eta_a \eta_b+1) G\left(-\frac{\eta_a+1}{\eta_a-1};1\right)}{4 (\eta_a-1) \eta_a (\eta_b-1) (1+\eta_b) (\eta_a+\eta_b) (\eta_a \eta_b-1)}+\frac{\eta_b G\left(-\frac{-\eta_a-\eta_b}{\eta_a-1};1\right)}{2 (\eta_a+\eta_b) \left(\eta_b^2+1\right)}+\frac{(1+\eta_a) \eta_b (\eta_a \eta_b+1) G\left(-\frac{-\eta_a-\eta_b+2}{\eta_a-1};1\right)}{8 \eta_a (\eta_b-1) (1+\eta_b) (\eta_a+\eta_b) (\eta_a \eta_b-1)}+\frac{\eta_b (\eta_a \eta_b+1) G\left(-\frac{\eta_b-1}{\eta_a-1};1\right)}{4 (1-\eta_b) (1+\eta_b) (\eta_a+\eta_b) (\eta_a \eta_b-1)}+\frac{(1+\eta_a) \eta_b (\eta_a \eta_b+1) G(-\eta_b;\eta_a)}{4 \eta_a (\eta_b-1) (1+\eta_b) (\eta_a+\eta_b) (\eta_a \eta_b-1)}+\frac{G\left(\frac{\eta_a+\eta_b}{(1-\eta_a) \eta_b};1\right)}{2 \eta_b (\eta_a+\eta_b)}+\frac{\left(2 \eta_b^2+\eta_a \eta_b-1\right) G\left(r_2;1\right) \left(1-\eta_a\right)r_2 }{4 \sqrt{\eta_a}\sqrt{\eta_b} (1-\eta_b)  (1+\eta_b) (\eta_a+\eta_b) (\eta_a \eta_b-1) \sqrt{\eta_b^2+\eta_a \eta_b-1}}-\frac{\eta_b (\eta_a \eta_b+1) G(-\eta_b;\eta_a)}{2 (1-\eta_b) (1+\eta_b) (\eta_a+\eta_b) (\eta_a \eta_b-1)}-\frac{\eta_b (\eta_a \eta_b+1) G(0;\eta_b)}{4 (1-\eta_b) (1+\eta_b) (\eta_a+\eta_b) (\eta_a \eta_b-1)}-\frac{(1+\eta_a) \eta_b (\eta_a \eta_b+1) G\left(-\frac{\eta_a+1}{\eta_a-1};1\right)}{4 (\eta_a-1) (\eta_b-1) (1+\eta_b) (\eta_a+\eta_b) (\eta_a \eta_b-1)}-\frac{(1+\eta_a) \eta_b (\eta_a \eta_b+1) G(1;\eta_a)}{4 \eta_a (\eta_b-1) (1+\eta_b) (\eta_a+\eta_b) (\eta_a \eta_b-1)}-\frac{(1+\eta_a) \eta_b (\eta_a \eta_b+1) G(1;\eta_b)}{4 \eta_a (\eta_b-1) (1+\eta_b) (\eta_a+\eta_b) (\eta_a \eta_b-1)}-\frac{(1+\eta_a) \eta_b (\eta_a \eta_b+1) G(-1;\eta_a)}{8 \eta_a (\eta_b-1) (1+\eta_b) (\eta_a+\eta_b) (\eta_a \eta_b-1)}-\frac{(1+\eta_a) \eta_b (\eta_a \eta_b+1) G(-1;\eta_b)}{8 \eta_a (\eta_b-1) (1+\eta_b) (\eta_a+\eta_b) (\eta_a \eta_b-1)}-\frac{(1+\eta_a) \eta_b (\eta_a \eta_b+1) G(0;\eta_a)}{8 \eta_a (\eta_b-1) (1+\eta_b) (\eta_a+\eta_b) (\eta_a \eta_b-1)}-\frac{\eta_b^2 G\left(\frac{1-\eta_a \eta_b}{(1-\eta_a) \eta_b};1\right)}{(1-\eta_b) (1+\eta_b) (\eta_a \eta_b-1) \left(\eta_b^2+1\right)}-\frac{(1+\eta_a) \eta_b (\eta_a \eta_b+1)}{8 \eta_a (\eta_b-1) (1+\eta_b) (\eta_a+\eta_b) (\eta_a \eta_b-1) (\zeta_b-1)}-\frac{\eta_b (\eta_a \eta_b+1)}{4 (1-\eta_b) (1+\eta_b) (\eta_a+\eta_b) (\eta_a \eta_b-1) \zeta_b}+\frac{\left(\eta_a^2-1\right) \eta_b (\eta_a \eta_b+1)}{8 \eta_a (\eta_b-1) (1+\eta_b) (\eta_a+\eta_b) (\eta_a \eta_b-1) (\zeta_b-1) (\eta_a+\eta_b-\eta_b \zeta_b+\zeta_b-2)}-\frac{\left(1-\eta_a\right) \left(2 \eta_b^2+\eta_a \eta_b-1\right) G\left(r_1;1\right)r_1}{4 \sqrt{\eta_a}\sqrt{\eta_b} (1-\eta_b) (1+\eta_b) (\eta_a+\eta_b) (\eta_a \eta_b-1) \sqrt{\eta_b^2+\eta_a \eta_b-1}}-\frac{\left(1-\eta_a\right) \left(\eta_a^2 \eta_b^2+3 \eta_b^2+\eta_a \left(\eta_b^3+\eta_b\right)-2\right) G\left(r_2;1\right)}{4 (\eta_a-1) \sqrt{\eta_a}\sqrt{\eta_b} (1-\eta_b)  (1+\eta_b) (\eta_a+\eta_b) (1-\eta_a \eta_b) \sqrt{\eta_b^2+\eta_a \eta_b-1}}-\frac{\left(\eta_a^2 \eta_b^2+3 \eta_b^2+\eta_a \left(\eta_b^3+\eta_b\right)-2\right) G\left(r_1;1\right) }{4  \sqrt{\eta_a}\sqrt{\eta_b} (1-\eta_b)  (1+\eta_b) (\eta_a+\eta_b) (1-\eta_a \eta_b) \sqrt{\eta_b^2+\eta_a \eta_b-1}}\Bigg\}$}$+\,\mathcal{O}(\eps^0)\,.$
\eleq
\noindent The functional form of the finite $\mathcal{O}(\eps^0)$ part in eq.~(\ref{eq:res1}) is much more complicated and hence omitted here. In particular, it is there that $\zeta_b$ also appears under the square root. Such terms can be simplified by constructing symmetric and anti-symmetric combinations of the form
\beq
\bsp
&r_{34}^{+} =\,r_3+r_4\,,\qquad
r_{34}^{-} =\,r_3-r_4\,,\\
&r_{56}^{+} =\,r_5+r_6,\,\qquad
r_{56}^{-} =\,r_5-r_6\,.
\esp
\eeq
The roots in eq.~(\ref{eq:roots1}) can then be replaced by
\beq
\bsp
\label{eq:newRepRoots}
&r_3\,=\,\frac{r_{34}^{+}+r_{34}^{-}}{2}\,,\qquad
r_4\,=\,\frac{r_{34}^{+}-r_{34}^{-}}{2}\,,\\
&r_5\,=\,\frac{r_{56}^{+}+r_{56}^{-}}{2}\,,\qquad
r_6\,=\,\frac{r_{56}^{+}-r_{56}^{-}}{2}\,.
\esp
\eeq
Since these roots only appear in weight-one GPLs, i.e. logarithms, the replacements in eq.~(\ref{eq:newRepRoots}) lead to a representation of the integrand $\Tilde{I}_1(\eta_a,\eta_b,\zeta_b;\eps)$ in which the symmetric part becomes free of square roots. The anti-symmetric part however still contains $\sqrt{\eta_a \eta_b-(\eta_b-1) (\zeta_b-1) (\eta_b \zeta_b+\eta_b-\zeta_b+1)}$. As the integration of the latter is more complicated, we will treat it separately. For this reason we decompose the integrand explicitly in terms of its symmetric part and anti-symmetric part,
\begin{equation}
    \Tilde{I}_1(\eta_a,\eta_b,\zeta_b;\eps) =\,\Tilde{I}_1^{\mathcal{S}}(\eta_a,\eta_b,\zeta_b;\eps)+\Tilde{I}_1^{\mathcal{A}}(\eta_a,\eta_b,\zeta_b;\eps)\,.
\end{equation}
We will compute the $\zeta_b$ integrals of both terms separately in the next two sections.

\subsubsection{Integrating the symmetric part}
We begin with the treatment of $\Tilde{I}_1^{\mathcal{S}}(\eta_a,\eta_b,\zeta_b;\eps)$. As before, we start by factorizing any higher-order polynomials in $\zeta_b$ that appear in the denominators. We find three quadratics which lead to the following six roots
\beq
\bsp
\label{eq:roots2}
&\Bigg\{r_{7/8}=\frac{1\pm i}{1-\eta_b},\,r_{9/10}=\,\frac{\sqrt{\eta_a}\pm i \,\sqrt{\eta_b}}{\sqrt{\eta_a} (1-\eta_b)},\,r_{11/12}=\,\frac{\sqrt{\eta_a}(1-\eta_b)\mp i\sqrt{\eta_b}(1-\eta_a)}{\sqrt{\eta_a} (1-\eta_b)}\Bigg\}\,.
\esp
\eeq
Note the explicit appearance of the imaginary unit $i$. Next we perform a partial fraction decomposition in $\zeta_b$. Finally, the integrand should be written in a fibration basis. This is complicated by the fact that the arguments of some of the GPLs are rational functions with non-linear denominators in $\zeta_b$. The latter should also be factorized, leading to two more roots,
\beq
\bsp
\label{eq:roots3}
\left\{r_{13/14}=\,\frac{\sqrt{\eta_a}\pm\sqrt{\eta_b} \sqrt{\eta_a^2+\eta_a \eta_b-1}}{\sqrt{\eta_a} (1-\eta_b)}\right\}\,.
\esp
\eeq
The logarithms can furthermore be simplified using the standard rules\footnote{Of course, when applying the rules in eq.~(\ref{eq:logrels}), one needs to be careful to take into account the proper factors of $2\pi i$ when any of the arguments become real and negative or complex.}
\beq
\label{eq:logrels}
\ln (x\,y) =\,\ln (x)+\ln (y)\,, \qquad \ln\left(\frac{x}{y}\right) =\,\ln (x)-\ln (y)\,. 
\eeq
We can now apply the {\tt GIntegrate} routine to compute the primitive function of $\Tilde{I}_1^{\mathcal{S}}(\eta_a,\eta_b,\zeta_b;\eps)$. The result of the integration, after removing the spurious divergences as discussed below eq.~(\ref{eq:roots1}), reads\newpage
\bleq{res2}
$\int_{0}^{1}\rd\zeta_b\,\Tilde{I}_1^{\mathcal{S}}(\eta_a,\eta_b,\zeta_b;\eps)=\dfrac{\eta_b (\eta_a \eta_b+1)}{2\eps^2 (\eta_b-1) (1+\eta_b) (\eta_a+\eta_b) (\eta_a \eta_b-1)}-\dfrac{1}{4\eps \left(\eta_b^4-1\right) \left(\eta_b \eta_a^2+\left(\eta_b^2-1\right) \eta_a-\eta_b\right) \eta_b}\Bigg\{\dfrac{\eta_b^{5/2} G\left(r_1;1\right) \eta_a^{3/2}}{\sqrt{\eta_b^2+\eta_a \eta_b-1}}+\dfrac{\eta_b^{9/2} G\left(r_1;1\right) \eta_a^{3/2}}{\sqrt{\eta_b^2+\eta_a \eta_b-1}}-\dfrac{\eta_b^{5/2} G\left(r_2;1\right) \eta_a^{3/2}}{\sqrt{\eta_b^2+\eta_a \eta_b-1}}-\dfrac{\eta_b^{9/2} G\left(r_2;1\right) \eta_a^{3/2}}{\sqrt{\eta_b^2+\eta_a \eta_b-1}}+2 \eta_b^5 G(0;\eta_a) \eta_a+2 \eta_b^3 G(0;\eta_a) \eta_a-2 \eta_b^5 G(0;\eta_b) \eta_a-2 \eta_b^3 G(0;\eta_b) \eta_a+4 \eta_b^5 G(1;\eta_a) \eta_a+4 \eta_b^3 G(1;\eta_a) \eta_a+4 \eta_b^5 G(1;\eta_b) \eta_a+4 \eta_b^3 G(1;\eta_b) \eta_a-\eta_b^3 G\left(r_1;1\right) \eta_a-\eta_b G\left(r_1;1\right) \eta_a-\eta_b^3 G\left(r_2;1\right) \eta_a-\eta_b G\left(r_2;1\right) \eta_a+\eta_b^5 G\left(\dfrac{1-\eta_b}{\eta_a-1};1\right) \eta_a+\eta_b^3 G\left(\dfrac{1-\eta_b}{\eta_a-1};1\right) \eta_a-\eta_b^5 G\left(\dfrac{\eta_a+\eta_b-2}{\eta_a-1};1\right) \eta_a-\eta_b^3 G\left(\dfrac{\eta_a+\eta_b-2}{\eta_a-1};1\right) \eta_a-2 \eta_b^5 G\left(\dfrac{\eta_a+\eta_b}{\eta_a-1};1\right) \eta_a+2 \eta_b^3 G\left(\dfrac{\eta_a+\eta_b}{\eta_a-1};1\right) \eta_a-\eta_b^5 G\left(\dfrac{1-\eta_a}{\eta_b-1};1\right) \eta_a-\eta_b^3 G\left(\dfrac{1-\eta_a}{\eta_b-1};1\right) \eta_a+\eta_b^5 G\left(\dfrac{\eta_a+\eta_b-2}{\eta_b-1};1\right) \eta_a+\eta_b^3 G\left(\dfrac{\eta_a+\eta_b-2}{\eta_b-1};1\right) \eta_a-4 \eta_b^5 G(-\eta_b;\eta_a) \eta_a-4 \eta_b^3 G(-\eta_b;\eta_a) \eta_a-2 \eta_b^5 G\left(\dfrac{\eta_a+\eta_b}{\eta_b-\eta_a \eta_b};1\right) \eta_a+2 \eta_b G\left(\dfrac{\eta_a+\eta_b}{\eta_b-\eta_a \eta_b};1\right) \eta_a-4 \eta_b^3 G\left(\dfrac{1-\eta_a \eta_b}{\eta_b-\eta_a \eta_b};1\right) \eta_a+2 \eta_b^2 (\eta_a \eta_b+1) \left(\eta_b^2+1\right) G(-1;\eta_a)+2 \eta_b^2 (\eta_a \eta_b+1) \left(\eta_b^2+1\right) G(-1;\eta_b)+2 \eta_b^4 G(0;\eta_a)+2 \eta_b^2 G(0;\eta_a)-2 \eta_b^4 G(0;\eta_b)-2 \eta_b^2 G(0;\eta_b)+4 \eta_b^4 G(1;\eta_a)+4 \eta_b^2 G(1;\eta_a)+4 \eta_b^4 G(1;\eta_b)+4 \eta_b^2 G(1;\eta_b)-2 \eta_b^4 G\left(r_1;1\right)-\eta_b^2 G\left(r_1;1\right)+G\left(r_1;1\right)-2 \eta_b^4 G\left(r_2;1\right)-\eta_b^2 G\left(r_2;1\right)+G\left(r_2;1\right)+\eta_b^2 G\left(\dfrac{1-\eta_b}{\eta_a-1};1\right)-\eta_b^4 G\left(\dfrac{\eta_a+\eta_b-2}{\eta_a-1};1\right)-\eta_b^2 G\left(\dfrac{\eta_a+\eta_b-2}{\eta_a-1};1\right)+2 \eta_b^4 G\left(\dfrac{\eta_a+\eta_b}{\eta_a-1};1\right)-2 \eta_b^2 G\left(\dfrac{\eta_a+\eta_b}{\eta_a-1};1\right)-\eta_b^4 G\left(\dfrac{1-\eta_a}{\eta_b-1};1\right)-\eta_b^2 G\left(\dfrac{1-\eta_a}{\eta_b-1};1\right)+\eta_b^4 G\left(\dfrac{\eta_a+\eta_b-2}{\eta_b-1};1\right)+\eta_b^2 G\left(\dfrac{\eta_a+\eta_b-2}{\eta_b-1};1\right)-4 \eta_b^4 G(-\eta_b;\eta_a)-4 \eta_b^2 G(-\eta_b;\eta_a)+2 \eta_b^4 G\left(\dfrac{\eta_a+\eta_b}{\eta_b-\eta_a \eta_b};1\right)-2 G\left(\dfrac{\eta_a+\eta_b}{\eta_b-\eta_a \eta_b};1\right)-4 \eta_b^4 G\left(\dfrac{1-\eta_a \eta_b}{\eta_b-\eta_a \eta_b};1\right)-\dfrac{\sqrt{\eta_a} \eta_b^{7/2} G\left(r_2;1\right)}{\sqrt{\eta_b^2+\eta_a \eta_b-1}}-\dfrac{\sqrt{\eta_a} \eta_b^{11/2} G\left(r_2;1\right)}{\sqrt{\eta_b^2+\eta_a \eta_b-1}}+\dfrac{\eta_b^{7/2} G\left(r_1;1\right) \sqrt{\eta_a}}{\sqrt{\eta_b^2+\eta_a \eta_b-1}}+\dfrac{\eta_b^{11/2} G\left(r_1;1\right) \sqrt{\eta_a}}{\sqrt{\eta_b^2+\eta_a \eta_b-1}}+\dfrac{\eta_b^{9/2} G\left(r_1;1\right)}{\sqrt{\eta_a} \sqrt{\eta_b^2+\eta_a \eta_b-1}}-\dfrac{\sqrt{\eta_b} G\left(r_1;1\right)}{\sqrt{\eta_a} \sqrt{\eta_b^2+\eta_a \eta_b-1}}-\dfrac{\eta_b^{9/2} G\left(r_2;1\right)}{\sqrt{\eta_a} \sqrt{\eta_b^2+\eta_a \eta_b-1}}+\dfrac{G\left(r_2;1\right) \sqrt{\eta_b}}{\sqrt{\eta_a} \sqrt{\eta_b^2+\eta_a \eta_b-1}}\Bigg\}+\mathcal{O}(\eps^0)\,.$
\eleq
As before, the finite piece is more involved and we do not present it here.

\subsubsection{Integrating the anti-symmetric part}
Next we compute the integral of $\Tilde{I}_1^{\mathcal{A}}(\eta_a,\eta_b,\zeta_b;\eps)$ with respect to $\zeta_b$. Although the denominators do not have new higher-order polynomials in the integration variable, $\zeta_b$ now appears under the square root in a non-trivial way. In particular, we find that the integrand contains the following two logarithms,
{\beq
\bsp
\label{eq:sqrts}
&\Bigg\{\ln \Bigg(\frac{\left(\sqrt{\eta_a \eta_b-(1-\eta_b) \zeta_b (\zeta_b-\eta_b (\zeta_b-2))}+\sqrt{\eta_a} \sqrt{\eta_b}\right)}{\left(\sqrt{\eta_a \eta_b-(1-\eta_b) \zeta_b (\zeta_b-\eta_b (\zeta_b-2))}-\sqrt{\eta_a} \sqrt{\eta_b}\right)}\\&\times\frac{\left(\sqrt{\eta_a} \sqrt{\eta_a \eta_b-(1-\eta_b) \zeta_b (\zeta_b-\eta_b (\zeta_b-2))}-\sqrt{\eta_b}\right)}{\left(\sqrt{\eta_a} \sqrt{\eta_a \eta_b-(1-\eta_b) \zeta_b (\zeta_b-\eta_b (\zeta_b-2))}+\sqrt{\eta_b}\right)}\Bigg),\\&\ln \Bigg(\frac{\left(\sqrt{\eta_a \eta_b-(1-\eta_b) \zeta_b (\eta_b (-\zeta_b)+2 \eta_b+\zeta_b)}-\sqrt{\eta_a} \sqrt{\eta_b}\right)}{\left(\sqrt{\eta_a \eta_b-(1-\eta_b) \zeta_b (\zeta_b-\eta_b (\zeta_b-2))}+\sqrt{\eta_a} \sqrt{\eta_b}\right)}\\&\times \frac{\left(-\eta_a \sqrt{\eta_b} (\zeta_b-1)+\sqrt{\eta_a} \sqrt{\eta_a \eta_b-(1-\eta_b) \zeta_b (\eta_b (-\zeta_b)+2 \eta_b+\zeta_b)}+\sqrt{\eta_b} \zeta_b\right)}{ \left(\eta_a \sqrt{\eta_b} (\zeta_b-1)+\sqrt{\eta_a} \sqrt{\eta_a \eta_b-(1-\eta_b) \zeta_b (\eta_b (-\zeta_b)+2 \eta_b+\zeta_b)}-\sqrt{\eta_b} \zeta_b\right)}\Bigg)\Bigg\}\,.
\esp
\eeq}
As we want the integration kernel to be linear to stay in the realm of GPLs, we need to \textbf{rationalize} the roots, which boils down to the construction of a suitable transformation of the integration variable to remove the root altogether. This can be done automatically using the {\tt RationalizeRoots} package \cite{Besier:2019kco}. For the expressions shown in eq.~(\ref{eq:sqrts}), it turns out that the appropriate transformation is
\begin{equation}
\label{eq:rationalize}
    \zeta_b\to \frac{2 \eta_b (\eta_a t+\eta_b-1)}{\eta_b \left(\eta_a t^2-2\right)+\eta_b^2+1}\,.
\end{equation}
Note that, while the integration over $\zeta_b$ simply ran between zero and one, the limits of the $t$-integration are non-trivial. In particular, the lower limit becomes
\begin{equation}
\label{eq:tmin}
    t_{\text{min}} =\,\frac{1-\eta_b}{\eta_a}
\end{equation}
while for the upper limit we actually find two distinct solutions
\begin{equation}
\label{eq:tmax}
    t_{\text{max}}^{\pm} =\,\frac{\eta_a \eta_b\pm\sqrt{\eta_a \eta_b \left(\eta_a \eta_b+\eta_b^2-1\right)}}{\eta_a \eta_b}\,.
\end{equation}
To decide which value to choose, we can perform a numerical test. For this, consider the test function
\begin{equation}
    f(\zeta_b,\eta_a,\eta_b) =\,\sqrt{\eta_a \eta_b-(1-\eta_b) \zeta_b (\zeta_b-\eta_b (\zeta_b-2))}\,.
\end{equation}
For the sake of argument, we choose $\eta_a=1/10$ and $\eta_b=1/5$. Numerically integrating $f$ between zero and one then gives\footnote{We just use the standard {\tt NIntegrate} command here without any further options.}
\begin{equation}
    \int_{0}^{1}\rd\zeta_b\,f(\zeta_b,1/10,1/5) =\,0.00540291\, +0.527791 i\,.
\end{equation} 
Next we can compute the same integral with the transformation in eq.~(\ref{eq:rationalize}). We find
\begin{equation}
    \int_{t_{\text{min}}}^{t_{\text{max}}^{+}}\rd t\,\mathcal{J}(t,1/10,1/5)f(t,1/10,1/5) =\,0.00540282\, -0.527791 i
\end{equation}
and
\begin{equation}
    \int_{t_{\text{min}}}^{t_{\text{max}}^{-}}\rd t\,\mathcal{J}(t,1/10,1/5)f(t,1/10,1/5) =\,0.00540282\, +0.527791 i\,.
\end{equation}
Here
\begin{equation}
   \mathcal{J}(t,\eta_a,\eta_b)=\frac{2 \eta_a \eta_b (\eta_b (t (2-\eta_a t)-2 \eta_b t+\eta_b-2)+1)}{\left(\eta_a \eta_b t^2+(\eta_b-1)^2\right)^2}
\end{equation}
is the Jacobian associated to the transformation in eq.~(\ref{eq:rationalize}). From these considerations it is clear that one should use
\begin{equation}
\label{eq:tmaxfin}
    t_{\text{max}}\equiv t_{\text{max}}^{-}=\frac{\eta_a \eta_b-\sqrt{\eta_a \eta_b \left(\eta_a \eta_b+\eta_b^2-1\right)}}{\eta_a \eta_b}\,.
\end{equation}
After applying the transformation in eq.~(\ref{eq:rationalize}) to our integrand, 
\begin{equation}
    \Tilde{I}_1^{\mathcal{A}}(\eta_a,\eta_b,\zeta_b;\eps)\to\mathcal{J}(t,\eta_a,\eta_b)\Tilde{I}_1^{\mathcal{A}}(\eta_a,\eta_b,t;\eps)\,,
\end{equation}
we can continue with the integration. First we need to partial fraction the integrand and identify any higher-order polynomials in $t$. We find six such polynomials, leading to the following additional roots
\beq
\bsp
\label{eq:roots4}
&\Bigg\{r_{15/16}=\,\frac{\sqrt{\eta_a} \sqrt{\eta_b}\mp\sqrt{\eta_a \eta_b+\eta_b^2-1}}{\sqrt{\eta_a} \sqrt{\eta_b}},\,r_{17/18}=\,\mp\frac{i (\eta_b-1)}{\sqrt{\eta_a} \sqrt{\eta_b}},\\&r_{19/20}=\,\mp\frac{(\eta_b-1) \left(\sqrt{\eta_a+\eta_b}+\sqrt{\eta_b}\right)}{\eta_a \sqrt{\eta_b}},\,r_{21/22}=\,\frac{(\eta_b-1) \left(\sqrt{\eta_a} \sqrt{\eta_b}\mp\sqrt{\eta_a \eta_b+\eta_b^2-1}\right)}{\sqrt{\eta_a} \sqrt{\eta_b} (1+\eta_b)}\Bigg\}\,.
\esp
\eeq
Furthermore, the logarithms contain two quadratic polynomials in $t$ as well, which should also be factorized. This gives an additional set of four roots
\beq
\bsp
\label{eq:roots5}
&\Bigg\{r_{23/24}=\-\frac{(\eta_b-1) \left(\sqrt{\eta_a} \sqrt{\eta_b}\pm\sqrt{\eta_a^2+\eta_a \eta_b-1}\right)}{(\eta_a-1) \sqrt{\eta_a} \sqrt{\eta_b}},\\&r_{25/26}=\,-\frac{(\eta_b-1) \left(\sqrt{\eta_a} \sqrt{\eta_b}\pm\sqrt{\eta_a^2+\eta_a \eta_b-1}\right)}{ (1+\eta_a) \sqrt{\eta_a}\sqrt{\eta_b}}\Bigg\}\,.
\esp
\eeq
It will be more convenient in what follows to integrate $t$ between zero and one. As such, we transform
\begin{equation}
    t\to (t_{\text{max}} - t_{\text{min}}) (1 - t) + t_{\text{min}}\,.
\end{equation}
Note that this introduces a Jacobian $t_{\text{max}} - t_{\text{min}}$ with $t_{\text{min}}$ given in eq.~(\ref{eq:tmin}) and $t_{\text{max}}$ in eq.~(\ref{eq:tmaxfin}). We can then finally apply {\tt GIntegrate} to compute the primitive function and take the difference of the latter evaluated in one and zero. As before, the coefficients of any divergent $G(a;a)$-type terms vanish as expected. Since the integration does not produce any poles in $\eps$, we do not show any explicit results here. We note though that the result contains GPLs of maximum weight two. In particular, introducing the shorthands
\beq
    \mathsf{r}_a =\, \sqrt{\eta_a \eta_b \left(\eta_a \eta_b+\eta_a^2-1\right)}\text{\,\,and\,\,}\mathsf{r}_b =\, \sqrt{\eta_a \eta_b \left(\eta_a \eta_b+\eta_b^2-1\right)}\,,
\eeq 
we find seven weight-one GPLs of the form $G(a;1)$ with
\beq
\bsp
a&\in\Bigg\{\frac{\mathsf{r}_b+\eta_a \eta_b}{\eta_b-\eta_b^2},-\frac{\mathsf{r}_b+\eta_a}{(\eta_b-1) (\eta_a+\eta_b)},-\frac{2 \eta_a \eta_b}{-\eta_a \eta_b+i \sqrt{\eta_a}\sqrt{\eta_b} (\eta_b-1) +\mathsf{r}_b},\\&\frac{2 \eta_a \eta_b}{\eta_a \eta_b+i \sqrt{\eta_a}\sqrt{\eta_b} (\eta_b-1) -\mathsf{r}_b},\frac{2\mathsf{r}_b}{\mathsf{r}_b-\eta_b (\eta_a+\eta_b-1)},\frac{2 \eta_a \eta_b+(\eta_b-1) \mathsf{r}_b-(1+\eta_b) \mathsf{r}_b}{(1+\eta_b) \left(\eta_b (\eta_a+\eta_b-1)-\mathsf{r}_b\right)},\\&-\frac{-2 \eta_a \eta_b+(\eta_b-1) \mathsf{r}_b+(1+\eta_b) \mathsf{r}_b}{(1+\eta_b) \left(\eta_b (\eta_a+\eta_b-1)-\mathsf{r}_b\right)}\Bigg\}\,
\esp
\eeq
and 57 weight-two GPLs of the form $G(b,c;1)$ with
\beq
\bsp
b,c&\in\Bigg\{1,\frac{\mathsf{r}_b+\eta_a \eta_b}{\eta_b-\eta_b^2},\frac{\mathsf{r}_b-\eta_b}{\mathsf{r}_b-\eta_b (\eta_a+\eta_b-1)},-\frac{\mathsf{r}_b+\eta_a}{(\eta_b-1) (\eta_a+\eta_b)},\\&-\frac{2 \eta_a \eta_b}{-\eta_a \eta_b+i \sqrt{\eta_a}\sqrt{\eta_b} (\eta_b-1) +\mathsf{r}_b},\frac{2 \eta_a \eta_b}{\eta_a \eta_b+i \sqrt{\eta_a}\sqrt{\eta_b} (\eta_b-1) -\mathsf{r}_b},\frac{2\mathsf{r}_b}{\mathsf{r}_b-\eta_b (\eta_a+\eta_b-1)},\\&\frac{2 \eta_a \eta_b+(\eta_b-1) \mathsf{r}_b-(1+\eta_b) \mathsf{r}_b}{(1+\eta_b) \left(\eta_b (\eta_a+\eta_b-1)-\mathsf{r}_b\right)},-\frac{-2 \eta_a \eta_b+(\eta_b-1) \mathsf{r}_b+(1+\eta_b) \mathsf{r}_b}{(1+\eta_b) \left(\eta_b (\eta_a+\eta_b-1)-\mathsf{r}_b\right)},\\&\frac{\eta_a^2 \eta_b+\eta_a \left(\eta_b^2-\mathsf{r}_b\right)-\mathsf{r}_a (\eta_b-1)}{(1+\eta_a) \left(\eta_b (\eta_a+\eta_b-1)-\mathsf{r}_b\right)}-\frac{\mathsf{r}_b}{(1+\eta_a) \left(\eta_b (\eta_a+\eta_b-1)-\mathsf{r}_b\right)},\\&\frac{\eta_a^2 \eta_b+\eta_a \left(\eta_b^2-\mathsf{r}_b\right)+\mathsf{r}_a (\eta_b-1)}{(1+\eta_a) \left(\eta_b (\eta_a+\eta_b-1)-\mathsf{r}_b\right)}-\frac{\mathsf{r}_b}{(1+\eta_a) \left(\eta_b (\eta_a+\eta_b-1)-\mathsf{r}_b\right)},-\frac{\mathsf{r}_b+\eta_b}{\eta_b (\eta_a+\eta_b-2)},\\&\frac{\eta_a^2 \eta_b+\eta_a (\eta_b-2) \eta_b-\eta_a \mathsf{r}_b-\mathsf{r}_a (\eta_b-1)}{(\eta_a-1) \left(\eta_b (\eta_a+\eta_b-1)-\mathsf{r}_b\right)}+\frac{\mathsf{r}_b}{(\eta_a-1) \left(\eta_b (\eta_a+\eta_b-1)-\mathsf{r}_b\right)},\\&\frac{\eta_a^2 \eta_b+\eta_a (\eta_b-2) \eta_b-\eta_a \mathsf{r}_b+\mathsf{r}_a (\eta_b-1)}{(\eta_a-1) \left(\eta_b (\eta_a+\eta_b-1)-\mathsf{r}_b\right)}+\frac{\mathsf{r}_b}{(\eta_a-1) \left(\eta_b (\eta_a+\eta_b-1)-\mathsf{r}_b\right)}\Bigg\}\,.
\esp
\eeq
Here of course $b\neq 1$.

\subsection{Analytic computation of $I_2(\eta_{a}, \eta_{b}; \eps)$}
\label{sec:compI2}
We now proceed with the computation of the second integral,
\begin{equation}
    I_2(\eta_{a}, \eta_{b}; \eps) =\,\int_{0}^{1} \rd \zeta_{a}\, \rd \zeta_{b}\, \frac{c^{(3)}_{17}}{\mathcal{D}_{7}(\eta_a,\eta_b,\zeta_a,\zeta_b)}\,
\end{equation}
with
\begin{align}
    \mathcal{D}_{7}(\eta_a,\eta_b,\zeta_a,\zeta_b)=\,&\, (\eta_a \zeta_a-\eta_a \zeta_b \zeta_a+\eta_a \eta_b \zeta_b \zeta_a-\eta_b \zeta_b \zeta_a+\zeta_b \zeta_a-\zeta_a+\eta_b \zeta_b-\zeta_b) (\eta_b \zeta_a^2 \eta_a^2\nn\\&-\eta_b \eta_a^2-\eta_b^2 \eta_a-2 \eta_b \zeta_a^2 \eta_a+\eta_b^2 \zeta_b^2 \eta_a-2 \eta_b \zeta_b^2 \eta_a+\zeta_b^2 \eta_a+2 \eta_b \zeta_a \eta_a\nn\\&+2 \eta_b \zeta_b \eta_a-2 \zeta_b \eta_a+\eta_a+\eta_b \zeta_a^2+\eta_b-2 \eta_b \zeta_a)\,.
\end{align}
We remind the reader that $c^{(3)}_{17}$ is simply a function of $\eta_a$ and $\eta_b$, cf.~eq.~(\ref{eq:c317}). After performing the sector decomposition, which was described at the end of sec.~\ref{sec:defIC}, we can simplify the problem by setting
\begin{align}
\zeta_a\to \frac{v_1-1}{\eta_a-1},\qquad\zeta_b\to \frac{v_2-1}{\eta_b-1}    
\end{align}
such that
\begin{equation}
    I_2(\eta_{a}, \eta_{b}; \eps) =\,-\int_{\eta_a}^{1} \rd v_1\, \int_{\eta_b}^{1}\rd v_2\, \frac{c^{(3)}_{17}}{(\eta_a-1) (\eta_b-1) (v_1 v_2-1) \left(\eta_a^2 \eta_b+\eta_a \left(\eta_b^2-v_2^2\right)-\eta_b v_1^2\right)}\,.
\end{equation}
Note that the last factor of the denominator is still quadratic in \textit{both} integration variables. We treat this in the following way. First, transform $v_1$ and $v_2$ as
\begin{equation}
    v_1\to v_1 \sqrt{\eta_a (\eta_a+\eta_b)},\qquad v_2\to v_2 \sqrt{\eta_b (\eta_a+\eta_b)}
\end{equation}
leading to
\begin{equation}
     I_2(\eta_{a}, \eta_{b}; \eps) =\,\int_{v_1^{\text{min}}}^{v_1^{\text{max}}} \rd v_1\, \int_{v_2^{\text{min}}}^{v_2^{\text{max}}}\rd v_2\, \frac{c^{(3)}_{17}}{(\eta_a-1) \sqrt{\eta_a}\sqrt{\eta_b} (\eta_b-1)  \left(v_1^2+v_2^2-1\right) \left(\sqrt{\eta_a} \sqrt{\eta_b} v_1 v_2 (\eta_a+\eta_b)-1\right)}\,.
\end{equation}
The integration limits are
\begin{align}
    &v_1^{\text{min}}=\frac{\eta_a}{\sqrt{\eta_a} \sqrt{\eta_a+\eta_b}}\,, \qquad
    v_1^{\text{max}}=\frac{1}{\sqrt{\eta_a} \sqrt{\eta_a+\eta_b}}\,;\\
    &v_2^{\text{min}}=\frac{\eta_b}{\sqrt{\eta_b} \sqrt{\eta_a+\eta_b}}\,,\qquad 
    v_2^{\text{max}}=\frac{1}{\sqrt{\eta_b} \sqrt{\eta_a+\eta_b}}\,.
\end{align}
To lighten the notation, we will factor out the part which is independent of $(v_1,v_2)$. For this we write
\begin{equation}
    I_2(\eta_{a}, \eta_{b}; \eps) =\,f(\eta_a,\eta_b)\Tilde{I}_2(\eta_a,\eta_b; \eps)
\end{equation}
with
\begin{equation}
\label{eq:overallfac}
    f(\eta_a,\eta_b) =\,\frac{2}{\sqrt{\eta_a} \sqrt{\eta_b} (\eta_a+\eta_b) (\eta_a \eta_b-1)}
\end{equation}
and
\begin{equation}
\label{eq:I2til}
    \Tilde{I}_2(\eta_a,\eta_b; \eps) =\,\int_{v_1^{\text{min}}}^{v_1^{\text{max}}} \rd v_1\, \int_{v_2^{\text{min}}}^{v_2^{\text{max}}}\rd v_2\,\frac{1}{\left(v_1^2+v_2^2-1\right) (v_1 v_2 \sqrt{\eta_a} \sqrt{\eta_b} (\eta_a+\eta_b)-1)}\,.
\end{equation}
The quadratic denominator in eq.~(\ref{eq:I2til}) can now be factorized as
\begin{equation}
    v_1^2+v_2^2-1\to \left(v_2+\sqrt{1-v_1^2}\right)\left(v_2-\sqrt{1-v_1^2}\right)
\end{equation}
leading to
\begin{equation}
    \Tilde{I}_2(\eta_a,\eta_b; \eps) =\,\int_{v_1^{\text{min}}}^{v_1^{\text{max}}} \rd v_1\, \int_{v_2^{\text{min}}}^{v_2^{\text{max}}}\rd v_2\,\frac{1}{\left(v_2+\sqrt{1-v_1^2}\right)\left(v_2-\sqrt{1-v_1^2}\right) (v_1 v_2 \sqrt{\eta_a} \sqrt{\eta_b} (\eta_a+\eta_b)-1)}\,.
\end{equation}
At this point the denominator only has linear factors in $v_2$, such that the integration over the latter can be performed. We find
\beq
\bsp
\label{eq:I2til2}
&\Tilde{I}_2(\eta_a,\eta_b; \eps) =\;-\frac{1}{2}\int_{v_1^{\text{min}}}^{v_1^{\text{max}}}\frac{\rd v_1}{\sqrt{1-v_1^2} \left(\eta_a \eta_b \left(v_1^2-1\right) v_1^2 (\eta_a+\eta_b)^2+1\right)}\\&\times\Bigg\{G\left(-\sqrt{1-v_1^2};\frac{\sqrt{\eta_b}}{\sqrt{\eta_a+\eta_b}}\right)-G\left(-\sqrt{1-v_1^2};\frac{1}{\sqrt{\eta_b} \sqrt{\eta_a+\eta_b}}\right)-G\left(\sqrt{1-v_1^2};\frac{\sqrt{\eta_b}}{\sqrt{\eta_a+\eta_b}}\right)\\&-\sqrt{\eta_a} \sqrt{\eta_b} v_1 \sqrt{1-v_1^2} (\eta_a+\eta_b) \Bigg[G\left(-\sqrt{1-v_1^2};\frac{\sqrt{\eta_b}}{\sqrt{\eta_a+\eta_b}}\right)-G\left(-\sqrt{1-v_1^2};\frac{1}{\sqrt{\eta_b} \sqrt{\eta_a+\eta_b}}\right)\\&+G\left(\sqrt{1-v_1^2};\frac{\sqrt{\eta_b}}{\sqrt{\eta_a+\eta_b}}\right)-G\left(\sqrt{1-v_1^2};\frac{1}{\sqrt{\eta_b} \sqrt{\eta_a+\eta_b}}\right)\\&-2\; G\left(\frac{1}{\sqrt{\eta_a} \sqrt{\eta_b} v_1 (\eta_a+\eta_b)};\frac{\sqrt{\eta_b}}{\sqrt{\eta_a+\eta_b}}\right)+2\;G\left(\frac{1}{\sqrt{\eta_a} \sqrt{\eta_b} v_1 (\eta_a+\eta_b)};\frac{1}{\sqrt{\eta_b} \sqrt{\eta_a+\eta_b}}\right)\Bigg]\\&+G\left(\sqrt{1-v_1^2};\frac{1}{\sqrt{\eta_b} \sqrt{\eta_a+\eta_b}}\right)\Bigg\}\,.
\esp
\eeq
To evaluate the $v_1$-integration we need to rationalize $\sqrt{1-v_1^2}$, which can be achieved by setting
\begin{equation}
\label{eq:transv}
    v_1\to \frac{1-t^2}{t^2+1}\,.
\end{equation}
The resulting $t$-integration then runs between $t^{\text{min}}$ and $t^{\text{max}}$ with
\begin{equation}
\label{eq:tlims}
    t^{\text{min}}=\sqrt{\frac{1-v_1^{\text{min}}}{1+v_1^{\text{min}}}}\,,\qquad t^{\text{max}}=\sqrt{\frac{1-v_1^{\text{max}}}{1+v_1^{\text{max}}}}\,.
\end{equation}
Next we need to factorize the higher-order polynomials in $t$ that appear in the denominators of the integrand. As the transformation in eq.~(\ref{eq:transv}) is quadratic in $t$, the integrand in eq.~(\ref{eq:I2til2}) now contains \textit{quartic} polynomials in $t$, which introduce the following roots
\beq
\bsp
\label{eq:qroots}
&\Bigg\{q_{1/2}=\,\frac{1}{2} \Big(\eta_a^{3/2} \sqrt{\eta_b}-\sqrt{\eta_a^3 \eta_b+2 \eta_a^2 \eta_b^2+\eta_a \eta_b^3-4}\\&\mp \sqrt{2(\eta_a+\eta_b)\sqrt{\eta_a} \sqrt{\eta_b} \left(\eta_a^{3/2} \sqrt{\eta_b}-\sqrt{\eta_a^3 \eta_b+2 \eta_a^2 \eta_b^2+\eta_a \eta_b^3-4}+\sqrt{\eta_a} \eta_b^{3/2}\right)}+\sqrt{\eta_a} \eta_b^{3/2}\Big),\\&q_{3/4}=\,\frac{1}{2} \Big(\eta_a^{3/2} \sqrt{\eta_b}+\sqrt{\eta_a^3 \eta_b+2 \eta_a^2 \eta_b^2+\eta_a \eta_b^3-4}\\&\mp \sqrt{2(\eta_a+\eta_b)\sqrt{\eta_a} \sqrt{\eta_b} \left(\eta_a^{3/2} \sqrt{\eta_b}+\sqrt{\eta_a^3 \eta_b+2 \eta_a^2 \eta_b^2+\eta_a \eta_b^3-4}+\sqrt{\eta_a} \eta_b^{3/2}\right)}+\sqrt{\eta_a} \eta_b^{3/2}\Big)\Bigg\}\,.
\esp
\eeq
Finally, there are still some quadratic polynomials in the weight-one GPLs, the roots of which are
\beq
\bsp
\label{eq:roots6}
&\Bigg\{r_{27/28}=\,\sqrt{\eta_b} \sqrt{\eta_a+\eta_b}\mp\sqrt{\eta_a \eta_b+\eta_b^2-1},\,r_{29/30}=\,\frac{\sqrt{\eta_a+\eta_b}\mp\sqrt{\eta_a}}{\sqrt{\eta_b}},\\&r_{31/32}=\,\mp\sqrt{\frac{\sqrt{\eta_a} \sqrt{\eta_a+\eta_b}-1}{\sqrt{\eta_a} \sqrt{\eta_a+\eta_b}+1}},\,r_{33/34}=\,\mp\sqrt{\frac{\sqrt{\eta_a} \eta_b \sqrt{\eta_a+\eta_b}-1}{\sqrt{\eta_a} \eta_b \sqrt{\eta_a+\eta_b}+1}}\Bigg\}\,.
\esp
\eeq
At this point we are ready to feed the integrand to {\tt GIntegrate}. Evaluating the resulting primitive function in the integration limits, cf.~eq.~(\ref{eq:tlims}), taking the difference and multiplying back the overall factor of eq.~(\ref{eq:overallfac}), we obtain the final analytic expression for $I_2(\eta_a,\eta_b;\eps)$. As the result does not have any poles in $\eps$, we do not present it here.

\subsection{Assembly of the full result}
\label{sec:fullres}
Having obtained the analytic expressions for $I_{1}(\eta_a,\eta_b;\eps)$ and $I_{2}(\eta_a,\eta_b;\eps)$ to $\mathcal{O}(\eps^0)$, we can construct the full result $I(\eta_a,\eta_b;\eps)$, cf.~eq.~(\ref{eq:Integral}). The latter is a complicated function of $\eta_a$ and $\eta_b$ with GPLs up to weight two. In particular, we find 307 GPLs of weight two and 49 GPLs of weight one. Furthermore, the corresponding arguments are highly non-trivial, involving (combinations of) the roots $\{r_1,\dots,r_{34}\}$ presented in eqs.~(\ref{eq:roots1}), (\ref{eq:roots2}),  (\ref{eq:roots3}), (\ref{eq:roots4}), (\ref{eq:roots5}), (\ref{eq:roots6}) and $\{q_1,\dots,q_4\}$ in eq.~(\ref{eq:qroots}). However, thorough numerical verification supports the correctness of the result. For example, for $\eta_a=1/10$ and $\eta_b=2/10$ we find
\beq
\bsp
 I(1/10,2/10;\eps) \,=\, &\,\frac{0.36139455782312925170068027210884353742}{\eps^2}\\&+\frac{8.119439712687562538858026197047126714}{\eps}\\&+45.186699361703037968513689684200284590\,,
\esp
\eeq
which agrees nicely with our original numerical evaluation in eq.~(\ref{eq:num1}). 

\subsection{Integration over $\eta_a$ and $\eta_b$}
\label{sec:distexp}
Finally, our analytic expression for $I(\eta_a,\eta_b;\eps)$ needs to be integrated over $\eta_a$ and $\eta_b$, cf.~eq.~(\ref{eq:fullint}),
\begin{equation}
    \mathcal{I} =\,\int_{0}^{1}\rd\eta_a\,\int_{0}^{1}\rd\eta_b\,\rd\sigma_{\ha{\ha{a}}\ha{\ha{b}}}(\eta_a p_a,\eta_b p_b)\,I(\eta_a,\eta_b;\eps)\,.
\end{equation}
As discussed above, this needs to be done carefully due to endpoint singularities, cf.~eq.~(\ref{eq:fulldistexp}). In particular, we need to set up an appropriate subtraction. For the integral under consideration, we have
\beq
\bsp
\label{eq:fulldistexpI}
&\mathcal{I} =\, \int_{0}^{1} \rd \eta_a\, \int_0^{1} \rd \eta_b\,\bigg\{
I(\eta_a,\eta_b;\eps)\rd\sigma_{\ha{\ha{a}}\ha{\ha{b}}}(\eta_a p_a,\eta_b p_b)
-
\La I(\eta_a,\eta_b;\eps)\rd\sigma_{\ha{\ha{a}}\ha{\ha{b}}}( p_a,\eta_b p_b)
\\&-
\Lb I(\eta_a,\eta_b;\eps)\rd\sigma_{\ha{\ha{a}}\ha{\ha{b}}}(\eta_a p_a, p_b)
-
\Big(\Lab I(\eta_a,\eta_b;\eps)-\La\Lab I(\eta_a,\eta_b;\eps)
-\Lb\Lab I(\eta_a,\eta_b;\eps)\Big)\rd\sigma_{\ha{\ha{a}}\ha{\ha{b}}}(p_a, p_b)
\\&+
[\La] I(\eta_a,\eta_b;\eps)\rd\sigma_{\ha{\ha{a}}\ha{\ha{b}}}( p_a,\eta_b p_b)
+
[\Lb] I(\eta_a,\eta_b;\eps)\rd\sigma_{\ha{\ha{a}}\ha{\ha{b}}}(\eta_a p_a, p_b)
+
\Big([\Lab]I(\eta_a,\eta_b;\eps)-[\La \Lab] I(\eta_a,\eta_b;\eps)\\&-[\Lb \Lab] I(\eta_a,\eta_b;\eps)\Big)\rd\sigma_{\ha{\ha{a}}\ha{\ha{b}}}( p_a,p_b)
\bigg\}\,.
\esp
\eeq
The asymptotic behavior of $I(\eta_a,\eta_b;\eps)$ in the various limits can be computed using the method of \textbf{expansion by regions} \cite{Beneke:1997zp}. The basic idea of the latter is to identify regions with non-trivial behavior when some asymptotic limit is approached. One then performs a Taylor expansion in each region in the appropriate variable, after which the expanded integrands are integrated over the \textit{full} integration range. The determination of all relevant regions is in general highly non-trivial.\footnote{See e.g. \cite{Ma:2025emu} for a recent review.} Here, we employ the algorithm developed by Pak and Smirnov, which is implemented in the {\tt Mathematica} package {\tt asy2.m} \cite{Pak:2010pt,Jantzen:2012mw}. In the context of loop integrals, the latter starts from the \textbf{Schwinger-representation} of the integral, which symbolically corresponds to
\begin{equation}
\label{eq:alpharep}
    I \sim \int \left(\prod_{j=1}^{n}\rd x_j\,x_j^{\nu_j}\right)\delta\left(1-\sum_{i=1}^{n}x_i\right)\mathcal{U}^{a}\mathcal{F}^{b}\,.
\end{equation}
Here $\mathcal{U}$ and $\mathcal{F}$ are the \textit{Symanzik polynomials}, which are homogeneous in the integration variables. Pak and Smirnov then consider the scaling behavior of the product polynomial $\mathcal{U}\mathcal{F}$ by mapping it to a \textit{geometric problem}, namely finding a \textbf{convex hull} for a set of points in a multi-dimensional vector space. The latter problem is well-known and solved by the divide-and-conquer {\tt quickhull} algorithm \cite{quickhull}.\newline

Note that, since our result for $I(\eta_a,\eta_b;\eps)$ is already expanded in $\eps$, we cannot directly perform the expansion by regions. Instead, we need to start from the unexpanded form of the integrand, cf.~eq.~(\ref{eq:unexpandedInt}). Furthermore, the algorithm implemented in {\tt asy2.m} expects the integration to run from zero to infinity. As such, we perform the following integral transformation
\begin{equation}
    \zeta_a\to \frac{t_1}{t_1+1},\qquad\zeta_b\to \frac{t_2}{t_2+1}\,.
\end{equation}
The integrand then becomes
\beq
\bsp
\label{eq:intasy}
&I(\eta_a,\eta_b,t_1,t_2;\eps) =\,(1-\eta_a)^{1-2 \eps} \eta_a^{-\eps} (1-\eta_b)^{-1-2\eps} \eta_b^{-\eps} t_1^{-\eps} (t_1+1)^{\eps} \left(\frac{t_2}{t_2+1}\right)^{-1-\eps} (\eta_a t_1+1)^{-\eps}\\&\times (\eta_a t_1+t_1+2)^{-\eps} (2 \eta_a t_1+\eta_a+1)^{-\eps} (\eta_b t_2+1)^{2-\eps} (\eta_b t_2+t_2+2)^{-1-\eps} (2 \eta_b t_2+\eta_b+1)^{-1-\eps} \\&\times(t_1 (\eta_a \eta_b t_2+\eta_a+t_2+1)+\eta_b t_2+t_2+2) (\eta_a (\eta_b (2 t_1 t_2+t_1+t_2+1)+t_1)+\eta_b t_2+1) \\&\times(\eta_a (t_2+1) (\eta_b (2 t_1 t_2+t_1+t_2+1)+t_1)+\eta_b^2 (t_1+1) (2 t_2+1)+\eta_b t_2 (-2 t_1+t_2-1)\\&-t_1+t_2) (\eta_b t_2 (2 \eta_a t_1+\eta_a+1)+\eta_a \eta_b t_1+\eta_a t_1+\eta_a+\eta_b)^{-1+2\eps} (t_1 (t_2 (\eta_a+\eta_b)+\eta_a+1)\\&+\eta_b t_2+t_2+2)^{-1+2\eps}\Big[(t_1 (\eta_a \eta_b t_2+\eta_a-t_2-1)+(\eta_b-1) t_2)(\eta_a^2 \eta_b (2 t_1+1) (t_2+1)^2\\&+\eta_a \left(\eta_b^2 (t_1+1)^2 (2 t_2+1)-2 \eta_b (t_1 t_2 (t_1+t_2+4)+t_1+t_2)-(t_1+1)^2\right)-\eta_b (t_2+1)^2)\Big]^{-1}\,.
\esp
\eeq
The computation of the appropriate limits of eq.~(\ref{eq:intasy}), and their corresponding integration, will be discussed in the following sections.

\subsubsection{Asymptotic behavior as $\eta_b\to 1$}
As {\tt asy2.m} expects the singularities to be at zero, we start by setting
\begin{equation}
\label{eq:etaToN}
    \eta_a\to 1-n_a,\qquad \eta_b\to 1-n_b\,.
\end{equation}
The $\eta_a$ transformation is not strictly necessary at this point, but merely done for later convenience. The integrand now takes the following form
\beq
\bsp
\label{eq:intasyTrans}
&I(n_a,n_b,t_1,t_2;\eps) =\,(1-n_a)^{-\eps} n_a^{1-2 \eps} (1-n_b)^{-\eps} n_b^{-1-2\eps} t_1^{-\eps} (t_1+1)^\eps t_2^{-1-\eps} (t_2+1)^{\eps+1}\\&\times (-2 n_a t_1-n_a+2 t_1+2)^{-\eps} (-n_a t_1+t_1+1)^{-\eps} (-n_a t_1+2 t_1+2)^{-\eps}\\&\times (-2 n_b t_2-n_b+2 t_2+2)^{-1-\eps} (-n_b t_2+t_2+1)^{2-\eps} (-n_b t_2+2 t_2+2)^{-1-\eps} \\&\times(n_a n_b t_1 t_2-n_a t_1 t_2-n_a t_1-n_b t_1 t_2-n_b t_2+2 t_1 t_2+2 t_1+2 t_2+2)\\&\times (2 n_a n_b t_1 t_2+n_a n_b t_1+n_a n_b t_2+n_a n_b-2 n_a t_1 t_2-2 n_a t_1-n_a t_2-n_a-2 n_b t_1 t_2\\&-n_b t_1-2 n_b t_2-n_b+2 t_1 t_2+2 t_1+2 t_2+2) (2 n_a n_b t_1 t_2^2+3 n_a n_b t_1 t_2+n_a n_b t_1+n_a n_b t_2^2\\&+2 n_a n_b t_2+n_a n_b-2 n_a t_1 t_2^2-4 n_a t_1 t_2-2 n_a t_1-n_a t_2^2-2 n_a t_2-n_a+2 n_b^2 t_1 t_2+n_b^2 t_1\\&+2 n_b^2 t_2+n_b^2-2 n_b t_1 t_2^2-5 n_b t_1 t_2-3 n_b t_1-2 n_b t_2^2-5 n_b t_2-3 n_b+2 t_1 t_2^2+4 t_1 t_2+2 t_1\\&+2 t_2^2+4 t_2+2) (-n_a t_1 t_2-n_a t_1-n_b t_1 t_2-n_b t_2+2 t_1 t_2+2 t_1+2 t_2+2)^{-1+2\eps}\\&\times (2 n_a n_b t_1 t_2+n_a n_b t_1+n_a n_b t_2-2 n_a t_1 t_2-2 n_a t_1-n_a t_2-n_a-2 n_b t_1 t_2-n_b t_1\\&-2 n_b t_2-n_b+2 t_1 t_2+2 t_1+2 t_2+2)^{-1+2\eps}\Big[(-n_a n_b t_1 t_2+n_a t_1 t_2+n_a t_1+n_b t_1 t_2+n_b t_2) \\&\times(2 n_a^2 n_b t_1 t_2^2+4 n_a^2 n_b t_1 t_2+2 n_a^2 n_b t_1+n_a^2 n_b t_2^2+2 n_a^2 n_b t_2+n_a^2 n_b-2 n_a^2 t_1 t_2^2-4 n_a^2 t_1 t_2\\&-2 n_a^2 t_1-n_a^2 t_2^2-2 n_a^2 t_2-n_a^2+2 n_a n_b^2 t_1^2 t_2+n_a n_b^2 t_1^2+4 n_a n_b^2 t_1 t_2+2 n_a n_b^2 t_1+2 n_a n_b^2 t_2\\&+n_a n_b^2-2 n_a n_b t_1^2 t_2-2 n_a n_b t_1^2-2 n_a n_b t_1 t_2^2-8 n_a n_b t_1 t_2-6 n_a n_b t_1-2 n_a n_b t_2^2-6 n_a n_b t_2\\&-4 n_a n_b+2 n_a t_1 t_2^2+4 n_a t_1 t_2+2 n_a t_1+2 n_a t_2^2+4 n_a t_2+2 n_a-2 n_b^2 t_1^2 t_2-n_b^2 t_1^2-4 n_b^2 t_1 t_2\\&-2 n_b^2 t_1-2 n_b^2 t_2-n_b^2+2 n_b t_1^2 t_2+2 n_b t_1^2+4 n_b t_1 t_2+4 n_b t_1+2 n_b t_2+2 n_b)\Big]^{-1}\,.
\esp
\eeq
The non-trivial sectors as $\eta_b\to 1$ can be determined using the {\tt WilsonExpand} command of {\tt asy2.m}. The latter expects two polynomials, corresponding to $\mathcal{F}$ and $\mathcal{U}$ in eq.~(\ref{eq:alpharep}). As such, we turn eq.~(\ref{eq:intasyTrans}) into a polynomial by turning off all exponents, which then corresponds to our $\mathcal{F}$,
\beq
\bsp
\label{eq:Fpoly}
&\mathcal{F} =\,(1-n_a) n_a (1-n_b)n_b t_1 (t_1+1) t_2 (t_2+1)\\&\times (-2 n_a t_1-n_a+2 t_1+2) (-n_a t_1+t_1+1) (-n_a t_1+2 t_1+2)\\&\times (-2 n_b t_2-n_b+2 t_2+2) (-n_b t_2+t_2+1) (-n_b t_2+2 t_2+2)\\&\times(n_a n_b t_1 t_2-n_a t_1 t_2-n_a t_1-n_b t_1 t_2-n_b t_2+2 t_1 t_2+2 t_1+2 t_2+2)\\&\times (2 n_a n_b t_1 t_2+n_a n_b t_1+n_a n_b t_2+n_a n_b-2 n_a t_1 t_2-2 n_a t_1-n_a t_2-n_a-2 n_b t_1 t_2\\&-n_b t_1-2 n_b t_2-n_b+2 t_1 t_2+2 t_1+2 t_2+2) (2 n_a n_b t_1 t_2^2+3 n_a n_b t_1 t_2+n_a n_b t_1+n_a n_b t_2^2\\&+2 n_a n_b t_2+n_a n_b-2 n_a t_1 t_2^2-4 n_a t_1 t_2-2 n_a t_1-n_a t_2^2-2 n_a t_2-n_a+2 n_b^2 t_1 t_2+n_b^2 t_1\\&+2 n_b^2 t_2+n_b^2-2 n_b t_1 t_2^2-5 n_b t_1 t_2-3 n_b t_1-2 n_b t_2^2-5 n_b t_2-3 n_b+2 t_1 t_2^2+4 t_1 t_2+2 t_1\\&+2 t_2^2+4 t_2+2) (-n_a t_1 t_2-n_a t_1-n_b t_1 t_2-n_b t_2+2 t_1 t_2+2 t_1+2 t_2+2)\\&\times (2 n_a n_b t_1 t_2+n_a n_b t_1+n_a n_b t_2-2 n_a t_1 t_2-2 n_a t_1-n_a t_2-n_a-2 n_b t_1 t_2-n_b t_1\\&-2 n_b t_2-n_b+2 t_1 t_2+2 t_1+2 t_2+2)\Big[(-n_a n_b t_1 t_2+n_a t_1 t_2+n_a t_1+n_b t_1 t_2+n_b t_2) \\&\times(2 n_a^2 n_b t_1 t_2^2+4 n_a^2 n_b t_1 t_2+2 n_a^2 n_b t_1+n_a^2 n_b t_2^2+2 n_a^2 n_b t_2+n_a^2 n_b-2 n_a^2 t_1 t_2^2-4 n_a^2 t_1 t_2\\&-2 n_a^2 t_1-n_a^2 t_2^2-2 n_a^2 t_2-n_a^2+2 n_a n_b^2 t_1^2 t_2+n_a n_b^2 t_1^2+4 n_a n_b^2 t_1 t_2+2 n_a n_b^2 t_1+2 n_a n_b^2 t_2\\&+n_a n_b^2-2 n_a n_b t_1^2 t_2-2 n_a n_b t_1^2-2 n_a n_b t_1 t_2^2-8 n_a n_b t_1 t_2-6 n_a n_b t_1-2 n_a n_b t_2^2-6 n_a n_b t_2\\&-4 n_a n_b+2 n_a t_1 t_2^2+4 n_a t_1 t_2+2 n_a t_1+2 n_a t_2^2+4 n_a t_2+2 n_a-2 n_b^2 t_1^2 t_2-n_b^2 t_1^2-4 n_b^2 t_1 t_2\\&-2 n_b^2 t_1-2 n_b^2 t_2-n_b^2+2 n_b t_1^2 t_2+2 n_b t_1^2+4 n_b t_1 t_2+4 n_b t_1+2 n_b t_2+2 n_b)\Big]\,.
\esp
\eeq
As we do not have a second polynomial at hand, we simply set $\mathcal{U}=1$. Feeding our expressions for $\mathcal{F}$ and $\mathcal{U}$ to {\tt WilsonExpand}, we find the following three non-trivial regions
\begin{align}
\label{eq:transbReg1}
    &\text{Region 1:\,}\, t_2\to n_b^0\,t_2\,, t_1\to n_b^{-1}\,t_1\,,\\
\label{eq:transbReg2}    
    &\text{Region 2:\,}\, t_2\to n_b^0\,t_2\,, t_1\to n_b^{0}\,t_1\,,\\
\label{eq:transbReg3}    
    &\text{Region 3:\,}\, t_2\to n_b^0\,t_2\,, t_1\to n_b^{1}\,t_1\,.
\end{align}
In each of these we now need to perform a series expansion and then integrate the result. To keep the dependence on $n_b$ intact we remap 
\begin{equation}
    n_b\to\rho\, n_b\,,
\end{equation}
such that now $\rho$ plays the r\^ole of the small parameter. We will present the treatment of each region separately. As the second region leads to the most complicated integration, we first discuss the other two. \newline

\noindent\textbf{\underline{Region 1}}\newline

\noindent Following eq.~(\ref{eq:transbReg1}) we map $t_1\to n_b^{-1}\,t_1$ and $t_2\to n_b^0\,t_2$. The result is then expanded in $\rho$ which gives
\beq
\bsp
\label{eq:intReg1}
&\mathcal{I}_b^{(1)}(t_1,t_2,n_a;\eps) =\,\frac{2^{-2 - \eps} \left(\frac{n_a-1}{n_a-2}\right)^{-\eps} n_a^{-2 \eps} t_1^{-1 + \eps} t_2^{-1 - \eps} (1 + t_2)^{1 + 2 \eps}}{n_a + t_1 + n_a t_2}\,.
\esp
\eeq
As $\rho$ is simply a bookkeeping parameter, we put it to one after the expansion. Furthermore, the $\eta_b$ dependent factor, which is $(1-\eta_b)^{-1-3\eps}$, was factored out and will be reinstated at the end of the computation. We now need to integrate eq.~(\ref{eq:intReg1}) over the full integration range of $t_1$ and $t_2$, which is $[0,\infty]\times[0,\infty]$. As we prefer to integrate between zero and one we set
\begin{equation}
    t_1\to \frac{x_1}{1-x_1},\qquad t_2\to \frac{x_2}{1-x_2}\,,
\end{equation}
which turns the integrand in eq.~(\ref{eq:intReg1}) into
\beq
\bsp
\label{eq:finIntegrand1}
&\mathcal{I}_b^{(1)}(x_1,x_2,n_a;\eps) =\,\frac{2^{-2 - \eps} (1 - n_a)^{-\eps} (2 - n_a)^\eps n_a^{-2 \eps} (1 - x_1)^{-\eps} x_1^{-1 + \eps} (1 - x_2)^{-1 - \eps} x_2^{-1 - \eps}}{n_a + x_1 - x_1(n_a + x_2)}\,.
\esp
\eeq
The integration of eq.~(\ref{eq:finIntegrand1}) is straightforward and can be done using the standard {\tt Integrate} command of {\tt Mathematica}. Expanding the result in $\eps$ yields
\begin{equation}
\label{eq:intresReg1}
    \mathcal{I}_b^{(1)}(\eta_a;\eps) =\, -\frac{3}{8 \eps (1 - \eta_a)}\left(\frac{1}{\eps}+\ln\left(\frac{1 + \eta_a}{2 \eta_a (1 - \eta_a)}\right)\right)+O(\eps^0)
\end{equation}
in terms of the original $\eta_a$ variable. While we only show the $\eps$-poles in eq.~(\ref{eq:intresReg1}), we have computed the $\eps$-expansion to $\mathcal{O}(\eps)$.\newline

\noindent\textbf{\underline{Region 3}}\newline

\noindent The treatment of the third region, cf.~eq.~(\ref{eq:transbReg3}), is identical to that of the first one above. In fact, we find that the results are exactly the same, i.e., 
\begin{equation}
\label{eq:intresReg3}
    \mathcal{I}_b^{(3)}(\eta_a;\eps) =\, \mathcal{I}_b^{(1)}(\eta_a;\eps)\,.
\end{equation}

\noindent\textbf{\underline{Region 2}}\newline

\noindent Finally we consider the second region, cf.~eq.~(\ref{eq:transbReg2}). Following the same steps as above, we find that the function to integrate is
\beq
\bsp
\mathcal{I}_b^{(2)}(x_1,x_2,\eta_a,\eta_b;\eps) &=\, 2^{-2 - 2 \eps} (1 - \eta_a)^{-1 - 2 \eps} \eta_a^{-\eps} (1 - x_1)^{-1 - \eps} x_1^{-1 - 
  \eps} (1 - (1 - \eta_a) x_1)^{-\eps}\\&\quad\times (2 - (1 - \eta_a) x_1)^\eps (1 + 
   \eta_a - (1 - \eta_a) x_1)^\eps (1 - x_2)^{-1 - \eps} x_2^{-1 - \eps}\,,
\esp
\eeq
where as before both $x_1$ and $x_2$ run between zero and one and the $\eta_b$-dependence, $(1-\eta_b)^{-1-2 \eps}$, was divided out. The $x_2$-integration is straightforward and we find
\beq
\bsp
\mathcal{I}_b^{(2)}(x_1,\eta_a,\eta_b;\eps) &=\, \frac{4^{-1-\eps}\,\Gamma(-\eps)^2}{\Gamma(-2\eps)}(1 - \eta_a)^{-1 - 2 \eps} \eta_a^{-\eps} (1 - x_1)^{-1 - \eps} x_1^{-1 - 
  \eps} (1 - x_1 + \eta_a x_1)^{-\eps}\\&\quad\times (2 - x_1 + \eta_a x_1)^\eps (1 + \eta_a - x_1 + \eta_a x_1)^\eps\,.
\esp
\eeq
The integration over $x_1$ however can no longer be handled with the built-in {\tt Mathematica} command {\tt Integrate}. Hence we turn to {\tt GIntegrate} of {\tt PolyLogTools}. As the latter cannot treat symbolic powers, we first expand in $\eps$. This in turn means we need to treat the singularities at zero and one, which we do by subtraction. Next, we go to a fibration basis with respect to $x_1$ and perform a partial fraction decomposition. The integration then yields
\begin{equation}
\label{eq:intresReg2}
   \mathcal{I}_b^{(2)}(\eta_a;\eps) =\, \frac{1}{\eps(1-\eta_a)}\left(\frac{1}{\eps}+\ln\left(\frac{1 + \eta_a}{2 \eta_a (1 - \eta_a)^2}\right)\right)+O(\eps^0)\,.
\end{equation}
Again we only show the poles in $\eps$. A priori, the $\mathcal{O}(\eps)$ part of $\mathcal{I}_b^{(2)}(\eta_a;\eps)$ is a complicated expression of about 200 terms, containing GPLs up to weight three. It turns out however that significant simplification is possible. A particularly useful tool for this is {\tt GatherObjectsMatrix}, which is a part of {\tt PolyLogTools}. The latter, when acting on some expression, returns a list of lists containing the GPLs with their corresponding (rational) coefficients. Such a list is typically easier to manipulate than the expression itself. For example, while acting with {\tt Simplify} on the full expression is typically quite time-consuming, it is fast after application of {\tt GatherObjectsMatrix}. One can then re-multiply the GPLs with their coefficients and apply the standard GPL relations, such as the shuffle relation in eq.~(\ref{eq:shuffle}), to obtain further simplifications. After this, the $\mathcal{O}(\eps)$ part of $\mathcal{I}_b^{(2)}(\eta_a;\eps)$ becomes much more manageable, containing less than 50 terms. In particular, the $\eta_a$-dependent part of the alphabet of the GPLs reduces from $\left\{ \frac{4}{1 - \eta_a}, \frac{2}{1 - \eta_a}, \eta_a, -\frac{2 \eta_a}{1-\eta_a}, -\frac{
 4 \eta_a}{1-\eta_a}, \frac{2 (1 + \eta_a)}{1-\eta_a}, 2 - \frac{4}{1-\eta_a} \right\}$ to simply $\left\{\eta_a \right\}$.

We can now reconstruct the full behavior of the integrand in eq.~(\ref{eq:intasy}) as $\eta_b\to 1$ by combining the results in eqs.~(\ref{eq:intresReg1}), (\ref{eq:intresReg3}) and (\ref{eq:intresReg2}) with the appropriate powers of $1-\eta_b$. We find
\begin{equation}
\label{eq:Lb}
    \Lb I(\eta_a,\eta_b;\eps) =\, 2(1-\eta_b)^{-1-3\eps}\,\mathcal{I}_b^{(1)}(\eta_a;\eps)+(1-\eta_b)^{-1-2\eps}\,\mathcal{I}_b^{(2)}(\eta_a;\eps)\,.
\end{equation}

\subsubsection{Asymptotic behavior as $\eta_a\to 1$}
To study the asymptotic behavior of the integrand as $\eta_a\to 1$, we can again start from the form presented in eq.~(\ref{eq:intasyTrans}). In particular, the $\mathcal{F}$ polynomial needed for {\tt WilsonExpand} is the same as before, cf.~eq.~(\ref{eq:Fpoly}). The regions are the same as well, cf.~eq.~(\ref{eq:transbReg1})-(\ref{eq:transbReg3}), with $t_1\leftrightarrow t_2$ and $n_a\leftrightarrow n_b$, i.e.,
\begin{align}
\label{eq:transReg1}
    &\text{Region 1:\,}\, t_1\to n_a^0\,t_1\,, t_2\to n_a^{-1}\,t_2\,,\\
\label{eq:tranReg2}    
    &\text{Region 2:\,}\, t_1\to n_a^0\,t_1\,, t_2\to n_a^{0}\,t_2\,,\\
\label{eq:transReg3}    
    &\text{Region 3:\,}\, t_1\to n_a^0\,t_1\,, t_2\to n_a^{1}\,t_2\,.
\end{align}
Following the same steps as for the $\eta_b\to 1$ limit above we find
\begin{equation}
\label{eq:La0a}
    \La I(\eta_a,\eta_b;\eps) =\, 2(1-\eta_a)^{3\eps}\,\mathcal{I}_a^{(1)}(\eta_b;\eps)+(1-\eta_a)^{1-2\eps}\,\mathcal{I}_a^{(2)}(\eta_b;\eps)\,.
\end{equation}
Note that, at the end of the day, we are only interested in the \textit{singular} behavior as $\eta_a\to 1$. In fact, we defined the formal limit operator $\La$ as discarding non-singular and regular terms as $\eta_a\to 1$. So, instead of using eq.~(\ref{eq:La0a}), we actually set
\begin{equation}
\label{eq:La0}
    \La I(\eta_a,\eta_b;\eps) =\, 0\,.
\end{equation}
Of course we can reach this conclusion immediately after determining the sectors and \textit{before} going through any explicit integrations. But, then again, where would be the fun in that?

\subsubsection{Asymptotic behavior as $\eta_{a,b}\to 1$}
Finally, we consider the asymptotic behavior of the integrand as both $\eta_a$ and $\eta_b$ approach one at the same time. Again we can start from the expression presented in eq.~(\ref{eq:intasyTrans}). To make the double limit explicit, we furthermore rescale both $n_a$ and $n_b$ as
\begin{equation}
    n_a\to \lambda n_a,\qquad n_b\to \lambda n_b
\end{equation}
such that $\lambda$ plays the r\^ole of the small parameter. The $\mathcal{F}$ polynomial now takes on the following form
\beq
\bsp
&\mathcal{F} =\, \lambda n_a n_b t_1 (t_1+1) t_2 (t_2+1) (1-\lambda n_a) (1-\lambda n_b) (-\lambda n_a t_1+t_1+1) (t_1 (2-\lambda n_a)+2)\\&\times (2 (t_1+1)-\lambda (2 n_a t_1+n_a)) (-\lambda n_b t_2+t_2+1) (t_2 (2-\lambda n_b)+2)\\&\times (2 (t_2+1)-\lambda (2 n_b t_2+n_b)) (n_a t_1 (-\lambda n_b t_2+t_2+1)+n_b (t_1+1) t_2) \\&\times(-\lambda t_1 (t_2 (n_a+n_b)+n_a)+t_2 (2-\lambda n_b)+2 t_1 (t_2+1)+2) \\&\times\left(\lambda^2 n_a n_b (2 t_1 t_2+t_1+t_2)-\lambda (n_a (2 t_1+1) (t_2+1)+n_b (t_1+1) (2 t_2+1))+2 (t_1+1) (t_2+1)\right)\\&\times \left(\lambda^2 n_a n_b (2 t_1 t_2+t_1+t_2+1)-\lambda (n_a (2 t_1+1) (t_2+1)+n_b (t_1+1) (2 t_2+1))+2 (t_1+1) (t_2+1)\right)\\&\times (n_a (t_1+1) \left(\lambda^2 n_b^2 (t_1+1) (2 t_2+1)-2 \lambda n_b (t_2+1) (t_1+t_2+2)+2 (t_2+1)^2\right)\\&+\lambda n_a^2 (2 t_1+1) (t_2+1)^2 (\lambda n_b-1)-n_b (t_1+1)^2 (\lambda (2 n_b t_2+n_b)-2 (t_2+1)))\\&\times \left(t_1 \left(\lambda^2 n_a n_b t_2-\lambda (t_2 (n_a+n_b)+n_a)+2 (t_2+1)\right)+t_2 (2-\lambda n_b)+2\right)\\&\times (\lambda^2 n_b (n_a (t_2+1) (2 t_1 t_2+t_1+t_2+1)+n_b (t_1+1) (2 t_2+1))-\lambda (t_2+1) \\&\times(n_a (2 t_1+1) (t_2+1)+n_b (t_1+1) (2 t_2+3))+2 (t_1+1) (t_2+1)^2)\,.
\esp
\eeq
Using {\tt WilsonExpand} we then find a lonely non-trivial region, namely
\begin{equation}
\label{eq:regionAB}
     t_1\to \lambda^0\,t_1\,, t_2\to \lambda^0\,t_2\,.
\end{equation}
After performing the series expansion the integrand becomes
\begin{equation}
\label{eq:intrepLab}
    \mathcal{I}_{ab}^{(1)}(x_1,x_2,n_a,n_b;\eps) =\, \frac{(1-x_1)^{-\eps} x_1^{-\eps} (1-x_2)^{-1-\eps} x_2^{-1-\eps}}{4 (n_a-n_b x_2+n_b-n_a x_1) (n_a x_1+n_b x_2)}
\end{equation}
where, as before, $x_1$ and $x_2$ run between zero and one. Note that the integrand now still depends on both $n_a$ and $n_b$. However, as we will see below, only their \textit{ratio} will end up in the final expression. Finally, an overall factor of 
\begin{equation}
\label{eq:facAB}
    \lambda^{-2-4\eps}\,n_a^{1-2\eps}\,n_b^{-1-2\eps}
\end{equation}
was divided out and will be reinstated when reconstructing the full limit formula. Before we can integrate $\mathcal{I}_{ab}^{(1)}(x_1,x_2,n_a,n_b;\eps)$ over $x_1$ and $x_2$, we need to treat the overlapping singularities by applying sector decomposition. After the decomposition we can perform the integration with the help of {\tt GIntegrate}. The second integration step introduces spurious divergences of the form $G(a;a)$, which however cancel as expected. Finally, we need to expand in $\eps$. Contrary to the single limits above, this expansion now needs to go to $\mathcal{O}(\eps^2)$. The result is a complicated function of $n_a$ and $n_b$ with about 2000 terms and GPLs up to weight four. However, we can simplify the result significantly by introducing the ratio
\begin{equation}
\label{eq:defRat}
    R \equiv \frac{n_b}{n_a}
\end{equation}
and substituting
\begin{equation}
    n_b =\,R\, n_a\,.
\end{equation}
If we then go to a fibration basis with respect to $R$, all GPLs will be free of $n_a$. After applying the shuffle relation of eq.~(\ref{eq:shuffle}) and performing a partial fraction decomposition with respect to $R$, the expression compactifies significantly, with less than 50 terms remaining. Moreover, the weight-four GPLs all simplify to classical and Nielsen polylogarithms in $R$, cf.~eqs.~(\ref{eq:Classpolylogs})-(\ref{eq:Nielsenpolylogs}), in combination with Riemann zeta-values $\zeta_n$ up to $n=4$. Explicitly we find
\beq
\bsp
\label{eq:intresab}
    &\mathcal{I}_{ab}^{(1)}(R;\eps) =\, \frac{1}{4(R+1)}\Bigg\{\frac{1}{\eps^2}+\frac{1}{\eps}\Big[3 \ln (R)-2 \ln (R+1)\Big]-\frac{1}{2}\Big[8 \Li_{2}(-R)+3 \ln ^2(R)\\&-4 \ln ^2(R+1)+4 \ln (R+1) \ln (R)+10 \zeta_{2}\Big]+\frac{\eps}{6}\Big[24 \Li_{3}(-R)+48 \Li_{3}(R+1)\\&-48 \ln (R+1) \Li_{2}(R+1)-18 \zeta_{2} \ln (R)+36 \zeta_{2} \ln (R+1)+3 \ln ^3(R)-8 \ln ^3(R+1)\\&+6 \ln (R+1) \ln ^2(R)+12 \ln ^2(R+1) \ln (R)-24 \ln (-R) \ln ^2(R+1)-36 \zeta_{3}\Big]\\&+\frac{\eps^2}{24}\Big[-192 \zeta_{2} \Li_{2}(-R)+96 \Li_{2}^2(-R)+96 \Li_{4}(-R)-384 \Li_{4}(R+1)\\&-192 \ln ^2(R+1) \Li_{2}(R+1)-384 \ln (R) \Li_{3}(R+1)+192 \ln (R+1) \ln (R) \Li_{2}(R+1)\\&+384 \ln (R+1) \Li_{3}(R+1)-576 S_{2,2}(-R)-36 \zeta_{2} \ln ^2(R)+144 \zeta_{2} \ln ^2(R+1)\\&+144 \zeta_{2} \ln (R+1) \ln (R)+816 \zeta_{3} \ln (R)-96 \zeta_{3} \ln (R+1)+3 \ln ^4(R)-16 \ln ^4(R+1)\\&+8 \ln (R+1) \ln ^3(R)+32 \ln ^3(R+1) \ln (R)-64 \ln (-R) \ln ^3(R+1)\\&+24 \ln ^2(R+1) \ln ^2(R)-450 \zeta_{4}\Big]\Bigg\}\,.
\esp
\eeq
We have divided out an overall factor of $1/n_a^2$, which will be reinstated later. Note that, while the result might look complex with the appearance of quantities like $\ln(-R)$, it is actually real, which can be checked numerically. Finally, restoring all overall factors, we find that the asymptotic behavior of $I(\eta_a,\eta_b;\eps)$ as both $\eta_a\to 1$ and $\eta_b\to 1$ is characterized by
\begin{equation}
\label{eq:Lab}
    \Lab I(\eta_a,\eta_b;\eps) =\, \lam^{-2-4\eps}(1-\eta_a)^{-1-2\eps}(1-\eta_b)^{-1-2\eps}\,\mathcal{I}_{ab}^{(1)}\left(\frac{1-\eta_b}{1-\eta_a};\eps\right)\,
\end{equation}
in which we restored the dependence of $R$ on $\eta_a$ and $\eta_b$, cf.~eq.~(\ref{eq:defRat}). As the prefactor genuinely diverges as $\eta_{a,b}\to 1$ and $\eps\to 0$, we need to keep this term for the subtraction.

\subsubsection{Asymptotic behavior of $\Lab I(\eta_a,\eta_b;\eps)$}

As discussed above, we also require the behavior of $\Lab I(\eta_a,\eta_b;\eps)$ as either $\eta_a\to 1$ or $\eta_b\to 1$. To compute these limits, we start from the integral representation of the double limit formula, cf.~eq.~(\ref{eq:intrepLab}),
\begin{equation}
    \mathcal{I}^{(1)}_{ab}(t_1,t_2,n_a,n_b;\eps) =\, \frac{n_a^{1-2 \eps} n_b^{-1-2\eps} t_1^{-\eps} (t_1+1)^{2 \eps} t_2^{-1-\eps} (t_2+1)^{2+2\eps}}{4 (n_a t_2+n_a+n_b t_1+n_b) (n_a t_1 (t_2+1)+n_b (t_1+1) t_2)}\,.
\end{equation}
In order to have the integration range be $[0,\infty]\times[0,\infty]$, we transformed
\begin{equation}
    x_1\to\frac{t_1}{t_1+1}\,,\qquad x_2\to\frac{t_2}{t_2+1}\,.
\end{equation}
The $\mathcal{F}$ polynomial which we feed to {\tt WilsonExpand} for the determination of the regions is then
\begin{equation}
    \mathcal{F} =\,\frac{1}{4} n_a n_b t_1 (t_1+1) t_2 (t_2+1) (n_a t_2+n_a+n_b t_1+n_b) (n_a t_1 (t_2+1)+n_b (t_1+1) t_2)\,.
\end{equation}

\noindent{\underline{$\bm{\eta_{a}\to 1}$}}\newline

\noindent In the limit that $\eta_a\to 1$ we find three distinct regions, corresponding to
\begin{align}
    &\text{Region 1:\,}\, t_1\to n_a^0\,t_1\,, t_2\to n_a^{-1}\,t_2\,,\\
    &\text{Region 2:\,}\, t_1\to n_a^0\,t_1\,, t_2\to n_a^{0}\,t_2\,,\\  
    &\text{Region 3:\,}\, t_1\to n_a^0\,t_1\,, t_2\to n_a^{1}\,t_2\,.
\end{align}
Note that these are exactly the same regions as the ones we found in the computation of $\La I(\eta_a,\eta_b;\eps)$, cf.~eqs.~(\ref{eq:transReg1})-(\ref{eq:transReg3}). In particular, the $\eta_a$ behavior is also the same as before, i.e.
\begin{align}
    &\text{Region 1:\,}\, (1-\eta_a)^{-3\eps},\\
    &\text{Region 2:\,}\, (1-\eta_a)^{1-2\eps}\,,\\  
    &\text{Region 3:\,}\, (1-\eta_a)^{-3\eps}\,.
\end{align}
As these are all non-singular in four dimensions, we do not need any subtractions, and hence we can simply set
\begin{equation}
\label{eq:LaLab}
    \La \Lab I(\eta_a,\eta_b;\eps) =\, 0\,.
\end{equation}

\noindent{\underline{$\bm{\eta_{b}\to 1}$}}\newline

\noindent Also in this limit we find three regions, which are the same as those in the computation of $\Lb I(\eta_a,\eta_b;\eps)$,
\begin{align}
    &\text{Region 1:\,}\, t_2\to n_b^0\,t_2\,, t_1\to n_b^{-1}\,t_1\,,\\
    &\text{Region 2:\,}\, t_2\to n_b^0\,t_2\,, t_1\to n_b^{0}\,t_1\,,\\
    &\text{Region 3:\,}\, t_2\to n_b^0\,t_2\,, t_1\to n_b^{1}\,t_1\,.
\end{align}
The corresponding $\eta_b$ behavior is
\begin{align}
    &\text{Region 1:\,}\, (1-\eta_b)^{-1-3\eps},\\
    &\text{Region 2:\,}\, (1-\eta_b)^{-1-2\eps}\,,\\  
    &\text{Region 3:\,}\, (1-\eta_b)^{-1-3\eps}\,.
\end{align}
As such, all regions need to be taken into account in the subtraction. The integration over $t_1$ and $t_2$ is straightforward and we find
\beq
\bsp
\label{eq:LbLab}
    \Lb \Lab I(\eta_a,\eta_b;\eps) &=\, (1-\eta_a)^{-1-\eps} (1-\eta_b)^{-1-3\eps} \left(-\frac{3}{4 \eps^2}+\frac{3 \zeta_{2}}{4}+\frac{9\zeta_{3} \eps }{2}+\frac{171}{16}\zeta_{4} \eps^2 \right)\\&+(1-\eta_a)^{-1-2\eps} (1-\eta_b)^{-1-2\eps} \left(\frac{1}{\eps^2}-2 \zeta_{2}-4\zeta_{3} \eps -2\zeta_{4} \eps^2\right )\,.
\esp
\eeq

\noindent\underline{\textbf{Cross-checks}}\newline

\noindent We can check our results for the overlapping limits, $\La \Lab$ and $\Lb \Lab$, by matching different limits in different regimes. In particular, the following conditions must hold:
\begin{itemize}
    \item When $\eta_a\to 1$, the expression for $\Lab I(\eta_a,\eta_b;\eps)$ must match the one for $\La \Lab I(\eta_a,\eta_b;\eps)$,
    \item when $\eta_a\to 1$, the expression for $\Lb I(\eta_a,\eta_b;\eps)$ must match the one for $\Lb \Lab I(\eta_a,\eta_b;\eps)$,
    \item when $\eta_b\to 1$, the expression for $\Lab I(\eta_a,\eta_b;\eps)$ must match the one for $\Lb \Lab I(\eta_a,\eta_b;\eps)$ and
    \item when $\eta_b\to 1$, the expression for $\La I(\eta_a,\eta_b;\eps)$ must match the one for $\La \Lab I(\eta_a,\eta_b;\eps)$.
\end{itemize}
We have explicitly verified that our analytic results satisfy these four matching conditions.

\subsubsection{Integrals of the limit formul\ae\,}
Next, we compute the integrals of all limit formul\ae\, presented above over the appropriate variables, cf.~eqs.~(\ref{eq:intlimsin})-(\ref{eq:intlimsfin}).\newline

\noindent\underline{${[\La]  }I(\eta_a,\eta_b;\eps)$}\newline

\noindent As $\La I(\eta_a,\eta_b;\eps)=0$, we naturally also have
\begin{equation}
\label{eq:ILa}
    [\La]  I(\eta_a,\eta_b;\eps) =\,\int_{0}^{1}\rd \eta_a \,\La I(\eta_a,\eta_b;\eps) =\,0\,.
\end{equation}

\noindent\underline{${[\Lb]  I(\eta_a,\eta_b;\eps)}$}\newline

\noindent Next we compute
\begin{equation}
    [\Lb]  I(\eta_a,\eta_b;\eps) =\,\int_{0}^{1}\rd \eta_b\, \Lb I(\eta_{a},\eta_b;\eps)\,.
\end{equation}
From eq.~(\ref{eq:Lb}) it is clear that this integration is completely trivial, and we find
\beq
\bsp
\label{eq:ILb}
    [\Lb]  I(\eta_a,\eta_b;\eps) =\,& -\frac{2}{3\eps}\mathcal{I}_b^{(1)}(\eta_a;\eps)-\frac{1}{2\eps}\mathcal{I}_b^{(2)}(\eta_a;\eps) \\=\,&\,\frac{1}{4(\eta_a-1)}\Bigg\{\frac{1}{ \eps^3}+\frac{-3 \ln (1-\eta_a)+\ln\left(\frac{\eta_a+1}{2\eta_a}\right)}{ \eps^2 }+\frac{1}{6 \eps }\Bigg(24 \text{Li}_2\left(\frac{1-\eta_a}{2}\right)+24 \text{Li}_2\left(\frac{\eta_a+1}{2}\right)\\&-3 \ln  ^2(\eta_a+1)+3 \left(7 \ln  ^2(1-\eta_a)+\ln  ^2(\eta_a)-2 (\ln  (2)-3 \ln  (\eta_a)) \ln  (1-\eta_a)\right)\\&+6 \ln  \left(\frac{1-\eta_a}{8}\right) \ln  (\eta_a+1)-7 \pi ^2+21 \ln  ^2(2)\Bigg)\Bigg\}+\mathcal{O}(\eps^0)\,.
\esp
\eeq
The $\mathcal{O}(\eps^0)$ part is computed as well but not presented here. It has a simple functional form though, with only logarithms and classical polylogarithms up to weight three with simple $\eta_a$-dependent arguments.\newline

\noindent\underline{${[\La \Lab] I(\eta_a,\eta_b;\eps)}$}\newline

\noindent As $\La \Lab I(\eta_a,\eta_b;\eps)=0$, we naturally also have
\begin{equation}
\label{eq:ILaLab}
    [\La  \Lab] I(\eta_a,\eta_b;\eps) =\,\int_{0}^{1}\rd \eta_a \,\La \Lab I(\eta_a,\eta_b;\eps) =\,0\,.
\end{equation}

\noindent\underline{${[\Lb \Lab] I(\eta_a,\eta_b;\eps)}$}\newline

\noindent Also the integration of $\Lb \Lab I(\eta_a,\eta_b;\eps)$ over $\eta_b$ is straightforward. From eq.~(\ref{eq:LbLab}) we find
\beq
\bsp
\label{eq:ILbLab}
    [\Lb \Lab] I(\eta_a,\eta_b;\eps) =\,&\, \frac{1}{4(\eta_a-1)}\Bigg\{\frac{1}{\eps^3}-\frac{3 \ln (1-\eta_a)}{\eps^2}+\frac{7 \ln ^2(1-\eta_a)-6 \zeta_2}{2 \eps}\\&+\frac{1}{2} \left(14 \zeta_2 \ln (1-\eta_a)-5 \ln ^3(1-\eta_a)-4 \zeta_3\right)\Bigg\}\,.
\esp
\eeq

\noindent\underline{${[\Lab]  I(\eta_a,\eta_b;\eps)}$}\newline

\noindent Finally we consider
\begin{equation}
    [\Lab]  I(\eta_a,\eta_b;\eps) =\,\int_{0}^{1}\rd \eta_a\, \int_{0}^{1}\rd \eta_b\, \Lab I(\eta_a,\eta_b;\eps)
\end{equation}
with $\Lab I(\eta_a,\eta_b;\eps)$ given by eq.~(\ref{eq:Lab}). As the dependence of $\Lab I(\eta_a,\eta_b;\eps)$ is quite complicated, the integration is non-trivial. First there are overlapping singularities which need to be treated using sector decomposition. The double integral can then be computed by an iterative application of partial fractioning, setting up a fibration basis in the integration variable under consideration and applying {\tt GIntegrate}. The result is
\beq
\bsp
    [\Lab]  I(\eta_a,\eta_b;\eps) =\, \frac{1}{16\eps^4}-\frac{\zeta_2}{4\eps^2}+\frac{\mathcal{E}}{\eps}+\mathcal{O}(\eps^0)
\esp
\eeq
with
\beq
\bsp
\label{eq:epsm1}
\mathcal{E} =\,& -\frac{1}{16} \ln ^3(2)+\frac{3}{16} G(0;2) \ln ^2(2)-\frac{1}{16} G(2;1) \ln ^2(2)-\frac{7}{24} G(0,0;2) \ln (2)\\&+\frac{1}{2} \zeta_2 \ln (2)-\frac{1}{6} G\left(-\frac{1}{2};1\right) G(0;2) G(2;1)+\frac{1}{8} G(-2;1) G(0;2) G\left(-\frac{1}{2};1\right)\\&-\frac{1}{8} G(0;2) G(-2,-2;1)-\frac{1}{48} G(0;2) G(-2,0;1)+\frac{1}{6} G(0;2) G(-2,2;1)\\&-\frac{1}{8} G\left(-\frac{1}{2};1\right) G(-2,2;1)-\frac{1}{24} G(0;2) G(0,-2;1)+\frac{1}{48} G(-2;1) G(0,0;2)\\&+\frac{1}{24} G(2;1) G(0,0;2)-\frac{1}{48} G\left(-\frac{1}{2};1\right) G(0,0;2)-\frac{1}{24} G(0;2) G(0,2;1)\\&+\frac{1}{12} G\left(-\frac{1}{2};1\right) G(0,2;1)+\frac{1}{8} G(2;1) G\left(0,-\frac{1}{2};1\right)+\frac{1}{6} G(0;2) G(2,-2;1)\\&-\frac{1}{24} G(0;2) G(2,0;1)+\frac{1}{12} G\left(-\frac{1}{2};1\right) G(2,0;1)-\frac{1}{8} G(0;2) G(2,2;1)\\&+\frac{1}{24} G\left(-\frac{1}{2};1\right) G(2,2;1)+\frac{1}{48} G(0;2) G\left(-\frac{1}{2},0;1\right)+\frac{1}{24} G(2;1) G\left(-\frac{1}{2},0;1\right)\\&+\frac{1}{8} G(-2,-2,2;1)+\frac{1}{48} G(-2,0,0;1)-\frac{1}{24} G(-2,0,2;1)-\frac{1}{24} G(-2,2,0;1)\\&-\frac{1}{24} G(-2,2,2;1)+\frac{1}{24} G(0,-2,0;1)-\frac{1}{12} G(0,-2,2;1)+\frac{1}{12} G(0,0,-2;1)\\&+\frac{3}{16} G(0,0,0;2)-\frac{1}{12} G(0,0,2;1)-\frac{1}{12} G(0,2,-2;1)-\frac{1}{8} G(0,2,-1;1)\\&+\frac{1}{24} G(0,2,0;1)+\frac{1}{8} G(0,2,2;1)-\frac{1}{24} G(2,-2,0;1)-\frac{1}{24} G(2,-2,2;1)\\&-\frac{1}{12} G(2,0,-2;1)-\frac{1}{8} G(2,0,-1;1)+\frac{1}{24} G(2,0,0;1)+\frac{1}{8} G(2,0,2;1)\\&+\frac{1}{12} G(2,2,-2;1)+\frac{1}{8} G(2,2,-1;1)-\frac{1}{12} G(2,2,2;1)-\frac{1}{48} G\left(-\frac{1}{2},0,0;1\right)\\&+\frac{1}{8} G(0,2;1) \ln (3)+\frac{1}{8} G(2,-1;1) \ln (3)-\frac{1}{8} G(2,2;1) \ln (3)-\frac{1}{8} G(2;1) \Li_2\left(-\frac{1}{2}\right)\\&-\frac{35}{48} G(0;2) \zeta_2-\frac{17}{48} G(2;1) \zeta_2-\frac{1}{16} \ln (3) \zeta_2-\frac{21 \zeta_3}{16}\,.
\esp
\eeq
The expression for the $1/\eps$-pole in eq.~(\ref{eq:epsm1}) looks quite daunting. Then again, it is just a combination of \textit{numbers} up to weight three. The question is then whether we can find a more compact representation for these numbers. As the reader will no doubt be happy to read, the answer is in fact affirmative. We can find this new and improved representation by application of the \textbf{PSLQ algorithm} \cite{pslq}. For this, we evaluate $\mathcal{E}$ to high precision, say to 100 digits. Then we apply PSLQ, using
\begin{equation}
    \left\{\zeta_3,\Li_3\left(\frac{1}{2}\right),\ln ^3(2),\ln ^2(2) \ln (3),\ln(2) \ln ^2(3),\ln ^3(3),\zeta_2 \ln (2),\zeta_2 \ln (3)\right\}
\end{equation}
as a basis. We find that the mess of eq.~(\ref{eq:epsm1}) reduces to a \textit{single} term, namely\footnote{In fact, this is the very last term on the right-hand side of eq.~(\ref{eq:epsm1}), meaning that the remaining terms exactly cancel among each other.}
\begin{equation}
\label{eq:epsm1fin}
    \mathcal{E} =\,-\frac{21\zeta_3}{16}\,.
\end{equation}
We have checked that eqs.~(\ref{eq:epsm1}) and (\ref{eq:epsm1fin}) agree numerically up to (at least) 200 digits. The same reasoning can now also be applied to the $\mathcal{O}(\eps^0)$ part, which a priori contains GPLs up to weight four with arguments living in
\begin{equation}
    \left\{-2,-1,-\frac{1}{2},0,1,2,4\right\}\,.
\end{equation}
Using
\beq
\bsp
\Bigg\{&\zeta (4),\Li_4\left(\frac{1}{2}\right),\ln ^4(2),\ln ^3(2) \ln (3),\ln ^2(2) \ln ^2(3),\ln (2) \ln ^3(3),\ln ^4(3),\zeta_3 \ln (2),\\&\zeta_3 \ln (3),\zeta_2 \ln ^2(2),\zeta_2 \ln ^2(3),\zeta_2 \ln (2) \ln (3)\Bigg\}
\esp
\eeq
as a basis for PSLQ we find that the full expression takes on the form
\beq
\bsp
\label{eq:ILab}
    [\Lab]  I(\eta_a,\eta_b;\eps) =\, \frac{1}{16\eps^4}-\frac{\zeta_2}{4\eps^2}-\frac{21\zeta_3}{16\eps}-5 \Li_4\left(\frac{1}{2}\right)+\frac{19 \zeta_4}{32}+\frac{5}{4} \zeta_2 \ln ^2(2)-\frac{35}{8} \zeta_3 \ln (2)-\frac{5}{24} \ln ^4(2)\,.
\esp
\eeq


\subsection{Hadron-level integration}
\label{sec:hadronint}
Up to now, we have been discussing the contribution of the $\frac{1}{1-x_{\ha{a},\ha{r}}}\frac{1}{1-x_{a,s}}$ integral to the full integrated counterterm, cf.~eq.~(\ref{eq:omxm1omxHm1}). Recall though that the latter gets contributions from 25 additional integrals, cf.~eq.~(\ref{eq:intStr}). Mercifully, the integration procedure to get to an expression similar to that in eq.~(\ref{eq:fulldistexpI}) turns out to be the same as the procedure described above.
To obtain the final result, we still need to transform from the partonic variables to the hadronic ones. In practice, this means replacing the partonic momenta $p_{a/b}$ by the hadronic ones, $x_{a/b}\,p_{A/B}$, and performing the convolution with the PDFs. The complete integrated subtraction term then takes the following form
\beq
\bsp
\label{eq:hadrint0}
&\left[ \IcC{asr}{IFF(0,0)}\IcC{as}{IF}\right]\otimes\rd\sigma_{AB}=\,\sum_{a,b}\frac{\omega(as)}{\omega(asr)}\int_{0}^{1} \rd x_a\,\rd x_b\, f_{a/A}(x_a)f_{b/B}(x_b) \int_0^{1} \rd \eta_a\,\rd \eta_b\\&\times\bigg\{
[\IcC{asr}{IFF(0,0)}\IcC{as}{IF}(\eta_a,\eta_b;\eps)]\rd\sigma_{\ha{\ha{a}}\ha{\ha{b}}}(\eta_a x_a p_A,\eta_b x_b p_B)
-
\La[ \IcC{asr}{IFF(0,0)}\IcC{as}{IF}(\eta_a,\eta_b;\eps)]\rd\sigma_{\ha{\ha{a}}\ha{\ha{b}}}( x_a p_A,\eta_b x_b p_B)
\\&-
\Lb[ \IcC{asr}{IFF(0,0)}\IcC{as}{IF}(\eta_a,\eta_b;\eps)]\rd\sigma_{\ha{\ha{a}}\ha{\ha{b}}}(\eta_a x_a p_A, x_b p_B)
-
\Big(\Lab[ \IcC{asr}{IFF(0,0)}\IcC{as}{IF}(\eta_a,\eta_b;\eps)]\\&-\La\Lab[ \IcC{asr}{IFF(0,0)}\IcC{as}{IF}(\eta_a,\eta_b;\eps)]-\Lb\Lab[ \IcC{asr}{IFF(0,0)}\IcC{as}{IF}(\eta_a,\eta_b;\eps)]\Big)\rd\sigma_{\ha{\ha{a}}\ha{\ha{b}}}(x_a p_A, x_b p_B)
\\&+
[\La][  \IcC{asr}{IFF(0,0)}\IcC{as}{IF}(\eta_a,\eta_b;\eps)]\rd\sigma_{\ha{\ha{a}}\ha{\ha{b}}}( x_a p_A,\eta_b x_b p_B)
+
[\Lb][  \IcC{asr}{IFF(0,0)}\IcC{as}{IF}(\eta_a,\eta_b;\eps)]\rd\sigma_{\ha{\ha{a}}\ha{\ha{b}}}(\eta_a x_a p_A, x_b p_B)
\\&+
\Big([\Lab][ \IcC{asr}{IFF(0,0)}\IcC{as}{IF}(\eta_a,\eta_b;\eps)]-[\La \Lab][ \IcC{asr}{IFF(0,0)}\IcC{as}{IF}(\eta_a,\eta_b;\eps)]\\&-[\Lb \Lab][ \IcC{asr}{IFF(0,0)}\IcC{as}{IF}(\eta_a,\eta_b;\eps)]\Big)
\bigg\}\,.
\esp
\eeq
Finally, we can further simplify the last steps of integration by evaluating the reduced differential cross section in a unique argument, $\rd\sigma_{\ha{\ha{a}}\ha{\ha{b}}}(x_a p_A,x_b p_B)$. For this, we replace the integration variables $x_a$ and $x_b$ by $x_a/\eta_a$ and $x_b/\eta_b$ respectively whenever appropriate. Note that this way the action of the distributions is transferred to the product of PDFs. Our final expression for the integrated counterterm then becomes
\beq
\bsp
\label{eq:hadrintfin}
\left[ \IcC{asr}{IFF(0,0)}\IcC{as}{IF}\right]\otimes\rd\sigma_{AB}&=\,\sum_{a,b}\frac{\omega(as)}{\omega(asr)}
    \int_{0}^{1}\rd x_a\,\rd x_b\,\rd\sigma_{\ha{\ha{a}}\ha{\ha{b}}}(x_a p_A,x_b p_B)\,\int_{0}^{1}\rd \eta_a\,\rd \eta_b\\&\times\Bigg\{\IcC{asr}{IFF(0,0)}\IcC{as}{IF}(\eta_a,\eta_b;\eps\,|\,\eta_a,\eta_b)\frac{f_{a/A}(x_a/\eta_a)}{\eta_a}\frac{f_{b/B}(x_b/\eta_b)}{\eta_b}\\&+\IcC{asr}{IFF(0,0)}\IcC{as}{IF}(\eta_a,\eta_b;\eps\,|\,1,\eta_b)f_{a/A}(x_a)\frac{f_{b/B}(x_b/\eta_b)}{\eta_b}\\&+\IcC{asr}{IFF(0,0)}\IcC{as}{IF}(\eta_a,\eta_b;\eps\,|\,\eta_a,1)\frac{f_{a/A}(x_a/\eta_a)}{\eta_a}f_{b/B}(x_b)\\&+\IcC{asr}{IFF(0,0)}\IcC{as}{IF}(\eta_a,\eta_b;\eps\,|\,1,1)f_{a/A}(x_a)f_{b/B}(x_b)\Bigg\}
\esp
\eeq
in terms of the \textbf{coefficient functions} of the \textit{parton-level} integrated counterterm which are defined as
\begin{align}
    &\IcC{asr}{IFF(0,0)}\IcC{as}{IF}(\eta_a,\eta_b;\eps\,|\,\eta_a,\eta_b) =\, (\eta_a \eta_b)^{2\eps}[\IcC{asr}{IFF(0,0)}\IcC{as}{IF}(\eta_a,\eta_b;\eps)]\,,\\
    &\IcC{asr}{IFF(0,0)}\IcC{as}{IF}(\eta_a,\eta_b;\eps\,|\,1,\eta_b) =\, \eta_b^{2\eps}(-\bom{L}_a+[\bom{L}_a])[\IcC{asr}{IFF(0,0)}\IcC{as}{IF}(\eta_a,\eta_b;\eps)]\,,\\
    &\IcC{asr}{IFF(0,0)}\IcC{as}{IF}(\eta_a,\eta_b;\eps\,|\,\eta_a,1) =\, \eta_a^{2\eps}(-\bom{L}_b+[\bom{L}_b])[\IcC{asr}{IFF(0,0)}\IcC{as}{IF}(\eta_a,\eta_b;\eps)]\,,\\
    &\IcC{asr}{IFF(0,0)}\IcC{as}{IF}(\eta_a,\eta_b;\eps\,|\,1,1) =\, (-\bom{L}_{ab}+\bom{L}_a\bom{L}_{ab}+\bom{L}_b\bom{L}_{ab}+[\bom{L}_{ab}]-[\bom{L}_a\bom{L}_{ab}]\nonumber\\&\qquad\qquad\qquad\qquad\qquad\qquad\qquad-[\bom{L}_b\bom{L}_{ab}])[\IcC{asr}{IFF(0,0)}\IcC{as}{IF}(\eta_a,\eta_b;\eps)]\,.
\end{align} 
Note the appearance of additional factors of $(\eta_a \eta_b)^{2\eps}$, $\eta_a^{2\eps}$ and $\eta_b^{2\eps}$, which are generated when the integral transformations act on the factor of $\left(\frac{\mu^2}{x_a x_b s_{AB}}\right)^{2\ep}$ in the reduced differential cross section, cf.~eq.~(\ref{eq:redXSapp}). Of course, these final integrals involving the PDFs are to be computed numerically.\footnote{The authors wish to congratulate those brave readers who made it to the end of this appendix.}

\resumetoc

\renewcommand{\theequation}{\ref{appx:GPLs}.\arabic{equation}}
\setcounter{equation}{0}
\renewcommand{\thefigure}{\ref{appx:GPLs}.\arabic{figure}}
\setcounter{figure}{0}
\renewcommand{\thetable}{\ref{appx:GPLs}.\arabic{table}}
\setcounter{table}{0}
\section{Properties of GPLs}
\label{appx:GPLs}
In this appendix we briefly review some useful properties of generalized polylogarithms (GPLs). This is by no means meant to be an extensive overview. GPLs are defined recursively as \cite{Goncharov:1998kja}
\begin{equation}
    G(a_1,\dots,a_n;z) =\,\int_{0}^{z}\frac{\rd t}{t-a_1}G(a_2,\dots,a_n;t)\,, \qquad G(z) \equiv G(;z) =1\,.
\end{equation}
Here $n\geq 0$ is called the \textit{weight} of the GPL. For example, the weight-one GPLs simply correspond to the \textit{logarithm},
\begin{equation}
\label{eq:GtoLog}
     G(0;z) =\,\ln z\,, \qquad G(a;z) =\,\ln \left(1-\frac{z}{a}\right)\,.
\end{equation}
Other important special cases of GPLs are the \textit{classical polylogarithms}
\begin{equation}
\label{eq:Classpolylogs}
    \Li_n(z) =\, \int_{0}^{z}\frac{\rd t}{t}\Li_{n-1}(t) =\, -G(\underbrace{0,\dots,0}_{n-1},1;z)\,,
\end{equation}
with $\Li_{1}(x)\equiv\,-\ln(1-x)$ and the \textit{Nielsen generalized polylogarithms}
\begin{equation}
\label{eq:Nielsenpolylogs}
    S_{n,p}(z) =\, \frac{(-1)^{n+p-1}}{(n-1)!p!}\int_{0}^{1}\frac{\rd t}{t}\ln^{n-1}(t)\ln^{p}(1-zt) =\, (-1)^{p}\,G(\underbrace{0,\dots,0}_{n},\underbrace{1,\dots,1}_{p};z)\,.
\end{equation}
For higher-weight GPLs it is convenient to introduce the weight vector $\Vec{a}_n =\,(a_1,\dots,a_n)$. Each separate $a_i$ is called a \textit{letter} while some combination of $a$'s is a \textit{word}. A particularly useful property of GPLs is that they constitute a \textit{shuffle algebra},
\begin{equation}
\label{eq:shuffle}
    G(\Vec{a}_m;z)G(\Vec{b}_n;z) =\,\sum_{\Vec{c}_{m+n}=\Vec{a}_m\shuffle\Vec{b}_n} G(\Vec{c}_{m+n};z)\,.
\end{equation}
Here the shuffle product $\shuffle$ gives the sum of all possible permutations of the letters while keeping the order of the letters in each word unchanged. For example,
\begin{equation}
    G(a;z)G(b,c;z) =\,G(a,b,c;z)+G(b,a,c;z)+G(b,c,a;z)\,.
\end{equation}
In certain applications, such as analytic integration, it is also useful to write a given GPL in a so-called \textit{fibration basis}, meaning it is written as some linear combination
\begin{equation}
\label{eq:fibre}
    \sum_{I=\{i_1,\dots,i_n\}}c_I\,G(\Vec{a}_{1,i_1};x_1)\dots G(\Vec{a}_{n,i_n};x_n)
\end{equation}
in which each weight-vector $\Vec{a}_{k,i_k}$ is independent of $x_k$. For example, with respect to the ordering $(x,y)$ we have
\begin{equation}
    G(1+x;1-y)\xrightarrow[]{(x,y)}-G(-1;x)+G(0;y)+G(-y;x)
\end{equation}
while the reverse ordering gives
\begin{equation}
    G(1+x;1-y)\xrightarrow[]{(y,x)}-G(-1;x)+G(0;x)+G(-x;y)\,.
\end{equation}

\renewcommand{\theequation}{\ref{appx:mommaps}.\arabic{equation}}
\setcounter{equation}{0}
\renewcommand{\thefigure}{\ref{appx:mommaps}.\arabic{figure}}
\setcounter{figure}{0}
\renewcommand{\thetable}{\ref{appx:mommaps}.\arabic{table}}
\setcounter{table}{0}

\section{Review of momentum mappings}
\label{appx:mommaps}
Here we give an overview of the different types of momentum mappings, which were described in detail in \cite{DelDuca:2025yph}.

\subsection{Soft mapping}
\label{sec:softmap}
The single soft momentum mapping, $(p_a,p_b;\mom{}_{X+m+2}) \xrightarrow{S_r} (\ti{p}_a,\ti{p}_b;\momt{}_{X+m+1})$, is defined as \cite{DelDuca:2019ctm}
\beq
\bsp
\ti{p}_a^\mu &=\,\lambda_r p_a^\mu\,,
\\
\ti{p}_b^\mu &=\,\lambda_r p_b^\mu\,,
\\
\ti{p}_n^\mu &=\,{\Lambda(P,\ti{P})^\mu}_{\!\nu}\, p_n^\nu\,, \qquad n\in F\,,n\neq r\,,
\\
\ti{p}_X^\mu &=\,{\Lambda(P,\ti{P})^\mu}_{\!\nu}\, p_X^\nu\,,
\esp
\label{eq:softmap}
\eeq
where ${\Lambda(P,\ti{P})^\mu}_{\!\nu}$ is a proper Lorentz transformation that 
takes the massive momentum $P$ into a momentum of the same mass, $\ti{P}$. One specific representation is
\beq
{\Lambda(P,\ti{P})^\mu}_{\!\nu} =\,
	{g^\mu}_{\!\nu} - \frac{2(P+\ti{P})^\mu (P+\ti{P})_\nu}{(P+\ti{P})^2} + \frac{2\ti{P}^\mu P_\nu}{P^2}\,.
\label{eq:Lambda_munu}
\eeq
The value of $\lam_r$ is fixed by requiring that $P^2=\ti{P}^2$,
\beq
    \lam_r =\, 1-\frac{s_{r Q}}{s_{ab}}\,.
\eeq

\subsection{Initial-state collinear mapping}
The initial-state collinear momentum mapping, $(p_a,p_b;\mom{}_{X+m+2}) \cmap{ab,r}{II,F} (\ha{p}_a,\ha{p}_b;\momh{}_{X+m+1})$, is defined as
\beq
\bsp
\ha{p}_a^\mu &=\,\xi_{a,r} p_a^\mu\,,
\\
\ha{p}_b^\mu &=\,\xi_{b,r} p_b^\mu\,,
\\
\ha{p}_n^\mu &=\,{\Lambda(P,\ha{P})^\mu}_{\!\nu}\, p_n^\nu\, \qquad {\text{with}}\, n\in F\,, n\ne r\,,
\\
\ha{p}_X^\mu &=\,{\Lambda(P,\ha{P})^\mu}_{\!\nu}\, p_X^\nu\,.
\esp
\label{eq:IFmap}
\eeq
Here ${\Lambda(P,\ha{P})^\mu}_{\!\nu}$ is the same Lorentz transformation as in the soft mapping, cf.~eq.~(\ref{eq:Lambda_munu}), while
\beq
\xi_{a,r} = \sqrt{\frac{s_{ab} - s_{br}}{s_{ab} - s_{ar}}
	\frac{s_{ab} - s_{r Q} }{s_{ab}}}\,,
\qquad
\xi_{b,r} = \sqrt{\frac{s_{ab} - s_{ar}}{s_{ab} - s_{br}}
	\frac{s_{ab} - s_{r Q}}{s_{ab}}}\,.
\label{eq:xiars_xibrs_def}
\eeq

\subsection{Final-state collinear mapping}
The final-state collinear momentum mapping, $(p_a,p_b;\mom{}_{X+m+2}) \cmap{ir}{FF} (\ha{p}_a,\ha{p}_b;\momh{}_{X+m+1})$, is defined as
\beq
\bsp
\ha{p}_a^\mu &=\,(1-\al_{ir})p_a^\mu\,,
\\
\ha{p}_b^\mu &=\,(1-\al_{ir})p_b^\mu\,,
\\
\ha{p}_{ir}^\mu &=\, p_{i}^\mu+p_{r}^\mu - \al_{ir} Q^\mu\,,
\\
\ha{p}_n^\mu &=\,p_n^\mu\, \qquad {\text{with}}\, n\in F\,, n \ne i,r\,,
\\
\ha{p}_X^\mu &=\,p_X^\mu\,.
\esp
\label{eq:FFmap}
\eeq
The value of $\al_{ir}$ is fixed by requiring that the parent momentum, 
$\ha{p}_{ir}$, be massless, $\ha{p}_{ir}^2 =\,0$,
\beq
\al_{ir} =\,\frac{1}{2}\left[\frac{s_{(ir)Q}}{s_{ab}} - \sqrt{\frac{s_{(ir)Q}^2}{s_{ab}^2} - \frac{4s_{ir}}{s_{ab}}}\,\right]\,.
\label{eq:Cir0FF_al}
\eeq

\renewcommand{\theequation}{\ref{appx:Ipoles}.\arabic{equation}}
\setcounter{equation}{0}
\renewcommand{\thefigure}{\ref{appx:Ipoles}.\arabic{figure}}
\setcounter{figure}{0}
\renewcommand{\thetable}{\ref{appx:Ipoles}.\arabic{table}}
\setcounter{table}{0}

\section{Pole parts of the coefficient functions}
\label{appx:Ipoles}
In this appendix we present the pole parts of the coefficient functions of the $A_{12}$ insertion operator at NNLO accuracy, cf. eqs.~(\ref{eq:I12-in})-(\ref{eq:I12-out}). We focus on the purely gluonic subprocess of Higgs production in the HEFT approximation, discarding light-quark contributions (i.e. $n_f=\,0$). Due to the fully symmetric nature of this process, the coefficient functions are not all independent. In particular, some are related by a simple $\eta_a\leftrightarrow\eta_b$ exchange, namely
\begin{align}
    I_{12;gg,gg}(\eta_a,\eta_b;\eps\,|\,\eta_a,1) =\,&\, I_{12;gg,gg}(\eta_b,\eta_a;\eps\,|\,1,\eta_b)\,,\\
    I_{12;gg,gg}(\eta_a,\eta_b;\eps\,|\,\eta_b,\eta_b) =\,&\,I_{12;gg,gg}(\eta_b,\eta_a;\eps\,|\,\eta_a,\eta_a)\,.
\end{align}
Defining
\beq
\bI_{12}(\ep) = \left(\frac{\al_s}{2\pi}\right)^2\frac{e^{2\eps \gamma_E}}{\Gamma^2(1-\eps)} \CA^2\, \bom{\bar{I}}_{12}(\ep)\,,
\eeq
the remaining coefficient functions  then read\footnote{Note that, in principle, the coefficient functions are to be thought of as piecewise functions with distinct expressions for $\eta_a<\eta_b$ and $\eta_b<\eta_a$. This is also how they are implemented in the {\tt NNLOCAL} code. Here however, we simply introduce the appropriate theta-functions (or, more accurately, distributions) for ease of notation.}
\begin{align}
    \bar{I}_{12;gg,gg}(\eta_a,\eta_b;\eps\,|\,\eta_a,\eta_b)=\,&\,\frac{4}{\eps^2}\Bigg\{\theta(\eta_a-\eta_b)\Bigg(\eta_a^2 \eta_b^2-\frac{\eta_a^2-\eta_a+2}{2 (1-\eta_b)}-\frac{\eta_a^2+\eta_a+2}{2 (1+\eta_b)}+\eta_a^2-\eta_a \eta_b+\frac{\eta_a}{\eta_b}\nn\\&+\frac{\eta_b}{\eta_a-\eta_b}-\frac{1}{2 (1-\eta_b) (\eta_a-\eta_b)}+\frac{1}{2 (1+\eta_b) (\eta_a-\eta_b)}+\frac{1}{2 \eta_a (1-\eta_b)}+\frac{1}{\eta_a \eta_b}\nn\\&-\frac{1}{2 \eta_a (1+\eta_b)}+\eta_b^2+1\Bigg)+\theta(\eta_b-\eta_a)[\eta_a\leftrightarrow\eta_b]
    +6 \eta_a^2 \eta_b^2-2 \eta_a^2 \eta_b\nn\\&-\frac{2 \Big(\eta_a^2-\eta_a+2\Big)}{\eta_b}-\frac{2 \Big(\eta_a^2+\eta_a+2\Big)}{1+\eta_b}+8 \eta_a^2-2 \eta_a \eta_b^2-\frac{4 \Big(\eta_b^2-\eta_b+2\Big)}{1-\eta_a}\nn\\&-\frac{2 \Big(\eta_b^2+\eta_b+2\Big)}{1+\eta_a}-\frac{2 \Big(\eta_b^2-\eta_b+2\Big)}{\eta_a}-2 \eta_a \eta_b+\frac{4}{(1-\eta_a) (1-\eta_b)}+\frac{4}{(1-\eta_a) \eta_b}\nn\\&-\frac{2}{(1+\eta_a) \eta_b}+\frac{2}{(1+\eta_a) (1+\eta_b)}+\frac{4}{\eta_a (1-\eta_b)}+\frac{6}{\eta_a \eta_b}-\frac{2}{\eta_a (1+\eta_b)}-4 \eta_a\nn\\&-\frac{4 \Big(\eta_a^2-\eta_a+2\Big)}{1-\eta_b}+8 \eta_b^2-4 \eta_b+16\Bigg\}\nn\\&+\frac{4}{\eps}\Bigg\{\theta(\eta_a-\eta_b)\Bigg(-\frac{11 \Big(\eta_a^2-\eta_a+2\Big)}{6 (1-\eta_b)}-\frac{3 \eta_b-8}{2 (\eta_a+\eta_b-2)}-\frac{\eta_b^2-3}{3 (\eta_a+\eta_b)^2}\nn\\&-\frac{11 \Big(\eta_b^2-\eta_b+2\Big)}{6 (1-\eta_a)}-\frac{11 \Big(\eta_b^2+\eta_b+2\Big)}{12 (1+\eta_a)}+\frac{3 \eta_b^2-6 \eta_b+8}{2 (\eta_a+\eta_b-2)^2}-\frac{11 \eta_b^2+13 \eta_b+19}{12 \eta_a}\nn\\&+\frac{1}{12} \Big(22 \eta_b^2 \eta_a^2+22 \eta_a^2-22 \eta_b \eta_a+22 \eta_b^2+85\Big)+\ln (1-\eta_a)\Bigg(\eta_b^4-\eta_a \eta_b^3-\eta_a^2 \eta_b^2\nn\\&-\eta_b^2+\eta_a \eta_b-\frac{\Big(\eta_b^4+\eta_b^2+2\Big) \eta_b}{\eta_a+\eta_b}-\frac{\eta_b}{(\eta_a+\eta_b) (\eta_a \eta_b-1)}-\eta_a^2-\frac{2 (2 \eta_b+1)}{\eta_a-\eta_b}\nn\\&-\frac{\eta_b^2-\eta_b-1}{\eta_a}-\frac{\eta_b^2-\eta_b+2}{1-\eta_a}+\frac{3 \Big(\eta_a^2-\eta_a+2\Big)}{2 (1-\eta_b)}-\frac{3}{2 (1-\eta_a) (1-\eta_b)}\nn\\&-\frac{3}{2 \eta_a (1-\eta_b)}+\frac{3}{(1-\eta_a) (\eta_a-\eta_b)}+\frac{\eta_a^2+\eta_a+2}{2 (1+\eta_b)}+\frac{1}{2 \eta_a (1+\eta_b)}\nn\\&-\frac{1}{(\eta_a-\eta_b) (1+\eta_b)}+\frac{1}{\eta_a (\eta_a+\eta_b)}-\frac{1}{2 (1-\eta_a) (1-\eta_a \eta_b)}+\frac{1}{2 (1+\eta_b) (1-\eta_a \eta_b)}\nn\\&-\frac{1}{2 (1-\eta_a \eta_b)}-1+\frac{\eta_a^2-3 \eta_a-1}{\eta_b}+\frac{1}{(1-\eta_a) \eta_b}-\frac{2}{\eta_a \eta_b}\Bigg) \nn\\&+\ln (\eta_a)\Bigg(-\eta_b^4+\eta_a \eta_b^3+\frac{\Big(\eta_b^4+\eta_b^2+2\Big) \eta_b}{\eta_a+\eta_b}+\frac{2 \eta_b}{\eta_a-\eta_b}-\frac{\eta_b}{(\eta_a+\eta_b) (1-\eta_a \eta_b)}\nn\\&+\frac{\eta_b^2-\eta_b+2}{2 (1-\eta_a)}+\frac{\eta_b^2+\eta_b+2}{2 (1+\eta_a)}-\frac{1}{(1-\eta_a) (\eta_a-\eta_b)}+\frac{1}{(1+\eta_a) (\eta_a-\eta_b)}\nn\\&+\frac{1}{2 (1-\eta_a) (1-\eta_a \eta_b)}+\frac{1}{2 (1+\eta_a) (1-\eta_a \eta_b)}-2+\frac{2 \eta_a}{\eta_b}-\frac{1}{2 (1-\eta_a) \eta_b}\nn\\&+\frac{1}{2 (1+\eta_a) \eta_b}+\frac{1}{(\eta_a+\eta_b) \eta_b}\Bigg) +\ln (1+\eta_a)\Bigg(\eta_b^4-\eta_a \eta_b^3-\eta_a^2 \eta_b^2-\eta_b^2+\eta_a \eta_b\nn\\&-\frac{\Big(\eta_b^4+\eta_b^2+2\Big) \eta_b}{\eta_a+\eta_b}-\frac{2 \eta_b}{\eta_a-\eta_b}-\frac{\eta_b}{(\eta_a+\eta_b) (\eta_a \eta_b-1)}-\eta_a^2+\frac{\eta_a^2-\eta_a+2}{2 (1-\eta_b)}\nn\\&-\frac{1}{2 \eta_a (1-\eta_b)}+\frac{\eta_a^2+\eta_a+2}{2 (1+\eta_b)}+\frac{1}{2 \eta_a (1+\eta_b)}-\frac{1}{(\eta_a-\eta_b) (1+\eta_b)}\nn\\&-\frac{1}{(1-\eta_b) (\eta_b-\eta_a)}+\frac{1}{\eta_a (\eta_a+\eta_b)}+\frac{1}{2 (1-\eta_b) (1-\eta_a \eta_b)}+\frac{1}{2 (1+\eta_b) (1-\eta_a \eta_b)}\nn\\&-\frac{1}{1-\eta_a \eta_b}-\frac{2 \eta_a}{\eta_b}-\frac{2}{\eta_a \eta_b}\Bigg) +\ln (1-\eta_b)\Bigg(-\eta_b^4+\eta_a \eta_b^3-2 \eta_a^2 \eta_b^2-2 \eta_b^2+2 \eta_a \eta_b\nn\\&+\frac{\Big(\eta_b^4+\eta_b^2+2\Big) \eta_b}{\eta_a+\eta_b}+\frac{\eta_b}{(\eta_a+\eta_b) (\eta_a \eta_b-1)}-2 \eta_a^2+\frac{2 (1+\eta_b)}{\eta_a-\eta_b}+\frac{\eta_b^2-\eta_b-1}{\eta_a}\nn\\&+\frac{3 \Big(\eta_b^2-\eta_b+2\Big)}{2 (1-\eta_a)}+\frac{\eta_b^2+\eta_b+2}{2 (1+\eta_a)}-\frac{2}{(1-\eta_a) (\eta_a-\eta_b)}+\frac{\eta_a^2+\eta_a+2}{1+\eta_b}\nn\\&+\frac{1}{\eta_a (1+\eta_b)}-\frac{1}{2 (1+\eta_a) (1+\eta_b)}+\frac{1}{2 (1-\eta_a) (1-\eta_a \eta_b)}-\frac{1}{2 (1+\eta_b) (1-\eta_a \eta_b)}\nn\\&+\frac{1}{2 (1-\eta_a \eta_b)}-3-\frac{\eta_a^2-\eta_a-1}{\eta_b}-\frac{3}{2 (1-\eta_a) \eta_b}-\frac{2}{\eta_a \eta_b}+\frac{1}{2 (1+\eta_a) \eta_b}\nn\\&+\frac{1}{(\eta_a+\eta_b) \eta_b}\Bigg) +\ln (\eta_a-\eta_b)\Bigg(-\frac{2 \eta_a}{\eta_b}-\frac{2 \eta_b}{\eta_a-\eta_b}-\frac{\eta_b^2-\eta_b+2}{1-\eta_a}-\frac{\eta_b^2+\eta_b+2}{1+\eta_a}\nn\\&+\frac{1}{(1-\eta_a) (\eta_a-\eta_b)}-\frac{1}{(1+\eta_a) (\eta_a-\eta_b)}+\frac{1}{(1-\eta_a) \eta_b}-\frac{1}{(1+\eta_a) \eta_b}+2\Bigg)\nn\\& +\ln (\eta_b)\Bigg(\eta_b^4-\eta_a \eta_b^3+\eta_a^2 \eta_b^2+\eta_b^2-\eta_a \eta_b-\frac{\Big(\eta_b^4+\eta_b^2+2\Big) \eta_b}{\eta_a+\eta_b}+\frac{2 \eta_b}{\eta_a-\eta_b}\nn\\&+\frac{\eta_b}{(\eta_a+\eta_b) (1-\eta_a \eta_b)}+\eta_a^2+\frac{\eta_b^2-\eta_b+2}{1-\eta_a}+\frac{\eta_b^2+\eta_b+2}{1+\eta_a}-\frac{\eta_a^2-\eta_a+2}{2 (1-\eta_b)}\nn\\&+\frac{1}{2 (1-\eta_a) (1-\eta_b)}+\frac{1}{2 \eta_a (1-\eta_b)}-\frac{1}{(1-\eta_a) (\eta_a-\eta_b)}+\frac{1}{(1+\eta_a) (\eta_a-\eta_b)}\nn\\&-\frac{\eta_a^2+\eta_a+2}{2 (1+\eta_b)}-\frac{1}{2 \eta_a (1+\eta_b)}+\frac{1}{2 (1+\eta_a) (1+\eta_b)}+\frac{1}{\eta_a (\eta_a+\eta_b)}\nn\\&-\frac{1}{2 (1-\eta_a) (1-\eta_a \eta_b)}-\frac{1}{2 (1+\eta_a) (1-\eta_a \eta_b)}+\frac{2 \eta_a}{\eta_b}-\frac{1}{(1-\eta_a) \eta_b}\nn\\&+\frac{1}{(1+\eta_a) \eta_b}\Bigg) +\ln (1+\eta_b)\Bigg(-\eta_b^4+\eta_a \eta_b^3-2 \eta_a^2 \eta_b^2-2 \eta_b^2+2 \eta_a \eta_b\nn\\&+\frac{\Big(\eta_b^4+\eta_b^2+2\Big) \eta_b}{\eta_a+\eta_b}-\frac{\eta_b}{(\eta_a+\eta_b) (1-\eta_a \eta_b)}-2 \eta_a^2+\frac{\eta_b^2-\eta_b+2}{2 (1-\eta_a)}+\frac{\eta_b^2+\eta_b+2}{2 (1+\eta_a)}\nn\\&+\frac{\eta_a^2-\eta_a+2}{1-\eta_b}-\frac{1}{(1-\eta_a) (1-\eta_b)}-\frac{1}{\eta_a (1-\eta_b)}+\frac{\eta_a^2+\eta_a+2}{1+\eta_b}+\frac{1}{\eta_a (1+\eta_b)}\nn\\&-\frac{1}{(1+\eta_a) (1+\eta_b)}-\frac{1}{\eta_a (\eta_a+\eta_b)}+\frac{1}{2 (1-\eta_a) (1-\eta_a \eta_b)}+\frac{1}{2 (1+\eta_a) (1-\eta_a \eta_b)}\nn\\&-4-\frac{1}{2 (1-\eta_a) \eta_b}-\frac{1}{\eta_a \eta_b}+\frac{1}{2 (1+\eta_a) \eta_b}\Bigg) +\ln (\eta_a+\eta_b)\Bigg(3 \eta_b^2 \eta_a^2+3 \eta_a^2-3 \eta_b \eta_a\nn\\&+3 \eta_b^2-\frac{3 \Big(\eta_b^2-\eta_b+2\Big)}{2 (1-\eta_a)}-\frac{3 \Big(\eta_b^2+\eta_b+2\Big)}{2 (1+\eta_a)}-\frac{3 \Big(\eta_a^2-\eta_a+2\Big)}{2 (1-\eta_b)}\nn\\&+\frac{3}{2 (1-\eta_a) (1-\eta_b)}+\frac{3}{2 (1-\eta_a) \eta_b}-\frac{3}{2 (1+\eta_a) \eta_b}-\frac{3 \Big(\eta_a^2+\eta_a+2\Big)}{2 (1+\eta_b)}\nn\\&+\frac{3}{2 (1+\eta_a) (1+\eta_b)}+6+\frac{3}{2 (1-\eta_b) \eta_a}+\frac{3}{\eta_b \eta_a}-\frac{3}{2 (1+\eta_b) \eta_a}\Bigg) \nn\\&+\ln (1-\eta_a \eta_b)\Bigg(-\eta_b^2 \eta_a^2-\eta_a^2+\eta_b \eta_a-\eta_b^2+\frac{\eta_b^2-\eta_b+2}{2 (1-\eta_a)}+\frac{\eta_b^2+\eta_b+2}{2 (1+\eta_a)}\nn\\&+\frac{\eta_a^2-\eta_a+2}{2 (1-\eta_b)}-\frac{1}{2 (1-\eta_a) (1-\eta_b)}-\frac{1}{2 (1-\eta_a) \eta_b}+\frac{1}{2 (1+\eta_a) \eta_b}+\frac{\eta_a^2+\eta_a+2}{2 (1+\eta_b)}\nn\\&-\frac{1}{2 (1+\eta_a) (1+\eta_b)}-2-\frac{1}{2 (1-\eta_b) \eta_a}-\frac{1}{\eta_b \eta_a}+\frac{1}{2 (1+\eta_b) \eta_a}\Bigg) \nn\\&+\frac{11}{6 (1-\eta_a) (1-\eta_b)}+\frac{11}{6 \eta_a (1-\eta_b)}-\frac{11 \eta_a^2+13 \eta_a+19}{12 \eta_b}+\frac{11}{6 (1-\eta_a) \eta_b}\nn\\&+\frac{11}{6 \eta_a \eta_b}-\frac{11}{12 (1+\eta_a) \eta_b}-\frac{11 \Big(\eta_a^2+\eta_a+2\Big)}{12 (1+\eta_b)}-\frac{11}{12 \eta_a (1+\eta_b)}\nn\\&+\frac{11}{12 (1+\eta_a) (1+\eta_b)}+\frac{2}{\eta_a (2-\eta_a-\eta_b)}+\frac{2}{\eta_b (2-\eta_a-\eta_b)}+\frac{\eta_b}{3 (\eta_a+\eta_b)}\nn\\&-\frac{5}{3 (\eta_a+\eta_b)^2 (1-\eta_a \eta_b)}-\frac{4}{3 \eta_a (2-\eta_a-\eta_b)^2}-\frac{4}{3 \eta_b (2-\eta_a-\eta_b)^2}\nn\\&+\frac{2}{3 (\eta_a+\eta_b)^2 (1-\eta_a \eta_b)^2}\Bigg)+\theta(\eta_b-\eta_a)[\eta_a\leftrightarrow\eta_b]\nn\\&+\frac{112 (1-\eta_a) \eta_a^4}{\eta_b^6}-\frac{2 (1-\eta_a) (28 \eta_a-97) \eta_a^3}{\eta_b^5}-\frac{(1-\eta_a) \Big(338 \eta_a^2+263 \eta_a-565\Big) \eta_a^2}{3 \eta_b^4}\nn\\&+\frac{(1-\eta_a) \Big(120 \eta_a^4+201 \eta_a^3-757 \eta_a^2-280 \eta_a+318\Big) \eta_a}{3 \eta_b^3}+\frac{2 \Big(\eta_b^2-\eta_b-1\Big)}{1+\eta_a}\nn\\&-\frac{2 \Big(11 \eta_b^2-9 \eta_b+9\Big)}{3 (1-\eta_a)}+\frac{2}{3} \Big(144 \eta_a^4+60 \eta_b \eta_a^3-168 \eta_a^3+33 \eta_b^2 \eta_a^2-57 \eta_b \eta_a^2-11 \eta_a^2\nn\\&+60 \eta_b^3 \eta_a-57 \eta_b^2 \eta_a+39 \eta_b \eta_a+6 \eta_a+144 \eta_b^4-168 \eta_b^3-11 \eta_b^2+6 \eta_b+69\Big)\nn\\&+\ln (1-\eta_a)\Bigg(-12 \eta_b^2 \eta_a^2+4 \eta_b \eta_a^2-16 \eta_a^2+4 \eta_b^2 \eta_a+4 \eta_b \eta_a+8 \eta_a-16 \eta_b^2+8 \eta_b\nn\\&+\frac{8 \Big(\eta_b^2-\eta_b+2\Big)}{1-\eta_a}+\frac{4 \Big(\eta_b^2+\eta_b+2\Big)}{1+\eta_a}+\frac{8 \Big(\eta_a^2-\eta_a+2\Big)}{1-\eta_b}-\frac{8}{(1-\eta_a) (1-\eta_b)}\nn\\&+\frac{4 \Big(\eta_a^2-\eta_a+2\Big)}{\eta_b}-\frac{8}{(1-\eta_a) \eta_b}+\frac{4}{(1+\eta_a) \eta_b}+\frac{4 \Big(\eta_a^2+\eta_a+2\Big)}{1+\eta_b}\nn\\&-\frac{4}{(1+\eta_a) (1+\eta_b)}-32+\frac{4 \Big(\eta_b^2-\eta_b+2\Big)}{\eta_a}-\frac{8}{(1-\eta_b) \eta_a}-\frac{12}{\eta_b \eta_a}+\frac{4}{(1+\eta_b) \eta_a}\Bigg) \nn\\&+\ln (\eta_a)\Bigg(\frac{112 \eta_a^5}{\eta_b^7}+\frac{250 \eta_a^4}{\eta_b^6}-6 \eta_a^4-6 \eta_b \eta_a^3-\frac{6 \Big(25 \eta_a^2-46\Big) \eta_a^3}{\eta_b^5}+7 \eta_b^2 \eta_a^2-\eta_b \eta_a^2\nn\\&-\frac{4 \Big(87 \eta_a^2-46\Big) \eta_a^2}{\eta_b^4}+46 \eta_a^2-\eta_b^2 \eta_a-5 \eta_b \eta_a+\frac{2 \Big(24 \eta_a^4-171 \eta_a^2+40\Big) \eta_a}{\eta_b^3}-2 \eta_a\nn\\&+8 \eta_b^2-2 \eta_b-\frac{4 \Big(\eta_b^2-\eta_b+2\Big)}{1-\eta_a}-\frac{3 \Big(\eta_b^2+\eta_b+2\Big)}{1+\eta_a}-\frac{4 (1-\eta_a)^2}{1-\eta_b}\nn\\&+\frac{4}{(1-\eta_a) (1-\eta_b)}-\frac{2 \eta_a^5-102 \eta_a^3+\eta_a^2+37 \eta_a+2}{\eta_b}+\frac{4}{(1-\eta_a) \eta_b}-\frac{3}{(1+\eta_a) \eta_b}\nn\\&-\frac{(\eta_a+2) (3 \eta_a+1)}{1+\eta_b}+\frac{3}{(1+\eta_a) (1+\eta_b)}+\frac{2 \Big(60 \eta_a^4-88 \eta_a^2+11\Big)}{\eta_b^2}+16-\frac{\eta_b^2-\eta_b+2}{\eta_a}\nn\\&+\frac{4}{(1-\eta_b) \eta_a}+\frac{7}{\eta_b \eta_a}-\frac{3}{(1+\eta_b) \eta_a}\Bigg) +\ln (1+\eta_a)\Bigg(\frac{112 \eta_b^5}{\eta_a^7}+\frac{250 \eta_b^4}{\eta_a^6}-6 \eta_b^4\nn\\&-6 \eta_a \eta_b^3-\frac{6 \Big(25 \eta_b^2-46\Big) \eta_b^3}{\eta_a^5}-4 \eta_a^2 \eta_b^2-4 \eta_a \eta_b^2-\frac{4 \Big(87 \eta_b^2-46\Big) \eta_b^2}{\eta_a^4}+30 \eta_b^2\nn\\&+16 \eta_a^3 \eta_b-4 \eta_a^2 \eta_b+4 \eta_a \eta_b+\frac{2 \Big(24 \eta_b^4-171 \eta_b^2+40\Big) \eta_b}{\eta_a^3}-8 \eta_b+68 \eta_a^4-40 \eta_a^2\nn\\&-8 \eta_a+\frac{4 \Big(\eta_b^2+\eta_b+2\Big)}{1+\eta_a}+\frac{2 \Big(60 \eta_b^4-88 \eta_b^2+11\Big)}{\eta_a^2}\nn\\&-\frac{2 \Big(\eta_b^5-51 \eta_b^3+2 \eta_b^2+21 \eta_b+4\Big)}{\eta_a}+\frac{4 \Big(\eta_a^2+\eta_a+2\Big)}{1+\eta_b}+\frac{4}{\eta_a (1+\eta_b)}\nn\\&-\frac{4}{(1+\eta_a) (1+\eta_b)}-8+\frac{4 \Big(32 \eta_a^5-33 \eta_a^3-\eta_a^2+9 \eta_a-2\Big)}{\eta_b}-\frac{4}{\eta_a \eta_b}+\frac{4}{(1+\eta_a) \eta_b}\nn\\&+\frac{2 \Big(58 \eta_a^6-81 \eta_a^4+33 \eta_a^2-4\Big)}{\eta_b^2}-\frac{2 (1-\eta_a) \eta_a (1+\eta_a) \Big(20 \eta_a^4-14 \eta_a^2+1\Big)}{\eta_b^3}\Bigg) \nn\\&+ \ln (1-\eta_b)\Bigg(-12 \eta_b^2 \eta_a^2+4 \eta_b \eta_a^2-16 \eta_a^2+4 \eta_b^2 \eta_a+4 \eta_b \eta_a+8 \eta_a-16 \eta_b^2+8 \eta_b \nn\\&+\frac{8 \Big(\eta_b^2-\eta_b+2\Big)}{1-\eta_a}+\frac{4 \Big(\eta_b^2+\eta_b+2\Big)}{1+\eta_a}+\frac{8 \Big(\eta_a^2-\eta_a+2\Big)}{1-\eta_b}-\frac{8}{(1-\eta_a) (1-\eta_b)} \nn\\&+\frac{4 \Big(\eta_a^2-\eta_a+2\Big)}{\eta_b}-\frac{8}{(1-\eta_a) \eta_b}+\frac{4}{(1+\eta_a) \eta_b}+\frac{4 \Big(\eta_a^2+\eta_a+2\Big)}{1+\eta_b} \nn\\&-\frac{4}{(1+\eta_a) (1+\eta_b)}-32+\frac{4 \Big(\eta_b^2-\eta_b+2\Big)}{\eta_a}-\frac{8}{(1-\eta_b) \eta_a}-\frac{12}{\eta_b \eta_a}+\frac{4}{(1+\eta_b) \eta_a}\Bigg)\nn\\&+\ln (\eta_b)\Bigg(\frac{112 \eta_b^5}{\eta_a^7}+\frac{250 \eta_b^4}{\eta_a^6}-6 \eta_b^4-6 \eta_a \eta_b^3-\frac{6 \Big(25 \eta_b^2-46\Big) \eta_b^3}{\eta_a^5}+7 \eta_a^2 \eta_b^2-\eta_a \eta_b^2\nn\\&-\frac{4 \Big(87 \eta_b^2-46\Big) \eta_b^2}{\eta_a^4}+46 \eta_b^2-\eta_a^2 \eta_b-5 \eta_a \eta_b+\frac{2 \Big(24 \eta_b^4-171 \eta_b^2+40\Big) \eta_b}{\eta_a^3}-2 \eta_b\nn\\&+8 \eta_a^2-\frac{4 (1-\eta_b)^2}{1-\eta_a}-2 \eta_a-\frac{(\eta_b+2) (3 \eta_b+1)}{1+\eta_a}+\frac{2 \Big(60 \eta_b^4-88 \eta_b^2+11\Big)}{\eta_a^2}\nn\\&-\frac{2 \eta_b^5-102 \eta_b^3+\eta_b^2+37 \eta_b+2}{\eta_a}-\frac{4 \Big(\eta_a^2-\eta_a+2\Big)}{1-\eta_b}+\frac{4}{(1-\eta_a) (1-\eta_b)}\nn\\&+\frac{4}{\eta_a (1-\eta_b)}-\frac{3 \Big(\eta_a^2+\eta_a+2\Big)}{1+\eta_b}-\frac{3}{\eta_a (1+\eta_b)}+\frac{3}{(1+\eta_a) (1+\eta_b)}+16\nn\\&-\frac{\eta_a^2-\eta_a+2}{\eta_b}+\frac{4}{(1-\eta_a) \eta_b}+\frac{7}{\eta_a \eta_b}-\frac{3}{(1+\eta_a) \eta_b}\Bigg) \nn\\&+\ln (1+\eta_b)\Bigg(\frac{112 \eta_a^5}{\eta_b^7}+\frac{250 \eta_a^4}{\eta_b^6}-6 \eta_a^4-6 \eta_b \eta_a^3-\frac{6 \Big(25 \eta_a^2-46\Big) \eta_a^3}{\eta_b^5}-4 \eta_b^2 \eta_a^2\nn\\&-4 \eta_b \eta_a^2-\frac{4 \Big(87 \eta_a^2-46\Big) \eta_a^2}{\eta_b^4}+30 \eta_a^2+16 \eta_b^3 \eta_a-4 \eta_b^2 \eta_a+4 \eta_b \eta_a\nn\\&+\frac{2 \Big(24 \eta_a^4-171 \eta_a^2+40\Big) \eta_a}{\eta_b^3}-8 \eta_a+68 \eta_b^4-40 \eta_b^2-8 \eta_b+\frac{4 \Big(\eta_b^2+\eta_b+2\Big)}{1+\eta_a}\nn\\&-\frac{2 \Big(\eta_a^5-51 \eta_a^3+2 \eta_a^2+21 \eta_a+4\Big)}{\eta_b}+\frac{4}{(1+\eta_a) \eta_b}+\frac{4 \Big(\eta_a^2+\eta_a+2\Big)}{1+\eta_b}\nn\\&-\frac{4}{(1+\eta_a) (1+\eta_b)}+\frac{2 \Big(60 \eta_a^4-88 \eta_a^2+11\Big)}{\eta_b^2}-8-\frac{4}{\eta_b \eta_a}+\frac{4}{(1+\eta_b) \eta_a}\nn\\&+\frac{4 \Big(32 \eta_b^5-33 \eta_b^3-\eta_b^2+9 \eta_b-2\Big)}{\eta_a}+\frac{2 \Big(58 \eta_b^6-81 \eta_b^4+33 \eta_b^2-4\Big)}{\eta_a^2}\nn\\&-\frac{2 (1-\eta_b) \eta_b (1+\eta_b) \Big(20 \eta_b^4-14 \eta_b^2+1\Big)}{\eta_a^3}\Bigg) \nn\\&+\ln (\eta_a+\eta_b)\Bigg(-\frac{112 \eta_a^5}{\eta_b^7}-\frac{250 \eta_a^4}{\eta_b^6}-62 \eta_a^4-10 \eta_b \eta_a^3+\frac{6 \Big(25 \eta_a^2-46\Big) \eta_a^3}{\eta_b^5}+8 \eta_b^2 \eta_a^2\nn\\&+\frac{4 \Big(87 \eta_a^2-46\Big) \eta_a^2}{\eta_b^4}+10 \eta_a^2-10 \eta_b^3 \eta_a-8 \eta_b \eta_a-\frac{6 \Big(21 \eta_a^4-5 \eta_a^2-1\Big) \eta_a}{\eta_b}\nn\\&-\frac{2 \Big(20 \eta_a^6-10 \eta_a^4-156 \eta_a^2+39\Big) \eta_a}{\eta_b^3}-62 \eta_b^4+10 \eta_b^2-\frac{4 \Big(\eta_b^2-\eta_b+2\Big)}{1-\eta_a}\nn\\&-\frac{4 \Big(\eta_b^2+\eta_b+2\Big)}{1+\eta_a}-\frac{4 \Big(\eta_a^2-\eta_a+2\Big)}{1-\eta_b}+\frac{4}{(1-\eta_a) (1-\eta_b)}+\frac{4}{(1-\eta_a) \eta_b}\nn\\&-\frac{4}{(1+\eta_a) \eta_b}-\frac{4 \Big(\eta_a^2+\eta_a+2\Big)}{1+\eta_b}+\frac{4}{(1+\eta_a) (1+\eta_b)}+\frac{4}{(1-\eta_b) \eta_a}+\frac{8}{\eta_b \eta_a}\nn\\&-\frac{2 \Big(58 \eta_a^6-21 \eta_a^4-55 \eta_a^2+7\Big)}{\eta_b^2}+16-\frac{6 \eta_b \Big(21 \eta_b^4-5 \eta_b^2-1\Big)}{\eta_a}-\frac{4}{(1+\eta_b) \eta_a}\nn\\&-\frac{2 \Big(58 \eta_b^6-21 \eta_b^4-55 \eta_b^2+7\Big)}{\eta_a^2}-\frac{2 \eta_b \Big(20 \eta_b^6-10 \eta_b^4-156 \eta_b^2+39\Big)}{\eta_a^3}\nn\\&+\frac{4 \eta_b^2 \Big(87 \eta_b^2-46\Big)}{\eta_a^4}+\frac{6 \eta_b^3 \Big(25 \eta_b^2-46\Big)}{\eta_a^5}-\frac{250 \eta_b^4}{\eta_a^6}-\frac{112 \eta_b^5}{\eta_a^7}\Bigg) \nn\\&-\frac{2 \Big(11 \eta_a^2-9 \eta_a+9\Big)}{3 (1-\eta_b)}+\frac{607 \eta_a^5-1092 \eta_a^4-420 \eta_a^3+1160 \eta_a^2-99 \eta_a-356}{6 \eta_b}\nn\\&+\frac{22}{3 (1-\eta_a) \eta_b}+\frac{2}{(1+\eta_a) \eta_b}+\frac{2 \Big(\eta_a^2-\eta_a-1\Big)}{1+\eta_b}\nn\\&+\frac{1200 \eta_a^6-4092 \eta_a^5-2205 \eta_a^4+11200 \eta_a^3-4440 \eta_a^2-3480 \eta_a+1537}{30 \eta_b^2}\nn\\&+ \ln (2)\Bigg(8 \eta_b \eta_a^2+8 \eta_b^2 \eta_a+16 \eta_a+16 \eta_b+\frac{4 \Big(\eta_b^2-\eta_b+2\Big)}{1-\eta_a}-\frac{4 \Big(\eta_b^2+\eta_b+2\Big)}{1+\eta_a}\nn\\&+\frac{4 \Big(\eta_a^2-\eta_a+2\Big)}{1-\eta_b}-\frac{4}{(1-\eta_a) (1-\eta_b)}+\frac{8 \Big(\eta_a^2+2\Big)}{\eta_b}-\frac{4}{(1-\eta_a) \eta_b}-\frac{4}{(1+\eta_a) \eta_b}\nn\\&-\frac{4 \Big(\eta_a^2+\eta_a+2\Big)}{1+\eta_b}+\frac{4}{(1+\eta_a) (1+\eta_b)}+\frac{8 \Big(\eta_b^2+2\Big)}{\eta_a}-\frac{4}{(1-\eta_b) \eta_a}-\frac{4}{(1+\eta_b) \eta_a}\Bigg)\nn\\&+\frac{607 \eta_b^5-1092 \eta_b^4-420 \eta_b^3+1160 \eta_b^2-99 \eta_b-356}{6 \eta_a}+\frac{22}{3 (1-\eta_b) \eta_a}+\frac{26}{\eta_b \eta_a}\nn\\&+\frac{2}{(1+\eta_b) \eta_a}+\frac{1200 \eta_b^6-4092 \eta_b^5-2205 \eta_b^4+11200 \eta_b^3-4440 \eta_b^2-3480 \eta_b+1537}{30 \eta_a^2}\nn\\&+\frac{(1-\eta_b) \eta_b \Big(120 \eta_b^4+201 \eta_b^3-757 \eta_b^2-280 \eta_b+318\Big)}{3 \eta_a^3}-\frac{2 (1-\eta_b) \eta_b^3 (28 \eta_b-97)}{\eta_a^5}\nn\\&-\frac{(1-\eta_b) \eta_b^2 \Big(338 \eta_b^2+263 \eta_b-565\Big)}{3 \eta_a^4}+\frac{112 (1-\eta_b) \eta_b^4}{\eta_a^6}\Bigg\}+\mathcal{O}(\eps^0)\,,
\end{align}
\begin{align}
    \bar{I}_{12;gg,gg}(\eta_a,\eta_b;\eps\,|\,1,\eta_b)=\,&\,-\frac{9p_{gg}(\eta_b)}{\eps^3}+\frac{4}{\eps^2}\Bigg\{5 p_{gg}(\eta_b) \ln (1-\eta_b)-\frac{5}{2} p_{gg}(\eta_b) \ln (1+\eta_b)+2 \ln (2) p_{gg}(\eta_b)\nn\\&+\frac{4 \Big(\eta_b^2-\eta_b+2\Big)}{1-\eta_a}-\frac{4}{(1-\eta_a) (1-\eta_b)}-\frac{4}{(1-\eta_a) \eta_b}+\frac{1}{12} \Big(55 \eta_b^2-47 \eta_b+58\Big)\nn\\&+\ln (\eta_b)\Bigg(2 \eta_b^2-4 \eta_b-\frac{2}{1-\eta_b}-\frac{2}{\eta_b}+2\Bigg) -\frac{11}{12 (1-\eta_b)}-\frac{55}{12 \eta_b}\Bigg\}\nn\\&+\frac{4}{\eps}\Bigg\{\frac{\theta(\eta_a-\eta_b)p_{gg}(\eta_b)}{ (1-\eta_a)} \Bigg(- \ln (1-\eta_a)+ \ln (1-\eta_b)+ \ln (\eta_b)- \ln (1+\eta_b)\nn\\&-\frac{11}{6}\Bigg)+\ln ^2(1-\eta_b)\Bigg(6 \eta_b^2-6 \eta_b-\frac{6}{1-\eta_b}+12-\frac{6}{\eta_b}\Bigg) \nn\\&+\ln (1-\eta_b)\Bigg(\frac{1}{3} \Big(-33 \eta_b^2+29 \eta_b-40\Big)-\frac{8 \Big(\eta_b^2-\eta_b+2\Big)}{1-\eta_a}+\frac{8}{(1-\eta_a) (1-\eta_b)}\nn\\&+\frac{11}{3 (1-\eta_b)}+\frac{8}{(1-\eta_a) \eta_b}+\frac{11}{\eta_b}\Bigg) +\ln (\eta_b) \ln (1-\eta_b)\Bigg(-4 \eta_b^2+4 \eta_b+\frac{4}{1-\eta_b}\nn\\&-8+\frac{4}{\eta_b}\Bigg) +\ln (1+\eta_b) \ln (1-\eta_b)\Bigg(\frac{1}{2} \Big(-3 \eta_b^2+4 \eta_b-8\Big)+\frac{1}{1-\eta_b}+\frac{2}{\eta_b}\Bigg) \nn\\&+\ln (2) \ln (1-\eta_b)\Bigg(\frac{1}{1-\eta_b}-\frac{\eta_b^2}{2}\Bigg) +\ln ^2(\eta_b)\Bigg(\eta_b^2-2 \eta_b-\frac{1}{1-\eta_b}+1-\frac{1}{\eta_b}\Bigg)\nn\\& +\ln ^2(1+\eta_b)\Bigg(\frac{1}{2} \Big(\eta_b^2-\eta_b+2\Big)-\frac{1}{2 (1-\eta_b)}-\frac{1}{2 \eta_b}\Bigg) +\frac{2 \Big(11 \eta_b^2-9 \eta_b+9\Big)}{3 (1-\eta_a)}\nn\\&+\frac{1}{36} \Big(19 \eta_b^2-115 \eta_b+182\Big)+\pi ^2 \Bigg(\frac{1}{12} \Big(-19 \eta_b^2+30 \eta_b-24\Big)+\frac{5}{3 (1-\eta_b)}+\frac{3}{2 \eta_b}\Bigg)\nn\\&+\ln (1-\eta_a)\Bigg(-\frac{8 \Big(\eta_b^2-\eta_b+2\Big)}{1-\eta_a}+\frac{8}{(1-\eta_a) (1-\eta_b)}+\frac{8}{(1-\eta_a) \eta_b}\Bigg) \nn\\&+ \ln (\eta_b)\Bigg(\frac{4 (1-\eta_b)^2}{1-\eta_a}+\frac{1}{12} \Big(145 \eta_b^2+19 \eta_b+136\Big)-\frac{4}{(1-\eta_a) \eta_b}-\frac{11}{4 \eta_b}\nn\\&-\frac{4}{(1-\eta_a) (1-\eta_b)}-\frac{11}{12 (1-\eta_b)}\Bigg)+ \ln (2) \ln (\eta_b)\Bigg(-2 \eta_b^2+6 \eta_b+\frac{2}{1-\eta_b}\nn\\&+\frac{2}{\eta_b}\Bigg)+\ln (1+\eta_b)\Bigg(\frac{1}{12} \Big(-13 \eta_b^2-35 \eta_b+46\Big)+\frac{4 \Big(\eta_b^2-\eta_b+2\Big)}{1-\eta_a}\nn\\&-\frac{4}{(1-\eta_a) (1-\eta_b)}-\frac{11}{12 (1-\eta_b)}-\frac{4}{(1-\eta_a) \eta_b}+\frac{13}{12 \eta_b}\Bigg)\nn\\& +\ln (\eta_b) \ln (1+\eta_b)\Bigg(3 \eta_b^2-3 \eta_b-\frac{3}{1-\eta_b}+6-\frac{3}{\eta_b}\Bigg) \nn\\&+\ln (2) \ln (1+\eta_b)\Bigg(\frac{1}{2} \Big(-11 \eta_b^2+10 \eta_b-20\Big)+\frac{6}{1-\eta_b}+\frac{5}{\eta_b}\Bigg) \nn\\&+\Li_2\Bigg(\frac{1-\eta_b}{2}\Bigg)\Bigg(\frac{1}{2} \Big(9 \eta_b^2-8 \eta_b+16\Big)-\frac{5}{1-\eta_b}-\frac{4}{\eta_b}\Bigg) +\Li_2(-\eta_b)(4 \eta_b+4)\nn\\& -\Li_2(\eta_b)(4 \eta_b+4) +\Li_2\Bigg(\frac{1+\eta_b}{2}\Bigg)\Bigg(\frac{1}{2} \Big(9 \eta_b^2-8 \eta_b+16\Big)-\frac{5}{1-\eta_b}-\frac{4}{\eta_b}\Bigg) \nn\\&-\frac{67}{36 (1-\eta_b)}-\frac{22}{3 (1-\eta_a) \eta_b}-\frac{19}{36 \eta_b}+\ln ^2(2)\Bigg(\frac{1}{2} \Big(9 \eta_b^2-8 \eta_b+16\Big)-\frac{5}{1-\eta_b}\nn\\&-\frac{4}{\eta_b}\Bigg) +\ln (2)\Bigg(-\frac{4 \Big(\eta_b^2-\eta_b+2\Big)}{1-\eta_a}+\frac{1}{12} \Big(13 \eta_b^2+35 \eta_b-46\Big)\nn\\&+\frac{4}{(1-\eta_a) (1-\eta_b)}+\frac{11}{12 (1-\eta_b)}+\frac{4}{(1-\eta_a) \eta_b}-\frac{13}{12 \eta_b}\Bigg) \Bigg\}+\mathcal{O}(\eps^0)\,,
\end{align}
\begin{align}
    \bar{I}_{12;gg,gg}(\eta_a,\eta_b;\eps\,|\,\eta_a,\eta_a)=\,&\,\frac{1}{\eps^3}\Bigg\{-2\eta_a+\frac{1}{ (1-\eta_a)}-\frac{1}{ (1+\eta_a)}\Bigg\}\nn\\&+\frac{4}{\eps^2}\Bigg\{\theta(\eta_a-\eta_b)\Bigg(-\frac{\eta_b}{\eta_a-\eta_b}+\frac{1}{2 (1-\eta_a) (\eta_a-\eta_b)}-\frac{1}{2 (1+\eta_a) (\eta_a-\eta_b)}-1\Bigg)\nn\\&+\ln (1-\eta_a)\Bigg(\eta_a-\frac{1}{2 (1-\eta_a)}+\frac{1}{2 (1+\eta_a)}\Bigg) +\ln (\eta_a)\Bigg(-\eta_a+\frac{1}{2 (1-\eta_a)}\nn\\&-\frac{1}{2 (1+\eta_a)}\Bigg) +\ln (1+\eta_a)\Bigg(\eta_a-\frac{1}{2 (1-\eta_a)}+\frac{1}{2 (1+\eta_a)}\Bigg) \Bigg\}\nn\\&+\frac{4}{\eps}\Bigg\{\theta(\eta_a-\eta_b)\Bigg\{[\ln(1+\eta_a)+\ln(1-\eta_a)+\ln(\eta_a-\eta_b)]\Bigg(2+\frac{2 \eta_b}{\eta_a-\eta_b}\nn\\&-\frac{1}{(1-\eta_a) (\eta_a-\eta_b)}+\frac{1}{(1+\eta_a) (\eta_a-\eta_b)}\Bigg)+\ln(\eta_a)\Bigg(-4-\frac{4 \eta_b}{\eta_a-\eta_b}\nn\\&+\frac{2}{(1-\eta_a) (\eta_a-\eta_b)}-\frac{2}{(1+\eta_a) (\eta_a-\eta_b)}\Bigg)\Bigg\}\nn\\&+\Bigg[\frac{\ln^2(1-\eta_a)+\ln^2(1+\eta_a)+\ln^2(\eta_a)}{2}+\ln(1-\eta_a)\ln(1+\eta_a)\nn\\&-\ln(1-\eta_a)\ln(\eta_a)-\ln(1+\eta_a)\ln(\eta_a)+\frac{\pi^2-1}{12}\Bigg]\Bigg(\frac{1}{1-\eta_a}-2\eta_a-\frac{1}{1+\eta_a}\Bigg)\Bigg\}\nn\\&+\mathcal{O}(\eps^0)\,,
\end{align}
\begin{align}
     \bar{I}_{12;gg,gg}(\eta_a,\eta_b;\eps\,|\,1,1)=\,&\,\frac{2}{\eps^4}+\frac{4}{\eps^3}\Bigg\{\frac{1}{1-\eta_a}+\frac{1}{1-\eta_b}+\frac{11}{48}+\frac{\ln (2)}{4}\Bigg\}+\frac{1}{\eps^2}\Bigg\{-\frac{\theta(\eta_a-\eta_b)+\theta(\eta_b-\eta_a)}{ (1-\eta_a) (1-\eta_b)}\nn\\&+\frac{8}{(1-\eta_a) (1-\eta_b)}+\ln (2) \Big(\frac{2}{1-\eta_a}+\frac{2}{1-\eta_b}\Big)+\frac{11}{6 (1-\eta_a)}-\frac{9 \ln (1-\eta_a)}{1-\eta_a}\nn\\&+\frac{\ln (\eta_a)}{1-\eta_a}+\frac{11}{6 (1-\eta_b)}-\frac{9 \ln (1-\eta_b)}{1-\eta_b}+\frac{\ln (\eta_b)}{1-\eta_b}-\frac{\pi ^2}{6}+\frac{35}{18}-\ln ^2(2)\Bigg\}\nn\\&+\frac{1}{\eps}\Bigg\{\theta(\eta_a-\eta_b)\Bigg\{\ln (1-\eta_a)\Bigg(\frac{1}{(1-\eta_a) (2-\eta_a-\eta_b)}-\frac{4}{(1-\eta_a) (\eta_a-\eta_b)}\nn\\&+\frac{5}{(1-\eta_a) (1-\eta_b)}\Bigg) +\ln (1-\eta_b)\Bigg(-\frac{1}{(1-\eta_a) (2-\eta_a-\eta_b)}\nn\\&+\frac{4}{(1-\eta_a) (\eta_a-\eta_b)}-\frac{2}{(1-\eta_a) (1-\eta_b)}\Bigg) +\frac{\ln (2-\eta_a-\eta_b)}{(1-\eta_a) (1-\eta_b)}\nn\\&+\frac{2 \ln (2)}{(1-\eta_a) (1-\eta_b)}\Bigg\}+\theta(\eta_b-\eta_a)[\eta_a\leftrightarrow\eta_b]-\pi ^2 \Bigg(\frac{13}{6 (1-\eta_a)}+\frac{3}{2 (1-\eta_b)}+\frac{11}{36}\Bigg)\nn\\&-\ln ^2(2) \Bigg(\frac{2}{1-\eta_a}+\frac{2}{1-\eta_b}\Bigg)-\ln (1-\eta_a)\Bigg(\frac{16}{(1-\eta_a) (1-\eta_b)}+\frac{22}{3 (1-\eta_a)}\Bigg)\nn\\& -\ln (1-\eta_b)\Bigg(\frac{16}{(1-\eta_a) (1-\eta_b)}+\frac{22}{3 (1-\eta_b)}\Bigg) +\frac{73}{18 (1-\eta_a)}+\frac{11 \ln ^2(1-\eta_a)}{1-\eta_a}\nn\\&-\frac{\ln ^2(\eta_a)}{1-\eta_a}-\frac{2 \ln (\eta_a) \ln (1-\eta_a)}{1-\eta_a}-\frac{6 \ln (2) \ln (1-\eta_a)}{1-\eta_a}-\frac{2 \ln (2) \ln (\eta_a)}{1-\eta_a}\nn\\&+\frac{67}{18 (1-\eta_b)}+\frac{11 \ln ^2(1-\eta_b)}{1-\eta_b}-\frac{\ln ^2(\eta_b)}{1-\eta_b}-\frac{6 \ln (2) \ln (1-\eta_b)}{1-\eta_b}\nn\\&-\frac{2 \ln (1-\eta_b) \ln (\eta_b)}{1-\eta_b}-\frac{2 \ln (2) \ln (\eta_b)}{1-\eta_b}+\frac{37 \zeta (3)}{4}+\frac{116}{27}+\frac{2 \ln ^3(2)}{3}\nn\\&-\frac{\ln(2)}{3}(1+\pi^2)\Bigg\}+\mathcal{O}(\eps^0)\,.
\end{align}
Here
\beq
p_{gg}(\eta) =\, \frac{1}{1 - \eta} + \frac{1}{\eta} - 2 + \eta (1 - \eta)\,.
\eeq
The terms proportional to $\theta(\eta_a-\eta_b)$ and $\theta(\eta_b-\eta_a)$ are related by an exchange of $\eta_a$ and $\eta_b$, which we denote by $\theta(\eta_b-\eta_a)[\eta_a\leftrightarrow\eta_b]$. For purposes of illustration, we set $\mu_F^2=\,\mu_R^2=\,m_H^2$.

\bibliographystyle{JHEP}
\bibliography{colorful}


\end{document}